\documentclass[10pt,twoside,openright,bibliography=totoc]{scrbook}
\usepackage{amsmath,amssymb}
\usepackage{verbatim}
\usepackage{graphicx}
\usepackage{booktabs}
\usepackage[usenames, dvipsnames]{color}
\usepackage{multirow}
\usepackage{setspace}
\usepackage{rotating}
\usepackage[nocompress]{cite}
\usepackage[ pdftex, plainpages = false, pdfpagelabels, 
                 pdfpagelayout = useoutlines,
                 bookmarks,
                 bookmarksopen = true,
                 bookmarksnumbered = true,
                 breaklinks = true,
                 linktocpage=all,
                 pagebackref=false,
                 colorlinks = true,
                 linkcolor = BrickRed,
                 urlcolor  = blue,
                 citecolor = BrickRed,
                 anchorcolor = green,
                 hyperindex = true,
                 hyperfigures
                 ]{hyperref} 
\usepackage{xifthen} 

\newcommand{\ket}[1]{\ensuremath{\left|#1\right\rangle}}
\newcommand{\bra}[1]{\ensuremath{\left\langle #1\right|}}
\newcommand{\braket}[2]{\ensuremath{\left\langle #1 \middle| #2 \right\rangle}}
\newcommand{\braketop}[3]{\ensuremath{\left\langle #1\middle| #2 \middle| #3\right\rangle}}
\newcommand{\cket}[1]{\ensuremath{\left|#1\right)}}
\newcommand{\cbra}[1]{\ensuremath{\left(#1\right|}}
\newcommand{\cbraket}[2]{\ensuremath{\left( #1 \middle| #2 \right)}}
\newcommand{\project}[2]{\ensuremath{\left(#1\middle| #2\right\rangle}}
\newcommand{\projectop}[3]{\ensuremath{\left(#1\middle| #2 \middle| #3\right\rangle}}
\newcommand{\comm}[2]{\ensuremath{\left[#1,#2\right]}}

\newcommand{\hamil}{\ensuremath{\mathcal{H}}}

\newcommand{\tr}{\hbox{Tr}}

\newcommand{\ticket}[1]{\boxed{\hbox{#1}}}

\newcommand{\moment}[1]{\ensuremath{\left\langle #1 \right\rangle}}
\newcommand{\cumulant}[1]{\ensuremath{\left\langle #1
    \right\rangle_{c}}}

\newcommand{\pdf}[2]{\ensuremath{\frac{\partial #1}{\partial #2}}}
\newcommand{\pdfh}[3]{\ensuremath{\frac{\partial^{#3} #1}{\partial
      {#2}^{#3}}}}

\newcommand{\cA}{\mathcal{A}}

\newcommand{\cF}{\mathcal{F}}
\newcommand{\cG}{\mathcal{G}}
\newcommand{\cH}{\mathcal{H}}

\newcommand{\cO}{\mathcal{O}}
\newcommand{\cQ}{\mathcal{Q}}
\newcommand{\cS}{\mathcal{S}}
\newcommand{\cT}{\mathcal{T}}

\newcommand{\mui}{M}
\newcommand{\exA}{\textbf{A}}
\newcommand{\exB}{\textbf{B}}
\newcommand{\exC}{\textbf{C}}
\newcommand{\exD}{\textbf{D}}

\newcommand{\arxiv}[2][]{\ifthenelse{\isempty{#1}}{\href{http://arxiv.org/abs/#2}{{\tt arXiv:\allowbreak{}#2}}} {\href{http://arxiv.org/abs/#2}{{\tt arXiv:\allowbreak{}#2 [#1]}}}}

\begin{document}
\bibliographystyle{utphys}
\begin{titlepage}
\title{Multiscale Structure in Eco-Evolutionary Dynamics}
\author{Blake C.\ Stacey}
\date{\today}
\publishers{University of Massachusetts Boston\\Department of Physics}
\end{titlepage}
\maketitle

\vspace*{\fill}
\thispagestyle{empty}
\begin{center}
\baselineskip=12pt
\parbox{8.0cm}{

I've done this because I love the act of writing, which is to say I
love the act of discovery, of revelation, and then the attempt to
share that revelation in all its fullness and clarity.

You can never be sure how many people will want to
share in that feeling.
\\
\hspace*{\fill}---\href{http://www.theatlantic.com/education/archive/2014/10/continuing-education/381082/}{\textsc{Ta-Nehisi Coates} (2014)}} \nocite{coates2014}
\end{center}

\vspace*{\fill}

\begin{center}
\parbox{8.0cm}{ This is a slightly expanded version of my PhD thesis,
  which was accepted by Brandeis University's Martin A.\ Fisher School
  of Physics in May of 2015.  In turn, that thesis was based on research
  I began at the New England Complex Systems Institute.  }
\end{center}

\pagebreak

\tableofcontents

\chapter{Abstract}

In a complex system, the individual components are neither so tightly
coupled or correlated that they can all be treated as a single unit,
nor so uncorrelated that they can be approximated as independent
entities.  Instead, patterns of interdependency lead to structure at
multiple scales of organization.  Evolution excels at producing such
complex structures.  In turn, the existence of these complex
interrelationships within a biological system affects the evolutionary
dynamics of that system.  I present a mathematical formalism for
multiscale structure, grounded in information theory, which makes
these intuitions quantitative, and I show how dynamics defined in
terms of population genetics or evolutionary game theory can lead to
multiscale organization.  For complex systems, ``more is different,''
and I address this from several perspectives.  Spatial host--consumer
models demonstrate the importance of the structures which can arise
due to dynamical pattern formation.  Evolutionary game theory reveals
the novel effects which can result from multiplayer games, nonlinear
payoffs and ecological stochasticity.  Replicator dynamics in an
environment with mesoscale structure relates to generalized
conditionalization rules in probability theory.

The idea of natural selection ``acting at multiple levels'' has been
mathematized in a variety of ways, not all of which are equivalent.
We will face down the confusion, using the experience developed over
the course of this thesis to clarify the situation.

\vfill

\emph{Chapter \ref{ch:multiscale} applies the general abstract framework of multiscale structure to some geometrical examples, to build intuition for it, and then connects it with population genetics and network theory.  Chapter~\ref{ch:redgreen} studies emergent multiscale structure in a spatial evolutionary ecosystem.  Next, Chapter~\ref{ch:volunteer} takes a different approach to the notion of ``more is different,'' using both simulations and dynamical systems theory to understand evolutionary games in which the interactions do not resolve into pairs.}

\emph{I have set aside Chapter \ref{ch:prob} to summarize the parts of probability theory which will be necessary for the following two chapters, because I've yet to find a textbook which has the necessary stuff all in one place.  Chapter~\ref{ch:adaptive} is purely analytical: I break a theorem from the literature, show how to fix it and then point out where it will break again.  The goal for Chapter~\ref{ch:doi} is to provide analytical arguments for at least a few of the things seen in Chapters~\ref{ch:multiscale}, \ref{ch:redgreen} and \ref{ch:volunteer}.  Specifically, I aim to use universality to predict critical exponents for phase transitions.}

\emph{Chapters \ref{ch:invasion-fitness} and \ref{ch:price-mls} are mostly about explaining other people's work in a way I can understand.  Chapter~\ref{ch:speculations} is essentially a concept piece, intended to sketch out the possibility of new interesting problems.}

\chapter{Multiscale Structure}

\label{ch:multiscale}

\section{Introduction}

A century and odd years ago, the philosopher William James
asked~\cite{james1907},
\begin{quotation}
\noindent What shall we call a \emph{thing} anyhow? It seems quite arbitrary,
for we carve out everything, just as we carve out constellations, to
suit our human purposes.  For me, this whole `audience' is one thing,
which grows now restless, now attentive. I have no use at present for
its individual units, so I don't consider them. So of an `army,' of a
`nation.' But in your own eyes, ladies and gentlemen, to call you
`audience' is an accidental way of taking you. The permanently real
things for you are your individual persons. To an anatomist, again,
those persons are but organisms, and the real things are the organs.
Not the organs, so much as their constituent cells, say the
histologists; not the cells, but their molecules, say in turn the
chemists.
\end{quotation}
The Jamesian view is that none of these scientific disciplines ought
to be taken as more ``fundamental'' than another.  Each must prove its
own worth by way of its pragmatic utility; none is by necessity merely
the reduction of another to a special case.

In the study of complex systems, we face this directly.  A complex
system exhibits structure at many scales of organization.  For
example, one can study human beings at any magnification, from the
molecular level to the societal, and an entire science flourishes at
each level.  We have developed a formalism for making this intuition
mathematically precise and quantitatively useful, employing the tools
of information theory~\cite{allen2014, baryam2004a, baryam2004b,
baryam2004c, sgs2004, metzler2005}.  To explore how this formalism can
be used, and to make clear the intricacies of multiscale information
theory, we shall in this chapter apply that theory to an illustrative
class of geometrical problems.  Having done this, we will be in a good
position to use it to study collective behaviors in systems developed
in mathematical biology.

Thinking clearly about what we mean by ``complexity'' is important for
biology, and few problems bring this home more clearly than the
so-called \emph{C-value paradox.}  This is the puzzle that the sizes
of species' genomes do not correlate with any obvious, intuitive or
meaningful measure of organismal complicatedness~\cite{eddy2012}.  A
species' C-value is the characteristic amount of DNA which occurs in
one set of chromosomes within its nucleus.  It can be measured in
picograms, for a physical unit, or in base pairs, for a more
informational one.  (A trillionth of a gram roughly works out to a
billion base pairs.)  One might think that species with larger
C-values would be more ``complex'' by some fairly apparent standard.
However, nature has not turned out that way.  The domestic
onion, \emph{Allium cepa,} has approximately 16 billion base pairs on
one set of chromosomes.  This is roughly five times the total size of
the human genome~\cite{palazzo2014}.  And the problem goes beyond the
check to our pride: life forms which seem by all accounts to be
comparably complicated can have widely separated genome sizes.  For
example, lungfish can have genomes 350 times larger than those of
pufferfish~\cite{palazzo2014}.  Even close evolutionary proximity is
no guarantee that C-values will agree.  \emph{Zea mays,} the maize
plant, diverged from the teosinte grass \emph{Zea luxurians} about
140,000 years ago~\cite{rossibarra2009}, and in that time, its genome
size has increased by half~\cite{tenaillon2011, eddy2012}.

So, the puzzle: what does all that extra genetic information do?  The
answer, in brief terms, is \emph{basically nothing.}

Nor are we humans making much use of our own genomes, percentage-wise.
Eukaryotic DNA contains, as a general rule, vast supplies
of \emph{junk}~\cite{eddy2012, doolittle2013, graur2013, hurst2013,
niu2013, palazzo2014, palazzo2015, graur2015, elliott2015}.  Some DNA
specifies the sequences of proteins, and is designated ``coding'' DNA.
Other stretches of the double helix play a role in regulating which
genes are active and when.  Still other portions of the genome are
transcribed into the RNA components of cell\-ular machinery like
ribosomes.  But even with all these accounted for, there remain
sequences which are, identifiably, detritus.

The C-value ``paradox'' is not so paradoxical after all, then: the
variable amounts of genomic bloat due to nonfunctional DNA make
C-value variations a rather unsurprising phenomenon.  The presence of
a large quantity of nonfunctional DNA can have a biological effect,
since it takes up space and increases the resources required for cells
to replicate.  For example, salamanders carry a truly remarkable
amount of genetic information, with different species possessing
genomes four to thirty-five times the size of our
own~\cite{palazzo2014}.  Plainly, it is possible to make a salamander
using far less DNA than some species of them have.  Among salamander
species, larger genome size is correlated with slower regeneration of
lost limbs, suggesting that elevated genome size might be somewhat
costly~\cite{sessions1987}.  The essential points are, first, that
this cost is, if it exists, not on the whole deleterious enough for
natural selection to act strongly against it~\cite{palazzo2014}, and
second, that it is due to the \emph{quantity} of DNA present, not its
\emph{specific sequence.}

Moreover, the presence of DNA detritus suggets a way of thinking about
complexity more quantitatively.  The matter is one of \emph{effective
description.}  Let us focus, for the moment, on the complexity of a
genome itself, rather than of the body plan associated with it.  Our
intuition leads us to say that we require more information to describe
a more complicated genome.  However, large stretches of a eukaryotic
genome will be junk sequences, which can be switched with other,
equally nonfunctional sequences or even deleted entirely with little
or no effect.  This suggests a strategy: we can describe the
functional portion of the genome---the protein-coding genes, the
regulatory regions and so forth---faithfully, and then we can loosely
characterize the rest.  We take careful notes about the functional
parts, and then we fill in the rest with broad brush-strokes.  A
coarse-grained description of the nonfunctional portion is adequate,
because any other nucleotide sequence which satisfies the same
coarse-grained criteria could be swapped in for the actual junk.

In turn, we can apply the same method to the functional portion.
Multiple nucleotide sequences are translated to the same protein,
because multiple codons in the genetic code stand for the same amino
acid~\cite{stryer1988, sella2006}.  We can, therefore, exploit this
redundancy and use a smaller number of characters to
represent \emph{what it is most important to know} about each
protein-coding gene.  Then, multiple amino-acid sequences are often
biologically equivalent to one another, because the substitution of
one amino acid for a similar one does not drastically change the
resulting protein~\cite{guo2004}.  So, we can describe a protein in a
coarse-grained way, and so on.  Indeed, this plurality of possible
sequences compatible with the same coarse-grained description is
biologically essential: given the amount of DNA which human cells
carry, we would otherwise be ground under heel by the genetic load of
deleterious mutations.  It also affects the \emph{rate} of evolution,
since a population can more easily explore the space of possible
genomes when there is a network of neutral paths through
it~\cite{maynardsmith1970}.

The general lesson is that \emph{partial descriptions} of a system can
have increased utility if they can exploit patterns and redundancies.
Furthermore, the way the utility of a description increases as we
allow more information to be used tells us about the structure of the
system we are describing.  This approach is different from the way
information theory has typically been used in the past, because we are
considering \emph{scale} and information as \emph{complementary
quantities}~\cite{allen2014}.  We measure the effort which goes into a
description in units of information, whereas the effectiveness of that
description is the scale of what it captures.

We now turn to mathematizing this idea, following earlier work by the
author and others on multiscale structure and information theory.  The
resulting formalism will be applicable at levels from the
intracellular to the societal.  This will allow us to discuss
descriptions, utilities and related concepts beginning from an
axiomatic starting point (so that we will not need molecular biology
in order to define utility).  With these concepts developed and some
illustrative examples analyzed, we will then apply them to
evolutionary dynamics.

\subsection{Information-Theoretic Axioms for Structured Systems}

For convenience, we review the basic axioms of the multiscale
information formalism which we developed in earlier
work~\cite{allen2014}.  In this formalism, a \emph{system} is defined
by a set of components, $A$, and an \emph{information function,} $H$,
which assigns a nonnegative real number to each subset $U \subset A$.
This number $H(U)$ is the amount of information needed to describe the
components in $U$.  To qualify as an information function, $H$ must
satisfy two axioms:

\begin{itemize}
\item \emph{Monotonicity:} The information in a subset $U$ that is
  contained in a subset $V$ cannot have more information than $V$.
  That is, $H(U) \leq H(V)$.
\item \emph{Strong subadditivity:} Given two subsets, the information
  contained in both cannot exceed the information in each of them
  separately minus the information in their intersection:
\begin{equation}
H(U \cup V) \leq H(U) + H(V) - H(U \cap V).
\label{eq:strong-subadditivity}
\end{equation}
\end{itemize}
Given an information function $H$, we can construct functions which
express different kinds of possible correlations among a system's
components, such as the mutual information, which is the difference
between the total information of two components considered separately
and the joint information of those two components taken together:
\begin{equation}
I(a;b) = H(a) + H(b) - H(a,b).
\end{equation}
By extension, we can also define the tertiary mutual information
\begin{equation}
I(a;b;c) = H(a) + H(b) + H(c) - H(a,b) - H(b,c) - H(a,c) + H(a,b,c).
\end{equation}
This can be extended to higher scales in the same fashion, defining
\emph{shared information} for sets of four or more components.

One way to understand the meaning of our axioms is the following.
Take a set of questions, which are all mutually independent in the
sense that answering one doesn't help to answer any other, or any
combination of others.  Each question pertains to one or more
components of a system.  Components have nonzero shared information if
one or more questions pertain to both of them.  In other words, if
each question is represented by a point, then each component is a set
of points, and set-theoretic intersection defines shared information.

As mentioned earlier, this formalism treats scale and information as
complementary quantities.  Often, ``scale'' is thought of in terms of
length or time (for example, the James quote above organizes the
learned disciplines essentially by the geometrical dimensions of what
they study).  For an axiomatic development, a more general definition
is appropriate, and so for our purposes, ``scale'' will refer to
the \emph{number of components} within a system which are involved in
an interrelationship.

\subsection{Indices of Structure}
\label{sec:multiscale-indices}
To specify the structure of a system according to our definition, it
is necessary to specify the information content of each subset $U
\subset A$.  Because the number of such subsets grows exponentially
with the number of components, complete descriptions of structure are
impractical for large systems.  Therefore, we require statistics which
can convey the general character of a system's structure without
specifying it completely.  Using an \emph{index of structure,} we can
summarize how a system is organized and compare that pattern of
organization to the patterns manifested by other systems.

One such index of structure is the \emph{complexity profile,}
introduced in \cite{baryam2004b} to formalize the intuition that a
genuinely complex system exhibits structure at multiple scales.  The
complexity profile is a real-valued function $C(k)$ that specifies the
amount of information contained in interdependencies of scale $k$ and
higher.  $C(k)$ can be computed using a combinatorial formula which
takes as input the values of the information function $H$ on all
subsets $U \subset A$
\cite{allen2014, baryam2004a, baryam2004b, baryam2004c, sgs2004,
  metzler2005}.  First, we define the quantity $Q(j)$ as the sum of
the joint information of all collections of $j$ components:
\begin{equation}
Q(j) = \sum_{i_1,\ldots,i_j} H(a_{i_1},\ldots,a_{i_j}).
\label{eq:Q}
\end{equation}
The complexity profile can be computed using the formula
\begin{equation}
C(k) = \sum_{j = N-k}^{N-1} (-1)^{j+k-N} 
\binom{j}{j+k-N}
       Q(j+1),
\label{eq:CP}
\end{equation}
where $N=|A|$ is the number of components in the system.  Generally,
$C(k)$ captures the amount of information contained in
interrelationships of order $k$ and higher.  We shall illustrate this
in the next section with a few examples.

The complexity profile satisfies a \emph{conservation law}: the sum of
$C(k)$ over all scales $k$ is
\begin{equation}
\sum_k C(k) = \sum_{a\in A} H(a).
\end{equation}
That is, the sum of the complexity over all scales is given by the
individual information assigned to each component, regardless of the
components' interrelationships.

Another useful index of structure is the Marginal Utility of
Information, or MUI~\cite{allen2014}.  While the complexity profile
characterizes the amount of information that is present in the system
behavior at different scales, the MUI is based on descriptive utility
of limited information through its ability to describe behavior of
multiple components.  Informally speaking, we describe a system by
``investing'' a certain amount of information, and for any amount of
information invested, an \emph{optimal} description yields the best
possible characterization of the system.  The MUI expresses how the
usefulness of an optimal description increases as we invest more
information.  We can define the MUI precisely, starting with the basic
axioms of information functions, by using notions from linear
programming~\cite{allen2014}.

In general outline, one constructs the MUI as follows.  Let $\cA$ be a
system, defined per our formalism as a set of components $A$ and an
information function $H$.  Then, let $d$ be a \emph{descriptor,} an
entity which conveys information about the system $\cA$.  To express
this mathematically, we consider the new, larger system made by
conjoining $d$ with the set of components $A$ and defining an
information function on the subsets of this expanded set.  The
information function for the augmented system reduces to that of the
original for all those interdependencies which do not involve the
descriptor $d$, and it expresses the shared information between $d$
and the original system.  The \emph{utility} of~$d$ is the sum of the
shared information between $d$ and each component within~$\cA$:
\begin{equation}
u(d) = \sum_{a \in A} I(d;a).
\end{equation}
This counts, in essence, the total scale of the system's organization
that is captured by~$d$.  We define the \emph{optimal utility} $U(y)$
as the utility of the best possible descriptor having $H(d) = y$.  The
MUI is then the derivative of~$U(y)$.

How do these structure indices capture the organization of a system?
We can illustrate the general idea by way of a conceptual example.
Consider a crew of movers, who are carrying furniture from a house to
a truck.  They can be acting largely independently, as when each mover
is carrying a single chair, or they can be working in concert,
transporting a large item that requires collective effort to move,
like a grand piano.  In the former case, knowing what any one mover is
doing does not say much about what specific act any other mover is
engaged with at that time.  Information about the crew applies at the
scale of an individual mover.  By contrast, in the latter case, the
behavior of one mover can be inferred from that of another, and
information about their actions is applicable at a larger scale.  From
these general considerations, it follows that for a system of largely
independent movers, $C(k)$ is large at small $k$ and drops off
rapidly, whereas when the movers are working collectively, $C(k)$ is
small for low $k$ and remains nonzero for larger $k$.  When the movers
act mostly independently, we cannot do much better
at \emph{describing} their behavior than by specifying the behavior of
each mover in turn.  Therefore, as we invest more information into
describing the system, the gain in utility of our description remains
essentially constant.  For the case of independent movers, then, the
MUI curve is low and flat.  On the other hand, when the movers are
acting in concert, a brief description can have a high utility, so the
MUI curve is peaked at the origin and falls off sharply.
Heuristically speaking, we can in this example think of the complexity
profile and the MUI as \emph{reflections} of each other.  When we
develop these indices quantitatively, we find in fact that this is
exactly true in a broad class of systems.

Both the complexity profile and the MUI obey a convenient sum rule.
If a system separates into two independent subsystems, the complexity
profile of the whole is the sum of the profiles of the pieces, and
likewise for the MUI.  This property of both structure indices follows
from the basic information-function axioms~\cite{allen2014}.  In the
next section, we will see examples of systems which illustrate the sum
rule for both the MUI and the complexity profile.

\section{Examples}

\subsection{Three-Component Systems}

To explore the consequences of our definitions, it is helpful to begin
with simple examples.  Following the recent review article about the
multiscale complexity formalism~\cite{allen2014}, we study the
following four systems, each of which contains three binary variables.

\begin{itemize}
\item \emph{Example \exA: Three independent bits:} The
  system comprises three components, and knowing the state of any one
  bit provides no inference about the state of any other.  As a whole,
  the system can be in any one of eight possible configurations, with
  no preference given to any of the eight possibilities.

\item \emph{Example \exB: Three completely interdependent bits:}
 The system as a whole is either in state $000$ or state $111$, with
 no preference given to either option.  Knowing the value of any one
 bit allows the inference of both other bits.

\item \emph{Example \exC: Independent blocks of dependent bits:} Each
  component is equally likely to take the value 0 or 1; however, the
  first two components always take the same value, while the third can
  take either value independently of the coupled pair.

\item \emph{Example \exD: The $2+1$ parity bit system}: Three bits
  which can exist in the states 110, 101, 011, or 000 with equal
  probability.  Each of the three bits is equal to the parity (0 if
  even; 1 if odd) of the sum of the other two.  Any two of the bits
  are statistically independent of each other, but the three as a
  whole are constrained to have an even sum.
\end{itemize}

\begin{figure}[h]
\includegraphics[width=8cm]{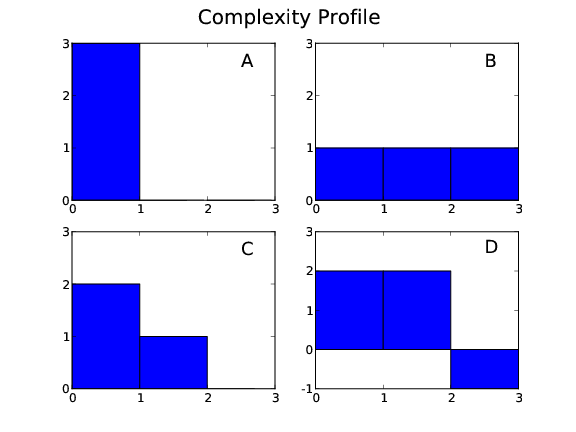}
\caption{\label{fig:cp-toy-examples} Complexity profiles for the
  three-component example systems \exA, \exB, \exC{} and \exD,
  computed using Eq.~(\ref{eq:CP}).  Examples \exA{} and \exB{}
  illustrate the general fact that highly interdependent systems have
  tall and narrow complexity profiles, whereas the profiles of systems
  with largely independent components are low and wide.  Example \exC,
  which we can think of as the combination of two independent
  subsystems, illustrates the complexity profile's sum rule.  Finally,
  example \exD, the parity-bit system, showcases the emergence of
  negative shared information.  Note that the total signed area
  bounded by each curve is 3 units.  (Figure reproduced
  from~\cite{allen2014}.)}
\end{figure}

\begin{figure}[h]
\includegraphics[width=8cm]{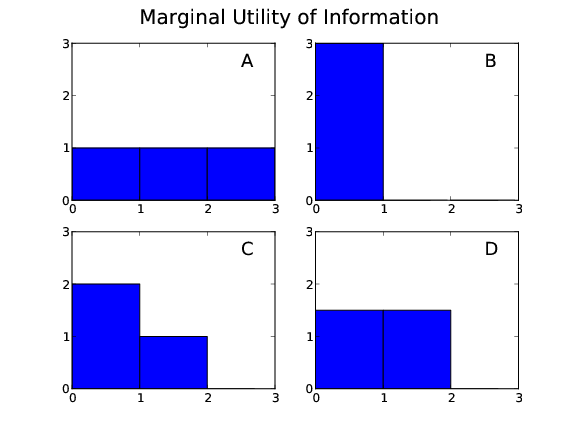}
\caption{\label{fig:mui-toy-examples} MUI plots for the
  three-component example systems \exA, \exB, \exC{} and \exD.  Note
  that, as with the complexity profiles in
  Figure~\ref{fig:cp-toy-examples}, the total area bounded by each
  curve is 3 units.  Furthermore, for examples \exA, \exB{} and \exC,
  the complexity profile and the MUI are reflections (generalized
  inverses) of each other.  This is generally true for systems which
  are the disjoint union of internally interdependent
  blocks~\cite{allen2014}, but it is not the case for the parity-bit
  system, example \exD.  (Figure reproduced from~\cite{allen2014}.)}
\end{figure}

Figures~\ref{fig:cp-toy-examples} and \ref{fig:mui-toy-examples} show
the complexity profiles and the MUI curves for these example systems.

\subsection{Minimal Incidence Geometry}
To develop additional intuition about our information-theoretic
formalism, and to build a bridge between different areas of
mathematics, we shall apply the information theory of multicomponent
systems to \emph{incidence geometry.} The premise of incidence
geometry is that one has a set of points and a set of lines which
connect them, satisfying some conditions which abstract basic notions
of geometry.  To wit, for any incidence geometry, every line contains
at least two distinct points, and for every line, there exist one or
more points not lying on that line.  We relate geometry to information
theory in the following way: Ascribe to each point 1 unit of
information, and define each line to be a system component.  Then for
any incidence geometry, the information ascribed to a component is
always greater than or equal to~2, and the information within the
whole system is always greater than~2.

The examples we shall consider from incidence geometry will illustrate
most of the key features of the multiscale information theory
formalism.  The noteworthy exception is that incidence geometry does
not provide examples of negative multivariate mutual information.
This is a subtlety which can arise when one considers dependencies
among three or more components~\cite{allen2014}, as we saw in
example \exD.  However, it will not be a major concern for the models
from mathematical biology which we will study later in this chapter.

\begin{figure}[h]
\includegraphics[width=8cm]{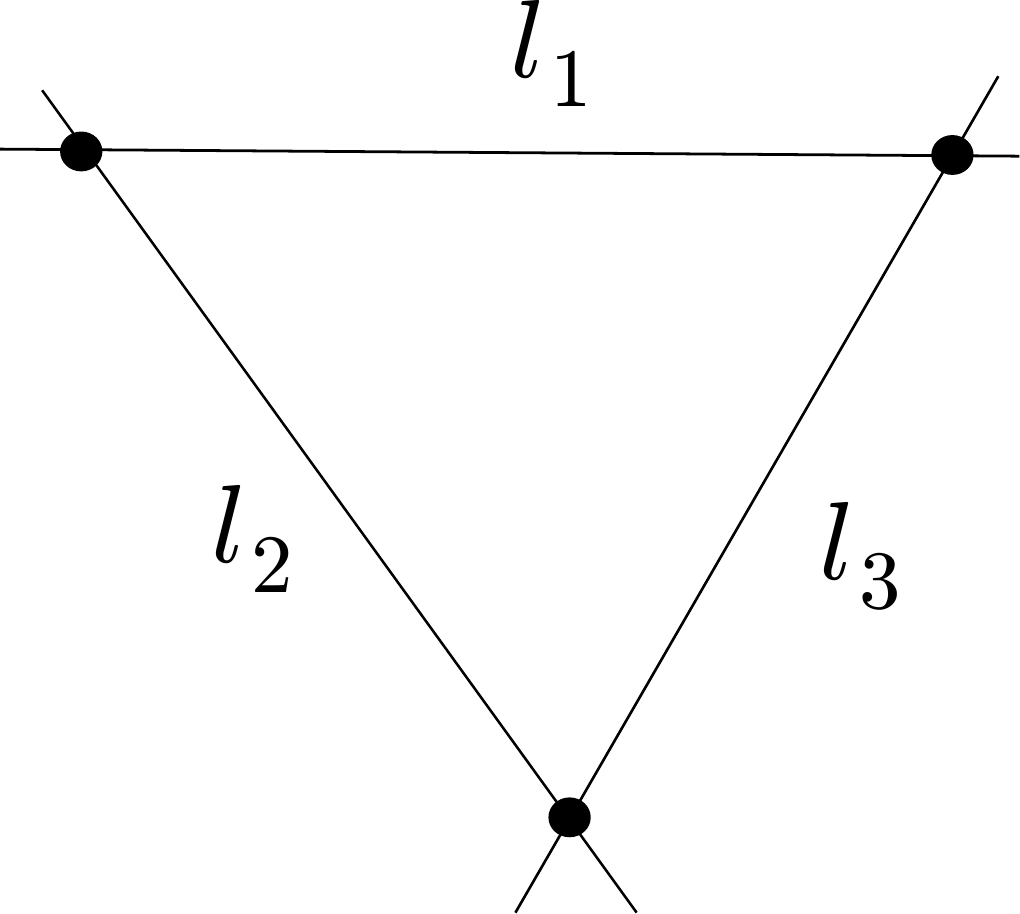}
\caption{\label{fig:minimal-incidence-geometry} The simplest possible
  incidence geometry.  Each of the three lines contains two distinct
  points, and for each line, there exists exactly one point which does
  not lie on that line.  When we associate a unit of information to
  each point, the shared information of any pair of lines is the
  information ascribed to their point of intersection.}
\end{figure}

The simplest possible incidence geometry contains 3 points and 3
lines.  We depict this construction in
Figure~\ref{fig:minimal-incidence-geometry}.

If we denote the three lines by $l_1$, $l_2$ and $l_3$, as in
Figure~\ref{fig:minimal-incidence-geometry}, then because each line
contains exactly two points, we have
\begin{equation}
H(l_1) = H(l_2) = H(l_3) = 2,
\end{equation}
while because any two lines intersect in exactly one point,
\begin{equation}
I(l_1;l_2) = I(l_2;l_3) = I(l_1;l_3) = 1.
\end{equation}
The joint information of all three components taken together is the
total number of points in the geometry:
\begin{equation}
H(l_1,l_2,l_3) = 3.
\end{equation}
From these three observations, we can deduce that the tertiary mutual
information of the three components vanishes:
\begin{equation}
C(3) = I(l_1;l_2;l_3) = 0.
\end{equation}
This is the information-theoretic restatement of the geometric fact
that the three lines do not all come together at a single point.  All
together, the complexity profile of the minimal incidence geometry is
given by
\begin{equation}
C(k) = \left\{
       \begin{array}{cc}
       3, & k \in \{1, 2\}, \\
       0, & k = 3.
       \end{array}
       \right.
\label{eq:minimal-geometry-CP}
\end{equation}

\begin{figure}[h]
\includegraphics[width=8cm]{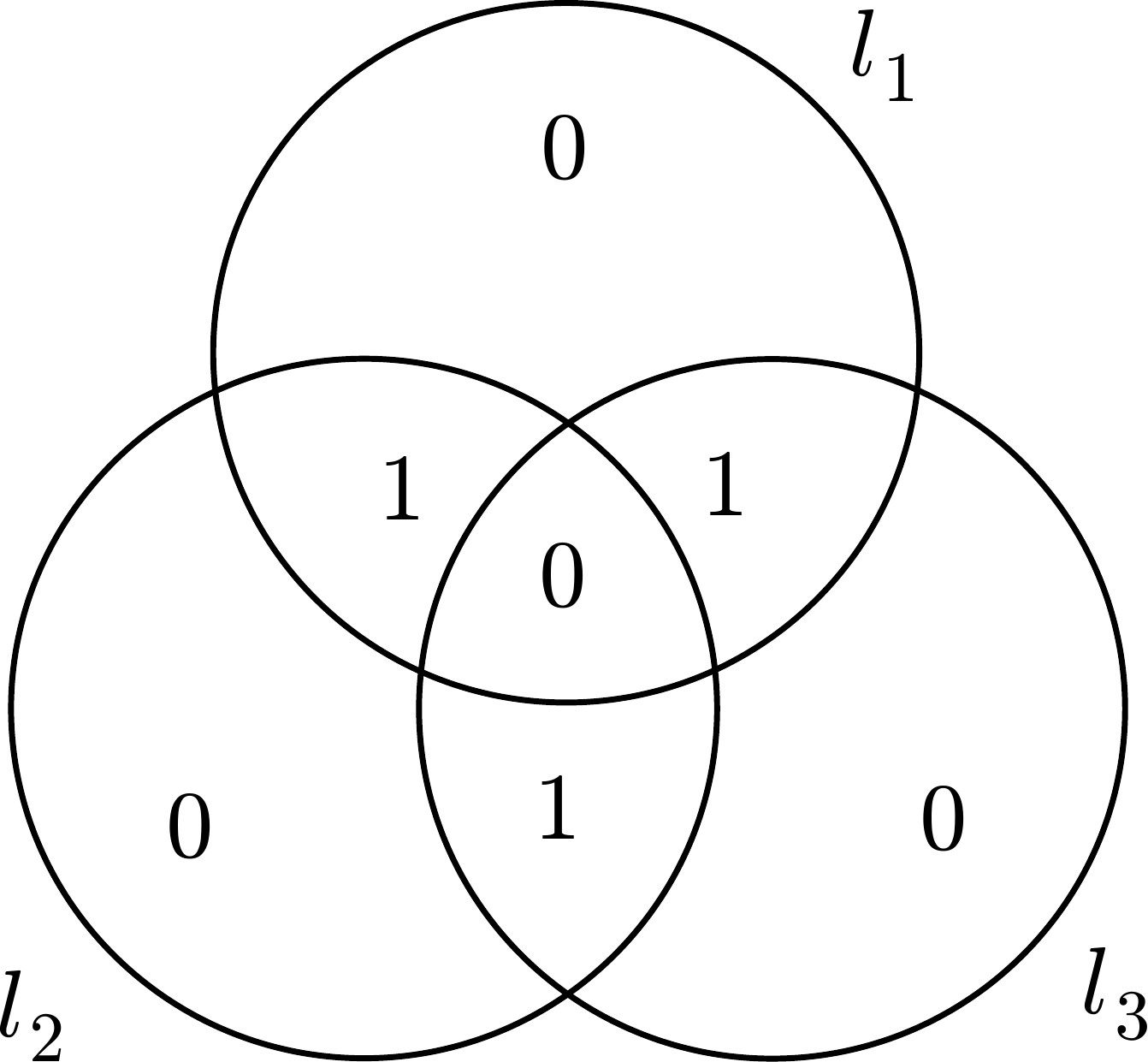}
\caption{\label{fig:venn-minimal-incidence-geometry} Information
  diagram for the minimal incidence geometry depicted in
  Figure~\ref{fig:minimal-incidence-geometry}.  Within each of the
  three circles, all of the regions which contain nonzero information
  are regions of overlap with another circle.  This is the
  information-theoretic consequence of the geometrical fact that each
  point belongs to more than one line.  Note that the central region,
  where all three circles overlap, contains no information.}
\end{figure}

The information-theoretic relationships among the three system
components $l_1$, $l_2$ and $l_3$ can be expressed in a three-circle
Venn diagram, which we depict in
Figure~\ref{fig:venn-minimal-incidence-geometry}.

The other structure index introduced above is the MUI.  We can deduce
the MUI curve of the minimal incidence geometry using the properties
of the MUI established in \cite{allen2014}.  First, a descriptor $d$
must have at least 3 units of information to capture all of the
information which was granted to the geometry.  Expanding on an
\emph{optimal} descriptor, one which wastes nothing, brings no benefit
beyond a descriptor length of $H(d) = 3$.  Therefore, the marginal
utility $\mui(y)$ will equal zero for $y \geq 3$.  In addition,
because the integral of the MUI curve is the utility of a full
description---that is, the total scale-weighted information of the
system---we know the integral of the MUI for this geometry will be 6.
Furthermore, we can constrain the height of the MUI curve, using the
following property:
\begin{itemize}
\item If there are no interactions or correlations of degree $k$ or
  higher---formally, if $I(a_1; \ldots; a_k)=0$ for all collections
  $a_1, \ldots, a_k$ of $k$ distinct components---then $\mui(y)\leq k$
  for all $y$ \cite[\S VII.B]{allen2014}.
\end{itemize}
Here, this means that $\mui(y) \leq 3$.  Furthermore, we know that
$\mui(y)$ is the derivative of a piecewise linear function, so
$\mui(y)$ is piecewise constant.  For the minimal geometry, then, we
expect the MUI should be
\begin{equation}
\mui(y) = \left\{
          \begin{array}{cc}
          2, & 0 \leq y < 3, \\
          0, & y \geq 3.
          \end{array}
          \right.
\label{eq:minimal-geometry-MUI}
\end{equation}
Note that the MUI curve is the reflection of the complexity profile
$C(k)$ in Eq.~(\ref{eq:minimal-geometry-CP}).

These properties hold for any incidence geometry: the MUI vanishes for
$y$ larger than the number of points used to build the geometry, the
integral of the MUI is the total scale-weighted information, and the
MUI is bounded above by one plus the maximal number of lines which
mutually intersect at a common point.  We can relate the MUI and the
complexity profile by noting that the areas bounded by both curves are
the same, and moreover, the width of the MUI curve $M(y)$ is the
height of the complexity profile $C(1)$, because both are given by the
number of points in the geometry.

\subsection{Fano Plane}

\begin{figure}[h]
\includegraphics[width=8cm]{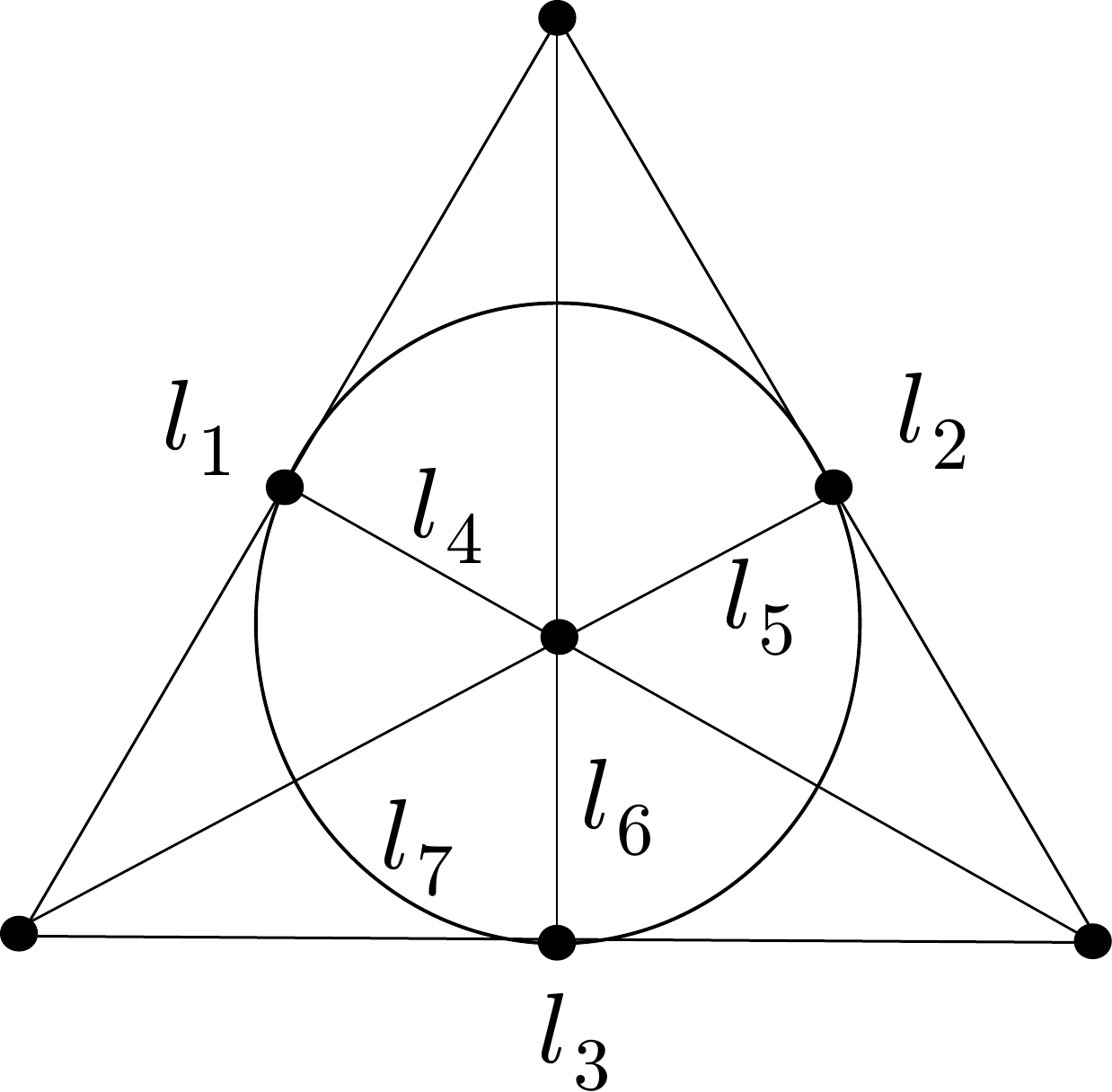}
\caption{\label{fig:fano} The Fano plane: a symmetrical arrangement of
seven points and seven lines.  The shared information between any two
lines is 1, but the shared information at higher scales depends on
which set of lines we choose.}
\end{figure}

The Fano plane, pictured in Figure~\ref{fig:fano}, has 7 points, 7
lines, 3 points on every line, and 3 lines through every point.  The
total scale-weighted information is the number of points per line
times the number of lines, or 21.  The information content of the
whole system is 7, while the mutual information between any line and
any other is again 1.

\begin{figure}[h]
\includegraphics[width=12cm]{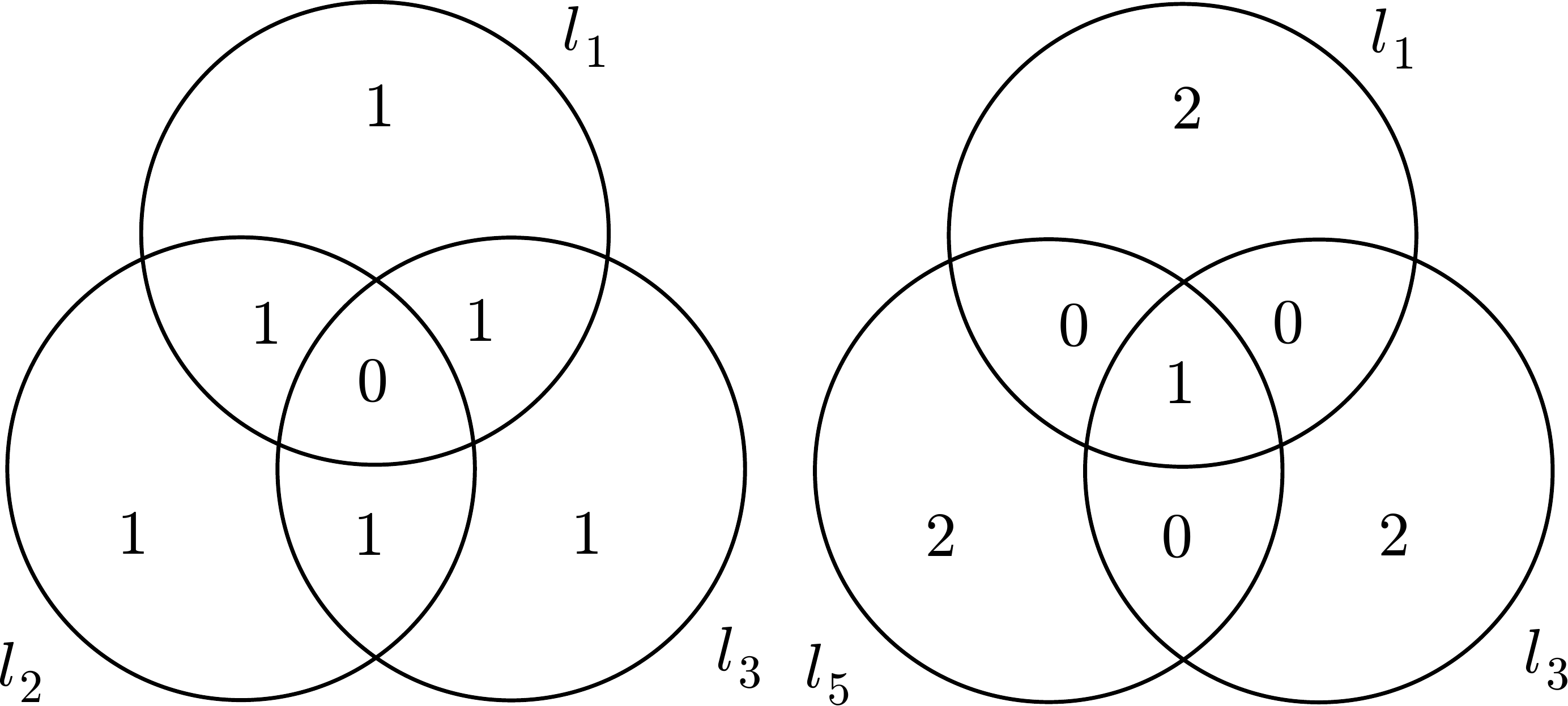}
\caption{\label{fig:venn-fano} Illustrative examples of the two
  possible three-circle Venn diagrams for three-component subsets of
  the Fano-plane system.  Note that the information in the central
  region is zero in one case but nonzero in the other.}
\end{figure}

For the Fano plane, there are two possible information-theoretic
scenarios involving three distinct lines.  If the three lines do not
all meet at a common point, as for example $l_1$, $l_2$ and $l_3$,
then the tertiary mutual information of those three components is
zero.  The other option is for the three lines to meet at a common
point, as with $l_1$, $l_3$ and $l_5$, in which case the tertiary
mutual information is 1.
\begin{equation}
I(l_1;l_2;l_3) = 0, \hbox{ but } I(l_1;l_3;l_5) = 1.
\end{equation}
The tertiary mutual information $I(l_i;l_j;l_k)$ can never be greater
than 1, because any two lines come together in one and exactly one
point.  This uniqueness of intersections also implies that, in the
three-circle Venn diagram, the lens-shaped regions where two circles
overlap always contain a total of 1 unit of information.  If the inner
region of triple overlap, corresponding to $I(l_i;l_j;l_k)$, contains
the value 1, then the outer region, $I(l_i;l_j|l_k)$, must contain the
value 0, and vice versa.  In addition, the total information content
enclosed by each of the three circles is always 3 units, the number of
points per line.  Together, these facts constrain the possible
three-circle Venn diagrams for subsets of the Fano-plane system,
leaving only the two possibilities depicted in
Figure~\ref{fig:venn-fano}.

We have in the Fano plane an elementary illustration of a commonplace
occurrence in complex systems: \emph{higher-order structure which
  cannot be resolved into lower-order interrelationships.}  In this
case, we see there exists a variety among triples of components which
cannot be inferred from considering pairs.  The Fano-plane system does
not fall within the particular special classes of systems studied as
examples in~\cite{allen2014}, since not all subsets of the same size
have the same information content.

We can deduce the complexity profile $C(k)$ of the Fano-plane system
by counting points lying on $k$ or more lines, or by computation using
Eq.~(\ref{eq:CP}).  If we take the latter approach, with 7 components
there are $2^7 - 1 = 127$ nonempty subsets $U$ for which we must find
the information content $H(U)$.  However, thanks to the high degree of
symmetry, the information content for the different possible subsets
of the component set is not that hard to work out.  Because there
exist 3 lines through any point, eliminating 1 line (component) can
only reduce the number of lines through any point down to 2.  So, the
information content of any six-component set is still 7.  To eliminate
all lines through a point, we must erase at least three components.
From considerations like these, we can deduce $C(k)$ by explicit
computation with Eq.~(\ref{eq:CP}).  For the Fano plane, the
complexity profile is given by
\begin{equation}
C(k) = \left\{
       \begin{array}{cc}
       7, & 1 \leq k \leq 3, \\
       0, & 4 \leq k \leq 7.
       \end{array}
       \right.
\end{equation}
We can also deduce this result quickly by recalling that, generally,
$C(k)$ captures the amount of information contained in
interrelationships of scale $k$ and higher.  In the context of
incidence geometry, the complexity profile $C(k)$ has the geometrical
interpretation as \emph{the number of distinct points which lie at the
  intersection of $k$ or more lines.}

Applying the same properties as we used for the minimal incidence
geometry, we can deduce that the MUI of the Fano plane vanishes for
descriptor lengths $y \geq 7$, that the integral of $\mui(y)$ is $7
\cdot 3 = 21$, and that $\mui(y) \leq 4$ for all $y$.

\section{The Dual of a Complex System}

Geometry makes much use of \emph{duality.}  The dual of a geometrical
arrangement is that arrangement which is found by interchanging the
roles of points and lines in the original.  For example, an
\emph{affine plane of order $n$} is a specialization of an incidence
geometry in which the following conditions hold \cite{wootters2006,
  ueberberg2010}:
\begin{itemize}
\item the geometry contains $n^2$ points in all;
\item the geometry contains $n(n+1)$ lines;
\item each line contains $n$ points;
\item each point lies on $n+1$ lines.
\end{itemize}
Interchanging points and lines yields a \emph{dual affine plane}
\cite{wootters2006, baron2009}, a geometry which meets the following
criteria:
\begin{itemize}
\item the geometry contains $n(n+1)$ points;
\item the geometry contains $n^2$ lines;
\item each line contains $n+1$ points;
\item each point lies on $n$ lines.
\end{itemize}
If we translate from geometry to information theory, what is the
meaning of duality?  We began by saying that each point in a geometry
corresponded to a unit of information, and each line was to become a
component in an information-theoretic system.  Applying the operation
of duality, we find that each line in the original geometry should be
ascribed a unit of information in the dual system, and each point in
the original geometry becomes a system component in the dual.

The complexity profile of the original system is defined for values of
$k$ from 1 to the number of lines in the original geometry.
Therefore, the complexity profile of the dual system is defined for
values of $k$ from 1 to the number of \emph{points} in the original
geometry.  The property of ``lines meeting at a common point'' in the
original becomes the property of ``points lying on a common line'' in
the dual.  Consequently, the complexity profile of the dual system can
be found from the original geometry.  The dual of $k$ lines
intersecting at a point is $k$ points sharing a common line.  We can
find the complexity profile of the dual system by counting the number
of distinct lines in the original system which contain $k$ or more
points.

For the minimal incidence geometry depicted in
Figure~\ref{fig:minimal-incidence-geometry} and the Fano plane
portrayed in Figure~\ref{fig:fano}, the duality exchange operation
does not change the complexity profile.  We can say that $C(k)$ for
those geometries is \emph{self-dual.}  Considering the affine planes
of order $n$ defined above, we know that each point lies on $n+1$
lines, so $C(k)$ is a rectangle of width $n+1$ and height $n^2$.  In a
dual affine plane of order $n$, each of the $n(n+1)$ points lie on $n$
lines, so $C(k)$ is a rectangle of width $n$ and height $n(n+1)$.  For
affine planes, the duality transformation preserves the \emph{area,}
but not the \emph{shape,} of the complexity profile.

One naturally wonders whether this property of the area under the
$C(k)$ curve is more general.  We can investigate this question using
the conservation law, proved in Allen \emph{et al.}~\cite{allen2014},
that the area under the complexity profile is always the total
scale-weighted information content of the system.  For the special
case of an incidence geometry, this means that
\begin{equation}
\sum_k C(k) = \sum_i H(l_i),
\end{equation}
where $i$ ranges from 1 to the total number of lines in the geometry.
Let $\{v_j\}$ be the set of all points in the geometry.  (For an affine
plane, $j$ thus ranges from 1 to $n^2$.)  The area bounded by the
complexity profile is then
\begin{equation}
\sum_i H(l_i) = \sum_i \sum_{\{j|v_j \in l_i\}} H(v_j).
\end{equation}
In this sum, the multiplicity of any $H(v_j)$ is the number of lines
which contain the point $v_j$.  Therefore,
\begin{equation}
\sum_i H(l_i)
 = \sum_j H(v_j)
   \cdot \left|
         \{ l_i | v_j \in l_i \}
         \right|.
\label{eq:area-under-profile}
\end{equation}
The components of the dual system are the points of the original
geometry, and the information content of each component in the dual
system is the number of lines which pass through that point in the
original geometry.  Summing over the components of the dual system to
find the total area under the dual complexity profile yields the same
sum as in Eq.~(\ref{eq:area-under-profile}).  In consequence, we can
say that \emph{the duality transformation generally preserves the area
  under the complexity profile.}

What can we deduce about the Marginal Utility of Information for
affine planes and their duals?  We recall that the MUI is, by
construction, a piecewise constant function.  The general properties
we deduced earlier imply that for an affine plane of order $n$,
$\mui(y)$ vanishes for $y \geq n^2$, and the integral of $\mui(y)$ is
the number of lines times the number of lines per point, or $n^2
(n+1)$.  Furthermore, because exactly $n+1$ lines meet at each point,
$M(y) \leq n + 2$ for all $y$.

The most straightforward way to meet these requirements is with the
following piecewise constant curve:
\begin{equation}
\mui(y) = \left\{
          \begin{array}{cc}
          n + 1, & 0 \leq y < n^2, \\
          0, & y \geq n^2.
          \end{array}
          \right.
\label{eq:affine-plane-MUI}
\end{equation}
For a \emph{dual} affine plane of order $n$, the integral of $\mui(y)$
is the same, $n^2(n+1)$.  The right-hand edge is instead at $n(n+1)$,
and the upper limit becomes $\mui(y) \leq n + 1$.  So,
\begin{equation}
\mui(y) = \left\{
          \begin{array}{cc}
          n, & 0 \leq y < n(n+1), \\
          0, & y \geq n(n+1).
          \end{array}
          \right.
\label{eq:dual-affine-plane-MUI}
\end{equation}
For both affine planes and their duals, $\mui(y)$ is the reflection of
the complexity profile $C(k)$.  And again, the duality transformation
preserves areas but not shapes.

These self-duality properties are the information-theoretic
consequences of the geometry theorem known as the \emph{principle of
  double counting} \cite{ueberberg2010}.  To wit: if $\{v_i\}$ are the
points of an incidence geometry and $\{l_j\}$ are its lines, then
\begin{equation}
\sum_{\{v_i\}}
  \left|\{
  l_j |\, \hbox{$v_i$ lies on $l_j$}
  \}\right|
=
\sum_{\{l_j\}}
  \left|\{
  v_i |\, \hbox{$l_j$ intersects $v_i$}
  \}\right|.
\label{eq:double-counting}
\end{equation}
We can prove this by observing that both sides of
Eq.~(\ref{eq:double-counting}) are equal to the size of the set of
ordered pairs defined by
\begin{equation}
S = \{(v_i, l_j) |\, \hbox{$v_i$ and $l_j$ are incident} \}.
\end{equation}

\vfill
\pagebreak

\section{Computation and Gammoids}

The idea of information is closely related to that
of \emph{computation,} and in this section, we will explore another
type of system for which we can define an information function,
thereby bringing our concepts of multiscale structure into a
computational context.

We can idealize a computation as a mathematical process that takes a
set of \emph{inputs} to a set of \emph{outputs.}  For example, if we
have a program that takes a number and returns its square, we can
represent that code pictorially as an arrow:
\begin{equation}
\includegraphics[height=3cm]{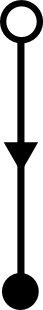}
\label{eq:gammoid1}
\end{equation}
We have drawn the output as a filled circle, and the input as an
unfilled one.  The picture (\ref{eq:gammoid1}) is a diagrammatic
representation for any function that acts on a single input to produce
one and only one output: the sine of an angle, the number of meters in
a distance specified in furlongs and so on.

Some functions take two inputs and return a single output.  Given two
numerical variables, for example, we could compute their sum, their
product, the logarithm of one to the base given by the other, and so
forth.  We can represent all of these functions pictorially with the
following diagram:
\begin{equation}
\includegraphics[height=2.5cm]{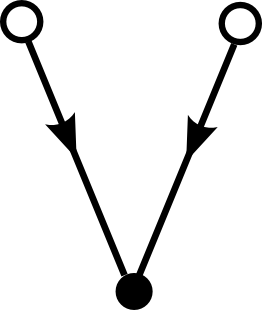}
\end{equation}

Alternatively, a single input might be used to create multiple
outputs.  For example, we might be given the time elapsed in seconds
since midnight, and calculate the current hour and the minute within
that hour:
\begin{equation}
\includegraphics[height=2.5cm]{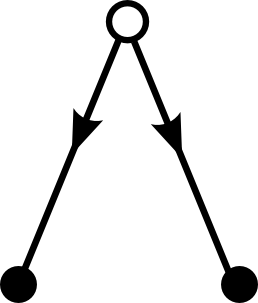}
\end{equation}
There is a sense in which these outputs are not independent, for they
derive from a common source.  This is true even if, as is the case for
separating clock time into minutes and hours, we cannot infer one
output from the other.

What other patterns of information flow might we encounter as we go
about our business?  Given two positive integers, we can find their
greatest common divisor and their least common multiple.  If a
function performs both of these tasks, then it has two outputs, both
of which depend on the two inputs:
\begin{equation}
\includegraphics[height=2.5cm]{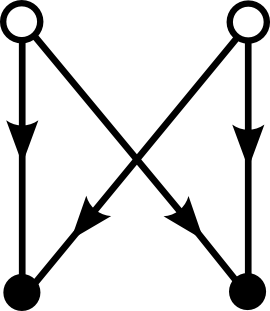}
\end{equation}

A computation may proceed by way of intermediate steps:
\begin{equation}
\includegraphics[height=4cm]{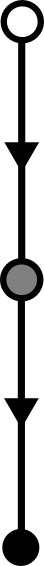}
\end{equation}

These intermediate stages may combine multiple inputs, and they can
yield multiple outputs.  In our diagrams, input vertices (unfilled
circles) have no incoming links, and output vertices (filled circles)
have no outgoing ones.  We introduce intermediate vertices with both
types of links, as in the following:
\begin{equation}
\includegraphics[height=3cm]{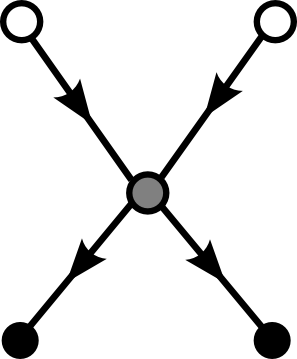}
\end{equation}
In this example, two pieces of input data are combined to yield an
intermediate result, which is then used to compute two output values.
If we saved this intermediate result, we could reconstruct the two
outputs without having the original input data.  It might be that we
cannot deduce the left-hand output from the right, but there is still
an interdependency between them, linking them by virtue of their
common past.

Consider a computation with two inputs and two outputs, one of which
is arrived at by means of an intermediate step:
\begin{equation}
\includegraphics[height=3cm]{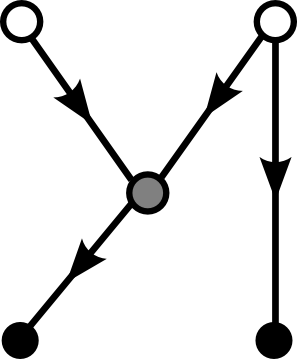}
\end{equation}
Both of the outputs derive ultimately from the same inputs.  However,
they are in a way independent of each other, because knowing the
result of the intermediate step lets us compute one output but not the
other.  Diagrams with multiple layers of intermediate steps are also possible,
and they have a straightforward interpretation in terms of
computations that proceed in stages.

These considerations motivate the following idea: The \emph{effective
amount of information} associated with a set of outputs is the number
of inputs and/or intermediate values which one must know in order to
produce those outputs.  We are, after a fashion, giving a mathematical
form to the notion that the difficulty of a computational task is how
arduous it would be to recover from a crash!

All of our diagrams have taken the form of \emph{directed graphs.}
Each one is a set of vertices, connected by a set of edges, where each
edge carries an indication of which vertex is its source and which is
its target.  In addition, we have distinguished subsets of the vertex
set, marking each vertex as input, intermediate or output.
Recognizing this, we can develop the idea more generally.

Let $G$ be a directed graph, or ``digraph'' for short.  Designate a
subset $S$ of its vertices as inputs, and select a subset $T$ of
vertices to be the outputs.  We can think of $S$ as the set of
``sources'' for information flow, and $T$ as the set of its
``targets.''  (In the digraphs we have drawn, the vertices in~$S$ have
no incoming edges, and those vertices in $T$ have no outgoing edges.
This is sensible, but it turns out not to be essential for proving the
key properties of the information function.)  The information content
$H(U)$ of a subset $U \subset T$ is the size of the smallest set of
vertices having the property that all paths from~$S$ to~$U$ must pass
through it.  The size of this ``minimal separating set'' tells us how
many intermediate variables we would need to save in order to compute
the outputs in~$U$.

It can be proven that the function $H$ satisfies
our axioms for being an information function.  Consequently, the
function $H$ yields sensible expressions for the interdependence of
output variables.  Having defined $H$, we can construct the indices of
multiscale structure $C(k)$ and $M(y)$ as we did before.

\vfill
\pagebreak

For example, take the graph
\begin{equation}
\includegraphics[height=3cm]{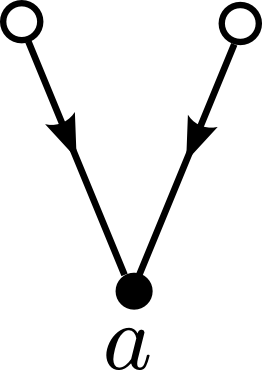}
\end{equation}
Here, we have one output, which we label $a$, and we see directly that
$H(a) = 2$.

A graph with two outputs might look like the following:
\begin{equation}
\includegraphics[height=3cm]{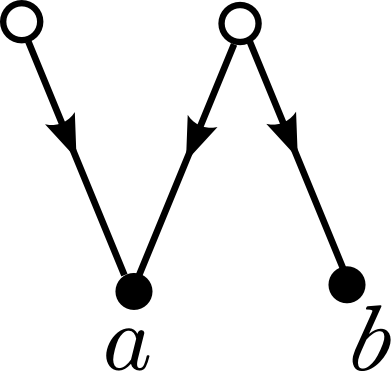}
\end{equation}
In this case, $H(a) = 2$, while $H(b) = 1$ and $H(a,b) = 2$.  The
shared information between outputs $a$ and $b$ is
\begin{equation}
I(a;b) = H(a) + H(b) - H(a,b) = 1,
\end{equation} 
corresponding nicely to the fact that $a$ and $b$ depend upon one
common input.

A more involved two-output graph could include an intermediate step:
\begin{equation}
\includegraphics[height=3cm]{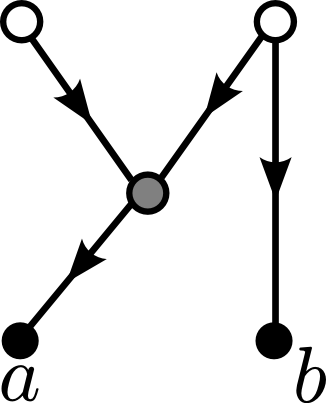}
\end{equation}
Here, $H(a) = H(b) = 1$, while $H(a,b) = 2$.  The shared information
between $a$ and $b$ is now $I(a;b) = 0$, because the intermediate
stage ``shields'' output $a$ from the original input.

What we have constructed here is known in pure mathematics as
a \emph{gammoid,} and our information function is the \emph{rank
function} of that gammoid~\cite{welsh1976}.  A gammoid is an example
of a \emph{matroid,} a structure defined as a set of elements $M$ and
a rank function that assigns a nonnegative integer to each subset
of~$M$.  Matroid rank functions are monotonic and strongly
subadditive, so they count as information functions.  As we have seen
in this section, matroid theory furnishes examples of mathematical
entities to which we can apply the ideas of multiscale structure.

\section{Network Dynamics}

We can apply the multiscale complexity formalism quantitatively to a
model which is an idealized representation of multiple interesting
biological scenarios.  Doing so requires relating the concepts of
probability and information; we provide a more detailed development of
the mathematical prerequisites in Chapter~\ref{ch:prob}.

One way to make progress on many biological problems is to make the
approximation that each component of a system can, at any given time,
be in one of two mutually exclusive states of being.  In essence, we
idealize a phenomenon by treating it as composed of binary random
variables, possibly correlated.  We might postulate that each organism
in a population can follow one of two survival strategies.  For
example, a male bower bird can maraud, attacking other birds' bowers,
or it can remain at its own bower, guarding its own mating display
from marauders~\cite{pruettjones1994, gonick1994}.  Or, we might
postulate that a gene comes in two variant forms.  Each instance of
the gene in the idealized population is then a binary random variable.
We can make an analogous approximation when modeling social and
economic systems.  For example, an individual voter can choose between
one of two political parties.  Or, in a simplified but still
instructive model for a stock market, the price of a company's stock
can be going either up or down~\cite{harmon2011}.

A specific implementation of this idea is the \emph{Moran model,}
which was originally formulated in biology but can be applied more
broadly~\cite{harmon2011}.  Consider a haploid population of
$N$ individuals, and a gene which comes in two alleles.  The genetic
character of the population can change as individuals are born and
die.  One simple dynamical model for this process picks an individual
at random with each tick of a discrete-time clock.  The chosen, or
\emph{focal,} individual mates with one of the other $N-1$ organisms
and produces an offspring, which then takes the place of the focal
individual.  The allele carried by the offspring is that carried by
one of its two parents, the choice of which parent being made randomly
with equal probability either way.

Reframing the Moran model in network-theory language turns out to be
convenient for developing extensions, such as treatments including
structured populations, wherein mating is not uniformly random.
Furthermore, doing so broadens the range of systems to which the
mathematics can be applied: moving away from the specifically
biological terminology makes it more explicit that the Moran model can
be applied equally well to biological evolution or to social
dynamics~\cite{harmon2011}.

The components of our system will be the $N$ nodes of a network.  Each
node is a random variable which can take the values 0 and 1.  In
addition to these $N$ nodes, we augment the system with a number $N_0$
of nodes whose states are all fixed at~0, and a quantity $N_1$ of
nodes whose states are fixed to be~1.

\begin{figure}[h]
\includegraphics[width=12cm]{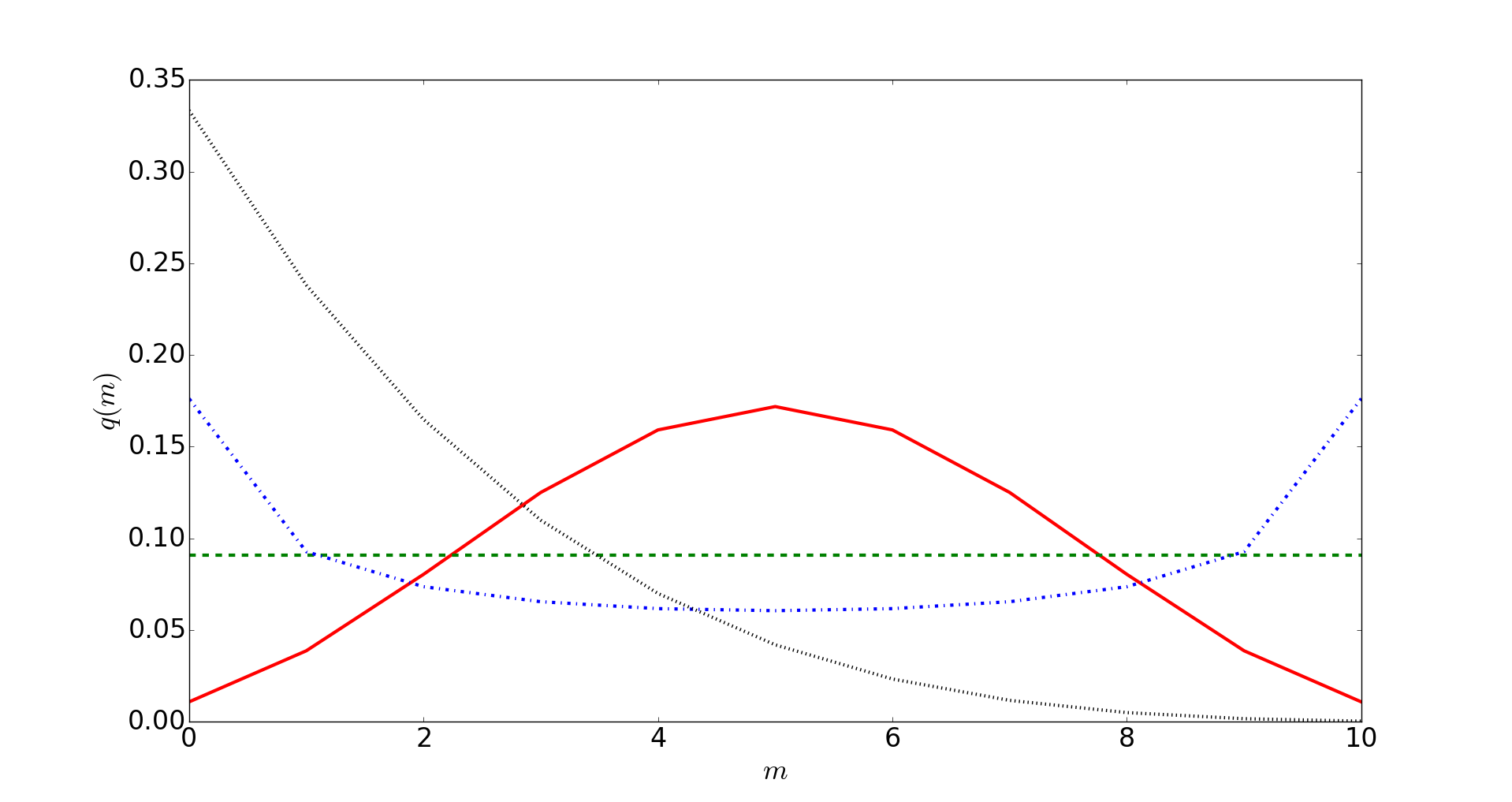}
\caption{\label{fig:network-imitation-dynamics-prob} Probability that
  exactly $m$ nodes out of 10 will be in the 1-state, for different
  values of $N_0$ and $N_1$. Red (solid): $N_0 = N_1 = 5$; green
  (dashed): $N_0 = N_1 = 1$; blue (dash-dotted): $N_0 = N_1 = 0.5$;
  black (dotted): $N_0 = 5$ and $N_1 = 1$.}
\end{figure}

At each time step, we pick one of the variable nodes at random.  We
then choose, stochastically, whether or not to change that node's
value.  With probability $p$, we keep the node value the same, and
with probability $1 - p$, we assign to it the value of another node,
chosen at random from a pool of candidates.  This pool contains both
the neighborhood of the node in the network topology and the $N_0 +
N_1$ fixed nodes.  In this way, the fixed nodes represent the
possibility of mutation: even if all the dynamical population has
allele 1, there remains the opportunity of picking up a 0, and vice
versa.  For a complete graph, the steady-state behavior of this
dynamical system can actually be found analytically~\cite{aguiar2011}.
The probability that exactly $m$ nodes out of the $N$ whose value can
vary will be in state 1 is
\begin{equation}
q(m) = A(N,N_0,N_1) \frac{\Gamma(N_1+m)\Gamma(N+N_0-m)}
                         {\Gamma(N-m+1)\Gamma(m+1)},
\label{eq:network-imitation-q}
\end{equation}
where the normalization constant $A$ is given by
\begin{equation}
A(N,N_0,N_1) = \frac{\Gamma(N+1)\Gamma(N_0+N_1)}
                    {\Gamma(N+N_0+N_1)\Gamma(N_0)\Gamma(N_1)}.
\end{equation}
We illustrate $q(m)$ for networks of $N = 10$ nodes and different
values of~$N_0$ and $N_1$ in
Figure~\ref{fig:network-imitation-dynamics-prob}.  The function $q(m)$
is an example of the \emph{beta-binomial distribution,} which is
significant in probability theory for reasons we will return to in
Chapter~\ref{ch:prob}.

Because the gamma function can take noninteger values, we can compute
the probability $q(m)$ even for nonintegral $N_0$ and $N_1$.  This is
useful if we wish to examine the low-mutation-rate limit.

If the network topology is that of a complete graph, then the system
has \emph{exchange symmetry,} an invariance under permutations which
simplifies the calculation of structure indices~\cite{allen2014}.
This simplification follows from the fact that if exchange symmetry
holds, all subsets having the same number of components can be taken
to contain the same quantity of information.  Formally, for each set
$U \subset A$, the information of $U$ is a function of the cardinality
$|U|$, which we can write as a subscript, $H(U) = H_{|U|}$.

\begin{figure}[h]
\includegraphics[width=12cm]{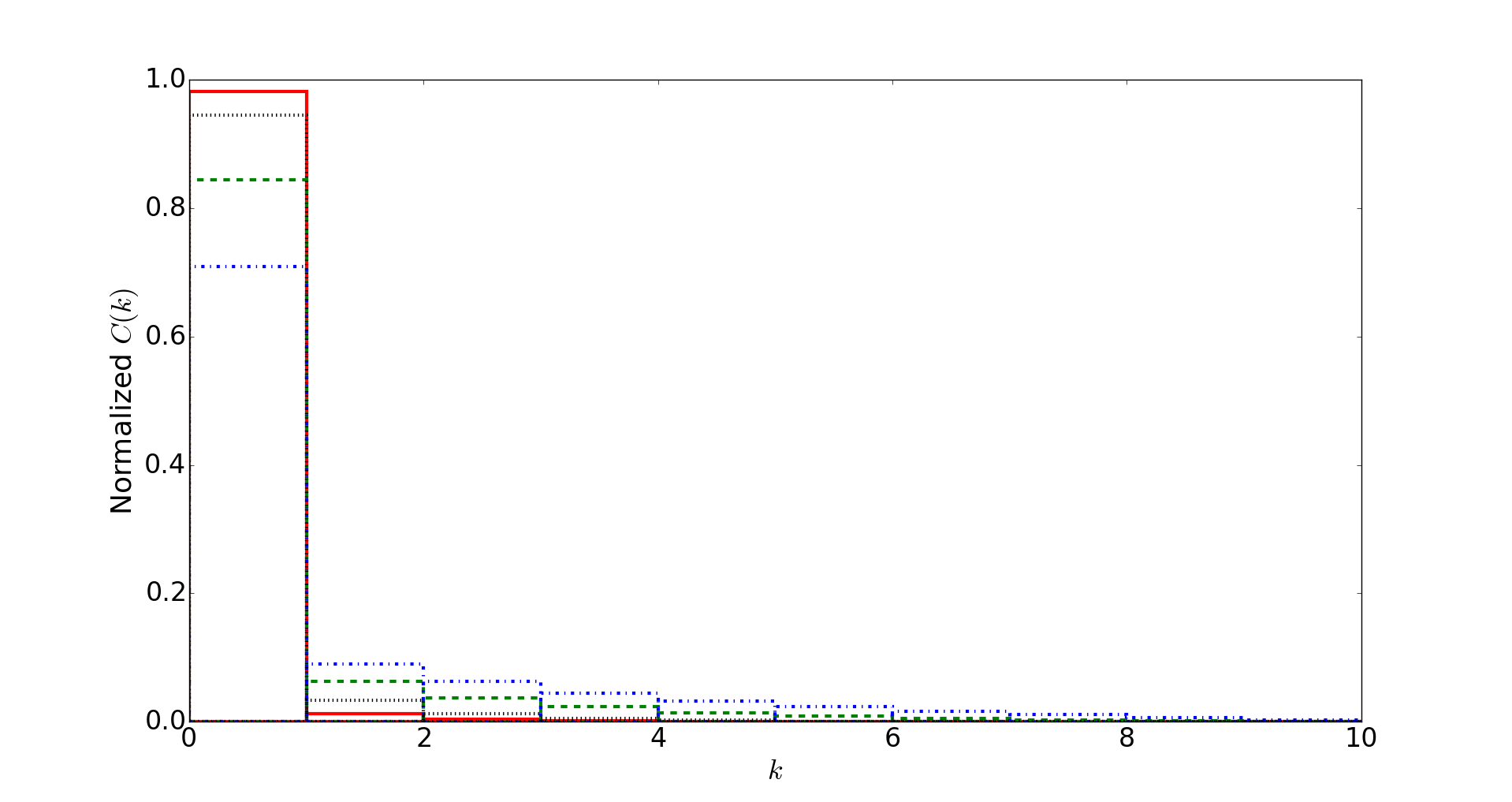}
\caption{\label{fig:network-imitation-dynamics-CP} Complexity profiles
  for the cases illustrated in
  Figure~\ref{fig:network-imitation-dynamics-prob}.  Each curve is
  normalized so that the total area under it is 1.  Elevated
  complexity $C$ at larger $k$ indicates collective behavior at larger
  scales.  Red (solid): $N_0 = N_1 = 5$; green (dashed): $N_0 = N_1 =
  1$; blue (dash-dotted): $N_0 = N_1 = 0.5$; black (dotted): $N_0 = 5$
  and $N_1 = 1$.}
\end{figure}

Recalling that the complexity profile $C(k)$ indicates the information
in dependencies of scale $k$ and higher, the information specific to
scale $k$ is
\begin{equation}
D(k) = C(k) - C(k+1).
\label{eq:D-of-k}
\end{equation}
The sum of $D(k)$ over all scales $k$ is $C(1)$.  For any fixed scale
$k$, the complexity $D(k)$ is (up to a prefactor) the binomial
transform of the sequence $a_l \equiv H_{l+N-k}$.
\begin{equation}
D(k) = \binom{N}{k} \sum_{l=0}^k (-1)^{l+1}
       \binom{k}{l} H_{l+N-k}.
\label{eq:D-of-k-binomial}
\end{equation}
We can calculate $H_n$ from the probability distribution $q(m)$, as
given by Eq.~(\ref{eq:network-imitation-q}).  Knowing $H_n$, we can
compute $D(k)$, from which we can easily find the complexity profile
$C(k)$.  Figure~\ref{fig:network-imitation-dynamics-CP} illustrates
the results.  We see that $C(k)$ depends upon the numbers of fixed
influence nodes, $N_0$ and $N_1$.  When $C(k)$ is concentrated at~$k =
1$, the nodes are changing their values almost independently of one
another.  This is the case, for example, when we set $N_0 = N_1 = 5$.
For those parameter values, the external influences are stronger than
those of the variable nodes upon each other, while being equally
balanced in both directions.  This creates a situation in which
knowing the status of any one variable node provides very little
information about the status of any other.  On the other hand, when
$C(k)$ is elevated at larger values of~$k$, then nonnegligible amounts
of information apply at higher scales.  This occurs when the external
influences are weaker than the internal dynamics, causing the variable
nodes to act collectively.

\section{Frequency-Dependent Moran Process}

One fundamental fact of evolutionary biology is that the environment
of an organism consists in large part of other organisms.  A simple,
albeit approximate, way to represent the configuration of an ecosystem
is by specifying the frequencies of abundance or population densities
for the species which are present.  (We will consider more
sophisticated approximations, and the hazards of oversimplified
representations, in later chapters.)  In this context, we can speak
of \emph{frequency-dependent fitness}: the success of an organism type
or an evolutionary strategy can be a function of the current
population densities.

The simplest kind of frequency dependence is a linear relationship
between population density and fitness.  As before, we consider two
varieties, and we keep the total population size constant, so the
frequencies of both types can be given in terms of a single variable
$x$.  We take the reproductive rates of type-0 and type-1 organisms to
be given by
\begin{align}
f_0(x) &= A_{01}x + A_{00}(1-x), \\
f_1(x) &= A_{11}x + A_{10}(1-x).
\end{align}
The coefficient $A_{ij}$ is the payoff which a type-$i$ player gains
by playing with a type-$j$ player.  Different values of the matrix $A$
represent different interactions between evolutionary strategies.

In order to apply Shannon information theory, we need a probability
distribution.  A convenient and meaningful one for these purposes is
the \emph{mutation-selection equilibrium} distribution, which
is the steady state of the frequency-dependent Moran process.  We can
find this distribution numerically by iterating the appropriate update
rule, which we can represent as multiplication by a transition
matrix.  The next step is to construct this matrix.  Having done so,
we will be able to compute $q(m)$ and thence obtain the complexity
profile, as before.  The result will typically depend both upon the
payoff matrix $A$ and on the mutation rate.

Let the total population size be $N$, and let $m$ denote the number of
type-1 individuals.  We suppose that reproduction is imperfect, with
mutations occurring at rate $u$.  That is, an offspring inherits its
parent's type with probability $1-u$, while with probability $u$, we
pick the offspring's type at random.  A nonzero mutation rate implies
that the population does not have to get stuck in a uniform
configuration: even if all individuals have the same type, an error in
reproduction can create an organism of the opposite type in the next
generation.  This is a necessary requirement for having a steady-state
probability distribution which is not concentrated entirely at~$m = 0$
or $m = N$.

To find the steady-state probability distribution for~$m$, we first
need to calculate the probabilities that $m$ will increase or
decrease.  In the frequency-dependent Moran process~\cite{allen2012}, the
probability that $m$ will decrease by~1 is
\begin{equation}
p_{m\to m-1} = \frac{m}{N}
 \left((1-u)\frac{(N-m)f_0\left(\frac{m}{N}\right)}
                 {mf_1\left(\frac{m}{N}\right)
                  + (N-m)f_0\left(\frac{m}{N}\right)}
        + \frac{u}{2}
 \right).
\end{equation}
And the probability of $m$ increasing by 1 is
\begin{equation}
p_{m\to m-1} = \frac{N-m}{N}
 \left((1-u)\frac{mf_1\left(\frac{m}{N}\right)}
                 {mf_1\left(\frac{m}{N}\right)
                  + (N-m)f_0\left(\frac{m}{N}\right)}
        + \frac{u}{2}
 \right).
\end{equation}
With these equations, we can find the steady-state probability
distribution $q(m)$, which will depend on the payoff matrix $A$ and
the mutation rate $u$.  (For the present purposes, a numerical
computation will suffice.)  Knowing $q(m)$, we can as before find the
complexity profile $C(k)$.  The resulting curve tells us about the
scales of organization which arise within the population as a
consequence of the evolutionary game dynamics.

To connect with the literature~\cite{allen2012}, we carry out this
calculation for the payoff matrix
\begin{equation}
A = \left(\begin{array}{cc}
          6 & 4 \\
          7 & 5
    \end{array}\right),
\label{eq:ch-ms-pd}
\end{equation}
which defines an instance of the \emph{Prisoner's Dilemma.}  One
application of this to biology is the case of the bower birds
mentioned earlier~\cite{pruettjones1994, gonick1994}.  Simplifying
somewhat, a male bower bird has two strategies available to it: to
guard its own bower, or to maraud and attempt to damage others.
Designate guarding as strategy 0 and marauding as strategy 1.  The
matrix element $A_{ij}$ denotes the payoff to a bird following
strategy $i$ against an opponent who plays strategy $j$.  In this
example, a guardian (row 0) who plays against a marauder (column 1)
obtains a score of~4.  The highest payoff is $A_{10}$, the score
obtained by a marauder who plays against a guardian.  In fact, it is
better to maraud than to guard, when facing either kind of foe:
\begin{equation}
A_{10} > A_{00}, \hbox{ and also } A_{11} > A_{01}.
\end{equation}
So far, it looks like the thing to do is to maraud.  However, the
payoff obtained when both birds follow this strategy is $A_{11}$,
which is \emph{less} than the payoff $A_{00}$ they would have obtained
if they had both stayed home.

The particular choice of numbers here is arbitrary, but the
relationships between the numbers are representative of typical
conditions in the wild.  As Gonick~\cite{gonick1994} summarizes,
``Seemingly forced by the game's logic into a hostile strategy, they
end up worse off than if they had only cooperated!''  A wide variety
of biological scenarios can be considered as examples of this
game~\cite{turner1999, grieg2004}.  A primary concern is to identify
the conditions under which cooperation (for example, both bower birds
guarding rather than marauding) is evolutionarily favorable.

This type of situation is designated a ``Prisoner's Dilemma'' because
it is usually introduced with an example of two people apprehended for
a crime and interrogated by the police.  Each player can choose to say
nothing, or to inform on the other player.  The payoff matrix is such
that it is better to inform than to stay silent, whatever option the
other player takes; however, if both players keep quiet, they fare
better than if they both inform on each other.

\begin{figure}[h]
\includegraphics[width=12cm]{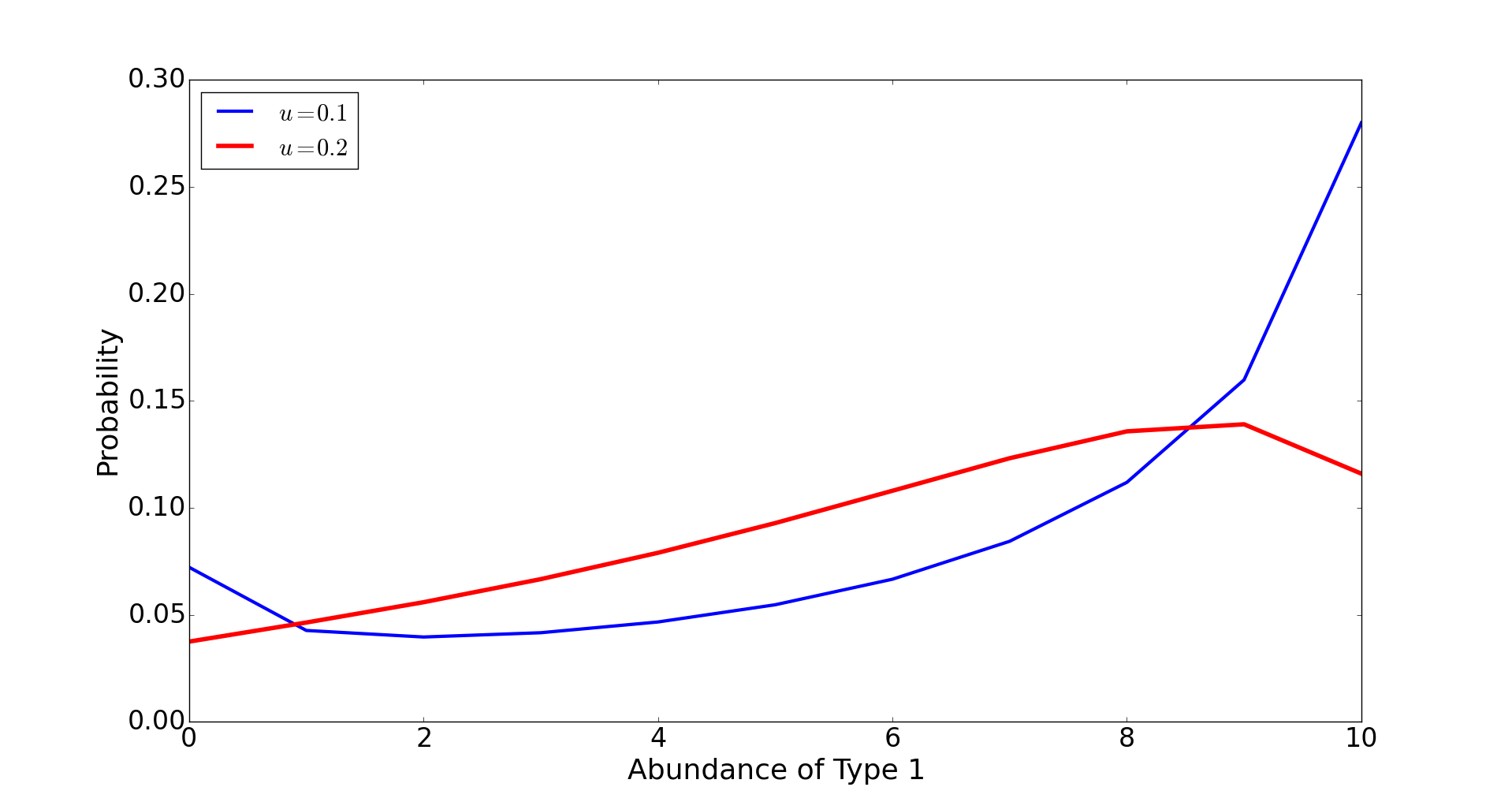}
\caption{\label{fig:moran-process-prob} Equilibrium probability that
  exactly $m$ agents out of 10 will be in the 1-state (marauding), for
  the Prisoner's Dilemma game defined by Eq.~(\ref{eq:ch-ms-pd}).
  Blue (thinner line): $u = 0.1$; red (thicker line): $u = 0.2$.}
\end{figure}

\begin{figure}[h]
\includegraphics[width=12cm]{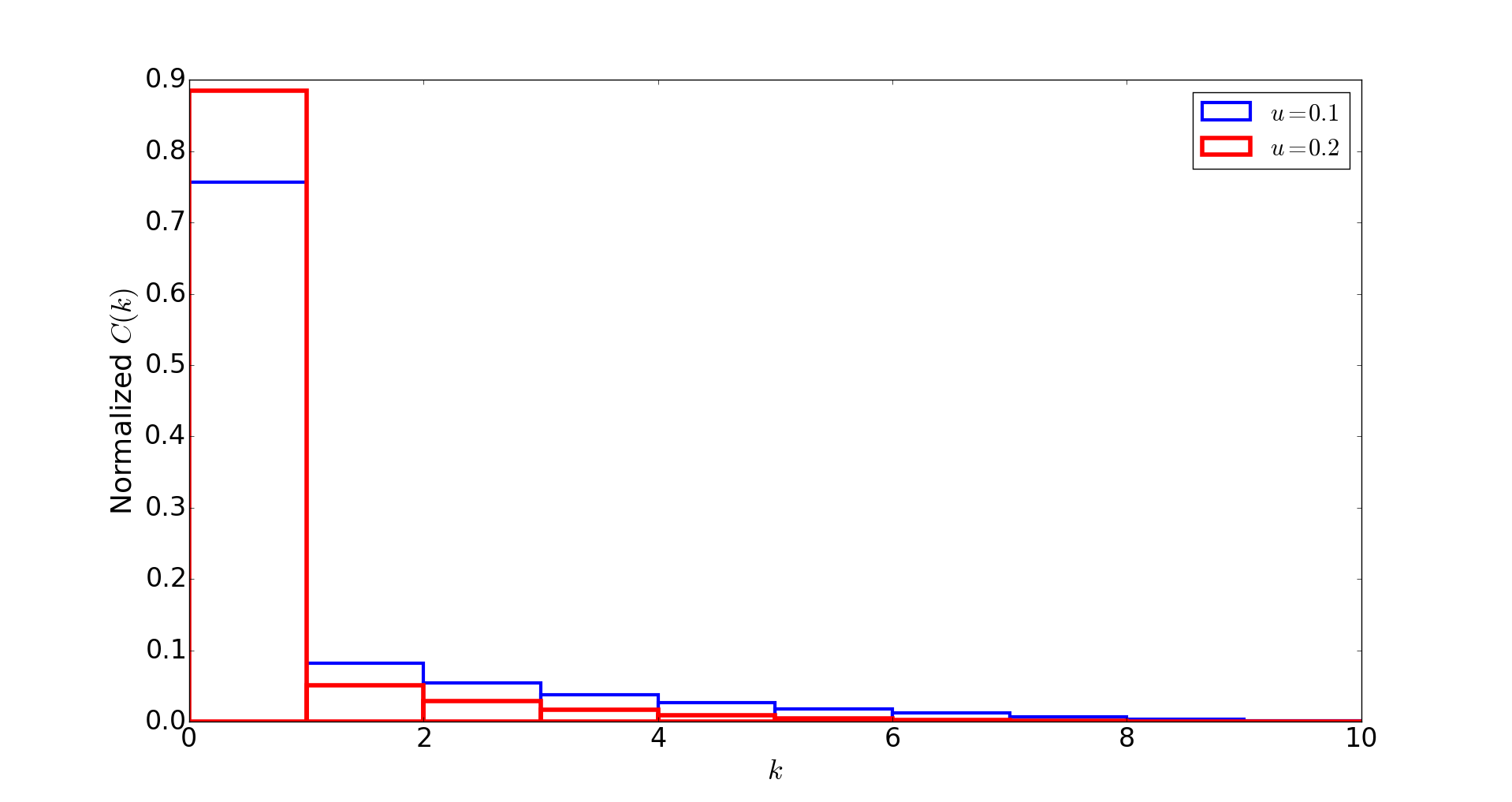}
\caption{\label{fig:moran-process-CP} Complexity profiles
  for the cases illustrated in Figure~\ref{fig:moran-process-prob}.
  Each curve is normalized so that the total area under it is 1.
  Elevated complexity $C$ at larger $k$ indicates collective behavior
  at larger scales.  Blue (thinner line): $u = 0.1$; red (thicker
  line): $u = 0.2$.}
\end{figure}

Figure~\ref{fig:moran-process-prob} shows the probability distribution
for the Moran process in mutation-selection equilibrium with this
payoff matrix, given two different mutation rates.  Note that the
effect of varying the mutation rate is quite dramatic.  As before, we
can compute the complexity profile, which we plot in
Figure~\ref{fig:moran-process-CP}.

\section{Multiscale Challenges and Evolution}

In the previous section, we considered the scales of organization
which can arise as an evolutionary process develops stochastically.
We can also apply our mathematical formalism of multiscale structure
to other aspects of evolutionary theory.  The following is excerpted
from an article by Allen, Bar-\!Yam and myself~\cite{allen2014}.

\begin{quotation}
The discipline of cybernetics, an ancestor to modern control theory,
used Shannon's information theory to quantify the difficulty of
performing tasks, a topic of relevance both to organismal survival in
biology and to system regulation in engineering.  Cyberneticist
W.\ Ross Ashby considered scenarios in which a regulator device must
protect some important entity from the outside environment and its
disruptive influences~\cite{ashby1956}.  In Ashby's examples, each
state of the environment must be matched by a state of the regulatory
system in order for it to be able to counter the environment's
influence on a protected component.  Successful regulation implies
that if one knows only the state of the protected component, one
cannot deduce the environmental influences; \emph{i.e.,} the job of
the regulator is to minimize mutual information between the protected
component and the environment.  This is an information-theoretic
statement of the idea of homeostasis.  Ashby's ``Law of Requisite
Variety'' states that the regulator's effectiveness is limited by its
own information content, or \emph{variety} in cybernetic terminology.
An insufficiently flexible regulator will not be able to cope with the
environmental variability.  A multiscale extension of Shannon
information theory provides a multiscale cybernetics, with which we
can study the scenarios in which ``that which we wish to protect'' and
``that which we must guard against'' are each systems of many
components, as are the tools we employ for regulation and
control~\cite{baryam2004a, baryam2004b, baryam2004c}.

Multiscale information theory enables us to overcome a key limitation
of the requisite variety concept.  In the examples of traditional
cybernetics~\cite{ashby1956}, each action of the environment requires
a specific, unique reaction on the part of the regulator.  This
neglects the fact that the impact which an event in the environment
has on the system depends upon the \emph{scale} of the environmental
degrees of freedom involved.  There is a great difference between
large-scale and fine-scale impacts.  Systems can deflect fine-scale
impacts without needing to specifically respond to them, while they
need to respond to large-scale ones or perish.  For example, a human
being can be indifferent to the impact of a falling raindrop, whereas
the impact of a falling rock is much more difficult to neglect, even
if specifying the state of the raindrop and the state of the rock
require the same amount of information.  An extreme case is the impact
of a molecule: air molecules are continually colliding with us, yet
the only effects we have to cope with actively are the large-scale,
collective behaviors like high-speed winds.  Ashby's Law does not make
this distinction.  Indeed, there is no framework for the discussion
due to the absence of a concept of scale in the information theory he
used: Each state is equally different from every other state and
actions must be made differently for each different environment.

Thus, in order to account for the real-world conditions, a multiscale
generalization of Ashby's Law is needed. According to such a Law, the
responses of the system must occur at a scale appropriate to the
environmental change, with larger-scale environmental changes being
met by larger-scale responses.  As with the case of raindrops
colliding with a surface, large-scale structures of a system can avoid
responding dynamically to small-scale environmental changes which
cause only small-scale fluctuations in the system.

Given a need to respond to larger-scale changes of the environment,
coarser-scale descriptions of that environment may suffice.  A
regulator that can marshall a \emph{large-scale response} can use a
coarse-grained description of the environment to counteract
large-scale fluctuations in the external conditions.  In this way,
limited amounts of information can still be useful.  To make requisite
variety a practical principle, one must recognize that information
applies to specific scales.

Ashby aimed to apply the requisite variety concept to biological
systems, as well as technological ones.  An organism which lacks the
flexibility to cope with variations in its environment dies.  Thus, a
mismatch in variety/complexity is costly in the struggle for survival,
and so we expect that natural selection will lead to organisms whose
complexity matches that of their environment.  However, ``the
environment'' of a living being includes other organisms, both of the
same species and of others.  Organisms can act and react in concert
with their conspecifics, and the effect of any action taken can depend
on what other organisms are doing at the same time~\cite{allen2014b}.
In some species, such as social insects~\cite{tschinkel2014}, distinct
scales of the individual, colony and species are key features
characterizing collective action.  This suggests a multiscale
cybernetics approach to the evolution of social behavior: We expect
that scales of organization within a population---the scales, for
example, of groups or colonies---will evolve to match the scales of
the challenges which the environment presents.  Furthermore, the
concept of multiscale response applies within the individual organism
as well.  Multiple scales of environmental challenges are met by
different scales of system responses.  To protect against infection,
for example, organisms have physical barriers (\emph{e.g.,} skin),
generic physiological responses (\emph{e.g.,} clotting, inflammation)
and highly specific adaptive immune responses, involving interactions
among many cell types, evolved to identify pathogens at the molecular
level. The evolution of immune systems is the evolution of separate
large- and small-scale countermeasures to threats, enabled by
biological mechanisms for information transmission and
preservation~\cite{stacey2008}.  As another example, the muscular
system includes both large and small muscles, comprising different
numbers of cells, corresponding to different scales of environmental
challenge (\emph{e.g.,} pursuing prey and escaping from predators
versus chewing food)~\cite{bar2003complexity}.
\end{quotation}

In Chapter~\ref{ch:volunteer}, we will use evolutionary game theory to
understand one example of multiscale requisite variety: a scenario in
which reproductive success depends on multiple organisms acting in
concert.

\chapter{Host--Consumer Evolution by Simulation}

\label{ch:redgreen}
\section{Introduction}
\label{sec:intro}
Spatial extent is a complicating factor in mathematical biology.  The
possibility that an action at point A cannot immediately affect what
happens at point B creates the opportunity for spatial nonuniformity.
This nonuniformity must change our understanding of evolutionary
dynamics, as the same organism in different places can have different
expected evolutionary outcomes. Since organism origins and fates are
both determined locally, we must consider heterogeneity explicitly to
determine its effects. We use simulations of spatially extended
host--pathogen and predator--prey ecosystems to reveal the limitations
of standard mathematical treatments of spatial heterogeneity. Our
model ecosystem generates heterogeneity dynamically; an adaptive
network of hosts on which pathogens are transmitted arises as an
emergent phenomenon. The structure and dynamics of this network differ
in significant ways from those of related models studied in the
adaptive-network field. We use a new technique, organism swapping, to
test the efficacy of both simple approximations and more elaborate
moment-closure methods, and a new measure to reveal the timescale
dependence of invasive-strain behavior.  Our results demonstrate the
failure not only of the most straightforward (``mean field'')
approximation, which smooths over heterogeneity entirely, but also of
the standard correction (``pair approximation'') to the mean field
treatment. In spatial contexts, invasive pathogen varieties can
prosper initially but perish in the medium term, implying that the
concepts of reproductive fitness and the Evolutionary Stable Strategy
have to be modified for such systems.

Mathematical modeling of biological systems involves a tradeoff
between detail and tractability.  Here, we consider evolutionary
ecological systems with spatial extent---a complicating factor.
Analytical treatments of spatial systems typically treat as equivalent
all configurations with the same overall population density, the same
allele frequencies, the same pairwise contact probabilities or the
like.  For ease of analysis, one seeks a simplified analytical model,
which {\em coarse-grains} ``microstates'' (the complete specification
of each organism) to ``macrostates'' (characterized by quantities like
average densities), allowing one to make useful predictions about the
model's behavior~\cite{levin1992,dieckmann2000}.  Corrections to
simple coarse-grainings can quickly generate an overbearing quantity
of algebra.  It is fairly well appreciated that the simplest
approximations break down in the spatial context.  What is less
acknowledged and not yet systematically understood is that the
extensions of the simpler approximations also fail.  Before exhausting
ourselves with ever-more-elaborate refinements, it would be useful to
have some understanding of when a particular series of approximations
is doomed to inadequacy.

In this chapter, we study the context in which commonly-used
coarse-grainings can be expected to fail at capturing the evolutionary
dynamics of an ecosystem, and in addition we provide a novel, direct
demonstration of that failure.  The fundamental issue is {\em spatial
  heterogeneity,} a long-recognized concern for mathematical
biology~\cite{wright1945, hartl2007}.  When does spatial heterogeneity
significantly impact the choice of appropriate mathematical treatment,
and when does a chosen mathematical formalism not capture the full
implications of spatial variability?  We show that one can test a
treatment of heterogeneity by transplanting organisms within a
simulated ecosystem in such a way that, were the treatment valid, the
modeled behavior of the ecosystem over time would remain essentially
unchanged.  We demonstrate situations where the system's behavior
changes dramatically and cannot be captured by a conventional
treatment.  The complications we explore imply that {\em short-term
  descriptions} of what is happening in an evolutionary ecological
model can be insufficient and, in fact, misleading, with regard not
just to quantitative details but also to qualitative characteristics
of ecological dynamics.

Many modeling approaches in mathematical biology which appear distinct
at first glance turn out to be describing the same phenomenon with
different equations~\cite{page2002, bijma2008, damore2011}.  What matters
for our purposes is not so much which technique is chosen, but whether
the underlying assumptions do, in fact, apply.

``Mean-field theory'' is a term from statistical
physics~\cite{baryam1999, kardar2007} which has been adopted in
ecology~\cite{goodnight2008, lion2008, givan2011}, referring to an
approximation in which each component of a system is modeled as
experiencing the same environment as any other.  This implies that the
probability distribution over all possible states of the system
factors into a product of probability distributions for individual
components.  An example in population genetics is the assumption that
a population is panmictic.  That is, if a new individual in one
generation has an equal chance of receiving an allele from any
individual in the previous generation, then we can approximate the
ecosystem dynamics using only the proportion of that allele, rather
than some more complicated representation of the population's genetic
makeup.  Modeling evolution of that population as ``change in allele
frequencies over time'' (per, {\em e.g.,}~\cite{williams1992,
  page2002}) is, implicitly, a mean-field
approximation~\cite{sayama2000}.  The mean-field approximation is also
in force if one postulates that an individual organism interacts with
some subset, chosen at random, of the total population, even if the
form and effect of interactions within that subset are complicated (as
in, {\em e.g.,}~\cite{vandyken2011, archetti2011}).

It is well known that real species are not necessarily panmictic.
However, many treatments which acknowledge this are still mean-field
models.  The textbook way of incorporating geographical distance into
a population-genetic model is to divide the system into $N$ local
subpopulations, ``islands,'' connected via migration~\cite{taylor1996,
  kokko2008, wild2009}.  Within each subpopulation, distance is
treated as negligible, and organisms are well mixed~\cite{levins1969,
  hartl2007}.  This approach makes a simplifying assumption that there
is a single distance scale below which panmixia
prevails~\cite{platt2010}, and it relies on well-defined boundaries
between panmictic subpopulations which persist over
time~\cite{levins1969}.  Furthermore, the connections among
subpopulations are frequently taken to have the topology of a complete
graph, {\em i.e.,} an organism in one subpopulation can migrate to any
other with equal ease~\cite{levins1969, hartl2007, kokko2008,
  wild2009}.  In this case, each of the $N$ subpopulations do
experience the same environment, to within one part in $N$.  Thus, the
mean-field approximation is in force at the island level, and the
island model incorporates spatial extent without incorporating a full
treatment of spatial heterogeneity.  For real
ecosystems~\cite{halley2004, scanlon2007, platt2010, reigada2012}, one
or more of these simplifying assumptions can fail.  Long-distance
migration is often thought to return a spatial ecosystem to a
well-mixed form, but if organisms' migration habits are themselves
adaptive, this is not necessarily so~\cite{ichinose2013}.  More
complicated population structures require more sophisticated
mathematical treatments of evolution, a fact which has mathematical
consequences, but more importantly has real-world implications for
practical issues like the evolution of drug-resistant
diseases~\cite{escalante2009}.

Where mean field approximations fail, ``higher order'' approximations
may be employed.  Rather than individual organisms or islands, a pair
approximation considers pairs of organisms or pairs of spatial regions
in average contexts.  However, this approximation can also fail when
local contexts of groups do not reflect the overall system behavior
due to heterogeneity across larger domains.  Patches of distinct
genetic composition in different parts of a spatial system that are
well separated cannot be treated correctly by such approximations.
Quantitative analyses confirm this inadequacy.  We introduce a new
approach to analyzing such approximations by swapping pairs of
organisms in a way that preserves the pair description. For spatial
systems, such swapping events violate the spatial separation between
patches and changes the evolutionary behavior of the system. The
swapping method therefore serves as a direct test of the (in)adequacy
of the pair approximation.  For evolution on random networks of sites
that do not embody large spatial distances, the pair approximation can
work and the swapping test does not change measures of evolutionary
dynamics.  However, such networks do not capture important properties
of spatial heterogeneity.

As one of the key properties of spatial extent is the propagation of
organisms from one part of the space to the other over long distances,
we show that important insights can be gained by considering models of
percolation.  Percolation describes the physical propagation of,
\emph{e.g.,} fluids through a random medium.  In certain limits the
evolutionary behavior of spatial systems can be mapped onto
percolation behavior, demonstrating that investigations of such
systems which go beyond mean-field or scaling studies are relevant to
evolutionary dynamics.  This and other advances that go beyond the
mean field are necessary to fully describe spatial evolutionary
dynamics as they are necessary for the description of many physical
systems of spatial extent.  The complexities of spatially extended
evolutionary dynamical systems beyond the prototypical problem of
percolation create new demands and opportunities for advancing our
insight into the dynamics of heterogenous systems and their
implications for evolution.

\section{Model and Methods}
\label{sec:methods}
We make the issue of spatial heterogeneity concrete by focusing on a
specific model of ecological and evolutionary interest.  We take a
model of hosts and consumers interacting on a 2D spatial lattice.
Each lattice site can be empty (0), occupied by a host ($H$) or
occupied by a consumer ($C$).  We use the term {\em consumer} as a
general label to encompass parasites, pathogens and predators.  Where
convenient for examples, we will specialize to one or another of these
terminologies.  Hosts reproduce into adjacent empty sites with some
probability $g$ per site, taken as a constant for all hosts.
Consumers reproduce into adjacent sites occupied by hosts, with
probability $\tau$ per host; sometimes $\tau$ is fixed for all
consumers, but we also consider cases in which it is a mutable
parameter passed from parent to offspring.  We will refer to~$\tau$ as
the \emph{transmissibility.}  Hosts do not die of natural causes,
while consumers perish with probability $v$ per unit time (leaving
empty sites behind).  Because consumers can only reproduce into sites
where hosts live, the effective graph topology of reproductively
available sites experienced by the consumers is constantly changing
due to their very presence.  This makes the ecosystem an
\emph{adaptive network,} a system in which the dynamics {\em of\,} a
network and the dynamics {\em on} that network can occur at comparable
timescales and reciprocally affect one another~\cite{gross2008,
gross2009, graeser2011, demirel2012, sayama2015}.  In this model,
dynamics can be highly complex, including spatial cascades of host and
consumer reproduction.  Even when a quasi-steady-state behavior
emerges, as we shall see, it is a consequence of fluctuations over
extended space and time intervals.

Several different types of biological interactions can be treated by
this modeling framework.  Hosts could represent regions inhabited by
autotrophs alone, while consumers represent regions containing a
mixture of autotrophs and the heterotrophs which predate upon
them~\cite{werfel2004}.  Alternatively, host agents could represent
healthy organisms, while consumers represent organisms infected with a
parasite or pathogen.  Thus, host--consumer models are closely related
to Susceptible--Infected--Recovered (SIR) models, which are
epidemiological models used to understand the spread of a disease
through a population.  SIR models describe scenarios in which each
individual in a network is either susceptible (S) to a pathogen,
infected (I) with it, or recovered (R) from it; susceptible nodes can
catch the disease from infected neighbors, becoming infected
themselves, while nodes which have become infected can recover from
the disease and are then resistant against further
infection. Susceptible, infected and recovered individuals roughly
correspond to hosts, consumers, and empty cells, respectively.  An
important difference between host--consumer models and epidemiological
models concerns the issue of {\em reinfection.}  In the host--consumer
model, an empty site left behind by a dead consumer can be reoccupied
by another consumer, but only if a host reproduces into it first.
Other research has considered models where R[ecovered] individuals can
also become I[nfected], with a different (typically lower) probability
than S[usceptible] ones, thereby incorporating imperfect immunity into
the model~\cite{jimenez2003,henkel2008}.  The degree of immunity is
independent of geography and the environment of the R[ecovered]
individual, unlike reoccupation in the host--consumer model.  Another
application is illustrated by the Amazon molly, {\em Poecilia
  formosa,} which is a parthenogenetic species: {\em P.~formosa,} all
of which are female, reproduce asexually but require the presence of
sperm to carry out egg development.  (This kind of sperm-dependent
parthenogenesis is also known as gynogenesis.)  {\em P.~formosa} are
thus dependent on males of other species in the same genus---usually
{\em P.~mexicana} or {\em P.~latipinna}---for reproduction.  Because
{\em P.~formosa} do not incur the cost of sex, they can outcompete the
species on which they rely, thereby possibly depleting the resource
they require for survival, {\em i.e.,} male fish~\cite{kokko2008,
  kokko2011}.  Thus, hosts could be regions containing sexual
organisms, with consumers standing for areas containing both sexual
and asexual individuals~\cite{kokko2008}.

\begin{figure}[h]
\includegraphics[width=4cm]{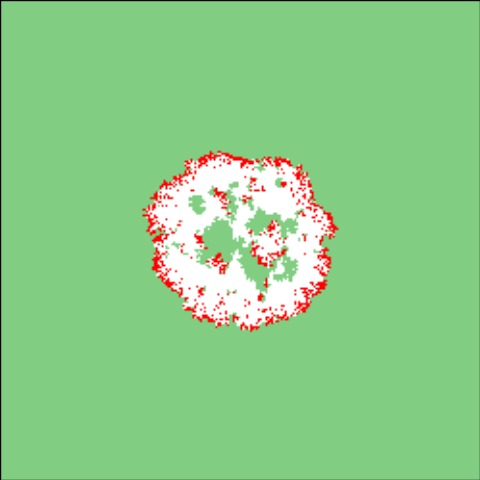}
\includegraphics[width=4cm]{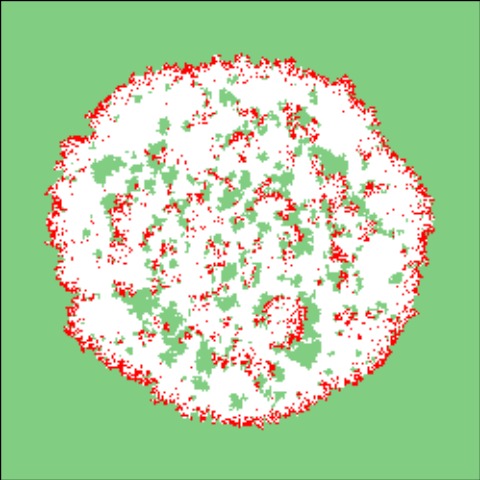}

\includegraphics[width=4cm]{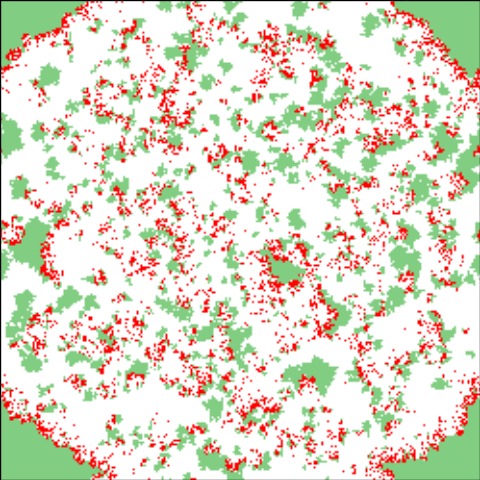}
\includegraphics[width=4cm]{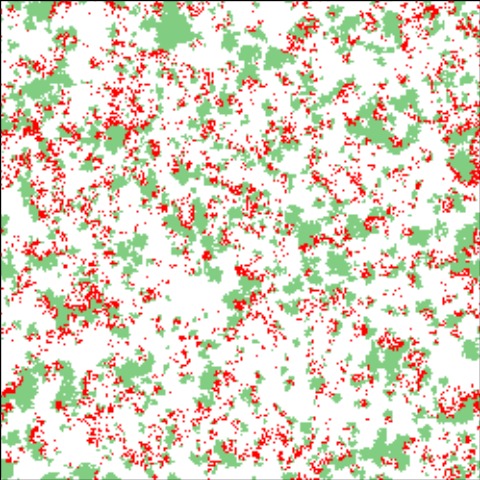}
\caption{\label{fig:animation} Snapshots of a simulated host--consumer
  ecosystem on a $250\times 250$ lattice, taken at intervals of 100
  generations.  Consumers are dark gray (red online), hosts are light
  gray (green online) and empty space is left white.  The simulation
  began with a single consumer at the center of the lattice, which
  gave rise to an expanding front of consumers.  The first image in
  this sequence shows the state of the ecosystem 100 generations into
  the simulation.  Hosts which survive the consumer wave recolonize
  the empty sites, leading to pattern formation.  Here, the host
  growth rate is $g = 0.1$, the consumer death rate is $v = 0.2$ and
  the consumer transmissibility is fixed at $\tau = 0.33$.}
\end{figure}

This host--consumer model displays waves of colonization, consumption
and repopulation.  Hosts reproduce into empty sites, and waves of
consumers follow, creating new empty regions open for host
colonization.  Therefore, clusters of hosts arise
dynamically~\cite{haraguchi2000, sayama2002, aguiar2003, sayama2003},
a type of pattern formation which can separate regions of the
resources available to pathogens into patches without the need for
such separation to be inserted manually.  Figure~\ref{fig:animation}
illustrates a typical example of this effect.  This is a specific
example of the general phenomenon of pattern formation in
nonequilibrium systems~\cite{sayama2000}.  Consumers are {\em
ecosystem engineers}~\cite{jones1994, strayer2006, post2009,
pringle2010, allen2013, callahan2014} which shape their local
environment: an excessively voracious lineage of consumers can deplete
the available resources in its vicinity, causing that lineage to
suffer a Malthusian catastrophe~\cite{haraguchi2000, cronin2005,
kerr2006, kokko2008, messinger2009, lion2010, messinger2012,
reigada2012}.  Because the ecology is spatially extended, this
catastrophe is a local niche annihilation, rather than a global
collapse~\cite{rauch2006}.  A mutant strain with a high
transmissibility can successfully invade in the short term but suffer
resource depletion in the medium term, meaning that in a population
where consumer transmissibilities evolve, averages taken over long
numbers of generations yield a moderate value~\cite{goodnight2008,
heilmann2010}.  This implies that an empirical payoff matrix or
reproduction ratio will exhibit nontrivial timescale
dependence~\cite{rauch2002, rauch2003, goodnight2008}.

This model is distinct from another approach to studying evolutionary
dynamics in spatial contexts, that of evolutionary game theory.
Game-theoretic models of spatially structured populations have been
explored at great length.  These investigations have found that
breakdowns of mean-field approximations are commonplace.  However,
evolutionary game theory has its own simplifying assumptions.  The
vast majority of studies consider only two-player games.  Population
size is usually taken to be constant, and population structure is
typically fixed in place.  In game-theoretic models, the benefits and
costs of different organism behavioral traits are parameters whose
values are chosen by the modeler.  By contrast, ``benefits'' and
``costs'' in host--consumer models are emergent properties which
depend on interactions over many generations.  Population size is not
fixed, and population structure is dynamical: the environment in which
different consumer varieties compete changes stochastically, in ways
affected by their presence.

\section{Results}

\label{sec:results}
\subsection{Evolution of Transmissibility}
\begin{figure}[h]
\includegraphics[width=6.5cm]{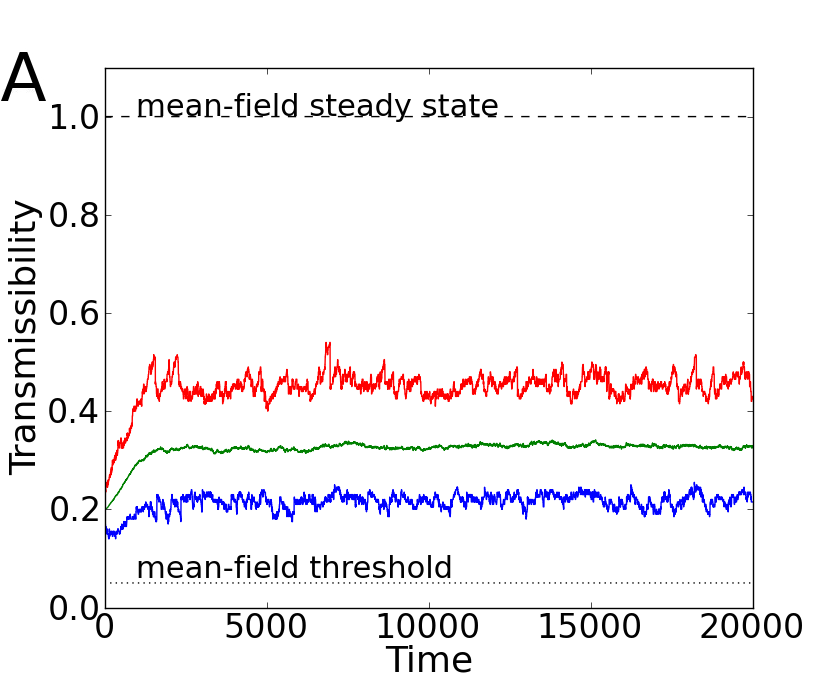}
\includegraphics[width=7cm]{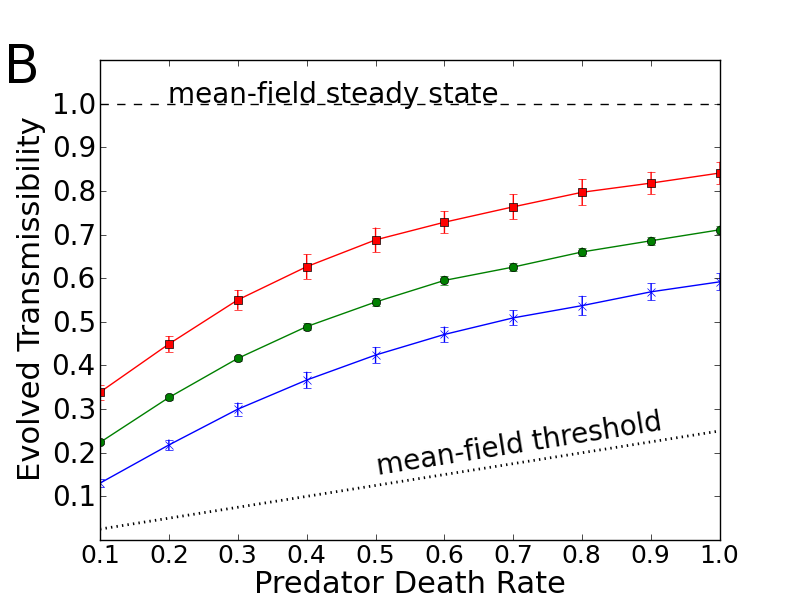}
\caption{\label{fig:minmax-6apr2011} {\bf (A)} Minimum (blue), average
  (green) and maximum (red) transmissibility $\tau$ for a consumer
  population over time, with~$g = 0.1$ and $v = 0.2$.  (The mutation
  rate is $\mu = 0.255$ and the step size is $\Delta\tau = 0.005$, as
  was used in reference~\cite{werfel2004}.)  The average $\tau$ tends
  to a quasi-steady-state value dependent on~$g$ and $v$; if the
  simulation is started with $\tau$ set to below this value, the
  average $\tau$ will increase, and likewise, the average $\tau$ will
  decrease if the consumer population is initialized with $\tau$ over
  the quasi-steady-state value~\cite{werfel2004}.  The horizontal
  dotted line indicates the threshold value of~$\tau$ which, in a
  mean-field model, is the smallest value at which a consumer
  population can sustain its numbers.  The dashed line indicates the
  value to which $\tau$ would trend in a well-mixed ecosystem.  {\bf
    (B)} Minimum, average and maximum $\tau$ as a function of~$v$,
  with~$g = 0.1$.  The dotted line shows the minimum sustainable
  $\tau$ as predicted by mean-field approximation.  Each point is
  found by averaging over 15,000 timesteps.  Error bars indicate one
  standard deviation.}
\end{figure}

We investigate evolution in the spatial host--consumer ecosystem
through simulation and analytic discussion.  If the transmissibility
$\tau$ is made a heritable trait, passed from a consumer to its
offspring with some chance of mutation, what effect will natural
selection have on the consumer population?
Figure~\ref{fig:minmax-6apr2011}(A) shows the average, minimum and
maximum values of the transmissibility $\tau$ observed in a population
over time.  The average $\tau$ tends to a quasi-steady-state value
dependent on the host growth rate $g$ and the consumer death rate $v$;
if the simulation is started with $\tau$ set to below this value, the
average $\tau$ will increase, and likewise, the average $\tau$ will
decrease if the consumer population is initialized with $\tau$ over
the quasi-steady-state value.  Even when the average $\tau$ has
achieved its quasi-steady-state value, the population displays a wide
spread of transmissibilities whose extremes fluctuate over
time~\cite{werfel2004}.

In a well-mixed ecosystem, the average $\tau$ of the population will
tend to 1, maximizing the reproductive rate of the individual
consumer.  This occurs because each consumer on average experiences
the same environment as any other, and thus has the same number of
hosts available to reproduce into.  A consumer with a higher $\tau$ has
a higher reproduction rate and therefore evolutionary dominance up to
the highest possible value, 1.  The observation of a
quasi-steady-state value below 1 is an important result.  This is the
first breakdown of the mean-field approximation, and it indicates the
inapplicability of traditional assumptions about fitness optimization,
with implications for the origins of reproductive restraint,
communication-based altruism and social behaviors in
general~\cite{rauch2002, rauch2003, werfel2004, aguiar2004, rauch2006,
  goodnight2008}.

One can avoid $\tau$ tending to 1 in a panmictic system by imposing
some extra constraint, such as a tradeoff between transmissibility and
lethality, where higher transmissibility becomes impossible due to
lethality that prevents transmission.  This tradeoff between
infectiousness and lethality can be considered as a within-host
version of resource overexploitation that here occurs at the
population level.  Such within-host tradeoffs are difficult to
establish empirically in living populations~\cite{froissart2010,
  asplen2012}.  Often, one lacks pertinent information, such as the
functional relationship between pathogen load and disease transmission
probability, or the extent to which empirical proxies for pathogen
load predict actual host mortality~\cite{hawley2013}.  An empirical
observation of low virulence should not by itself be taken as evidence
that a tradeoff exists: it may well be that another condition, such as
panmixia, fails to obtain.  The behavior of spatial models makes clear
that the relevant scale of the limiting factor is not necessarily
within the individual host.

Another difference between spatial and nonspatial host--consumer
systems is the rate at which consumers must reproduce in order to
sustain their population.  One can calculate the minimum sustainable
value of~$\tau$ in the mean-field approximation~\cite{mobilia2006b} by
balancing the birth and death rates.  If the host population is small
compared to the total ecosystem size, then the minimum sustainable
$\tau$ is the value which satisfies $k\tau = v$, where $k$ is the
number of neighbors adjacent to a site.  For the parameters used in
Figure~\ref{fig:minmax-6apr2011}(A), this value would be 0.05, which
is substantially smaller---by a factor of~4---than the lowest $\tau$
seen in the evolving spatial population.  Consumer populations
with~$\tau$ at the mean-field threshold are not sustainable in the
spatial case.  This is easily verified by numerical simulations or by
using the mean-field equations for the host--consumer
dynamics~\cite{aguiar2003b, aguiar2004, aguiar2003b-errata}.
Stochastic fluctuations suppress the active phase, \emph{i.e.,} the
range of parameter values which permit a living consumer population is
reduced~\cite{mobilia2006b}.

\begin{figure}[h]
\includegraphics[width=12cm]{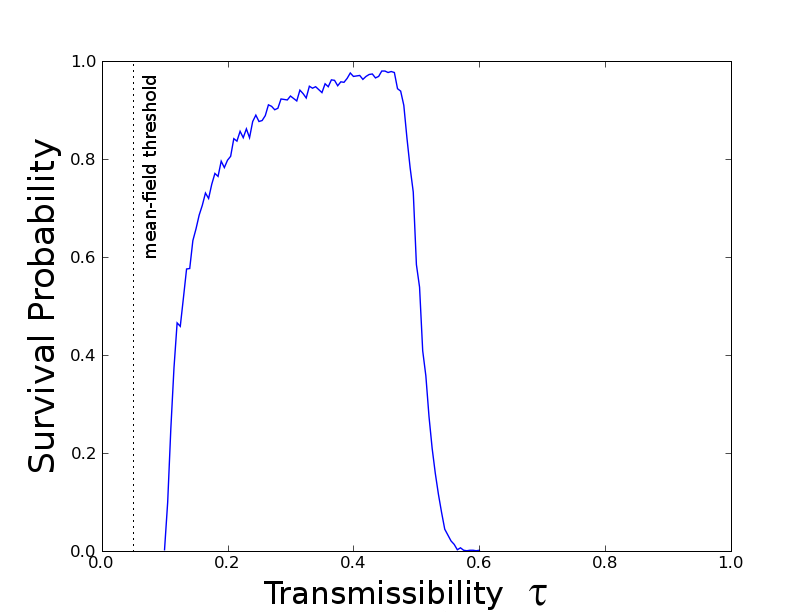}
\caption{\label{fig:survival-probability-v0p2-g0p1} Probability
  of a consumer strain surviving 2000 generations after injection at a
  single point in a $250\times 250$ lattice filled with hosts.
  (Computed with 1000 runs per point.  Host reproduction rate $g =
  0.1$, consumer death rate $v = 0.2$.)  Vertical dotted line shows
  the sustainability threshold found through the mean-field
  approximation.}
\end{figure}

To gain insight into this phenomenon, we study the case of fixed
$\tau$ by means of numerical simulations.  We fill the lattice with
hosts and inject a single consumer with a $\tau$ of our choice; then,
we observe how long the descendents of that consumer persist as a
function of~$\tau$.  The consumer population does not persist when
$\tau$ is either too low or too
high. Figure~\ref{fig:survival-probability-v0p2-g0p1} shows the
probability that a consumer strain will survive for a substantial
length of time (2000 generations) after injection into a lattice
filled with hosts.  This probability is hump-shaped, with an
asymmetric plateau bounded above and below by cutoffs.

To understand the upper cutoff visible in
Figure~\ref{fig:survival-probability-v0p2-g0p1}, {\em i.e.,} the value
of $\tau$ above which the consumer population again becomes
unsustainable, consider the limiting scenario where $g \approx 0$.  If
hosts do not reproduce into available empty sites, our system reduces
to an \emph{epidemic process} which has been studied
before~\cite{arashiro2006, henkel2008}.  Below the transition point at
$\tau = 0.5$, a consumer injected into a lattice of hosts will produce
a consumer strain (which we can think of as an infection) which
survives for a finite number of generations and then dies out, leaving
the lattice filled with hosts (susceptibles) marred by a small patch
of empty sites (recovered individuals).  Above the transition point, a
single consumer gives rise to an expanding wave of consumers which
propagates over the lattice, leaving empty sites in its wake, until it
consumes all the hosts in the ecosystem.  This regime is known
as \emph{annular growth.}  No finite ecosystem can sustain annular
growth indefinitely.  If the host growth rate $g$ is made nonzero,
then hosts can recolonize sites left empty by the expanding consumer
population, opening the possibility of host--consumer coexistence in
an ecosystem of dynamically formed and re-formed patches.
Figure~\ref{fig:animation} illustrates an example of this phenomenon.
This is a specific example of the general phenomenon that, even far
from phase transitions, nonequilibrium systems can display {\em
pattern formation}~\cite{sayama2000}.

We can, therefore, interpret the upper cutoff on consumer
sustainability as a Malthusian catastrophe due ultimately to the
limited amount of available hosts~\cite{aguiar2004}.  In physics
jargon, this cutoff is a {\em finite-size effect.} This is the key to
understanding what happens when multiple types of consumer are present
on the same lattice, and in particular the case we study in the next
section, where an invasive consumer variety is introduced to an
ecosystem where native hosts and consumers have already formed a
dynamic patch distribution.  The environment experienced by the
invasive variety is that formed by the native species, and the
``finite size'' of the resources available to the invasive variety is
not the size of the whole lattice, but that of a local
patch~\cite{aguiar2004}.

\subsection{Timescale Dependence of Invasion Success}
\label{sec:timescale}
\begin{figure}[h]
\includegraphics[width=6cm]{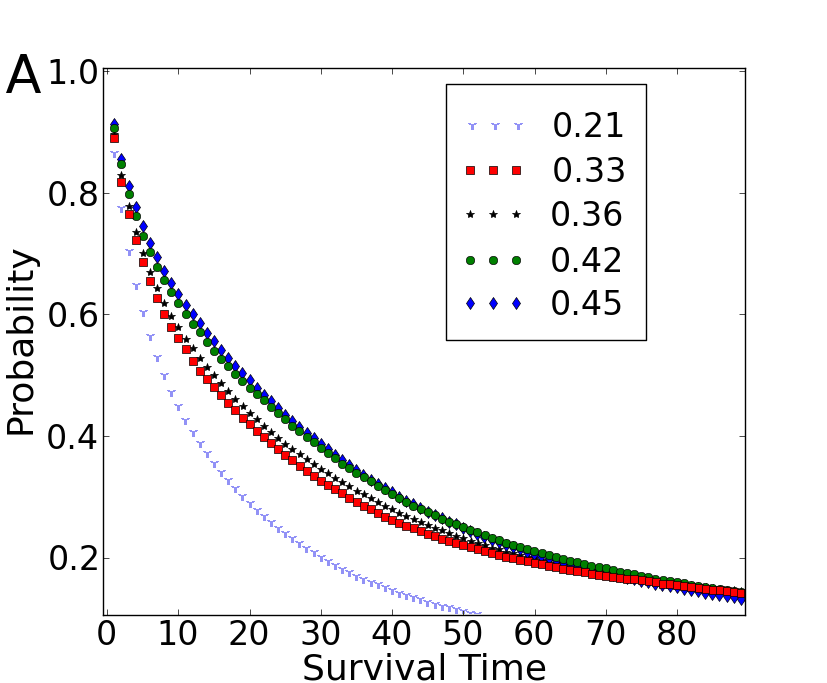}
\includegraphics[width=6cm]{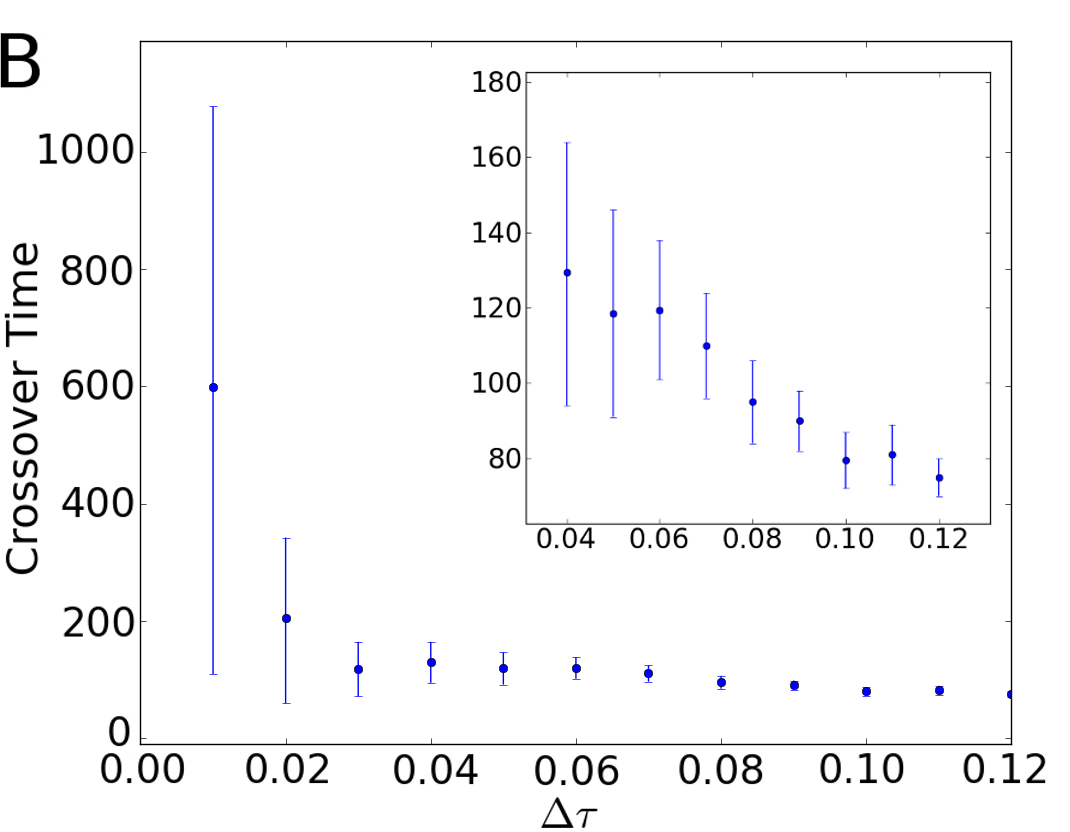}
\caption{\label{fig:survival-time-comparison} {\bf (A)} Survival
  probability as a function of time for five scenarios: injecting
  mutants with the same transmissibility as the native consumers,
  three examples of injecting mutants with transmissibility higher
  than the $\tau$ of the native consumers, and an example of injecting
  mutants with lower $\tau$ than the native population. {\bf (B)} Time
  intervals during which the survival-probability curves for the
  native and invasive strains overlap.  $\Delta\tau$ indicates the
  difference between the invasive and native transmissibilities.  The
  closer the mutant trait value is to the resident, the greater the
  duration of time over which the survival-probability curves for the
  native and mutant strains overlap.  Here, overlap is defined by
  probabilities being coincident at the 95\% confidence level; using
  other overlap criteria gives qualitatively the same results.  Inset:
  magnified view of the $\Delta\tau \geq 0.04$ region.}
\end{figure}

A key question about an ecological system is whether a new variety of
organism, having a different genetic character and phenotypic trait
values, can successfully invade a native population.  If a mutant
consumer strain with fixed transmissibility $\tau_m$ can successfully
invade a population of transmissibility $\tau_0 < \tau_m$, then we
expect the time-averaged value of~$\tau$ seen in the evolving system
to be larger than $\tau_0$.  To investigate this, we simulate
scenarios where the native population has $\tau$ close to the average
value seen in the evolutionary case described above.  We then inject a
mutant consumer strain with significantly larger $\tau$ and study the
results.  For a typical example, we see from
Figure~\ref{fig:minmax-6apr2011}(A) that when $g = 0.1$ and $v = 0.2$,
the average $\tau$ is approximately 0.33.  So, we simulate $\tau_m =
0.45$ mutants entering an ecosystem whose native population has
$\tau_0 = 0.33$.  Initially, the mutants prosper, but they ultimately
fail to invade.  As shown in
Figure~\ref{fig:survival-time-comparison}, the probability of a
$\tau_m = 0.45$ strain surviving for tens of generations after
injection is larger than that of a $\tau_0 = 0.33$ strain. That is,
mutants with the higher $\tau$ can out-compete the neutral case.
However, after $\approx\!74$ generations, the survival-probability
curves cross.  Observed over longer timescales, the mutant strain is
less successful than the native variety.  This pattern is consistent
for $\tau_m > \tau_0$: the average transmissibility seen in the
evolutionary case stands up to invasive varieties.  This key result
manifests the distinctive properties of the spatial structure of the
model.  The underlying reason for this result is that the mutants
encounter the resource limitations imposed by the patchy native
population.  Over short timescales, the mutant strain enjoys the
resources available within the local patch, consuming those resources
more rapidly than can be sustained once it encounters the limitations
of the local patch size.  In this way, the initial generations of the
mutant strain ``shade'' their descendants.  Thanks to
descendant-shading, short-term prosperity is not a guarantee of
medium- or long-term success.

This is to be contrasted with what happens in a well-mixed ecosystem.
In the well-mixed scenario, consumer strains with higher $\tau$
successfully invade and displace the native population with a high
probability.  The invasion success is consistent with the dynamics of
a continuously evolving ecosystem.  If $\tau$ is made an evolvable
trait in simulated panmictic systems, the average $\tau$ of the
population will tend to~1, as predicted by the mean-field analytic
proof.  There is no difference in a well-mixed scenario between
short-term and long-term success.  Descendant-shading does not occur
in the well-mixed case.  This follows from the lack of distinction
between local patches and large-scale structure.

One common measure of evolutionary success is the expected relative
growth rate of the number of offspring of a mutant individual within a
native population, \emph{i.e.,} the relative growth rate of a mutant
strain.  This rate, known as the {\em invasion fitness,} is often used
to investigate the stability of an evolutionary
ecosystem~\cite{vanbaalen1998, vanbaalen2000, lion2008}.  If the
invasion fitness is found to be positive, the native variety is judged
to be vulnerable to invasion by the mutant.  Conversely, if the
invasion fitness is found to be negative, the native variety is deemed
to be stable.  For the spatial host--consumer ecosystem, this method
gives qualitatively incorrect predictions for evolutionary dynamics.

Our investigation builds on earlier work which studied the timescale
dependence of fitness indicators in spatial host--consumer
ecosystems~\cite{rauch2002, rauch2003}. In this chapter we have
augmented the prior work by considering the survival probability to
show the effects of varying $\tau$.  We have also more systematically
shown the number of generations until dominance of the evolutionary
stable strain.  In addition, we reported the case of a mutant strain
invading a background population, clarifying the conceptual and
quantitative results of those earlier works, which considered instead
scenarios complicated by multiple ongoing mutations.

\subsection{Pair Approximations}

The inadequacy of mean-field treatments of spatial systems motivates
the development of more elaborate mathematical methods.  In this
section, we review one such methodology, based on augmenting
mean-field approximations with successively higher-order correlations,
and we test its applicability to our host--consumer spatial model.
The numerical variables used in this methodology are probabilities
which encode the state of the ecosystem and can change over time.  One
such variable is, for example, the probability $p_a$ that a lattice
site chosen at random contains an organism of type $a$.  Another is
$p_{ab}$, the probability that a randomly-chosen {\em pair} of
neighboring sites will have one member of type $a$ and the other
of type $b$.  The change of these quantities over time is usually
described by differential equations, for which analysis tools from
nonlinear dynamics are available~\cite{vanbaalen2000, haraguchi2000,
  aguiar2004, lion2008, rozhnova2009, allen2010, araujo2010}.

The importance of the joint probabilities $p_{ab}$ is that they
reflect correlations which mean-field approximations neglect.  To
understand the relevance of the joint probabilities $p_{ab}$, consider
a scenario where an invasive mutant variety forms a spatial cluster
near its point of entry.  Let $p_M$ be the probability that a lattice
site chosen at random contains a mutant-type organism, and let
$p_{MM}$ denote the probability that a pair of neighboring sites
chosen at random will both be occupied by mutant-type organisms.  Then
the average density of invasive mutants in the ecosystem, $p_M$, will
be low, while the conditional probability that a neighbor of an
invasive individual will also be of the invasive type, $q_{M|M} =
p_{MM} / p_{M}$, will be significantly higher.  (It is typical in
theoretical spatial ecology to denote conditional probabilities
with~$q$, rather than $p$~\cite{tgoei2000}.)  A discrepancy between
the conditional probability $q_{a|b}$ and the overall probability
$p_a$ can persist when the ecosystem has settled into a
quasi-steady-state behavior, and is then an indicator of spatial
pattern formation.  

\begin{figure}[!h]
\includegraphics[width=6cm]{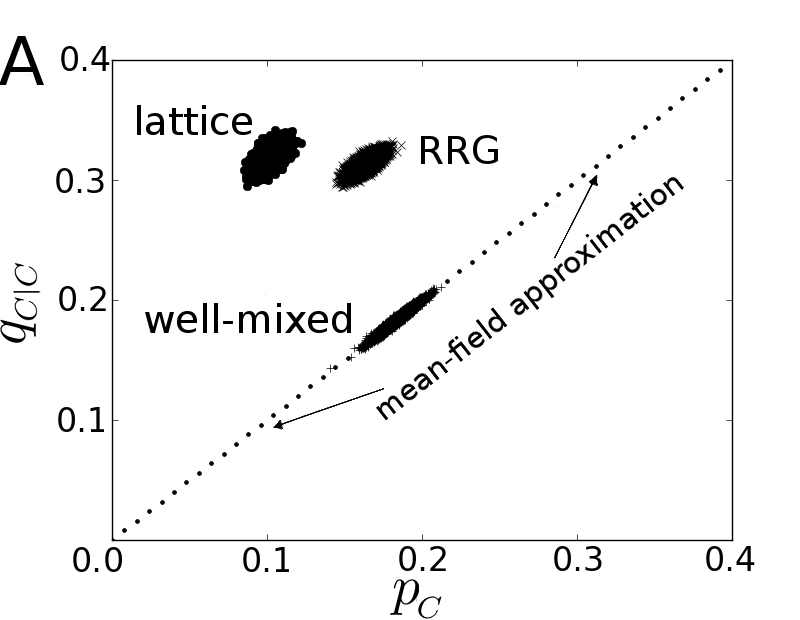}
\includegraphics[width=6cm]{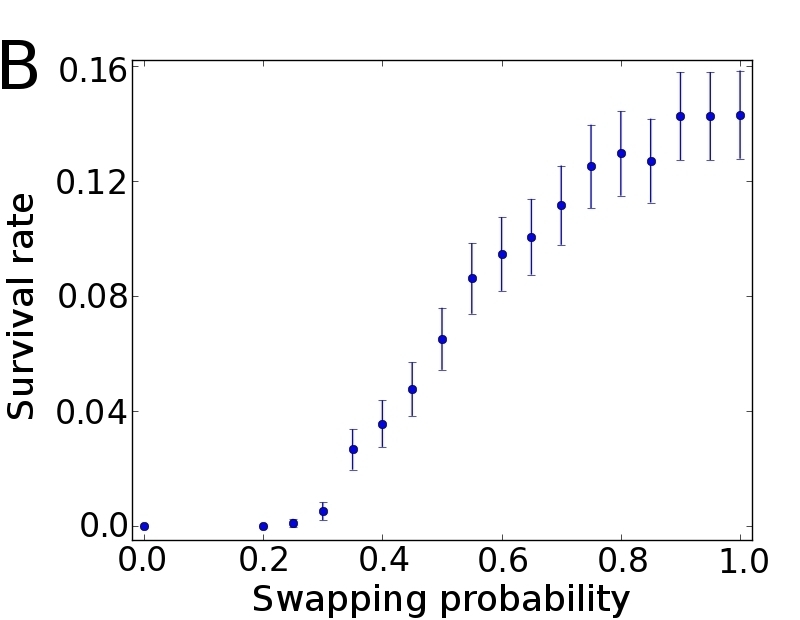}
\caption{\label{fig:qcc-over-pc} {\bf (A)} Pairwise conditional
  probability $q_{C|C}$ plotted against the average density of
  consumers, $p_C$, for three variations on the host--consumer model:
  a well-mixed case in which mean-field theory is applicable, a random
  regular graph (in which each site has exactly four neighbors) and a
  2D square lattice.  The dotted line, $p_C = q_{C|C}$, indicates the
  mean-field approximation.  $10^4$ timesteps were computed for each
  case.  The well-mixed case is simulated by dynamically rewiring
  sites at each time step, precluding the generation of spatial
  heterogeneity; consequently, the pairwise correlation $q_{C|C}$ is
  within statistical variation equal to~$p_C$ ($R^2 = 0.953$).  A
  random regular graph (RRG) with random but static connections does
  develop spatial heterogeneity so that $q_{C|C}$ is not the same
  as~$p_C$ ($R^2 = 0.581$). The discrepancy is even stronger in the
  lattice case ($R^2 = 0.304$).  {\bf (B)} Success rate of invasive
  mutant strains as a function of swapping probability.  Voracious
  mutant strains with $\tau = 0.45$ are introduced into a lattice
  ecosystem defined by a host growth rate of $g = 0.1$, a consumer
  death rate $v = 0.2$ (the same for both consumer varieties), and a
  native consumer transmissibility of $\tau = 0.33$.  Average success
  rates are found by simulating 2000 invasions per value of the
  swapping probability parameter; error bars indicate 95\%-confidence
  intervals.  Increasing the fraction of possible swaps which are
  actually performed makes the voracious invasive strain more likely
  to take over the ecosystem.}
\end{figure}

Applying this idea to the spatial host--consumer model, let $p_C$ be
the probability that a lattice site chosen at random contains a
consumer, and let $q_{C|C}$ denote the conditional probability that
lattice site adjacent to a consumer will also be occupied by a
consumer.  Figure~\ref{fig:qcc-over-pc}(A) shows $p_C$ and $q_{C|C}$
measured during the course of numerical simulations.  In a well-mixed
scenario (where we expect the mean-field approximation to be
applicable), the average consumer density $p_C$ and the
consumer--consumer pairwise correlation $q_{C|C}$ are essentially
equal over time.  In the spatial lattice scenario, $p_C$ and $q_{C|C}$
are noticeably different.

Treating the correlations $q_{a|b}$ as not wholly determined by the
probabilities $p_a$ is a way of allowing spatial heterogeneity to
enter an analytical model.  Whether it is a {\em sufficient} extension
in any particular circumstance is not, {\em a priori,} obvious.
Typically, the differential equations for the pair probabilities
$p_{ab}$ depend on triplet probabilities $p_{abc}$, which depend upon
quadruplet probabilities and so forth.  The standard procedure is to
truncate this hierarchy at some level, a technique known as {\em
  moment closure}~\cite{matsuda1992, vanbaalen1998, rozhnova2009,
  do2009, demirel2012}.  Moment closures constitute a series of
approximations of increasing intricacy~\cite{buice2009,dodd2009}.  The
simplest moment closure is the mean field approximation; going beyond
the mean field to include second-order correlations but neglecting
correlations of third and higher order constitutes a {\em pair
  approximation.}  These approximations do not incorporate all of the
information about spatial structure which may be necessary to account
for real-world ecological effects~\cite{vanbaalen1998}.

\subsection{Organism Swapping}
\label{sec:swapping}
Several factors have been identified which undermine pair
approximations~\cite{vanbaalen1998, aguiar2003b, aguiar2004,
  aguiar2003b-errata, werfel2004, szabo2007, allen2010, allen2011,
  demirel2012, smaldino2013}.  In our model, we can directly test the
efficacy of pair approximations in a completely general way.  The key
idea is to transplant individuals in such a way that the variables
used in the moment-closure analytical treatment remain unchanged.  At
each timestep, we look through the ecosystem for isolated consumers,
that is, for individual consumers surrounded only by a specified
number of hosts and empty sites.  We can exchange these individuals
without affecting the pairwise correlations.  For example, if we find
a native-type consumer adjacent to three hosts and one empty site, we
can swap it with an invasive-type consumer also adjacent to three
hosts and one empty site.  We can also exchange isolated pairs of
consumers in the same way.  The variables used in the moment-closure
treatment remain the same.  Were the moment-closure treatment valid,
we would expect the dynamics to remain unchanged when we perform such
exchanges.

When we perform the simulation, however, swapping strongly affects the
dynamics.  With this type of swapping in effect, mutants with higher
$\tau$ can invade a native population with lower $\tau$.  In one
typical simultation run with a native $\tau$ of~0.33 and an invasive
$\tau$ of~0.45, the invasive strain succeeded in~1,425 of~10,000
injections.  Without swapping, the number of successful invasions is
{\em zero.}

Swapping can be considered as creating a new ecosystem model with the
same moment-closure treatment as that of the original.  The behavior
of invasive strains is different, because transplanting organisms
allows invasive varieties to evade localized Malthusian catastrophes.
Swapping opens the ecosystem up to invasive strains, since, in
essence, it removes individuals from the ``scene of the crimes''
committed by their ancestors.

This type of swapping is, to our knowledge, a new test of
moment-closure validity.  Randomized exchanges have been incorporated
into computational ecology simulations for different purposes.  For
example, research on dispersal rates in an island model shuffled
individuals in such a way that the population size of each island was
held constant~\cite{poethke2007}.

If, instead of performing every permissible swap, we transplant
organisms with some probability between 0 and 1, we can interpolate
between the limit of no swapping, where invasions always fail, and the
case where pair approximation is most applicable and invasions succeed
significantly often.  The results are shown in
Figure~\ref{fig:qcc-over-pc}(B) and indicate that the impact of
swapping becomes detectable at a probability of $\approx\!0.25$ and
effectively saturates at a probability of $\approx\!0.9$.  

Our swapping method allows us to test the significance of
complications which can undermine pair approximation techniques or
make them impractical to apply, several of which have been identified.
First, introducing \emph{mutation} into a game-theoretic dynamical
system can make pair approximation treatments of that system give
inaccurate predictions~\cite{allen2010, allen2011}.  

Second, when the evolving population has a network structure, the
presence of
\emph{short loops} in the network often makes pair approximations
fail~\cite{szabo2007}.  For example, in a triangular lattice, one can
take a walk of three steps and return to one's starting point, whereas
on a hexagonal lattice, the shortest closed circuit is six steps long.
A pair approximation can work well for a dynamical system defined on
the hexagonal lattice but fail when the same dynamics are played out
on a triangular one.  This happens because the short loops provide
opportunities for contact which the coarse-graining necessary for a
pair approximation will miss.  (We will study this point in more
detail in Chapter~\ref{ch:volunteer}.)  This effect is amplified in
adaptive network models, where the underlying network changes
dynamically in response to the population living upon it.  In such
cases, even extending the moment closure to the triplet level brings
little improvement~\cite{demirel2012}.

Third, \emph{fluctuating population
  sizes} make pair approximations significantly more cumbersome to
construct, leading to systems of differential equations which are too
intricate to be significantly illuminating.  In a game-theoretic model
where a lattice is completely filled at all times with cooperators and
defectors, there is one independent population density variable and
three types of pairs.  By contrast, in an ecological model where two
consumer varieties are competing within an adaptive network of hosts,
a pair approximation requires nine independent
variables~\cite{aguiar2003b, aguiar2004, aguiar2003b-errata}.
Modeling phenomena of biological interest can easily increase the
complexity still more.  For example, if organism behavior changes in
response to social signals~\cite{werfel2004}, the number of possible
states per site, and thus the number of dynamical variables in a pair
approximation treatment, increases further.

Fourth, the pair-approximation philosophy of averaging over all pairs
in the system impedes the incorporation of \emph{environmental
heterogeneities.}  These include biologically crucial factors like
variable organism mobility, background toxicity or other localized
``costs of living,'' and resource availability~\cite{smaldino2013}.

Finally, \emph{dynamical pattern formation} creates spatial
arrangements which the pair approximation does not
describe~\cite{vanbaalen1998}.  This can be thought of as the
ecosystem generating its own environmental heterogeneities.

\subsection{Effect of Substrate Topology}
\label{sec:substrate}

It is instructive to compare the spatial lattice ecosystem with the
host--consumer model defined on a random regular graph (RRG).  In an
RRG, each node has the same number of neighbors, as they do in a
lattice network, but the connections are otherwise random.  RRGs have
been used as approximations to incorporate the effects of spatial
extent into population models, as they make for more tractable
mathematical treatments, although they are typically less realistic
than spatial lattices~\cite{lion2009}.  The network structure is set
at the beginning of a simulation and does not change over time.  The
important aspect of this network as compared to the spatial case is
that there exist short paths of links that couple all nodes of the
network.  This is quite different from the spatial case, where strains
in one part of the network cannot reach another in only a few
generations due to the need to traverse large numbers of spatially
local links.

When we simulate our host--consumer ecosystem on an RRG, we find that
an invasive consumer strain with higher transmissibility $\tau$ can
out-compete and overwhelm a native consumer population with lower
$\tau$.  In one typical simulation run, using the native and invasive
$\tau$ values of 0.33 and 0.45 respectively, 2,233 out of 10,000
invasions were successful, whereas on the lattice {\em no} invasion
succeeded using the same parameters.  Thus, the RRG does not capture
the essential features of the spatial scenario.  In particular, our
results show that the RRG case is more like the well-mixed case than
the spatial lattice, as far as stability against invasion is
concerned.

Our swapping test provides insight into the utility of the pair
approximation, which can be effective for the RRG even though it is
not for the spatial case.  Consider the pairwise correlation value
$q_{C|C}$, which would be a variable for a pair approximation
treatment.  On an RRG, the underlying network topology provides enough
locality that $p_C$ and $q_{C|C}$ are unequal, distinct from the well
mixed case as shown in Figure~\ref{fig:qcc-over-pc}(A).  This means
that the pair approximation is nontrivial for the RRG as it
incorporates the difference between~$q_{C|C}$ and $p_C$, which would
not be contained in a mean-field treatment.  We can also implement
swapping on the RRG, where invasions can succeed without it; as
expected, swapping does not affect the success rate on the RRG.  With
10,000 simulated invasions for each case, the 95\%-confidence interval
for the difference in success rates between full swapping and none is
$0.004 \pm 0.01$.  Thus, the pair approximation may be successful in
this network topology.  However, this does not mean that the RRG or
the pair approximation capture the full significance of a spatial
system, because the RRG network does not embody essential properties
of spatial extent---separation by potentially large distances.

\subsection{Percolation}
\label{sec:percolation}
In order to obtain quantitatively or even qualitatively correct
predictions for spatial host--consumer evolutionary dynamics,
different approaches are needed.  Having encountered the limitations
of moment closures, we now demonstrate a change of perspective which
yields quantitatively useful results.  In certain situations, the
process of pathogen propagation through the host population
distributed in space can be mapped onto a {\em percolation} problem. A
topic widely investigated in mathematics, percolation theory deals
with movement though a matrix of randomly placed obstacles. A
prototypical percolation problem is a fluid flowing downhill through a
regular lattice of channels, with some of the lattice junction points
blocked at random. The key parameter is the fraction of blocked
junction points. If this fraction is larger than a certain threshold
value, the fluid will be contained in a limited part of the
system. However, if the blocking fraction is below the threshold, the
fluid can percolate arbitrarily far from its starting point. This is a
\emph{phase transition,} a shift from one regime of behavior to
another, in this case between a phase in which fluid flow can continue
indefinitely and one in which flow always halts. Similar issues arise
when a pathogen propagates by cross-infection through a set of
spatially arranged hosts. Sufficiently many hosts in mutual contact
are required for the pathogen to propagate successfully. Pathogen
strains therefore survive or die out over time depending on whether
percolation is or is not possible~\cite{mobilia2006b, arashiro2006,
  davis2008, salkeld2010, givan2011, neri2011}.

One important goal of studying host--pathogen models is knowing the
pathogen properties that enable its survival in a population, or
equivalently what prevents it from persisting in a population. The
growth rate of a pathogen in a population can be an important public
health concern. We therefore focus on analyzing the minimum value of
the transmissibility that enables a pathogen population to persist,
and the growth dynamics of population sizes near that transition.

Of essential importance to the quantitative theoretical and empirical
analysis is the recognition that infected population growth can be
described by power laws $n \sim t^z$, with an exponent that differs from
that of the mean field. Identifying the value of the power $z$ is
important to practical projections of the number of infected
individuals. The initial growth curve of infected populations can be
correctly extrapolated if the exponent is known, guiding public health
responses.  Knowing what impediments are needed to prevent further
propagation can even better guide public health intervention
strategies.

\begin{figure}[h]
\includegraphics[width=6cm]{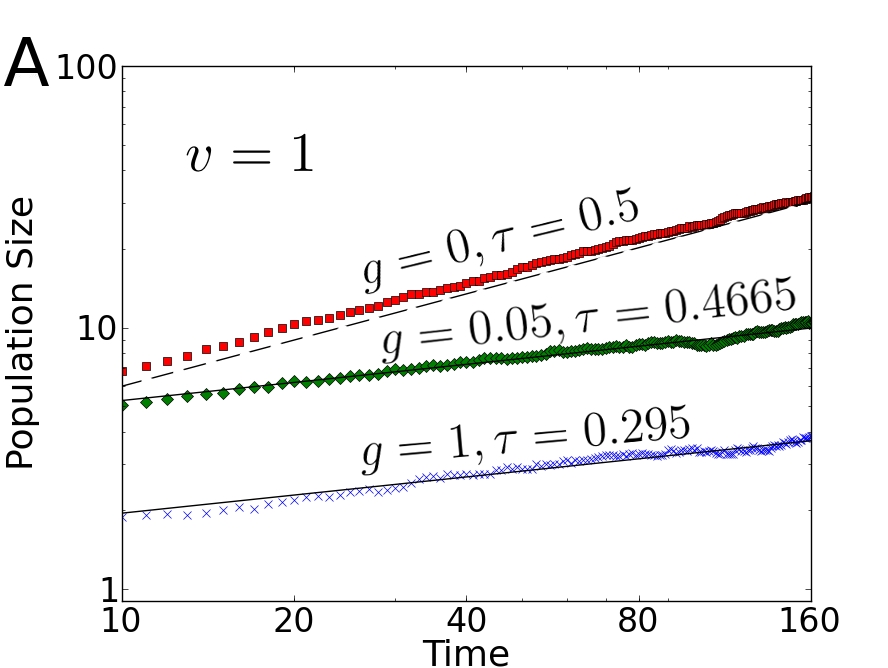}
\includegraphics[width=6cm]{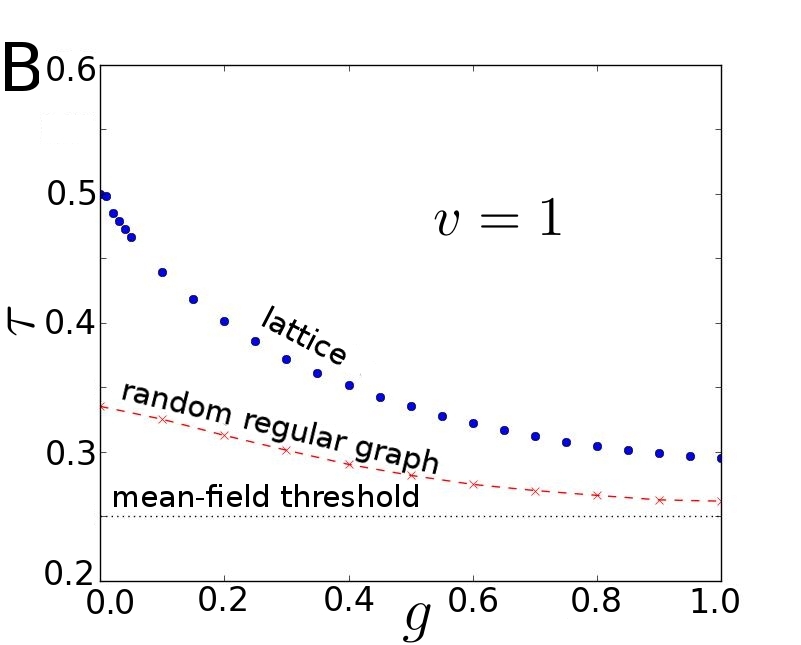}
\caption{\label{fig:percolation} {\bf (A)} Population size as a
  function of time, averaged over $10^3$ simulation runs, for $\tau$
  values near the transition points at $g = 0$ and $g = 0.05$ on the
  spatial lattice, with $v = 1$.  Dashed and solid lines indicate the
  population growth for systems at dynamic percolation and directed
  percolation transitions respectively, showing that these transitions
  have the characteristic properties of those universality classes.
  {\bf (B)} Critical $\tau$ for the host--consumer ecosystem with $v =
  1$.  The transition line crosses over from the dynamic percolation
  universality class at $g = 0$ to directed percolation between $g =
  0.015$ and $g = 0.02$.  Red Xs indicate the transition curve for the
  host-consumer dynamics on a random regular graph (RRG) of
  uniform degree 4; the dashed line connecting them is to guide the
  eye.  The RRG transition is neither directed percolation nor dynamic
  percolation.  }
\end{figure}

We show in Figure~\ref{fig:percolation} the results of numerical
simulations which indicates that the consumer extinction transition,
when the transmissibility $\tau$ becomes just large enough that the
consumer population sustains itself, lies in the {\em directed
  percolation universality class}~\cite{hinrichsen2000, henkel2008,
  saif2010, wendykier2011, szolnoki2011}.  A similar result has been
reported for related models~\cite{mobilia2006b, arashiro2006},
consistent with those models being in the same universality class.
The directed percolation universality class is a large set of models,
all of which exhibit a phase transition between two regimes of
behavior, and all of which behave in essentially the same way near
their respective transition points.  The scenario of fluid flow
through a random medium considered above is a classic example of a
directed percolation-class model, but many others exist as
well~\cite{hinrichsen2000, henkel2008}.  The {\em critical exponents}
describe how properties of the modeled system vary over time or as a
function of how far the control parameter is from the critical point.
They are the same for all systems in the universality class.  Other
universality classes exist as well, with different classes having
different quantitative values for the critical exponents.  Identifying
the universality class a system belongs to enables us to study a
complicated phenomenon by examining a simpler representative of its
class instead.  This is convenient, because regions of parameter space
near phase transitions are precisely where mean-field and
moment-closure approximations are least reliable, even for
short-timescale modeling.  Near the phase transition, stochastic
fluctuations create dynamical patterns with a wide range of sizes.  In
Figure~\ref{fig:percolation}(A), we see that percolation theory gives
quantitatively correct predictions for the growth of consumer
population sizes in the spatial host--consumer model.

We can understand the $g = 0$ and $g = 1$ extremes by mapping the
host--consumer model onto other stochastic models for which exact or
approximate results are available.  When $g = 0$, the host--consumer
model maps onto the SIR epidemic process~\cite{henkel2008}.  In turn
the SIR model on the square lattice can be understood in terms of bond
percolation on the square lattice~\cite{grassberger1997}, for which
the transition point is known exactly~\cite{kesten1980, grimmett1999}.
We can therefore predict analytically that the critical $\tau$ on the
square lattice is~0.5.  Percolation theory also gives a prediction for
the critical $\tau$ on an RRG: it should be approximately
$1/3$~\cite{newman2002}.  These both match the simulation results seen
in Figure~\ref{fig:percolation}(B).

In contrast, when $g = 1$, empty sites are filled as quickly as
possible, so the behavior of the host--consumer model should resemble
that of an epidemic model with only Susceptible and Infected sites.
In this case, the transition point of the epidemic model on the square
lattice is only known numerically~\cite{grassberger1997}.  The
numerical value, $\approx\!0.29$, does agree with the critical $\tau$
found by simulating the host--consumer model at~$g = 1$.

Thus, in the limiting cases of $g = 0$ and $g = 1$, the host--consumer
model is roughly equivalent to the SIR and SIS epidemic models.
However, a host--consumer model with~$0 < g < 1$ has dynamical
behavior distinct from an epidemic model which allows reinfection of
Recovered sites.  The key difference is that reoccupying an empty site
with a consumer requires prior recolonization by a host, whereas the
vulnerability of a R[ecovered] individual to becoming I[nfected] is
defined as an intrinsic property of the R[ecovered] type.  This
changes the role of ecology: both models incorporate space, but the
effect of spatial extent is different.  This manifests as a change in
the shape of the critical-threshold curve, as well as a change in
universality class~\cite{grassberger1997, henkel2008}.

Furthermore, when we use an RRG topology instead, comparing
host--consumer dynamics at~$g = 1$ with an SIS epidemic model reveals
their transitions to take place at different thresholds.  For the
host-consumer model, the critical $\tau$ on an RRG is approximately
0.2615, while the SIS threshold is approximately
$1/3$~\cite{gleeson2011, boguna2013, mata2014}.

The physics analysis of percolation behavior near the transition point
maps directly onto the critical public health problem of the growth of
infected populations, and more generally onto the dynamics of
evolutionary systems.  For these systems the mean field treatment
fails and the standard transmission of infectious diseases in a
population need not apply.  Applications to real world systems must
accommodate the actual network of connectivity. This network can also
be modified by intervention strategies.

\subsection{Patch Size and Structure}
\label{sec:patchsize}
\begin{figure}[h]
\includegraphics[width=12cm]{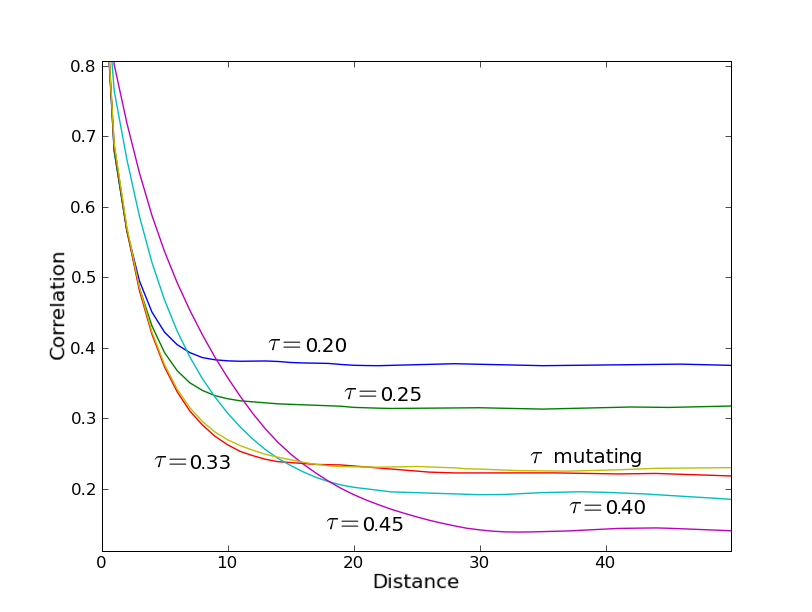}
\caption{\label{fig:correlation-curves-v0p2-g0p1} The decay of
host-host autocorrelation as a function of distance, for different
choices of~$\tau$ (fixing $v = 0.2$, $g = 0.1$).  Note that the
correlation curve for~$\tau = 0.33$ essentially coincides with the
curve found when $\tau$ is a mutable trait.}
\end{figure}

Since it is the size of a host patch which determines the amount of
resources available for consumers, we now investigate patch sizes in
detail.  One way to test if the host patches have a characteristic
size is to take snapshots of the dynamics in its quasi-steady-state
regime (\emph{e.g.,} the fourth panel of Figure~\ref{fig:animation})
and compute its \emph{autocorrelation.}  We can do this by running the
snapshot through a filter that produces a binary matrix whose entries
are 1 in locations occupied by hosts and 0 otherwise.  Applying FFTs
and the Wiener--Khinchin theorem then yields a 2D autocorrelation
matrix, which by averaging we can collapse to a function of distance.
If this autocorrelation function has a characteristic distance
scale---for example, if it decays exponentially---then we can use that
distance as the \emph{correlation length.}  We present examples
confirming this in Figure~\ref{fig:correlation-curves-v0p2-g0p1}.

\begin{figure}[h]
\includegraphics[width=6cm]{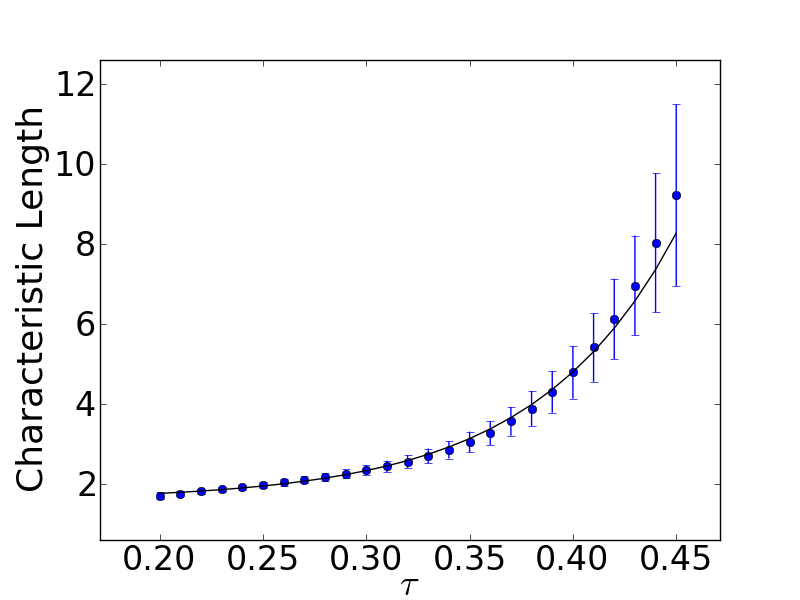}
\includegraphics[width=6cm]{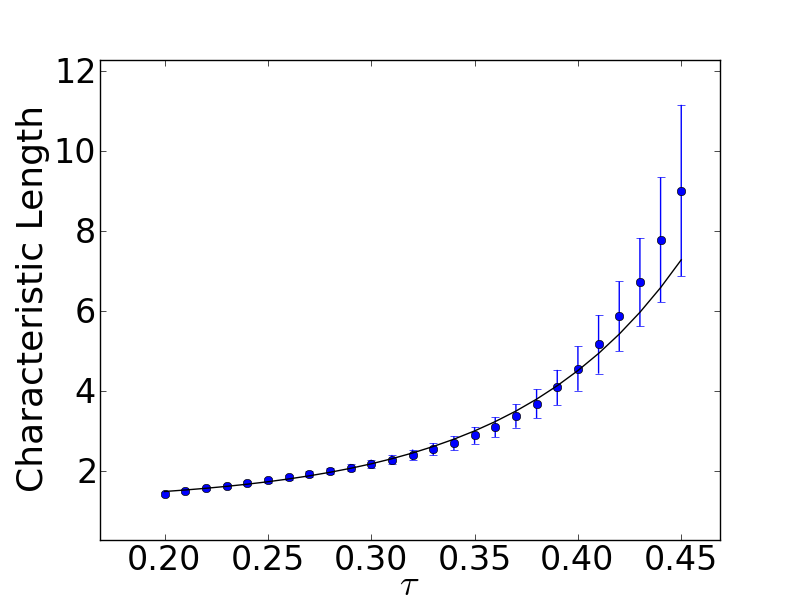}
\caption{\label{fig:correlation-lengths-v0p2-g0p1} \textbf{(Left)} 
  Correlation length of hosts for different values of~$\tau$ ($g =
  0.1$, $v = 0.2$). The solid line is the curve $\xi_H
  = \alpha \times \exp(\tau / \beta) + \gamma$, where the parameters
  found by least-squares fitting are $\alpha = 0.009 \pm 0.002$,
  $\beta = 0.068 \pm 0.002$ and $\gamma = 1.59 \pm
  0.02$.  \textbf{(Right)} Correlation length of empty space under the
  same conditions. The solid line is the curve $\xi_0
  = \alpha \times \exp(\tau / \beta) + \gamma$, with $\alpha =
  0.026 \pm 0.007$, $\beta = 0.082 \pm 0.004$ and $\gamma = 1.20 \pm
  0.05$.}
\end{figure}

Next, Figure~\ref{fig:correlation-lengths-v0p2-g0p1} summarizes the
results of this procedure for host--host correlations as we vary
$\tau$, holding the other parameters fixed at~$g = 0.1$, $v = 0.2$.
We see that the correlation length of the host distribution, which we
can regard as the characteristic size of host patches, increases
with~$\tau$.  By choosing a different filter, we can apply the same
procedure to find the characteristic size of empty regions.  The right
panel of Figure~\ref{fig:correlation-lengths-v0p2-g0p1} summarizes the
results.

We consider the following scenario: a native population of consumers,
with transmissibility $\tau_{\rm nat}$, is dynamically forming and
re-forming host patches.  Into this ecosystem, a mutant consumer with
transmissibility $\tau_{\rm mut}$ is introduced.  Call the
characteristic size of host patches $\xi_H$ and the characteristic
separation between them $\xi_0$.  Both of these length scales will
depend on $g$, $v$ and $\tau$.  The length $\xi_H(g,v,\tau = \tau_{\rm
mut})$ is the typical size of host patches in which the mutant variety
``expects'' to live.  If this length is too much larger than
$\xi_H(g,v,\tau = \tau_{\rm nat})$, the size of the patches which
exist due to the native population, then the mutant variety is likely
to suffer a Malthusian catastrophe.

\begin{figure}[!h]
\includegraphics[width=10cm]{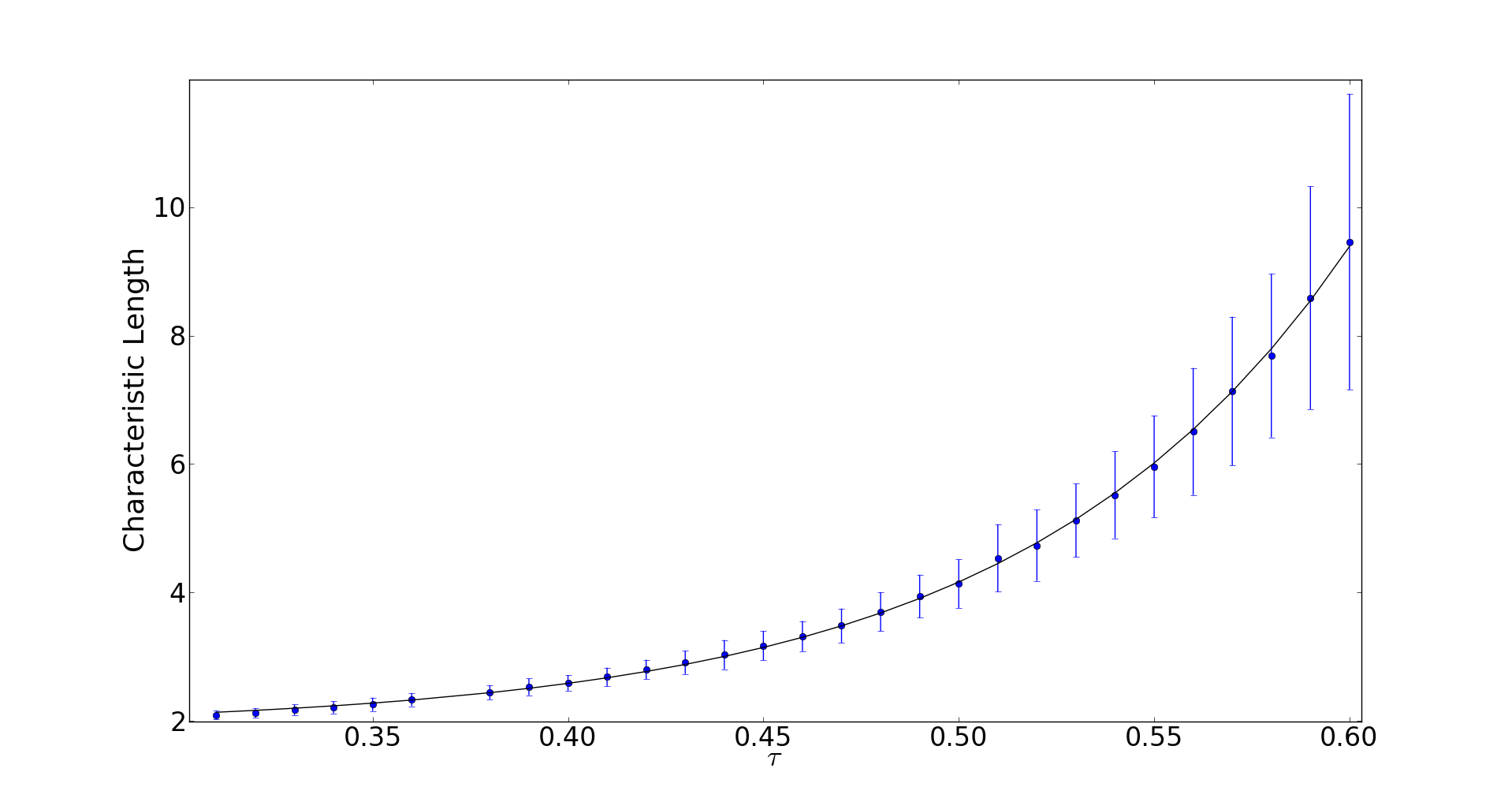}

\includegraphics[width=10cm]{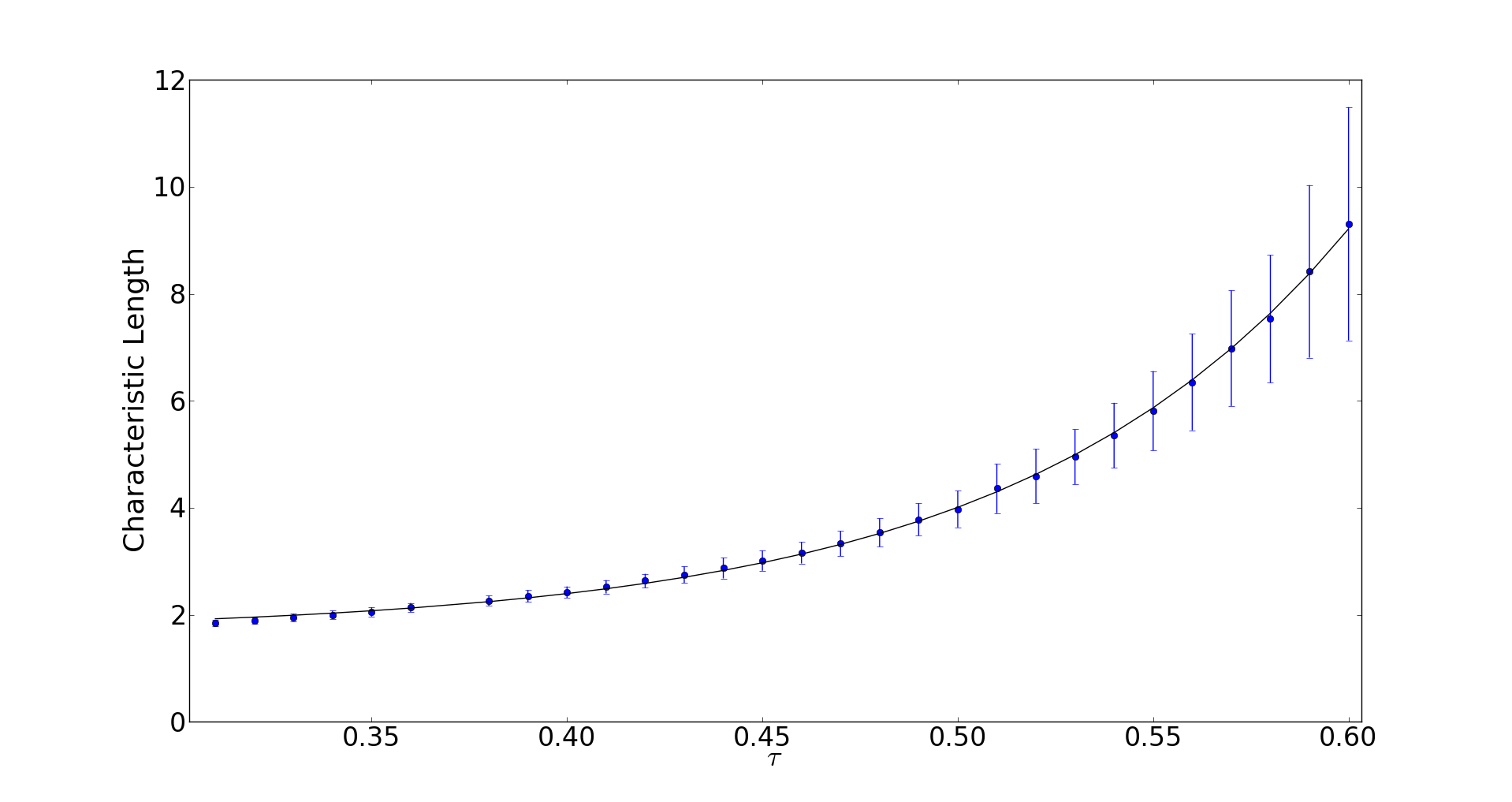}
\caption{\label{fig:correlation-lengths-v0p4-g0p1} \textbf{(Top)} 
  Host-host correlation length $\xi_H$ for different values of~$\tau$
  ($g = 0.1$, $v = 0.4$).  The solid line is the curve $\xi_H
  = \alpha \times \exp(\tau / \beta) + \gamma$, where the parameters
  found by least-squares fitting are $\alpha = 0.0089 \pm 0.0005$,
  $\beta = 0.0893 \pm 0.0009$ and $\gamma = 1.81 \pm
  0.01$.  \textbf{(Bottom)} Void-void correlation length $\xi_0$ under
  the same conditions.  Here, $\alpha = 0.014 \pm 0.001$, $\beta =
  0.096 \pm 0.002$ and $\gamma = 1.51 \pm 0.02$.}
\end{figure}

When we find the correlation lengths via simulations, as seen in
Figures~\ref{fig:correlation-lengths-v0p2-g0p1}
and \ref{fig:correlation-lengths-v0p4-g0p1}, we see that both $\xi_H$
and $\xi_0$ increase with~$\tau$.  Assuming that the deleterious
effect of ``expecting'' larger host patches grows with the discrepancy
between ``expected'' and actual patch sizes, the distribution
of~$\tau$ in the consumer population will accumulate in the region
where the $\xi_H$ curve starts to take off.

The missing piece is an \emph{analytical} expression for~$\xi_H$ and
$\xi_0$ in terms of the parameters $g$, $v$ and $\tau$, when all these
parameters are fixed.  What is the functional form of
\begin{equation}
\xi_H = \xi_H(g,v,\tau) \hbox{ and }
\xi_0 = \xi_0(g,v,\tau)\ ?
\label{eq:xi-missing-piece}
\end{equation}
Figure~\ref{fig:correlation-lengths-v0p2-g0p1} suggests that both
correlation lengths depend roughly exponentially on~$\tau$, but the
offsets and growth rates will depend on~$g$ and $v$.

One way in which we might try to estimate the length scale $\xi_H$ and
see, at least qualitatively, how it depends on the system parameters,
is to use a moment closure.  This method turns out not to work, and we
now investigate why.  Consider a square lattice of size
$L \times L$, and assume for simplicity that it contains some number
$K$ of host patches, all of a comparable size which we take to be the
characteristic length $\xi_H$.  The population density of hosts is
\begin{equation}
p_H = \frac{K\xi_H^2}{L^2}.
\end{equation}
The quantity $q_{H|H}$ used in a pair approximation denotes the
probability that a randomly chosen neighbor site of a randomly chosen
host will also contain a host.  If we imagine a patch of hosts in an
otherwise empty lattice (for convenience, suppose the patch is roughly
circular), then sites in the interior of the patch have all their
neighboring lattice sites occupied by hosts, while sites on the edge
have some empty space adjacent.  Only the sites on the perimeter can
be responsible for decreasing $q_{H|H}$ below unity.  Writing $P$ for
perimeter and $A$ for area, we have
\begin{equation}
q_{H|H} = 1 - \frac{KfP}{A},
\end{equation}
where $f$ is a constant we can take to be roughly one-half.  If, for
simplicity, we assume ordinary Euclidean scaling of perimeters and
areas,
\begin{equation}
P \sim \xi_H,\ A \sim \xi_H^2,
\end{equation}
then we have that
\begin{equation}
q_{H|H} = 1 - \frac{Kf'}{\xi_H}.
\end{equation}
We have two quantities, $p_H$ and $q_{H|H}$, which we can compute by
numerically integrating a set of coupled differential equations.  And
we have two unknowns: the patch size $\xi_H$ and the number of patches
$K$.  We solve for~$\xi_H$, obtaining
\begin{equation}
\xi_H^3 = f'L^2 \frac{p_H}{1 - q_{H|H}}.
\end{equation}
As the conditional probability $q_{H|H}$ approaches unity, the patch
size increases.

Unfortunately, when we compute $p_H$ and $q_{H|H}$ by numerically
iterating the dynamical equations (see \cite{aguiar2003b, aguiar2004,
aguiar2003b-errata}), we find that the inferred $\xi_H$
value \emph{decreases} with~$\tau$.  The direction of the change, as
predicted by pair approximation, is incorrect!  This is a sign that
the pair approximation is losing important information about
larger-scale structures.

Another possible way to predict how $\xi_H$ and $\xi_0$ depend on $g$,
$v$ and $\tau$ comes from the theory of spatial stochastic processes.
For some models, we can derive an analytical expression for the power
spectrum of fluctuations, as a function of both spatial and temporal
oscillation frequencies.  A peak in this power spectrum indicates a
characteristic length or time scale~\cite{lugo2008, biancalani2010,
butler2011}.  Because the host--consumer lattice model differs in a
few significant ways from those treated by these methods before, we
defer a detailed exploration of this idea until a later chapter.  The
upshot appears to be that the characteristic length scale increases in
the proper direction, that is, with increasing $\tau$, although thanks
to the differences hinted at, the concavity may not be correct.

\begin{figure}[h]
\includegraphics[width=12cm]{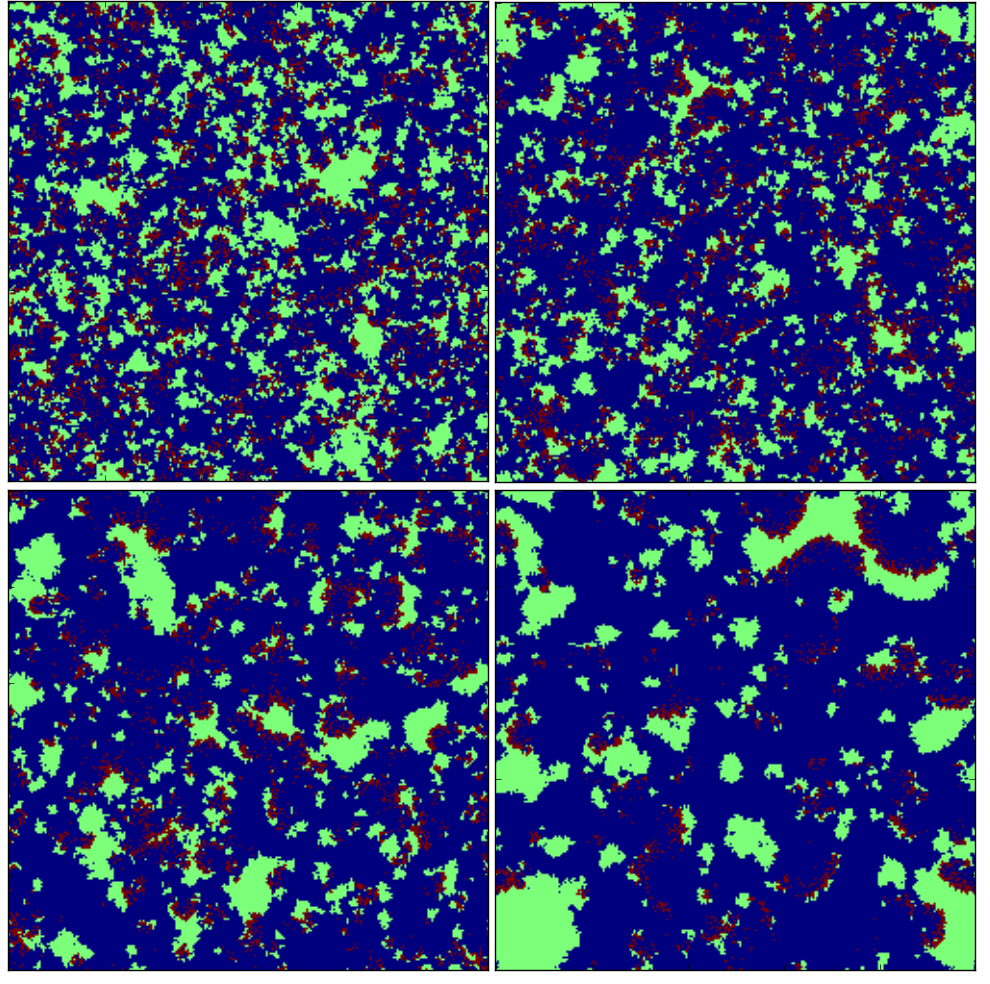}
\caption{\label{fig:snapshot-g0p1-v0p2} Typical frames from
host--consumer simulations run with different values of the
transmissibility $\tau$ (the other parameters are fixed at~$g = 0.1$,
$v = 0.2$).  {\bf (Top left)} $\tau = 0.30$; {\bf (top right)} $\tau =
0.35$; {\bf (bottom left)} $\tau = 0.40$; {\bf (bottom right)} $\tau =
0.45$.}
\end{figure} 

At this point, we might be concerned that computing a correlation
length in this fashion could be smearing over too much.  If there are
many small patches and a few large ones (and
Figure~\ref{fig:snapshot-g0p1-v0p2} makes this look plausible), then
does reducing the pattern to a single correlation length lose any
important information?  To cross-check the idea that host patch size
controls the evolved transmissibility, we therefore measure patch size
in another way.

The most straightforward way to assign a size to a patch is to count
the number of lattice sites it contains.  This has the advantage that
while we are computing it, we can also count the number of sites in
the patch which border on empty space.  So, we can see how the
perimeters of host patches relate to their areas.  If host patches
were Euclidean circles, then the perimeter would scale as the area to
the one-half power.  On the other hand, a patch which consists of a
single site in a discrete lattice is all boundary:  its perimeter and
its area are equal.  Our patches live on a square lattice, so a
cluster of hosts must contain at least five individuals for its area
to be greater than its perimeter.  The more hosts are contained within
the cluster, the more likely the cluster is to have an interior.  We
therefore expect a crossover:  below some value of the area, the
maximum observed perimeter will be equal to the area.  Above the
crossover point, the maximum observed perimeter will grow more
slowly.  Noting that the patch dynamics are stochastic and their edges
irregular, we hypothesize that this growth will be faster than the
area-to-the-0.5 obtained for Euclidean circles.

\begin{figure}[h]
\includegraphics[width=14cm]{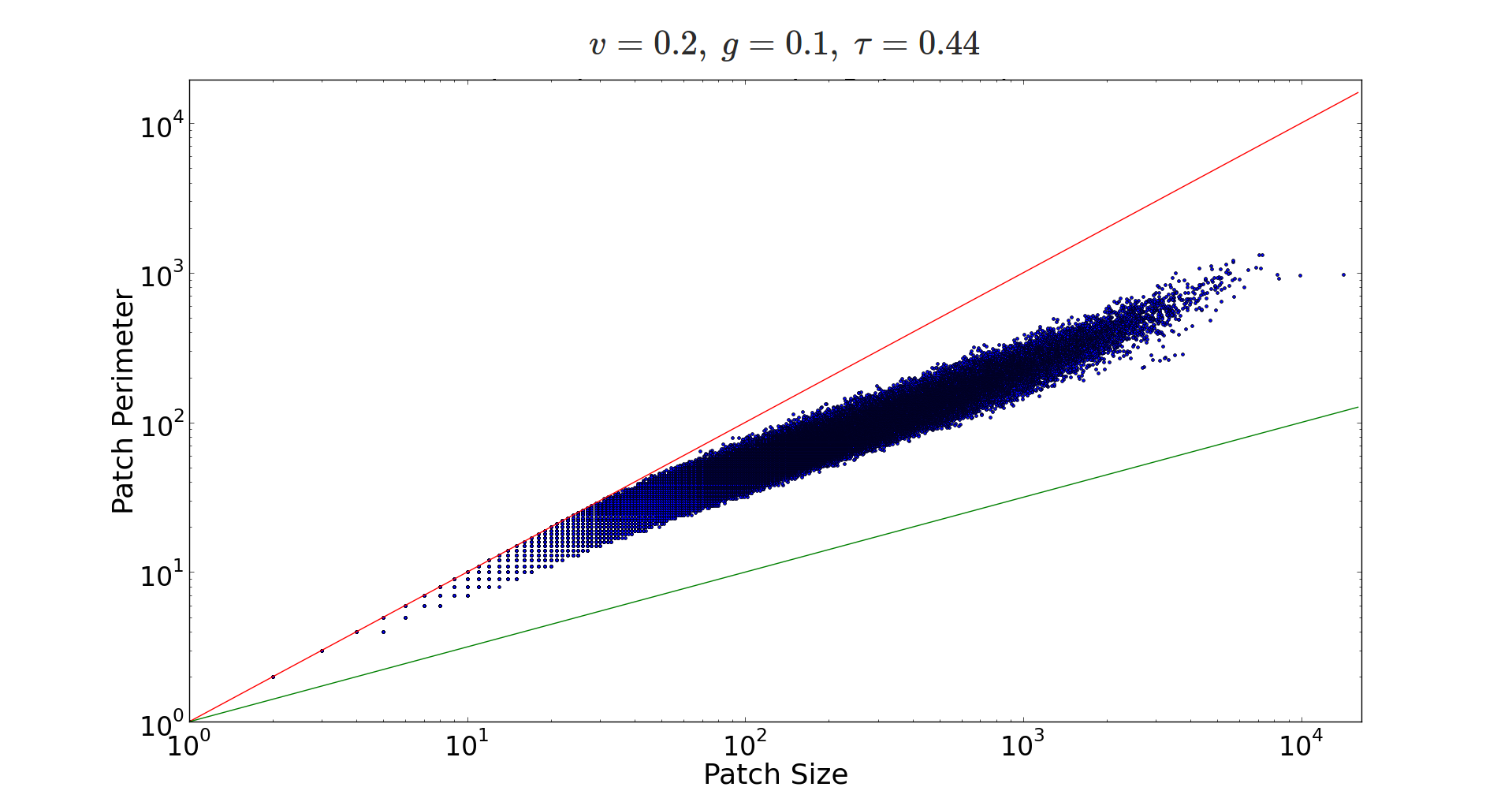}
\caption{\label{fig:patchperimeters-v0p2-g0p1-tau0p44} Perimeters of
host patches, plotted against their areas ($g = 0.1$, $v = 0.2$, $\tau
= 0.44$).  The upper (red) straight line indicates linear growth, and the lower
(green) line indicates square-root growth.  Note the crossover at area
$\approx\!30$ lattice sites.  Below the crossover point, the maximum
observed perimeter toes the line of linear growth, indicating that
host patches exist which have no interior.}
\end{figure}

Figure~\ref{fig:patchperimeters-v0p2-g0p1-tau0p44} bears this out.
When we plot the perimeters of host clusters against their areas, we
see a crossover at an area of $\approx\!30$ lattice sites (for $g =
0.1$, $v = 0.2$).  Furthermore, the increase of perimeter with area
above the crossover point is faster than the square root of the area.

Pascual \emph{et al.}~\cite{pascual2002} study the cluster-size
distribution for prey in a predator--prey model which is similar to
our host--consumer ecosystem.  However, they do not find a crossover:
instead, cluster perimeter scales smoothly and just barely sublinearly
with cluster area, across the whole range of observed areas.  This may
be due to an extra mixing effect which their model includes and ours
does not, an effect which tends to bring more of a cluster to its
perimeter.

\begin{figure}[h]
\includegraphics[width=14cm]{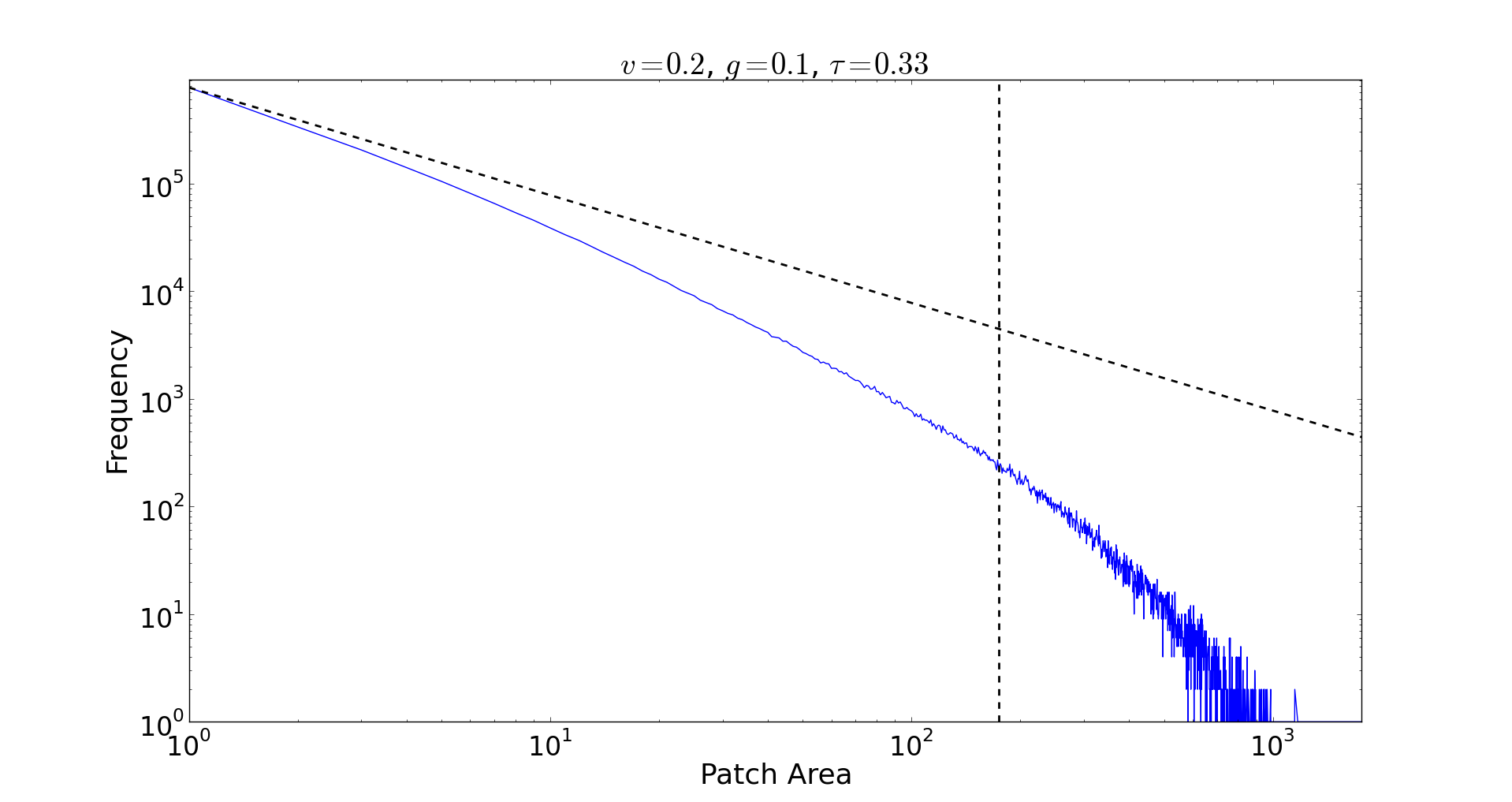}
\caption{\label{fig:loglog-area-counts-g0p1-v0p2-tau0p33} Number of
host patches observed as a function of patch area ($g = 0.1$, $v = 0.2$,
$\tau = 0.33$).  The vertical dashed line indicates the
99\textsuperscript{th} percentile, that is, the point at which 99\% of
the patches are that size or smaller.  The sloping dashed line
indicates a decay with the inverse of area, for comparison purposes.}
\end{figure}

Having measured the host patch sizes, we can investigate the size
distribution.  Figure~\ref{fig:loglog-area-counts-g0p1-v0p2-tau0p33}
shows the numbers of patches observed at different sizes.  The
fall-off of frequency with area is faster than an inverse-area
relationship, though not uniform.  At first blush, an appropriate
characterization would be a power-law decay with an exponential
cutoff.  Because this distribution is fairly broad, we use
a \emph{percentile} to characterize it: we find the value of the area
such that 99\% of the host patches are that size or smaller.  This is
indicated by the vertical dashed line in
Figure~\ref{fig:loglog-area-counts-g0p1-v0p2-tau0p33}.

\begin{figure}[h]
\includegraphics[width=14cm]{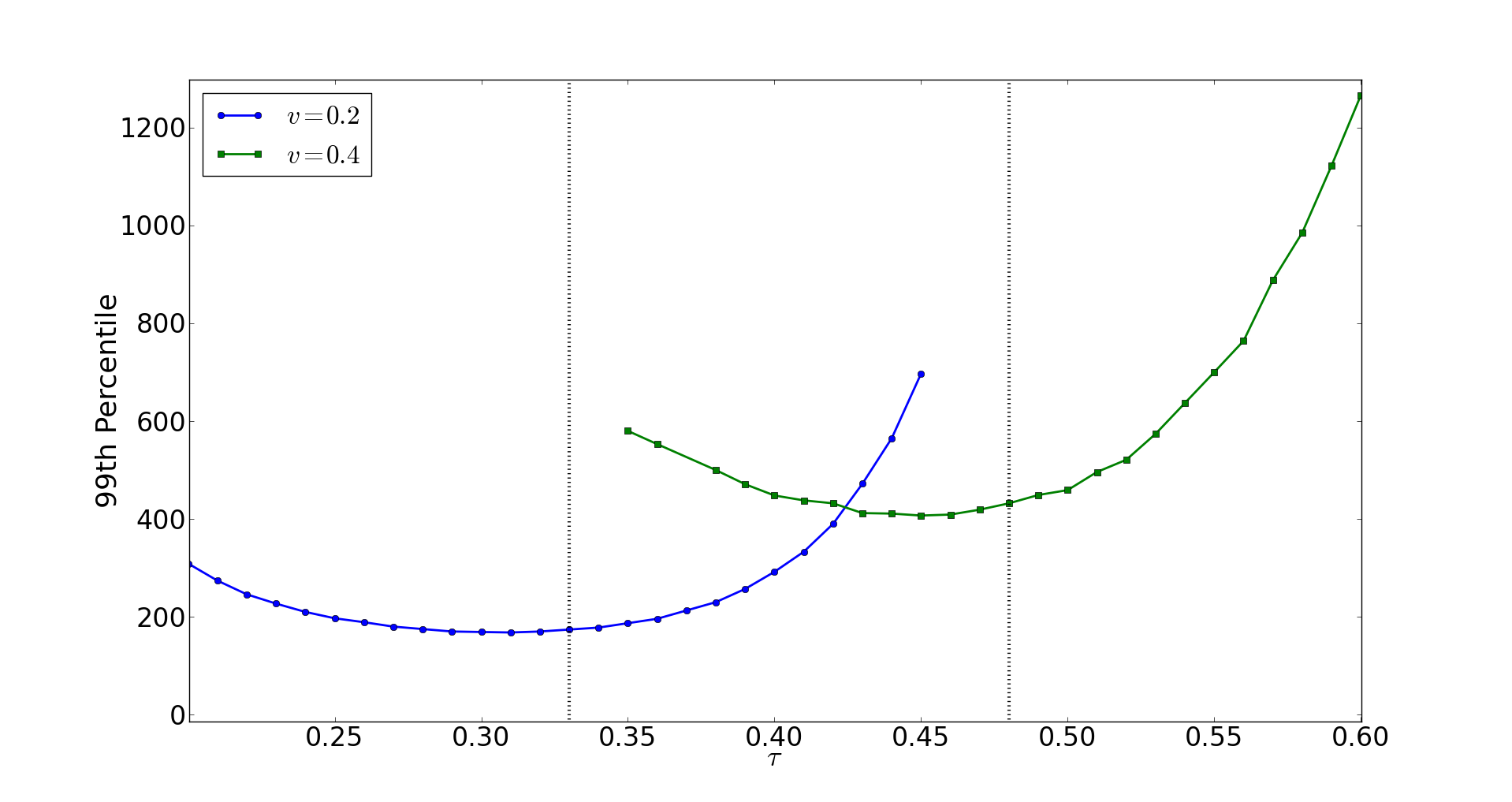}
\caption{\label{fig:areapercentile99-vs-tau-g0p1}
99\textsuperscript{th} percentile of host patch area versus consumer
transmissibility $\tau$, for $v = 0.2$ and $v = 0.4$ (computed with $g
= 0.1$).  The dashed vertical lines indicate the average
transmissibilities which evolve when $v$ and $g$ are fixed.}
\end{figure}

We see the results of repeating this calculation across a range of
$\tau$ values in Figure~\ref{fig:areapercentile99-vs-tau-g0p1}.
The \emph{minimum} of the 99\textsuperscript{th}-percentile curve
gives the value of the transmissibility which evolves when $\tau$ is a
mutable trait!

We can understand this, heuristically, using much the same argument we
made above.  A consumer strain which ``expects'' that large host
patches are available will fare poorly if they are not.  When $\tau$
is evolvable, the $\tau$ distribution of the consumer population will
tend to concentrate around the minimum of the
99\textsuperscript{th}-percentile curve.  In the short term, local
subpopulations with higher $\tau$ can blossom, causing occasional
upward shifts in the average $\tau$.  We expect, therefore, that the
average $\tau$ in the evolving population will lie somewhat to the
right of the curve's minimum point.

The hypothesis that these curves can be well approximated by
exponentially truncated power laws can be validated by standard
curve-fitting techniques~\cite{alstott2014}.  Applying methods
suitable to power-law analysis reveal that for all $\tau$ values, the
number of patches decays with roughly the 1.5 power of the patch
area.  The location of the cutoff determined by curve-fitting follows,
unsurprisingly, the relationship seen with the
99\textsuperscript{th}-percentile areas
(Figure \ref{fig:areapercentile99-vs-tau-g0p1}).  One should take
care, however, when applying these statistical methods and
interpreting the results, as they presume an independence among data
points which may or may not be applicable here.  If it is the case,
for example, that large patches are typically accompanied by smaller
patches nearby, then the assumption of statistical independence would
be invalid.

The qualitative shape of the area/abundance curve, that is, the
power-law dependence with an exponential cutoff, is also seen in the
results of \emph{coagulation and fragmentation
processes}~\cite{gueron1995, ma2011}.  Furthermore, such distributions
are known to be \emph{maximum entropy} distributions: they arise when
one maximizes the Shannon entropy (which we will discuss in
Chapter~\ref{ch:prob}), subject to a certain type of
constraint~\cite{visser2013}.  This suggests that the general
functional form is rather robust.

It is a good idea to try and understand the shape of the
99\textsuperscript{th}-percentile curves
(Figure \ref{fig:areapercentile99-vs-tau-g0p1}) heuristically, since
the minima of those curves indicate the values to which $\tau$ will
evolve.  Suppose, for concreteness, that the abundance as a function
of area \emph{is} a power-law decay with an exponential cutoff.  Why
should the cutoff decrease and then increase?  Can we think of this in
terms of countervailing influences?

First, we consider the low-$\tau$ regime.  When $\tau$ is small, the
cutoff should be large, because the system is near a critical point,
and critical points mean power laws.  The closer we move towards
criticality, the better the quality of the power law, and the less
noticeable any cutoff will be.  Therefore, as we increase $\tau$, we
move away from criticality, and the deviation from a clean power law
becomes more severe.  This qualitatively explains the decreasing part
of the curve in Figure~\ref{fig:areapercentile99-vs-tau-g0p1}.

\begin{figure}[h]
\includegraphics[width=14cm]{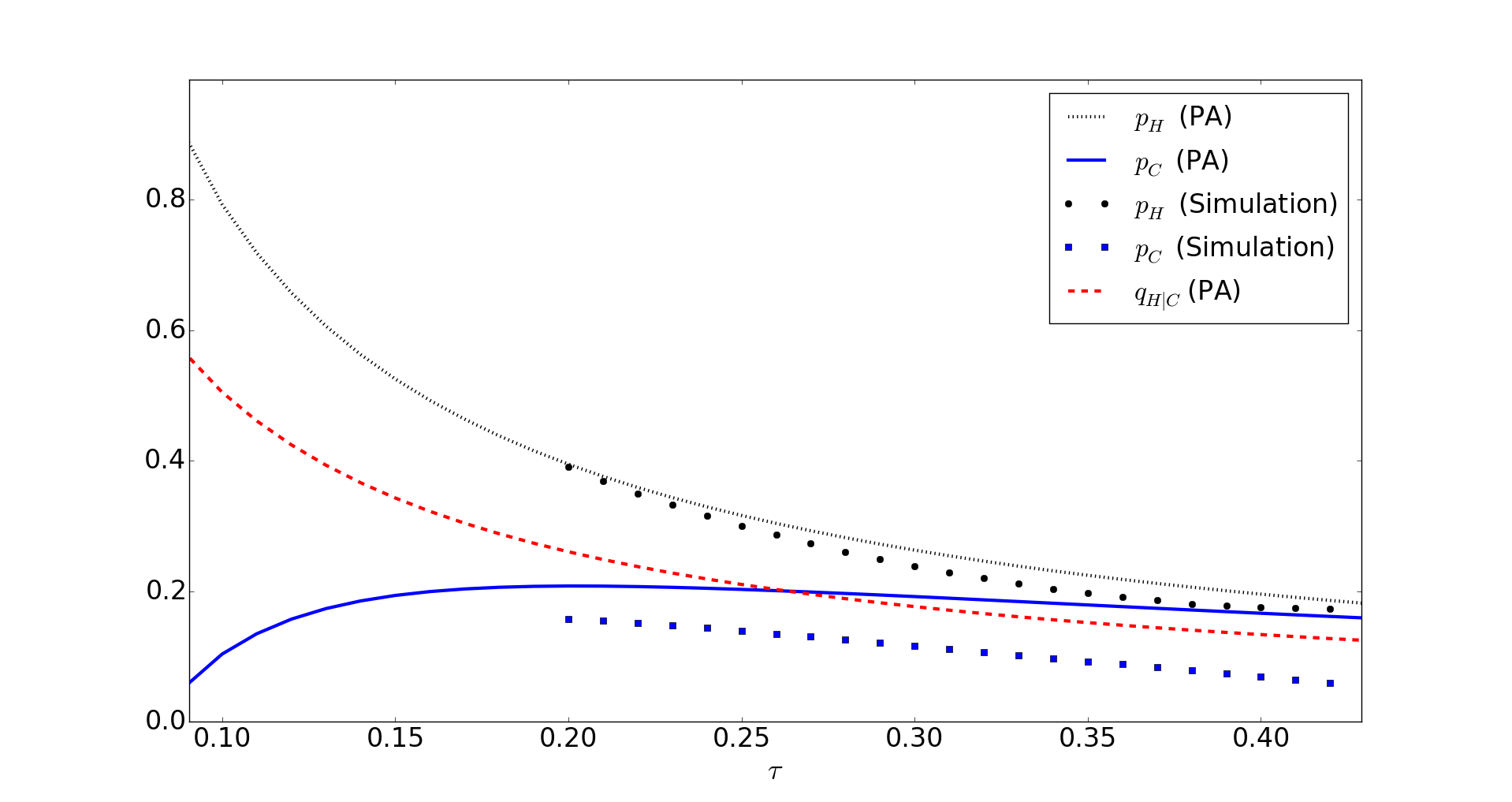}
\caption{\label{fig:hc-densities-PA-g0p1-v0p2} Host and consumer
densities as a function of~$\tau$, with $g = 0.1$ and $v = 0.2$.  Note
that the agreement between the simulation results and the pair
approximation is better for the host density than for the consumers.
The pair approximation for this ecosystem was developed by de
Aguiar \emph{et al.}~\cite{aguiar2003b, aguiar2004,
aguiar2003b-errata} and will be discussed in more detail in
Chapter~\ref{ch:invasion-fitness}.}
\end{figure}

What about when $\tau$ is large---say, when it is larger than the
value to which it would evolve when mutation is present?  Here, the
theory of critical points is no longer pertinent, and we find guidance
instead in the study of coagulation and fragmentation processes.
Numerical simulations indicate that for each value of~$v$ and $g$,
there is a region along the $\tau$ axis where the \emph{total host
population size} does not strongly depend upon~$\tau$ (see
Figure~\ref{fig:hc-densities-PA-g0p1-v0p2}).  That is, when $\tau$ is
large, we can adjust it and the host population size $P$ will not
change too much in response.  This suggests that we can construct a
simplified model in which $P$ is a parameter that we can vary
independently of the consumer transmissibility.

Gueron and Levin develop a model of group-size dynamics wherein a
population of fixed total size is divided into groups that can merge
and split stochastically~\cite{gueron1995}.  The rates of merging and
splitting are taken to be functions of the group sizes.  If $x$ and
$y$ are the sizes of two groups, then the probability per unit time
that those groups will merge is $m(x,y)$.  Similarly, the probability
per unit time that a group of size $x$ will fission into fragments of
sizes $y$ and $x-y$ is $s(x,y)$.  A convenient choice of functional
form that allows some analytical solutions is to take
\begin{equation}
m(x,y) = \mu a(x) a(y),
\end{equation}
where $a(x)$ is an increasing function of~$x$ and $\mu$ is a rate
parameter.  (For example, the probability of merging might increase
with the surface area of each group, implying a power-law dependence
on the group population.)  The splitting rate is taken to be
\begin{equation}
s(x,y) = 2\sigma a(x).
\end{equation}

This stochastic process has a stationary solution:  the distribution
of patch sizes follows the truncated decay
\begin{equation}
f(x) = \frac{2\sigma}{\mu} \left(\frac{1}{a(x)}\right) e^{-x/x_c}.
\end{equation}
The cutoff $x_c$ is fixed by the total population size:
\begin{equation}
P = \frac{2\sigma}{\mu} \int_0^\infty dz\, \frac{z}{a(z)} e^{-z/x_c}.
\end{equation}

In our host--consumer ecosystem, host patches split apart because they
are eaten into by consumers.  They can merge together on their own,
but their fission requires consumption.  Very crudely speaking, the
splitting rate should increase with the consumer population density.
(In a slightly more refined approximation, we could say that the
splitting rate should increase with the contact probability
$q_{H|C}$.)  This corresponds, in the Gueron--Levin model, to
increasing $\sigma$.  If we imagine that $P$ and $\mu$ are fixed,
increasing $\sigma$ must be balanced by decreasing the value of the
integral.  Likewise, lower consumer density corresponds to lower
$\sigma$ and thus a larger value for the integral.  The only way to
change the value of the integral, if the function $a(z)$ is a given,
is to alter the cutoff $x_c$.  To obtain a larger value, we move the
cutoff farther out.

We have, therefore, that if the host density is \emph{constant} with
respect to~$\tau$ but the consumer density \emph{falls,} then the
cutoff should be \emph{larger.}  When we measure the densities by
numerical simulation, these are the trends we find at larger $\tau$,
and so the upswing fits neatly into our picture.

\begin{figure}[h]
\includegraphics[width=14cm]{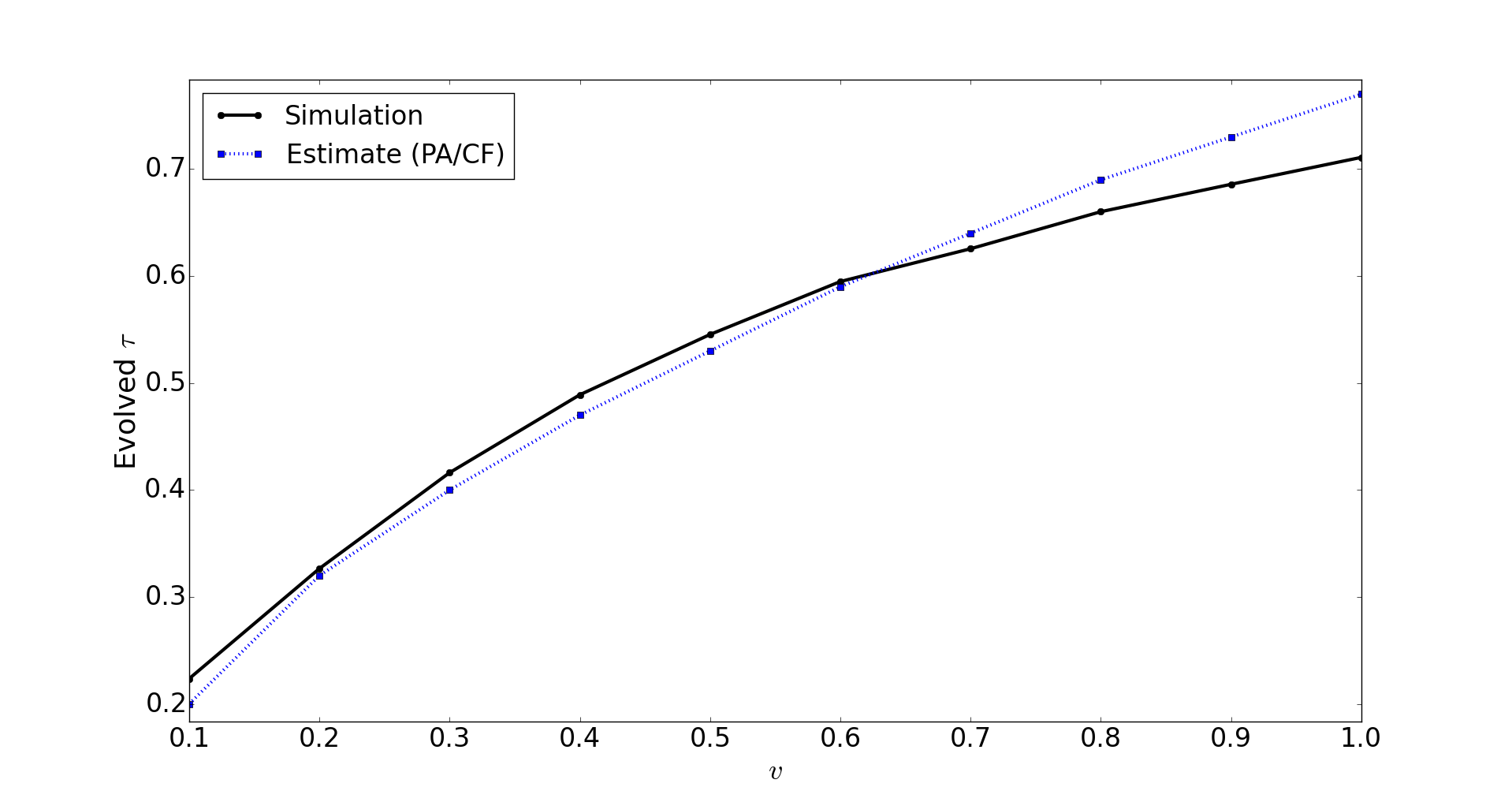}
\caption{\label{fig:evolved-tau-guesstimates-g0p1} Estimates of the
evolved transmissibility as a function of the host death rate $v$
(with $g = 0.1$).  These values were obtained by combining the results
of pair approximation with the insights from the Gueron--Levin
coagulation and fragmentation model and the idea that the
host-patch-size cutoff controls the evolution of~$\tau$.  By searching
in the pair approximation for a region where $dq_{H|C}/d\tau < 0$ and
$dp_H/d\tau > -C$, we identified that part of the $\tau$ axis where
patch splitting is suppressed and the cutoff elevated.  We set $C =
0.75$ based on the $p_H(\tau)$ curves computed for~$v = 0.2$ and $v =
0.4$.  This value, in turn, provides a reasonable estimate for the
evolved transmissibility obtained at other values of~$v$.}
\end{figure}

Figures \ref{fig:hc-densities-PA-g0p1-v0p2}
and \ref{fig:evolved-tau-guesstimates-g0p1} together indicate that
while pair approximation itself fails to capture the eco-evolutionary
dynamics of the spatial system, it can provide a qualitatively useful
guide when combined with a coagulation/fragmentation model and the
principle that host patch size controls consumer transmissibility by
way of localized Malthusian catastrophes.

The distribution of areas
(Figure~\ref{fig:loglog-area-counts-g0p1-v0p2-tau0p33}) and the
perimeter-area relationship
(Figure~\ref{fig:patchperimeters-v0p2-g0p1-tau0p44}) together provide
an approximation to the complexity profile of the host-patch system,
as we defined it back in Chapter~\ref{ch:multiscale}.  Crudely
speaking, the interesting part of a host patch is its boundary: in
order to say what the patch might do next, we need to know what is
happening at its edges.  Therefore, the effective information content
of a host patch should scale, roughly, with its perimeter.  If we
neglect the influences of one patch upon another, we can approximate
the host population as a collection of blocks, and we can invoke the
sum rule stated in \S\ref{sec:multiscale-indices}.  Each block
contributes a rectangle to the complexity profile.  The width of the
rectangle is the number of hosts in the patch, and the height is the
information content, which is given by the perimeter.  Alternatively,
by interchanging the axes, we can construct an MUI curve in the same
way.  In either case, the structure index so defined will only be an
approximation.

\section{Discussion}
\label{sec:discussion}

\begin{quotation}
\noindent But the problem, here, is that it's a form of adaptation that hasn't
been studied enough in animals and plants, which is that each change
in the species changes what we call the environment, so there is a
co-evolution of organism and environment. [\ldots] The organism by its
evolution changes the conditions of its life and changes what
surrounds it. Organisms are always creating their own hole in the
world, their own niche.

---Richard Lewontin~\cite{wilson2015}
\end{quotation}

Understanding the effects of spatial extent is a vital part of
evolutionary ecology.  Spatial extent changes the quantitative and
qualitative characteristics of a model's evolutionary behavior,
compared to well-mixed models.  The short-term success rate of novel
genetic varieties is not indicative of their long-term chance of
success relative to the prevalent type.  Standard stability criteria
fail to reflect the actual stability achieved over time.  We must
instead consider extended timescales because they are determined by
spatial patterns, whose ongoing formation is an intrinsic part of
nonequilibrium evolutionary dynamics.  Our analysis provides a clear
understanding of why there are dramatic differences between spatial
models and mean-field models, which simplify away heterogeneity
through mixing populations, averaging over variations or mandating a
globally connected patch structure.  We have further shown that
transplanting organisms dramatically changes the dynamics of spatial
systems, even when we preserve local correlations as would be
considered in a pair approximation treatment.  Our results prove that
any model striving to capture the effects of heterogeneity that does
not change its behavior with organism transplanting cannot fully
capture the dynamics of spatial evolution.  The following subsections
summarize the general conclusions we draw from these results.

\subsection{Defining Fitness}

In our host--consumer models, each individual either survives or it
does not, and any individual has a specific number of offspring and
survives over a certain amount of time; that is to say, an
``individual fitness'' (in the terminology of~\cite{orr2009}) is a
well-defined concept.  To find {\em expected} individual fitness, or
{\em average} individual fitness, we must define a set of individual
organisms over which to take an average, which is the very concept we
have established to be problematic.  Consequently, derived notions of
fitness, which depend on comparisons between such
averages~\cite{orr2009}, become elusive, context-dependent quantities.
The problem is both temporal and spatial: Average relative fitness in
one generation is not necessarily a good measure of the long-term
success of a strain in one, or a combination of, the broad variety of
dynamically-generated niches.  This problem is not, however, the same
as the traditional concept of variation of fitness across a static set
of niches, because the niche dynamics of our spatially explicit model
ensures that evolutionary outcomes are not reflected in any standard
definition of the average.

In the previous chapter, we examined the idea of ``frequency-dependent
fitness''~\cite{page2002,hartl2007}, and one might be inclined to
apply that term to voracious invasive strains in this system, as the
invasive strain is successful initially when rare but fails when it
becomes more common. The term ``frequency-dependent fitness'' is,
however, a misnomer in this context, because the organism type is rare
and successful when it is newly introduced, but as it declines to
extinction it becomes rare and unsuccessful.  Nor can we attribute the
decline to the frequency of hosts: the average population density of
hosts remains essentially unchanged, because the boom and the
following bust are localized.  {\em Frequency,} being defined by an
average over the whole ecosystem, is only a proper variable to use for
describing the ecosystem in the panmictic case.  One might attempt to
refine the concept of global frequency by including local frequencies.
However, the breakdown of moment-closure techniques implies that
defining fitness as a function of organism type together with average
local environment~\cite{smith2010} will, in many circumstances, not be
an adequate solution.

Consequently, we find that trying to assign a meaningful invasion
fitness value to an invasive variety of organism is too drastic a
simplification.  In turn, this implies that we cannot assign a fitness
value to a phenotypic or genetic characteristic such as infectiousness
or transmissibility.  To understand this point, we rephrase the
spatial host--consumer model in terms of alleles.  In an invasion
scenario, an individual consumer can have one of two possible alleles
of the ``transmissibility gene'', one coding for the native value
of~$\tau$ (\emph{e.g.,} $\tau = 0.33$) and the other for the invasive
value (\emph{e.g.,} $\tau = 0.45$).  A mean-field treatment would then
involve specifying the fraction of the population which carries the
native allele versus the fraction which carries the invasive variant.
We have seen, however, that the predictions based on such a heavily
coarse-grained caricature of the original model deviate from its
actual behavior.  In short, the evolutionary dynamics cannot be
characterized using the allele frequencies at a particular time.

If we can no longer summarize the genetic character of a population by
an allele frequency---or a set of allele frequencies for well-defined
local subpopulations---then computing the fitness of a genotype from
its generation-to-generation change in frequency is a fruitless task.
In a world which exhibits nonequilibrium spatial pattern formation,
allele frequencies are the wrong attribute for understanding the dynamics
of natural selection.  Formally, the conventional assumption that the
allele frequencies are a sufficient set of variables to describe
evolutionary dynamics is incorrect.  The spatial structure itself is a
necessary part of the system description at a particular time in order
to determine the subsequent generation outcomes, even in an average
sense.

The timescale-dependence issues which arise in spatial host--consumer
ecosystems exist in a wider context.  Multiple examples indicate that
initial success and eventual fixation are only two extremes of a
continuum which must be understood in its entirety to grasp the
stability of a system.  In the study of genetic drift, it has been
found that neutral mutations can fixate and beneficial mutations fail
to fixate due to stochasticity~\cite{hartl2007}.  Likewise, in the
study of clonal interference~\cite{fogle2008, maddamsetti2015}, one
beneficial mutation can out-compete another and prevent its fixation.
Closer to the theme of this chapter, recent work has also emphasized
that selection acting at multiple timescales is important for the
evolution of multicellularity~\cite{hammerschmidt2014, fox2015}.

Furthermore, classical genetics makes much use of the Price equation
for studying the change in a population's genetic composition over
time~\cite{bijma2008, damore2011}, and it is well known that analytic
models built using the Price equation lack ``dynamic
sufficiency''. That is, the equation requires more information about
the current generation than it produces about the
next~\cite{grafen1985, page2002, damore2011, simon2012, allen2012},
and so predictions for many-generation phenomena must be made
carefully, if they can be made at all.  Modeling approaches which are
fundamentally grounded in the Price equation, such as
``neighbor-modulated'' fitness calculations~\cite{reeve2007,
bijma2008, wild2009, lion2010, allen2013b, allen2013c} and their
``multilevel'' counterparts~\cite{reeve2007, bijma2008, wade2010,
vandyken2011, sfd2012, nielsen2012, campobello2015}, are not likely to
work well here, as the analyses in question draw conclusions only from
the short-timescale regime.  In addition, those particular analyses
which address host--consumer-like dynamics either rely on moment
closures~\cite{lion2010} or they assume a fixed, complete connection
topology of local populations which are internally
well-mixed~\cite{wild2009, wade2010}.  These simplified population
structures are quite unlike the dynamical patch formation seen in the
host--consumer lattice model.  (Wild and Taylor~\cite{wild2004}
demonstrate an equivalence between stability criteria defined via
immediate gains, or ``reproductive fitness'', and criteria defined
using fixation probability; however, their proofs are explicitly
formulated for the case of a well-mixed population of constant size,
neither assumption being applicable here.  Whether fixation
probability is equivalent to any other criterion of evolutionary
success generally depends on mutation rates, even in
panmixia~\cite{allen2012}.)

We will discuss invasion fitness and moment-closure calculations in
more mathematical detail in Chapter~\ref{ch:invasion-fitness}.  The
Price equation and the ideas which cluster around it will be our
subject in Chapter~\ref{ch:price-mls}.

In the adaptive dynamics literature, models have been studied in which
``the resident strikes back''~\cite{mylius2001, geritz2002,
  geritz2005}.  That is, an initially rare mutant variety $M$ can
invade a resident population of type $R$, but $M$ does not supplant
$R$ and become the new resident variety, even though a population full
of type $M$ is robust against incursions by type $R$.  This is often
considered a rare occurrence, requiring special conditions to
obtain~\cite{geritz2002, geritz2005}, though the theorems proved to
that effect apply to nonspatial models, and in adaptive dynamics, it
is standard to consider small differences between mutant and resident
trait values.  The spatial host--consumer ecosystem has the important
property that, if mutation is an ongoing process, the spatial extent
allows genetic diversity to grow.  We initialize the system with all
the consumers having the same trait value, but soon enough, different
local subpopulations have different trait values.  If the effects of
single mutations are small, then the different varieties arising have
roughly comparable survival probabilities, and so the distribution of
extant trait values can spread out.  However, the cumulative effect of
many mutations which happen to act in the same direction on a trait
such as transmissibility creates a variety which may engender its own
local Malthusian catastrophe.  So, the results of rare, big mutations
tell us about the spread of trait values we see in the case of
frequent, small mutations.

Chapter~\ref{ch:adaptive} will treat adaptive dynamics in greater
depth, exploring the simplifications which its approximations allow.
In essence, adaptive dynamics illustrates the maxim that when one can
apply a Taylor expansion, one can simplify.  One finds that reductions
in complexity depend upon assumptions of smoothness that are
remarkably convenient, but are not always applicable.

In our model, transmissibility and consumer death rate are
independently adjustable parameters. One can also build a model in
which one of these quantities is tied to the other, for example by
imposing a tradeoff between transmissibility and virulence of a
disease. Different functional forms of such a relationship are
appropriate for modeling different ecosystems: host/pathogen,
prey/predator, sexual/parthenogenetic and so forth. As long as spatial
pattern formation occurs and organism type impacts on the environment
of descendants via ecosystem engineering, the shortcomings of
mean-field theory are relevant, as are limitations of pair
approximations~\cite{messinger2012}.

\subsection{Pair Approximations and Stability}
\label{sec:discuss-pa}
Pair approximations have been used to test for the existence of an
{\em Evolutionary Stable Strategy} (ESS) in a system---that is, a
strategy which, when established, cannot be successfully replaced by
another~\cite{szabo2007}.  In addition to the limitations of pair
approximation for representing patch structure~\cite{vanbaalen2000},
as we saw in the previous section, the question of whether a mutant
strain can initially grow is distinct from the question of whether
that strain achieves fixation or goes extinct~\cite{doebeli1997,
  lessard2005, antal2005, strayer2006, paley2007, fogle2008,
  ponciano2009, heilmann2010, smaldino2013}.  The former is a question
about short-term behavior, and the latter concerns effects apparent at
longer timescales.  This distinction is often lost or obscured in
analytical treatments.  The reason is that one typically tests whether
a new type can invade by linearizing the corresponding differential
equations at a point where its density is negligible. However, this
only reveals the initial growth rate (see the fixed-point eigenvalue
analysis in~\cite{vanbaalen1998, vanbaalen2000, aguiar2004,
  lion2008}).

Our analysis implies that pair approximations are inadequate for
analysis of systems with spatial inhomogeneity.  Even including
including triple and other higher-order corrections does not suffice,
as this series approximation is poorly behaved at phase
transitions~\cite{buice2009}.  Such higher order terms continue to
reflect only the local structure of the system and not the existence
of well separated areas that diverge in their genetic composition.
Nonequilibrium pattern formation will necessarily also be poorly
described, at least until the order of expansion reaches the
characteristic number of elements in a patch, or an area that
encompasses any relevant heterogeneity.  Given the algebraic intricacy
of higher-order corrections to pair approximations~\cite{aguiar2004,
lion2009, raghib2010}, it is useful to know in advance whether such
elaborations have a chance at success.  As approximation techniques
based on successively refining mean-field treatments are blind to
important phenomena, we therefore need to build our analytical work on
a different conceptual foundation.

\subsection{Percolation}
\label{sec:discuss-percolation}
The mathematical connection between pathogen--host and percolation
problems can provide insight into the difficulties in analytical
treatment of the biological problem. Spatial heterogeneity gives rise
to failure of traditional analytic treatments of percolation and a
need for new methodologies. Since the pathogen problem maps onto the
percolation problem under some circumstances, the same analytic
problems must arise in the biological context.  While the presence of
a nonequilibrium transition point indicates that traditional analysis
techniques fail, it raises the possibility that new tools from the
theory of phase transitions~\cite{hinrichsen2000, janssen2005,
  henkel2008, buice2009} will become applicable.  For example, in
Section~\ref{sec:percolation}, we saw that percolation theory enables
us to make quantitatively accurate predictions of population growth
and of the critical parameter values which divide one ecological
regime from another.  Indeed, specific important problems in public
health, such as the growth in number of individuals infected in a
pandemic, can be considered directly within the context of
percolation. Simulations of propagation on approximations of real world
networks may help provide accurate predictions, but the general
properties of disease propagation can be understood analytically.

Later, in Chapter~\ref{ch:volunteer}, we will see that percolation
models are also helpful in evolutionary game theory, where they yield
quantitative results near phase transitions, as they do for
host--consumer models.  Chapter~\ref{ch:doi} will develop the
mathematics necessary to understand percolation phase transitions.
Pursuing this topic in depth will take us into the subject of
statistical field theory.

\subsection{Adaptive Networks}
\label{sec:adaptive-networks}
Our results also have significance in the context of adaptive-network
research.  This field studies systems in which a network's wiring
pattern and the states of its nodes change in interrelated ways.
Prior modeling efforts have considered epidemics on adaptive networks,
where the spread of the disease {\em through} the network changes the
connections {\em of} the network~\cite{gross2006, do2009, shaw2009,
shaw2010, kamp2010, vansegbroeck2010, wu2010, fehl2011}.  In such
models, if a susceptible node has an infected neighbor, it can break
that connection by rewiring to another susceptible node.  A key point
in the analysis is that the new neighbor is chosen at random from the
eligible population.  This choice of rewiring scheme is exactly what
makes a pair approximation work for that epidemic model, because it
eliminates higher-order correlations in the system~\cite{do2009}.
(Chapter~\ref{ch:invasion-fitness} will develop in detail how this
method of rewiring allows us to write differential equations for the
model.)  In our system, by contrast, hosts can form new connections by
reproducing into empty sites, but these contacts can only connect
geographically proximate individuals.

The difference we have seen between lattice behavior on one hand and
RRG or swapping-enabled behavior on the other emphasizes the need to
study the effect of spatial proximity on link rewiring.  While the
structure-erasing nature of unconstrained rewiring among susceptible
hosts has been acknowledged~\cite{kamp2010, vansegbroeck2010}, new
rewiring rules which reflect spatial and community structure have yet
to be systematically investigated.  The reason that they have not is
naturally related to the need for different analytic approaches.
``Myopic'' rewiring rules, such as restricting the set of eligible new
partners to the neighbors of a node's current partners, have on
occasion been considered, but in contexts other than epidemiology,
like evolutionary game theory~\cite{traulsen2009, holme2009}, making
the endeavour of exploring such rules in epidemic models all the more
worth pursuing.

\subsection{Conclusions}
Fisher \cite{fisher1930} introduced modern genetic theory in large
part motivated by the need to describe the existence of
biodiversity.\footnote{His \emph{Genetical Theory of Natural
Selection} also includes the interesting remark, ``No practical
biologist interested in sexual reproduction would be led to work out
the detailed consequences experienced by organisms having three or
more sexes; yet what else should he do if he wishes to understand why
the sexes are, in fact, always two?''  Biology replies, ``Always
two?''~\cite{ainsworth2015}} However, the expressions he described
which apply in panmictic populations and mean-field treatments lead to
a population genetics that rapidly converges to homogeneous
populations.  Spatial extents and their violation of the mean-field
approximations are a key to biodiversity in nature.  Their proper
theoretical treatment will be a large step forward for evolutionary
biology.

Most laboratory experiments, guided by traditional evolutionary
thinking, have used well-mixed populations.  The results obtained are
consistent with theoretical analysis precisely because the conditions
are consistent with those assumptions.  Such experiments do not
provide insight into the role of spatial extent and the implications
for real-world biological populations.  A growing number of
experiments today are going beyond such conditions and, as is to be
expected, are obtaining quite different results~\cite{kerr2006,
  xavier2007, davis2008, ponciano2009, heilmann2010, goenci2010,
  salkeld2010, reigada2012, wodarz2012, allen2013}.

Mean-field models are often helpful as a first step towards
understanding the behavior of systems, but we cannot trust them to
provide a complete story, and we should not let mean-field thinking
furnish all the concepts we use to reason about evolutionary dynamics.
Our analysis of transplanting organisms can be considered parallel to
real world concerns and manifest effects of invasive species
introduced by human activity and the impact of shipping and air
transportation on pathogen evolution~\cite{strayer2006, rauch2006}.
These are among the well-established examples of situations in which
spatial extent influences evolutionary dynamics~\cite{tgoei2000,
ponciano2009, aguiar2009, goenci2010, aguiar2011, martins2013,
vandyken2013, baptestini2013, allen2015, werfel2015a, werfel2015b}.
Identifying specific implications of the issues explored in this
chapter for particular biological systems~\cite{kerr2006, xavier2007,
davis2008, ponciano2009, heilmann2010, goenci2010, salkeld2010,
reigada2012, wodarz2012, allen2013, momeni2013, lambert2014,
vangestel2014, carlssongraner2015} requires field and laboratory work,
as well as theoretical insight to guide the questions that are being
asked.

\vfill

\emph{This chapter is based in part on a paper by myself, Andreas Gros and Yaneer
 Bar-\!Yam, originally published in October 2011 and updated at the
 beginning of 2014~\cite{stacey2014}.  The research reported in that
 paper was a project I initiated, building on earlier work by
 Bar-\!Yam, de Aguiar, Rauch, Sayama, Werfel and others.  I wrote the
 first version of the basic simulation code; Andi figured out how to
 distribute the work across multiple computers and implemented the
 organism-swapping idea, which I had after a conversation with him
 about something tangentially related.  The paper-writing process was
 a collaboration, in which I produced most of the words and my
 coauthors provided the Darwinian editorial pressure.}

\chapter{A Volunteer's Dilemma}

\label{ch:volunteer}
The ability of any organism to survive and produce viable offspring
depends on its environment, and that environment consists in
significant part other living beings.  In Chapter~\ref{ch:redgreen},
we studied this in an evolutionary system based on ecological
competition:  success required having food to eat.  Now, we turn to
evolutionary dynamics defined using \emph{game theory.}  Interactions
among organisms will be represented by mathematical games, and
reproductive success will depend on the numerical payoffs won when
individuals play those games together.

We can think of the evolutionary game which will be our focus for this
chapter as a realization of the general idea discussed at the end of
Chapter~\ref{ch:multiscale}.  This game, which we designate
the \emph{Volunteer's Dilemma,} is a scenario in which organisms must
act in concert to achieve success in the struggle for life.

\section{Well-Mixed Populations with Carrying Capacity}
\label{sec:vs-panmixia}
We introduce the Volunteer's Dilemma in the context of two species
living in the same environment.  The ecosystem is well-mixed, so we
can describe it by population densities.  The total number of
individuals is restricted, so that the sum of the population densities
is bounded by unity:
\begin{equation}
v + s \leq 1.
\end{equation}
Here, $v$ denotes the population density of Volunteers, and $s$ is the
population density of Slackers.  The game is played in the following
way: a group of $K \geq 3$ agents is assembled by randomly drawing
from the population.  Volunteers pay a cost, $c$.  If all the agents
in the group volunteer, then they each gain a benefit, $b$.  A single
agent slacking off deprives all the agents from gaining the benefit.
Note that we are here taking the ``benefit'' and ``cost'' of
strategies as given parameters of the model, rather than as emergent
consequences derived from some more fundamental dynamics.  This is
unlike the way in which the idea of competition is realized in the
spatial host--consumer system of Chapter~\ref{ch:redgreen}; we can
think of these two different styles of modeling as complementary.

This game is also known as an $n$-player \emph{stag
hunt}~\cite{skyrms2000, pacheco2009, vanveelen2012b}.  We will stick
with the Volunteer's Dilemma terminology in this chapter in order to
avoid ambiguity: the stag hunt game is typically defined as a
two-player interaction, and so ``$n$ people play a stag-hunt game''
could be interpreted as one person engaging in two-player stag hunts
with $n - 1$ others, and then receiving the total payoff from these $n
- 1$ separate games.

Reproduction in this environment depends upon the availability of
empty space, so the probability of a reproduction event diminishes as
the total population density $v + s$ increases.  We take a pessimistic
view of volunteerism: the cost of being a volunteer must be paid
whether or not reproduction happens.  This is an idealization of a
case where, for example, volunteering requires additional metabolic
products which demand heightened energy expenditure regardless of the
availability of space to reproduce into.  Consequently, we treat the
cost $c$ as augmenting the death rate $d$, while the benefit $b$
augments the growth rate, which is also modulated by the total
population density.  For convenience, we define $k = K - 1$ to be the
number of \emph{additional} Volunteers which a Volunteer must have in
its peer group in order to obtain the benefit.  We write the following
coupled equations for the time evolution of the ecosystem:
\begin{align}
\dot{v} &= -(d+c)v + \left[g + b\left(\frac{v}{v+s}\right)^k\right]
                     v(1 - v - s), \label{eq:vdot} \\
\dot{s} &= -ds + gs(1 - v - s). \label{eq:sdot}
\end{align}
We follow Durrett and Levin~\cite{durrett1994} in having the cost and
benefit parameters directly modify the relevant rates.  The parameters
$b$, $c$, $d$ and $g$ are all positive.  The increase in Volunteer
reproduction due to the benefit $b$ depends on how likely it is that a
pool of $k$ agents will contain only Volunteers.  A generalization is
possible to a case where the size $n$ of the interacting group is
larger than the critical number $k$ of Volunteers required to gain the
benefit.  This would introduce a term of the form
\begin{equation}
b\sum_{j=k}^n \binom{n}{j} \left(\frac{s}{v+s}\right)^{n-j}
             \left(\frac{v}{v+s}\right)^j.
\end{equation}

All along the line $v + s = 1$, the phase-space flow is inward,
because both $\dot{v}$ and $\dot{s}$ are negative.  Equations
(\ref{eq:vdot}) and (\ref{eq:sdot}) have an obvious fixed point at the
origin.  If $v = 0$, then Eq.~(\ref{eq:sdot}) reduces to
\begin{equation}
\dot{s} = -ds + gs(1 - s),
\end{equation}
which has a stable equilibrium at
\begin{equation}
s^* = 1 - \frac{d}{g}.
\end{equation}
Likewise, if $s = 0$, then the population is all Volunteers, and
\begin{equation}
\dot{v} = -(d+c)v + (g+b)v(1-v),
\end{equation}
which has a stable equilibrium located on the $v$-axis at
\begin{equation}
v^* = 1 - \frac{d+c}{g+b}.
\label{eq:vstar}
\end{equation}
When we move off the axes and consider the $v$--$s$ plane, then the
all-Slacker fixed point is a stable node, while the all-Volunteer
fixed point at~$(v^*,0)$ may or may not be stable.  Its stability
depends on whether or not there exists another fixed point in the
off-axis region.

If both $v$ and $s$ are nonzero, then setting $\dot{s} = 0$ in
Eq.~(\ref{eq:sdot}) shows that
\begin{equation}
(v^{**} + s^{**}) = 1 - \frac{d}{g}.
\end{equation}
Setting $\dot{v} = 0$ in Eq.~(\ref{eq:vdot}) and substituting in this
expression for the total population density yields
\begin{equation}
\left(\frac{v^{**}}{v^{**} + s^{**}}\right)^k
 = \frac{cg}{bd}.
\end{equation}
The left-hand side must be less than or equal to unity, so for this to
be valid, we must have $cg \leq bd$, or
\begin{equation}
\frac{b}{c} \geq \frac{g}{d}.
\label{eq:volunteer-boverc}
\end{equation}
If this is satisfied, then
\begin{equation}
v^{**} = \left(\frac{cg}{bd}\right)^{1/k}
        \left(1 - \frac{d}{g}\right).
\end{equation}

The fixed point at $(v^{**},s^{**})$ is a saddle point, and if it
exists, then the all-Volunteer equilibrium at~$(v^*,0)$ is a stable
node.  Otherwise, the equilibrium $(v^*,0)$ is a saddle point,
unstable to an off-axis push.  The location of the all-Volunteer
equilibrium on the $v$-axis is independent of the group size $k$;
however, its \emph{basin of attraction} is not.  The region within
which evolutionary trajectories go to~$(v^*,0)$ depends on the
location of the coexistence equilibrium, $(v^{**},s^{**})$, and as we
have just seen, both $v^{**}$ and $s^{**}$ depend on~$k$.  This is
another appearance of the theme we encountered in
Chapter~\ref{ch:redgreen}: the stability or instability of fixed
points does not tell us everything we need to know about evolutionary
dynamics.  Note that there is no assortment of like types in this
model: the groups within which individuals interact form at random,
without regard to types.  Nor can the effects of social behaviors on
individual fitness be decomposed into a linear combination of pairwise
interactions.  The success or failure of the Volunteer type is a
genuinely and irreducibly synergistic effect.

We will return to this point and see what it means for the traditional
language of population biology in Chapter~\ref{ch:price-mls}.  In that
context, we will see that the stability criterion we have here,
Eq.~(\ref{eq:volunteer-boverc}), has both familiar and surprising
features.  To put the matter briefly, that the rule takes the ratio of
a benefit parameter to a cost one is a commonplace
attribute~\cite{allen2015b}, but what the rule compares that ratio to
is not.  Furthermore, that the rule does not depend on any kind of
assortment among genetically similar individuals is, from a
traditional perspective, a surprising outcome.  We can interpret this
rule as saying that in an ecosystem containing one Slacker and many
Volunteers, if a group forms which contains the Slacker, that group
will on the whole perform worse than those comprising only Volunteers.
Consequently, the Slacker strategy will be penalized by natural
selection, if $b/c$ is sufficiently large.

\begin{figure}[h]
\includegraphics[width=6.5cm]{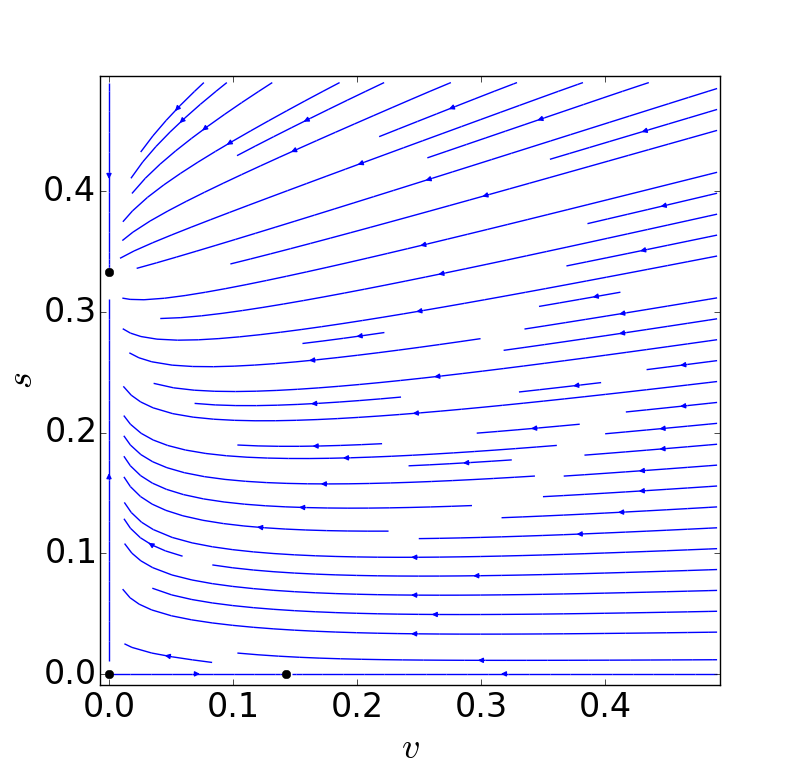}
\includegraphics[width=6.5cm]{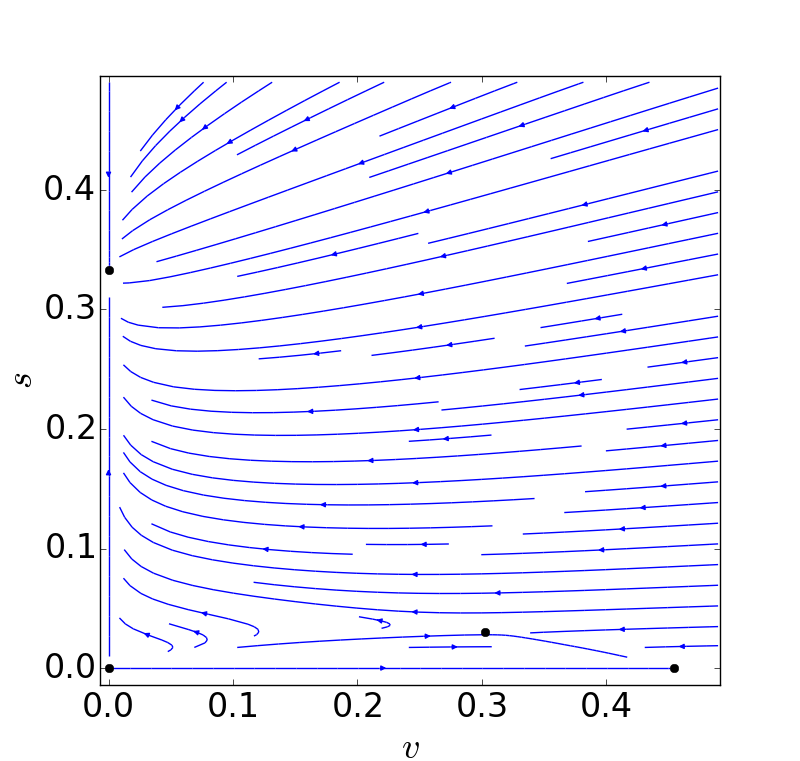}
\caption{\label{fig:vs-phaseplots} {\bf (Left)} Phase portrait for the
continuous, well-mixed Volunteer's Dilemma ecosystem, in the regime
where Volunteerism is an unstable strategy ($g = 0.15$, $d = 0.1$, $b
= c = 0.2$, $k = 3$).  Note that of the three fixed points, the origin
is unstable, the all-Slacker equilibrium is stable, and the
all-Volunteer equilibrium is only stable along the $v$ axis.  {\bf
(Right)} Phase portrait in the regime where Volunteerism is stable ($g
= 0.15$, $d = 0.1$, $b = 0.4$, $c = 0.2$, $k = 3$).  Note the presence
of an additional fixed point.}
\end{figure}

We now establish the stability conditions for the all-Volunteer
equilibrium explicitly, by linearizing the system's dynamics near that
fixed point.  For brevity, we define the abbreviations
\begin{equation}
x = v + s,\ p = \left(\frac{v}{v+s}\right)^k.
\end{equation}
The Jacobian of the dynamics defined by Eqs.~(\ref{eq:vdot}) and
(\ref{eq:sdot}) is, in terms of these new variables,
\begin{equation}
J = \left(\begin{array}{cc}
 -(d+c) + \frac{bkps}{x}(1-x) + (g+bp)(1-x-v)
 & -\frac{bkps}{x}(1-x) - v(g+bp) \\
 -gs
 &
 -d + g(1 - x - s)
\end{array}\right).
\end{equation}
At the all-Volunteer equilibrium, all terms proportional to~$s$ drop
out, and the Jacobian reduces to a conveniently upper-triangular form:
\begin{equation}
J = \left(\begin{array}{cc}
(d+c) - (g+b) & (d+c) - (g+b) \\
 0 & -d + \frac{g(d+c)}{g+b}
\end{array}\right).
\end{equation}
The eigenvalues of this matrix are
\begin{align}
\lambda_1 &= (d+c) - (g+b), \\
\lambda_2 &= -d + \frac{g(d+c)}{g+b}.
\end{align}
From Eq.~(\ref{eq:vstar}), we know that $d+c$ is always less than
$g+b$, if $v^*$ is a valid equilibrium distinct from the origin.
Therefore, $\lambda_1$ is guaranteed to be negative.  The condition
for $\lambda_2$ to be negative is exactly
Eq.~(\ref{eq:volunteer-boverc}).  As promised, the all-Volunteer
configuration is a stable one, provided that the ratio of the benefit
$b$ to the cost $c$ is greater than $g/d$, the ratio of the baseline
growth and death rates.  This rule has the appealing feature that the
game-payoff parameters appear on one side, and the baseline rates on
the other.

Contrast this with what happens if we define the dynamics using the
Prisoner's Dilemma instead.  This game, more widely studied than the
Volunteer's Dilemma, is a two-player game, rather than a game for an
arbitrarily large number of simultaneous players.  We recall its
definition from Chapter~\ref{ch:multiscale}: Again, there are two
strategies, which we can designate by \emph{Valiant}
and \emph{Slinker.}  (This terminology is nonstandard, but it lets us
keep the same variable names.)  Also as before, the interaction
payoffs depend upon two parameters, which we can call $b$ and $c$.
Valiant individuals pay a cost $c$ and gain a benefit $b$ if their
interaction partner is also Valiant.  Meanwhile, Slinkers gain the
same benefit from playing with a Valiant, but pay no cost.  A Slinker
who plays against another Slinker pays nothing and gains nothing.  If
individuals play with more than one partner during their lifespans,
the total benefit they accrue is the sum of the payoffs gained in each
instance of the game.  We can, therefore, introduce a group-size
parameter $k$, but unlike before, the dependence of growth rates
on~$k$ will be linear.

As before, we begin by defining a two-variable dynamical system, for
which we will then find equilibrium points.  Also as before, we let
the cost parameter modify the death rate, while the benefit parameter
modifies the growth rate.  The analogues to
Eqs.~(\ref{eq:vdot}) and (\ref{eq:sdot}) are
\begin{align}
\dot{v} &= -(d+c)v + \left(g + \frac{kbv}{v+s}\right)v(1-v-s), \\
\dot{s} &= -ds + \left(g + \frac{kbv}{v+s}\right)s(1-v-s).
\end{align}

This dynamical system has nonzero fixed points at
\begin{align}
v = 0, &\ s = 1 - \frac{d}{g}, \\
s = 0, &\ v = 1 - \frac{d+c}{g+kb}.
\end{align}
An equilibrium with both $v$ and $s$ nonzero is only possible if $c$
vanishes, which is an uninteresting case.

Can a population of Valiants withstand an intrusion by Slinkers?  At
the all-Valiant equilibrium, the Jacobian is, by straightforward
calculation,
\begin{equation}
J = \left(\begin{array}{cc}
          (d+c)-(g+kb) & (d+c)-(g+kb) \\
          0 & c
          \end{array}\right).
\end{equation}
The eigenvalues are $(d+c)-(g+kb)$, which is guaranteed to be negative,
and $c$, which is always positive.  Therefore, the all-Valiant
equilibrium is \emph{never} stable to off-axis perturbations.

Consequently, we see that the socially cooperative strategy is
\emph{always} vulnerable to invasion in the Prisoner's Dilemma,
whereas in the Volunteer's Dilemma, it can be stable if the costs and
benefits of playing the game satisfy a criterion which depends on the
death and growth rates.  This underlines the importance of studying a
variety of games, both two- and multiplayer, in evolutionary game
theory.  If we had confined our attention to the Prisoner's Dilemma,
we would have missed a dynamical system with interesting features.

\section{Volunteer's Dilemma in a Networked Population}
\label{sec:vs-lattice}
It has been suggested \cite{archetti2011,archetti2011b} that this kind
of nonlinearity can explain the kinds of evolutionary outcomes which
are usually taken as requiring ``assortment'' among genetically
similar individuals.  Spatial structure is one way to create such
assortment, since limited mobility means that geographically proximate
organisms are more likely to be genetically related as well.  However,
there is no fundamental reason why nature should not present us with
\emph{both} nonlinearity \emph{and} spatial structure, so it is only
natural to see what happens when both occur together.

There are multiple ways to incorporate spatial structure into the
evolutionary game dynamics we have defined here.  We will address two
possibilities in turn.  The first method essentially takes the
dynamical system defined by Eqs.~(\ref{eq:vdot}) and (\ref{eq:sdot})
and spreads it over a lattice.  We start with an $L \times L$ square
lattice, and we specify that each lattice site can be empty ($0$),
occupied by a Volunteer ($V$) or occupied by a Slacker ($S$).  Our
simulation proceeds in discrete time.  At each time step, we pick a
site at random.  If the chosen site is empty, we do nothing.  If it
contains a Slacker, we pick a neighboring site at random, and if that
site is empty, the Slacker can reproduce into it with probability $g$.
Volunteers reproduce likewise, except that their probability of
budding into an adjacent empty site depends on how many of their
neighbors are also Volunteers.  Specifically, if at least two of the
neighboring sites contain Volunteers, the baseline reproduction
probability $g$ is augmented by an amount $b$.  Next, if the
individual we chose is a Slacker, we kill it off with probability $d$,
and if it is a Volunteer, we kill it with probability $d + c$.  One
generation is defined to have passed when $L^2$ sites have been
sampled.

\begin{figure}[h]
\includegraphics[width=6cm]{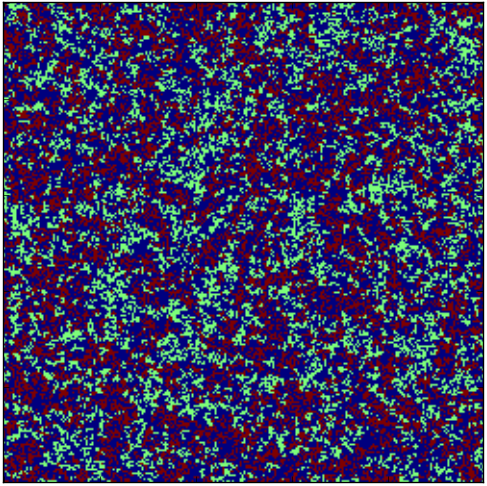}
\includegraphics[width=6cm]{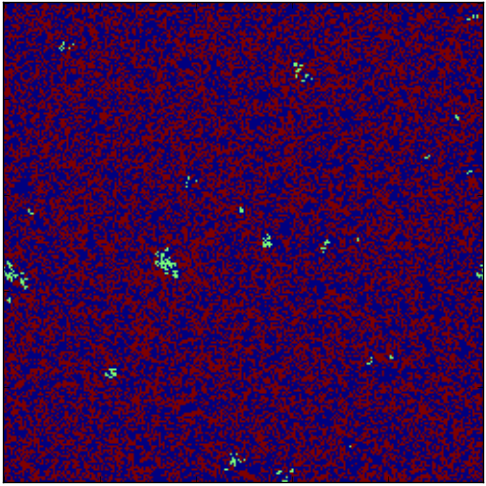}
\caption{\label{fig:vs-grid-L250-d0p1-b0p4-c0p2-g0p2} \textbf{(Left)}
  Snapshot of the Volunteer's Dilemma simulation on a $250 \times 250$
  lattice.  Red sites indicate Volunteers, cyan sites indicate
  Slackers and blue sites are empty space.  In this example, the death
  rate is $d = 0.1$, the baseline growth rate is $g = 0.2$, the cost
  of volunteering is $c = 0.2$ and the benefit is $b = 0.4$.  The
  sites were initialized at random for this illustration and the
  simulation run for 100 generations. \textbf{(Right)} Another
  snapshot of the Volunteer's Dilemma on a $250 \times 250$ lattice,
  taken $10^3$ generations into the simulation.  The parameters are
  the same as at left.  Note that Slackers are much less common, but
  have yet to disappear completely.}
\end{figure}

Figure~\ref{fig:vs-grid-L250-d0p1-b0p4-c0p2-g0p2} shows what can
happen when we implement these stochastic dynamical rules in a
simulation.  Having constructed the spatial model in this way, we can
analyze it following the ideas presented in the previous chapter.

\begin{figure}[h]
\includegraphics[width=12cm]{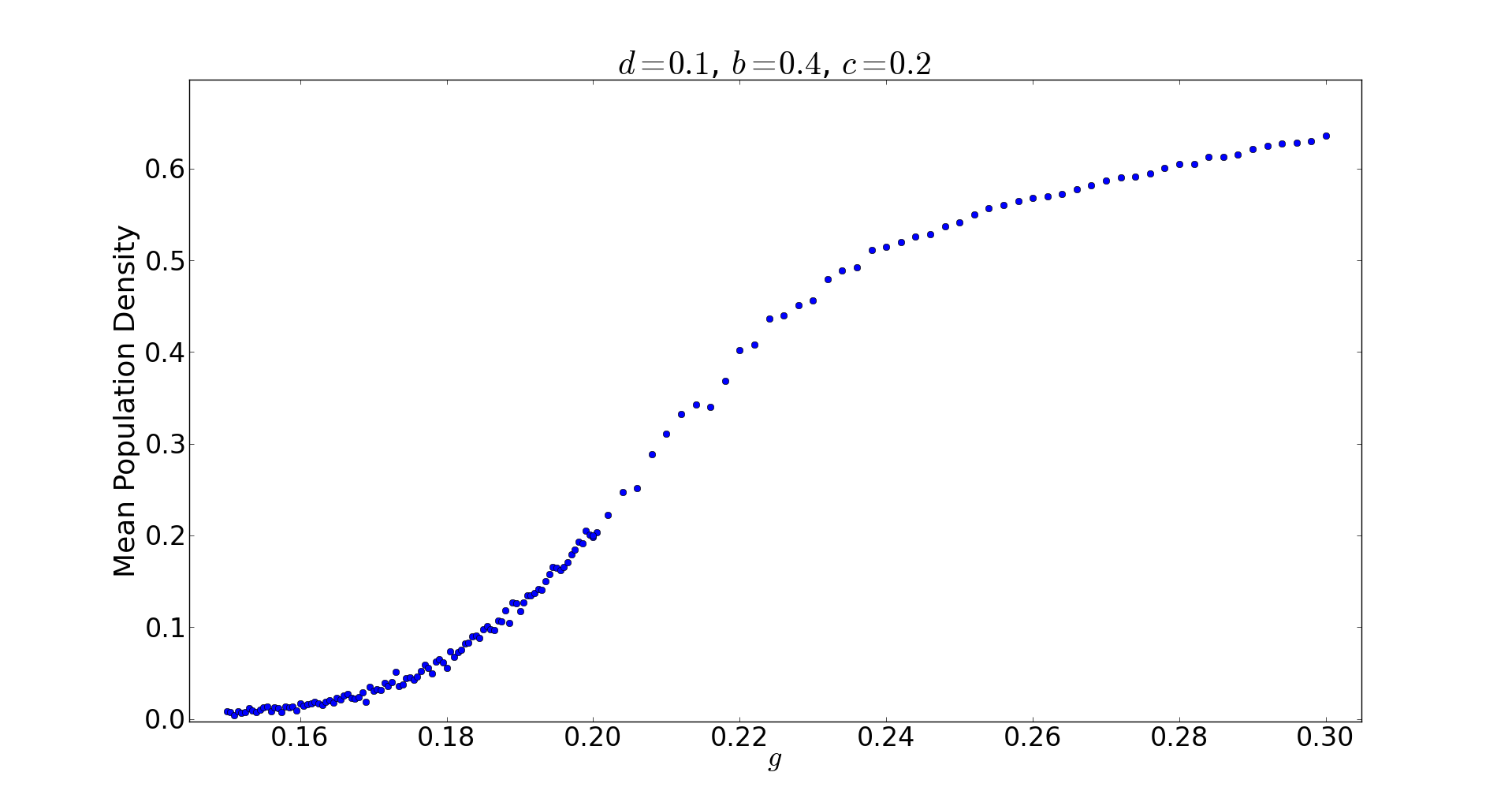}
\includegraphics[width=12cm]{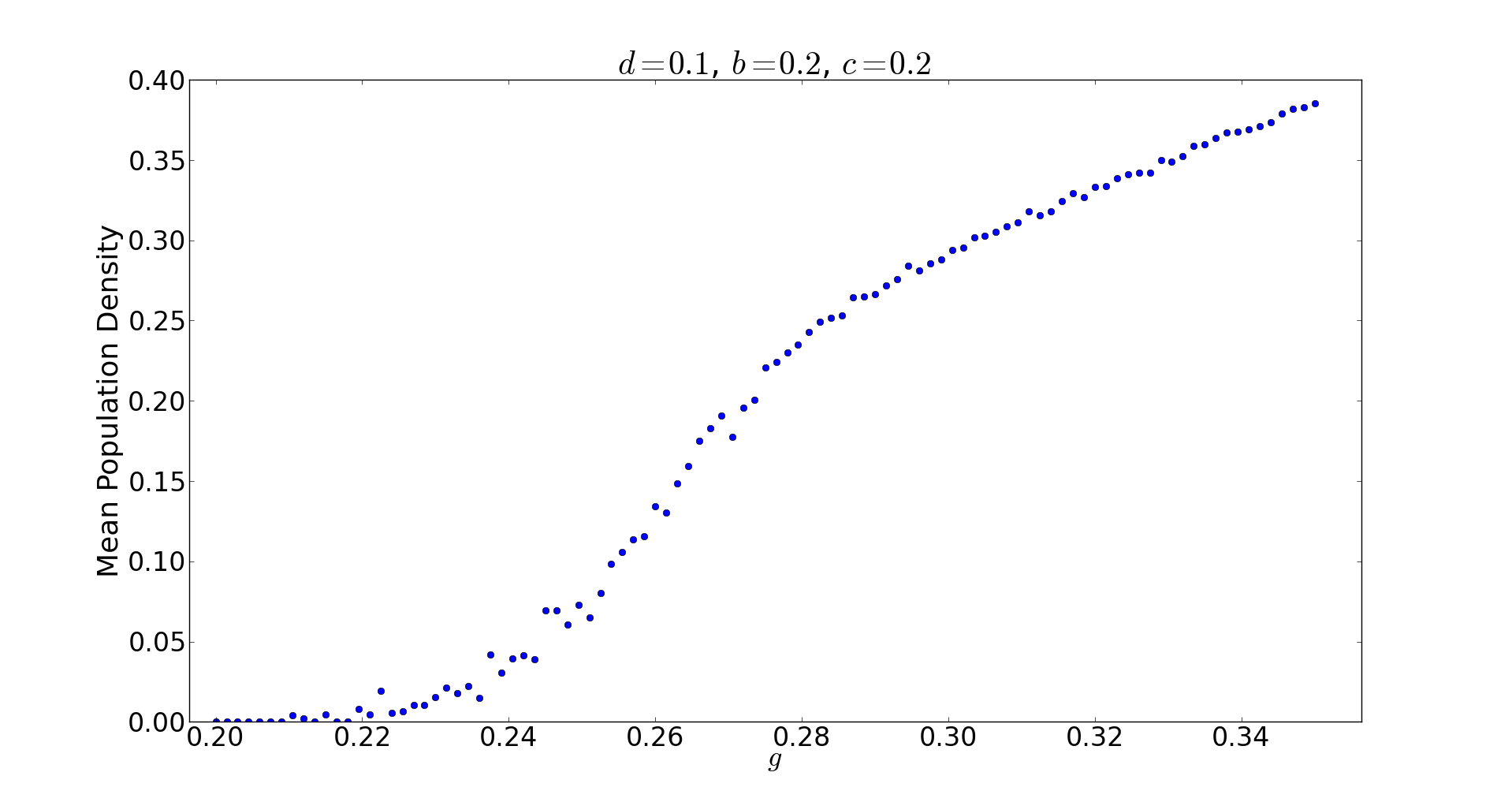}
\caption{\label{fig:pop-density-d0p1} \textbf{(Top)} Average Slacker density
  of surviving populations for $d = 0.1$ (and $b = 0.4$, $c = 0.2$,
  but those don't matter for Slackers). \textbf{(Bottom)} Average
  Volunteer density of surviving populations for $d = 0.1$, $b = 0.2$
  and $c = 0.2$.  In both panels, results were computed with 100
  trials per point; each trial was run for 500 generations.}
\end{figure}

Consider a single Slacker placed in an otherwise empty lattice.  It
will die, leaving an empty space behind, with probability $d$.  If the
population is not to vanish, the organism needs to reproduce before
then.  For budding to be more probable than dying, we must have $g >
d$.  We expect on general grounds that in practice the critical value
of~$g$ for having a sustainable population will be somewhat higher
than~$d$, thanks to the basic stochasticity of the system.  This is
known as ``fluctuations suppressing the active state.''  Simulation
results, illustrated in Figure~\ref{fig:pop-density-d0p1},
bear out this expectation.  In each trial, a Slacker population is
initialized with a single individual in an otherwise empty world.  The
number of trials in which the population survives for 500 generations
only exceeds 10\% for $g > 0.15$.  Furthermore, the population density
of those trials where the species does survive only grows to
significant amounts when $g$ is substantially greater than $d$.

This pattern repeats for Volunteers: a population grown from an
initial seed individual only becomes viable when $g + b$ is
significantly larger than $d + c$, as illustrated in
Figure~\ref{fig:pop-density-d0p1}.  Again, we see the effect
of stochastic fluctuations, which suppress the active phase of the
system.  Recall that thresholds were also elevated above mean-field
expectations in the spatial host--consumer system
(Chapter~\ref{ch:redgreen}, Figures \ref{fig:minmax-6apr2011} and
\ref{fig:percolation}).

\begin{figure}[h]
\includegraphics[width=12cm]{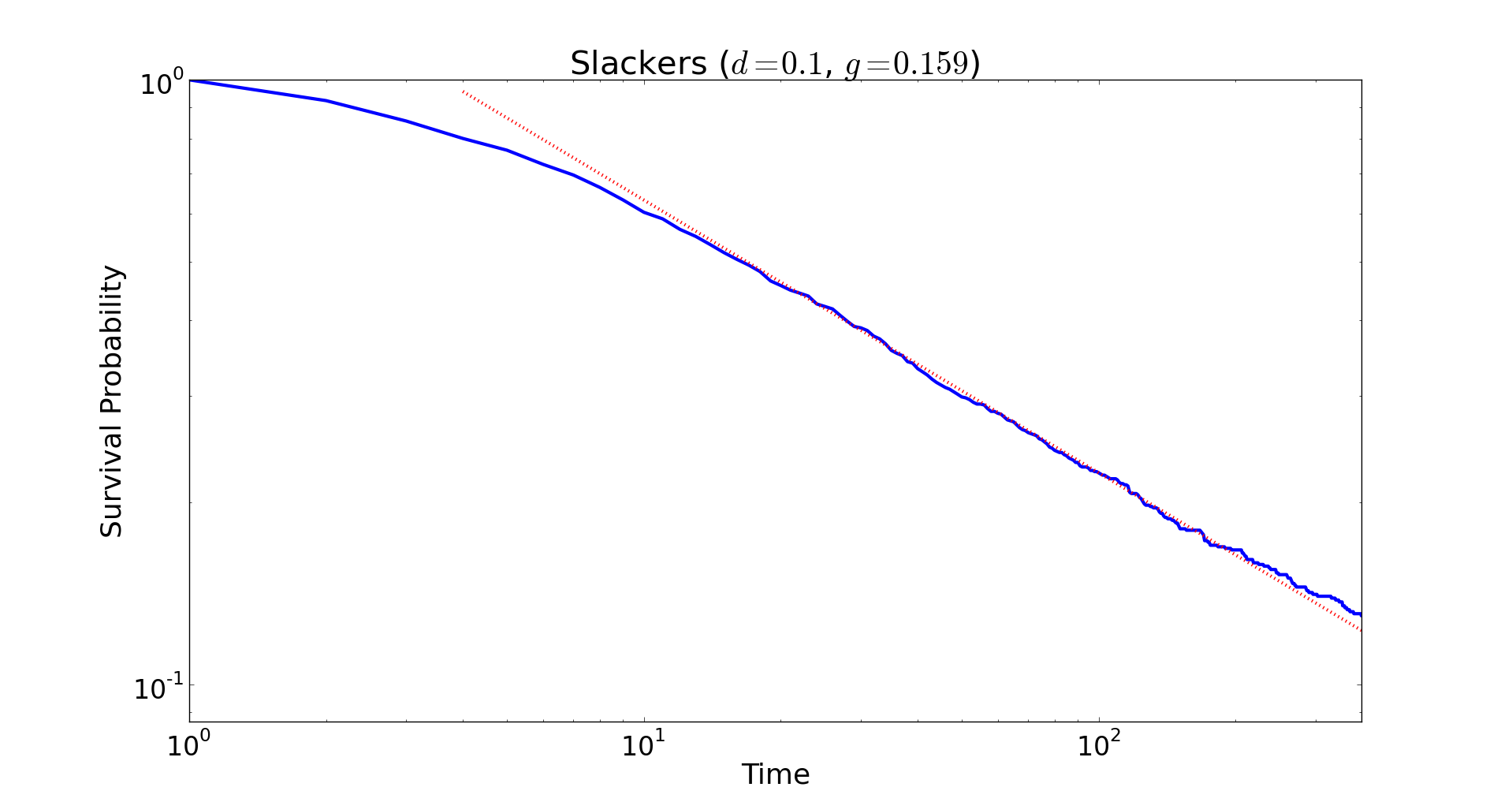}
\caption{\label{fig:survival-prob-slackers} Survival probability for a
  Slacker population initialized with a single seed on a $250 \times
  250$ lattice ($d = 0.1$, $g = 0.159$). The dotted line shows the
  characteristic fall-off expected for the directed-percolation
  universality class, a power-law decay with exponent $-0.451$. At
  longer times, an upward deviation indicates a finite-size effect.}
\end{figure}

\begin{figure}[h]
\includegraphics[width=12cm]{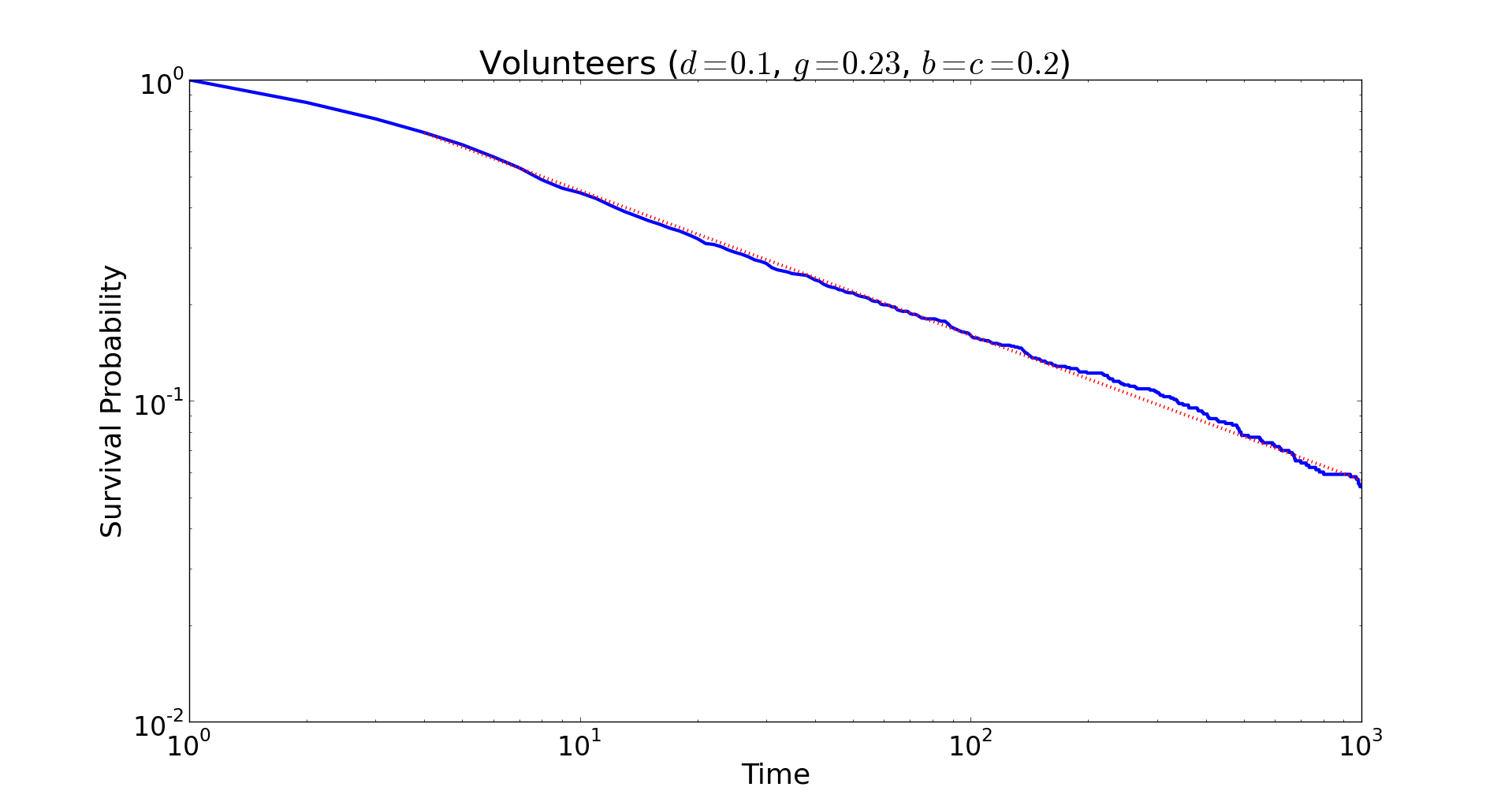}
\caption{\label{fig:survival-prob-volunteers} Survival probability for a
  Volunteer population initialized with a single seed on a $250 \times
  250$ lattice ($d = 0.1$, $g = 0.23$, $b = c = 0.2$). As with the
  Slacker population in Figure~\ref{fig:survival-prob-slackers}, the
  dotted line shows a fall-off characteristic of the
  directed-percolation universality class, a power-law decay with
  exponent $-0.451$.}
\end{figure}

Furthermore, we can understand the behavior near the transition point
using percolation theory.  The most convenient quantity for numerical
exploration is $P(t)$, the probability that a population grown
beginning with a single seed at time $0$ is still alive at time $t$.
We plot a survival probability curve for Slackers in
Figure~\ref{fig:survival-prob-slackers}.  The probability fall-off
matches that expected for a process which is near criticality in the
directed-percolation universality class.
Figure~\ref{fig:survival-prob-volunteers} shows that the same holds
true for a population of Volunteers.

\begin{figure}[h]
\includegraphics[width=12cm]{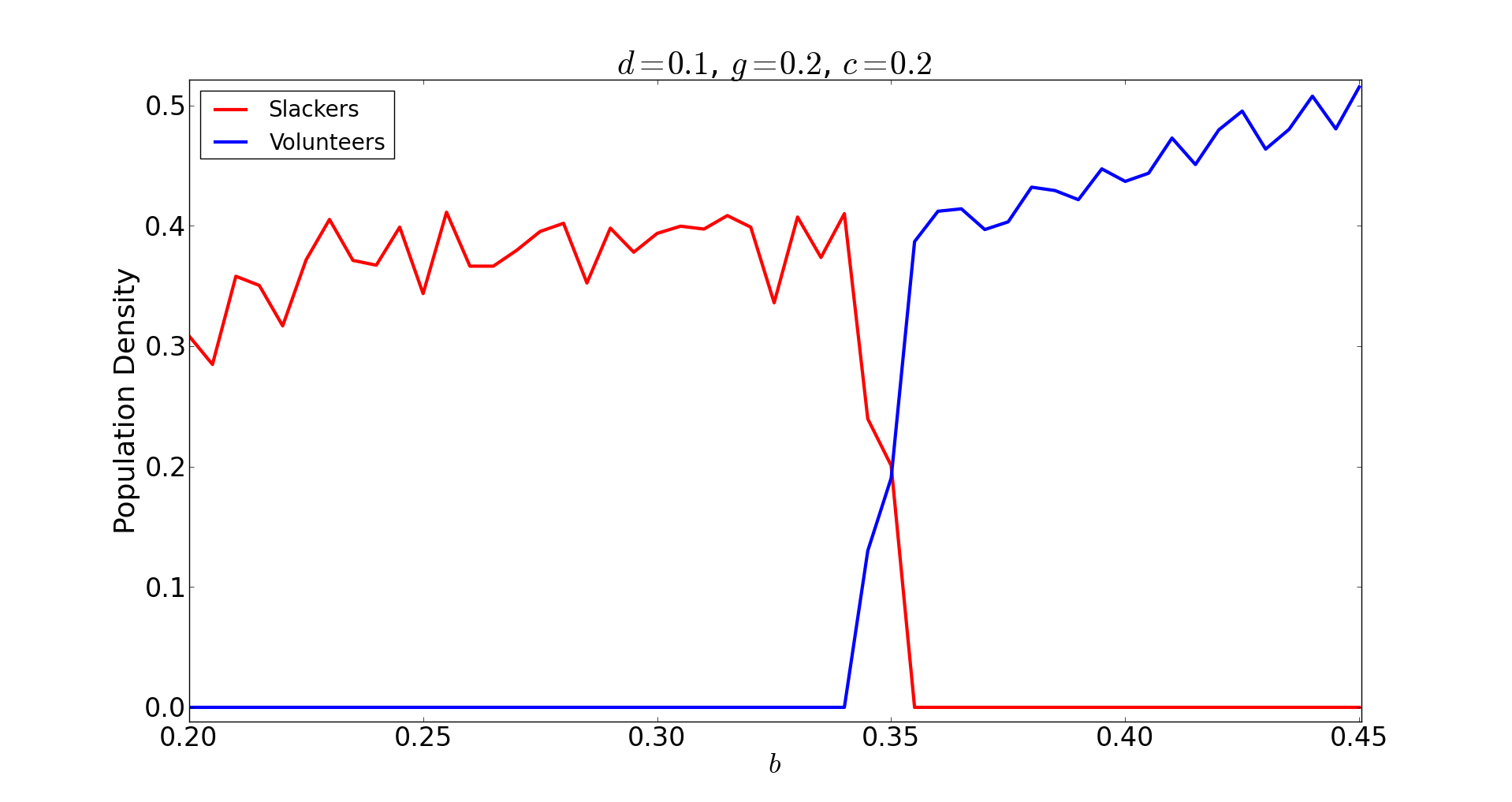}
\caption{\label{fig:population-densities-d0p1-g0p2-c0p2} Population
densities of Volunteers and of Slackers after 10,000 generations or
the extinction of the other species, whichever comes first. ($d =
0.1$, $g = 0.2$, $c = 0.2$.)}
\end{figure}

We can understand when Volunteerism is favored over Slacking by seeing
what happens when both species are present on the same lattice.  We
begin by filling the lattice at random with 5\% Slackers and 5\%
Volunteers, leaving it otherwise empty.
Figure~\ref{fig:population-densities-d0p1-g0p2-c0p2} indicates that as
the benefit parameter $b$ is increased, Volunteers go from unfavored
to favored.  Note that the crossover occurs when $b/c \approx 1.75$,
which is \emph{less} than the crossover point computed in the
mean-field model: per Eq.~(\ref{eq:volunteer-boverc}), the mean-field
critical ratio is given by $b/c = g/d = 2.0$.  Spatial structure also
promotes Volunteerism in another way.  In the well-mixed model, an
even balance between Slackers and Volunteers will evolve towards the
all-Slacker fixed point, even if the all-Volunteer fixed point is
stable, unless $b$ is increased still further.  That is, points on the
line $s = v$ can lie in the all-Slacker equilibrium point's basin of
attraction.  So, even though the all-Volunteer configuration is stable
against invasion, the Volunteer strategy does not predominate when
starting from an even balance.  However, in the spatial version, this
can happen easily.

\begin{figure}[h]
\includegraphics[width=12cm]{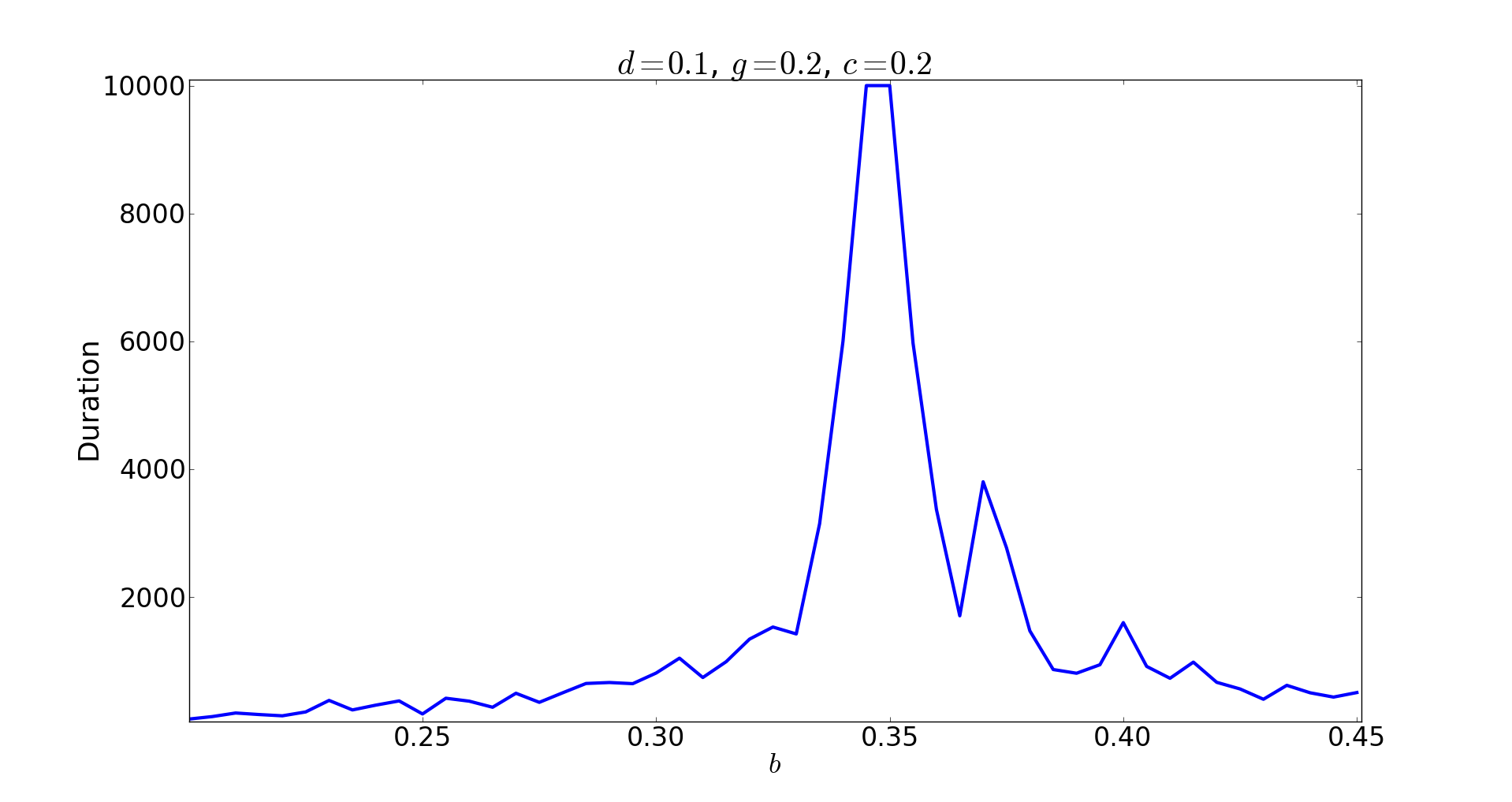}
\caption{\label{fig:durations-d0p1-g0p2-c0p2} Time until simulation
completion, set to be 10,000 generations or the extinction of a
species, whichever comes first. ($d = 0.1$, $g = 0.2$, $c = 0.2$.)}
\end{figure}

\begin{figure}[h]
\includegraphics[width=12cm]{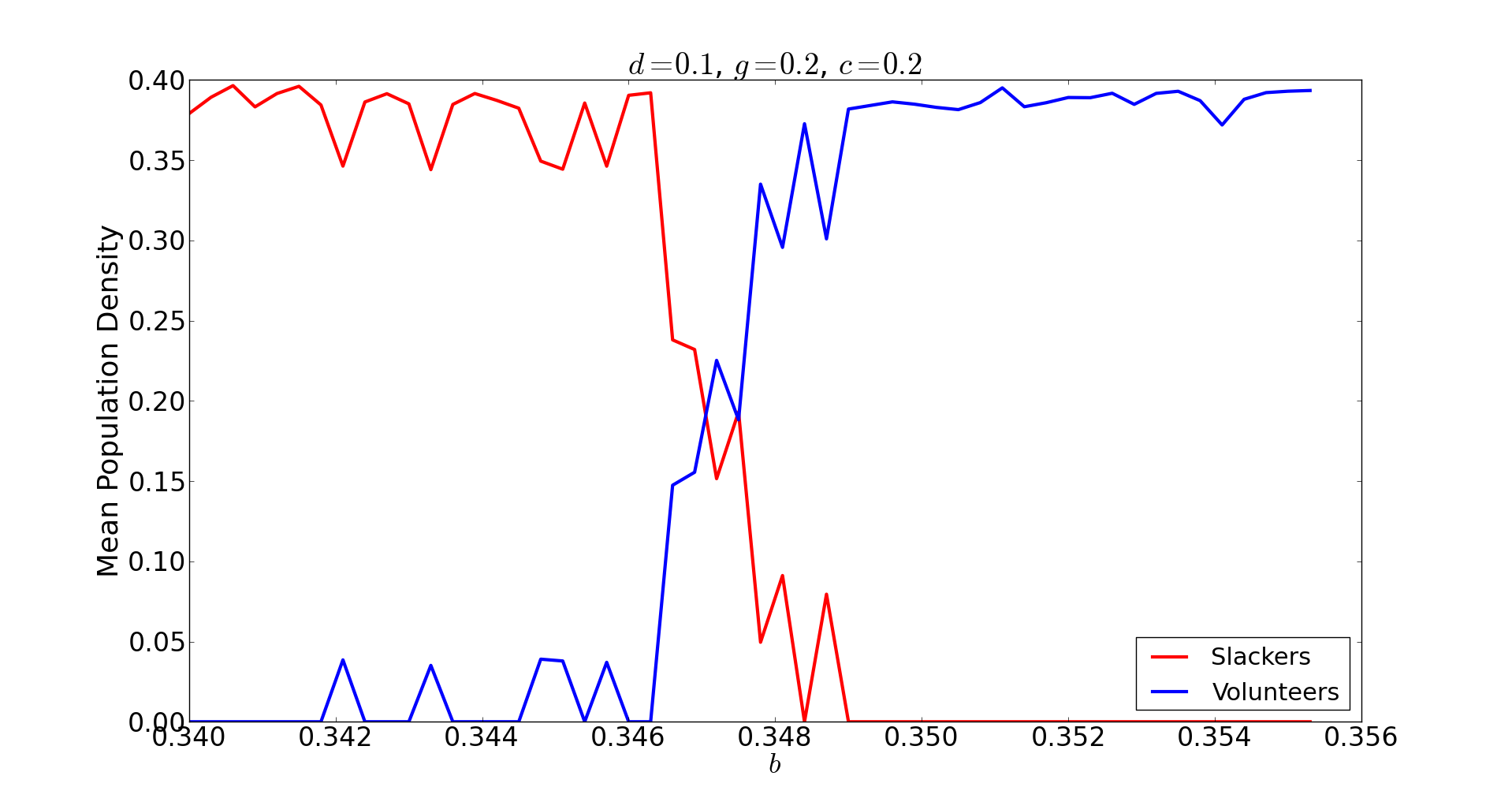}
\caption{\label{fig:pop-densities-d0p1-g0p2-c0p2-detail} Population
density of Slackers and of Volunteers, recorded when one of the two
populations goes extinct. Computed on a $50 \times 50$ lattice, with
10 trials for each value of~$b$. (The other parameters were fixed at
$d = 0.1$, $g = 0.2$, $c = 0.2$.)}
\end{figure}

\begin{figure}[h]
\includegraphics[width=12cm]{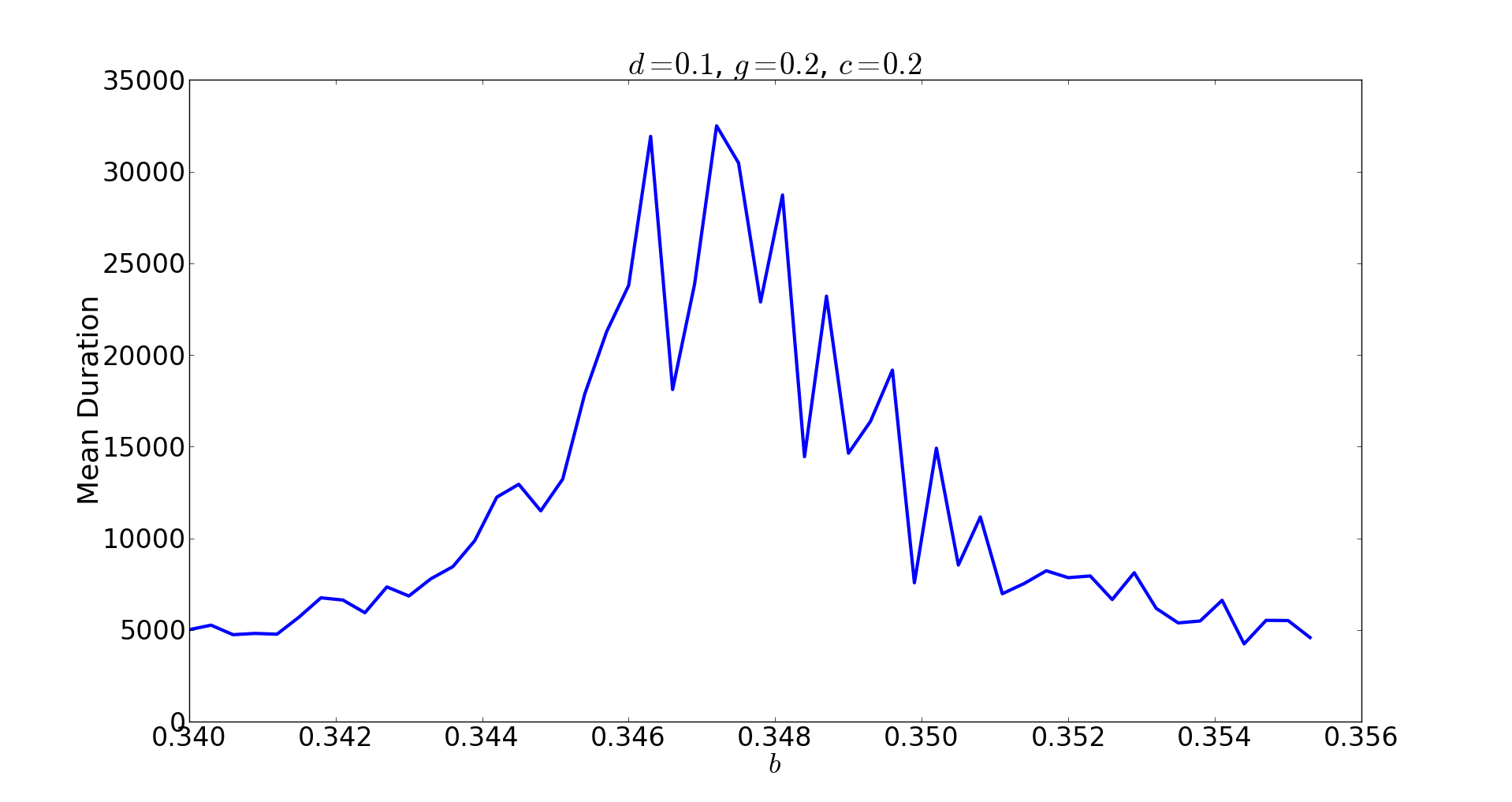}
\includegraphics[width=12cm]{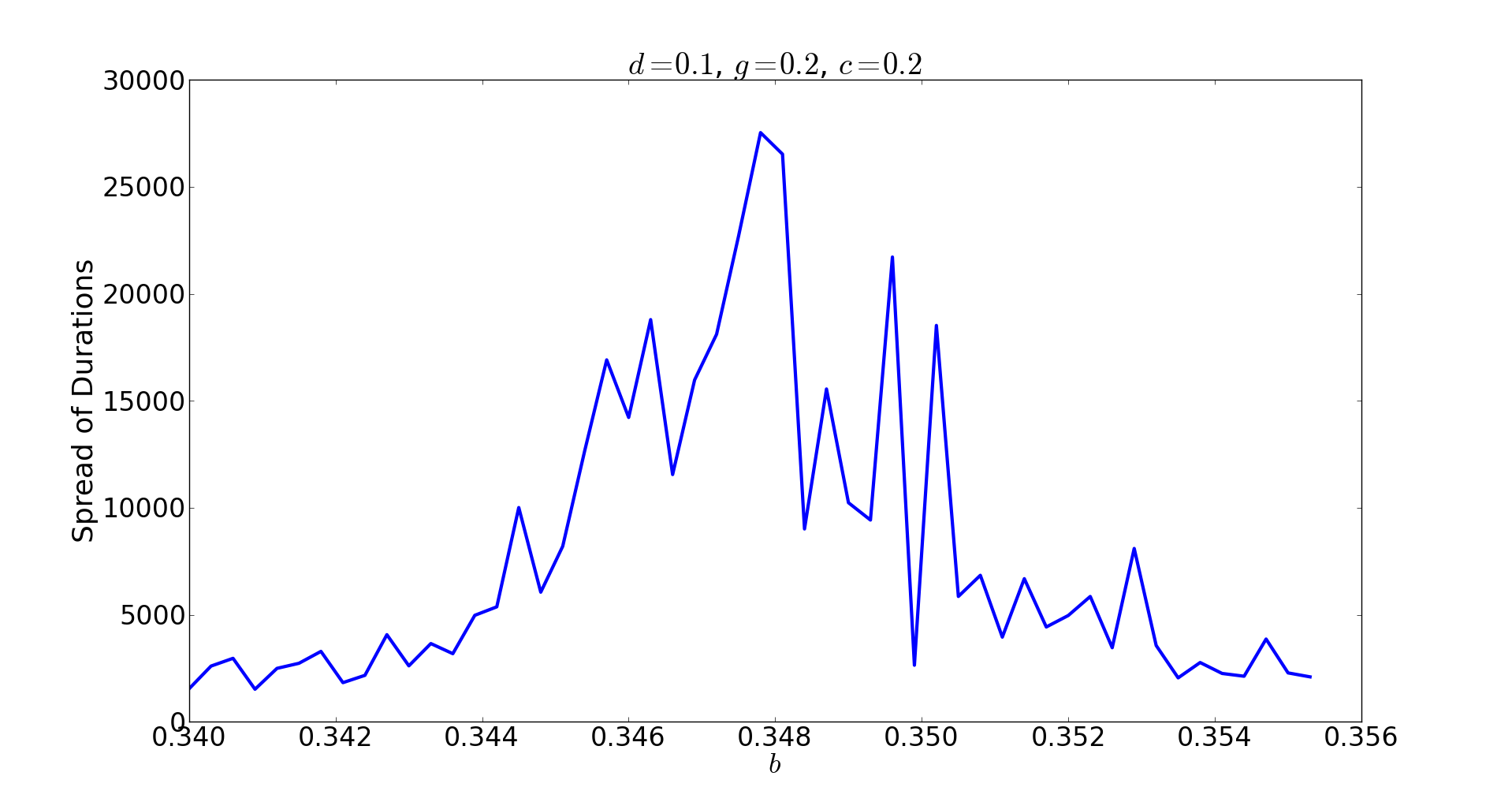}
\caption{\label{fig:durations-d0p1-g0p2-c0p2-detail} \textbf{(Top)}
Average time for a population to go extinct in the crossover region,
signaling the end of a simulation run.  \textbf{(Bottom)} Standard
deviation of simulation durations. Computed on a $50 \times 50$
lattice, with 10 trials for each value of~$b$. (The other parameters
were fixed at $d = 0.1$, $g = 0.2$, $c = 0.2$.)}
\end{figure}

In addition, as the two strategies become comparable in performance,
it takes longer for one to become predominant.  As
Figure~\ref{fig:durations-d0p1-g0p2-c0p2} illustrates, the time for
the losing strategy to vanish from the lattice diverges when $b$ is
near the critical transition value.  This is reminiscent of the
phenomenon known as \emph{critical slowing down} in statistical
physics~\cite{kardar2007b, chen2014}.  Examining the crossover region
in more detail (Figure~\ref{fig:pop-densities-d0p1-g0p2-c0p2-detail})
bears this out.  As the performance of the two varieties becomes
comparable, the average time required for one to drive the other to
extinction increases, and the variation in the time to extinction goes
up as well (Figure~\ref{fig:durations-d0p1-g0p2-c0p2-detail}).
Together, the characteristic exponents and the slowing-down effect
demonstrate the relevance of statistical physics to evolutionary game
theory.

\section{Fully Occupied Networks}

Another way of incorporating population structure into evolutionary
game theory is to put the dynamics on a network and keep each network
node occupied.  This means that the total population size is constant
and equal to the number of nodes in the network.  We take up this
approach next, as it connects with ways evolutionary game dynamics
have been studied before~\cite{nowak2010, allen2014b}.

Suppose that we're trying to figure out the fitness of the $i^{\rm
  th}$ individual, call it $f_i$, as a function of the genotypes of
the organisms with which it interacts, $\{s_j\}$.  To begin with a
simplified case, we might make the assumption that the total effect of
multiple causes taken together is the sum of the effects those causes
would have taken independently, and that the size of the effect grows
evenly with the size of the cause. So, we write a linear equation,
\begin{equation}
f_i = \beta^j_i s_j,
\label{eq:linear-payoff}
\end{equation}
where we've written the parameters with a $\beta$ as an homage to the
``regression coefficient'' jargon common to the art. The idea is that
if we had a whole bunch of measurements from a laboratory or a field
station, we could run a regression analysis and figure out what the
values of these coefficients should be. Of course, we can feed any set
of numbers we want into the our statistical software package; saying
that the results have any predictive value is a stronger statement,
which requires making a claim about the linearity of the interactions
at work.

An alternative way to measure genotype or trait values is to do so
relative to the social circle to which the individual $i$
belongs~\cite{bijma2008}. This analogous to the physicist's practice
of transforming from the laboratory frame to the centre-of-mass-frame,
except we consider instead a ``center-of-social-circle'' frame. This
coordinate change means that the parameters $\beta^j_i$ get mixed up
with each other, but since the equations for the next step of the
computations---figuring out how the genetic composition of the
population changes as a result of these fitness assignments---are also
linear, everything works out pretty simply.  We will return to this
point in Chapter~\ref{ch:price-mls}.

A {\em nonlinear} relationship between genotype and fitness is
interesting for both mathematical and biological
reasons~\cite{damore2011}.  One way to see why, which recalls our
discussion of Requisite Variety in Chapter~\ref{ch:multiscale}, is to
consider situations where success requires coordinated action.  For
example, suppose three graduate students are moving across town to a
new apartment.  They have to transport a heavy object, like a piano or
a drill press left in the living room.  To move the piano, all three
must heft simultaneously and walk in the same direction at the same
speed.  The payoff to flatmate $i$ is, using $s_i = 0$ to denote
``doing nothing'' and $s_i = 1$ to indicate ``hefting the piano'',
\begin{equation}
f_i = -cs_i + b\prod_j s_j.
\label{eq:volunteer-dilemma}
\end{equation}
The cost $c$ and benefit $b$ parameterize the situation, which is
another realization of the Volunteer's Dilemma game.  As
Nowak \emph{et al.}\ write of the payoff function
(\ref{eq:volunteer-dilemma}), ``Clearly in such a game one cannot
separate the effects of the action of the second player on the first
from the effects of the action of the first on
himself''~\cite{nowak2010}.  Nowak \emph{et al.}\ discuss the
three-player version of Eq.~(\ref{eq:volunteer-dilemma})
qualitatively, but they do not treat it in detail.  We now take up
that analysis.\footnote{They mention the three-player Volunteer's
Dilemma, using the stag-hunt terminology, in an appendix to a
paper~\cite{nowak2010} which provoked quite a bit of
controversy~\cite{allen2013c}.  Judging from the sound and fury, the
people who got upset about the paper didn't read the appendices.}

In this section, we will use three different types of connection
topology:
\begin{itemize}
\item A {\em regular graph,} in which each node has the same degree
  $k$, and the connection pattern is highly symmetric, as in a
  lattice;
\item A {\em random regular graph,} in which each node has the same
  degree $k$, but connections are otherwise unpatterned;
\item A {\em mixed graph,} which we generate anew at each time step
  of the simulation, keeping the node degrees fixed at $k$.
\end{itemize}
Simulations were carried out for $k = 3$, $k = 4$ and $k = 6$.  The
regular graph of degree $k = 4$ was a $20 \times 20$ lattice.  Both
periodic and truncated boundary conditions were studied.  For degree
$k = 3$, I used a hexagonal lattice of 391 nodes, and to test the
effects of boundary conditions, I compared against the results for
graph F400A in Foster's census of symmetric graphs \cite{foster2001}.
For $k = 6$, a trigonal planar lattice of 400 nodes was constructed,
and again both periodic and truncated boundary conditions were
studied.

We use a ``death-birth'' updating scheme \cite{allen2010}.  During
each generation of the simulation, we progress through the $N$ nodes
of the network in random order.  At each node, we compute the payoffs
of the neighbouring players, using Eq.~(\ref{eq:volunteer-dilemma}).
The focal node adopts the strategy of a neighbour chosen
stochastically with probabilities based on the neighbours' payoffs.
The probability of adopting the strategy of a player who obtained a
score $f$ is defined to be proportional to~$e^{wf}$, where the
parameter $w$ sets the strength of selection pressure.

Unlike the version of the Volunteer's Dilemma we studied in the
previous section, this version allows an individual of one species to
displace one of another.  Here, no network node is left empty, whereas
in the previous version, populations grow into available empty spaces
on the lattice.

\subsection{Initially Balanced Population}
\begin{figure}[ht]
\includegraphics[width=12cm]{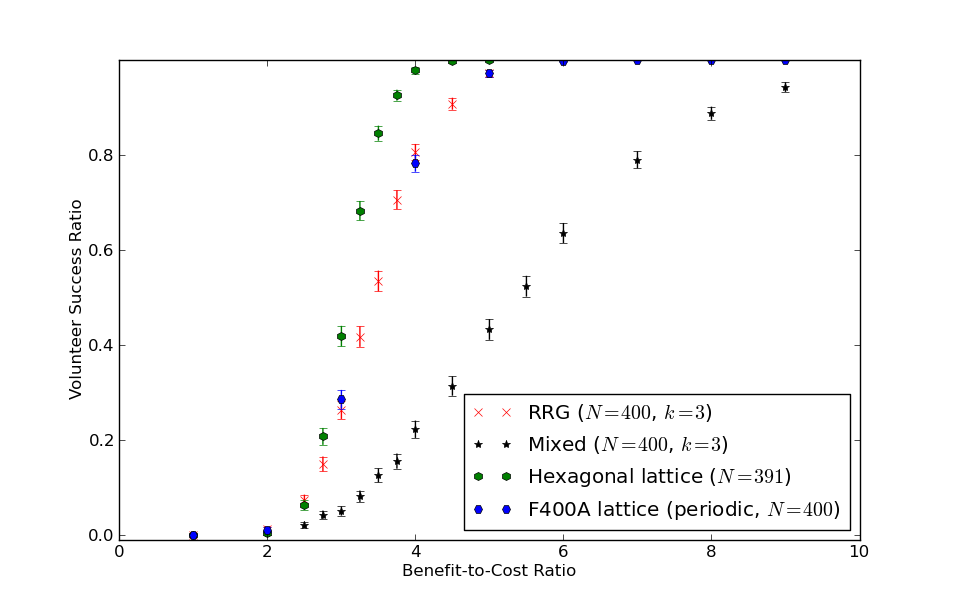}
\caption{\label{fig:Volunteer-success-vs-boverc-w0p1} 
  Volunteer success rate on various topologies with degree $k = 3$, as
  a function of benefit $b$ (with cost $c = 1.0$ and strength of
  selection $w = 0.1$). }
\end{figure}

\begin{figure}[ht]
\includegraphics[width=12cm]{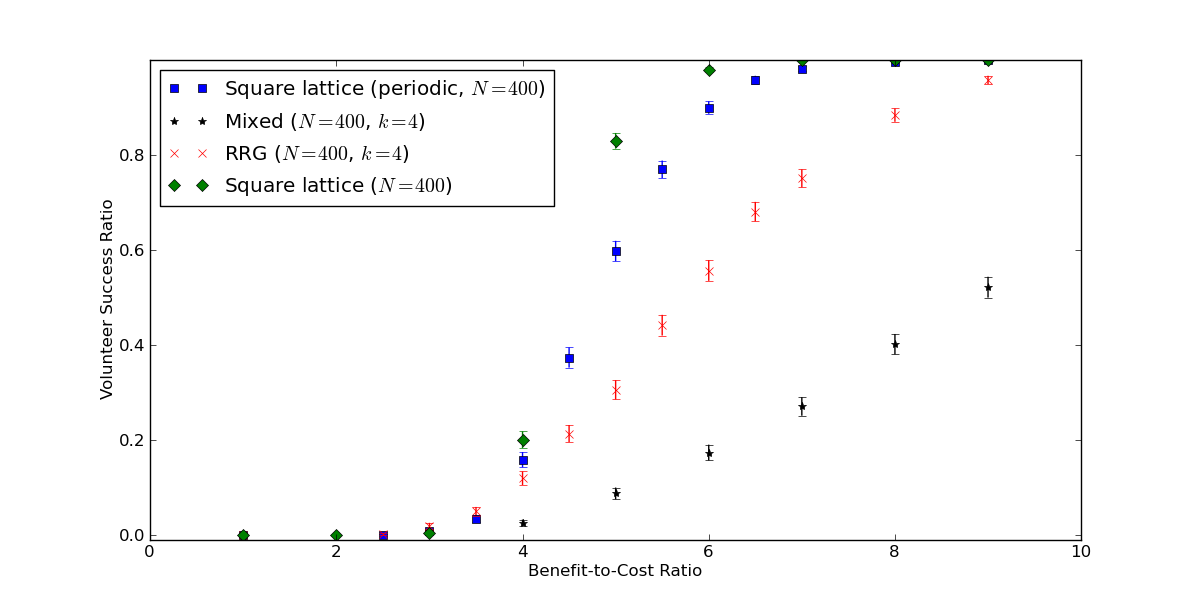}

\includegraphics[width=12cm]{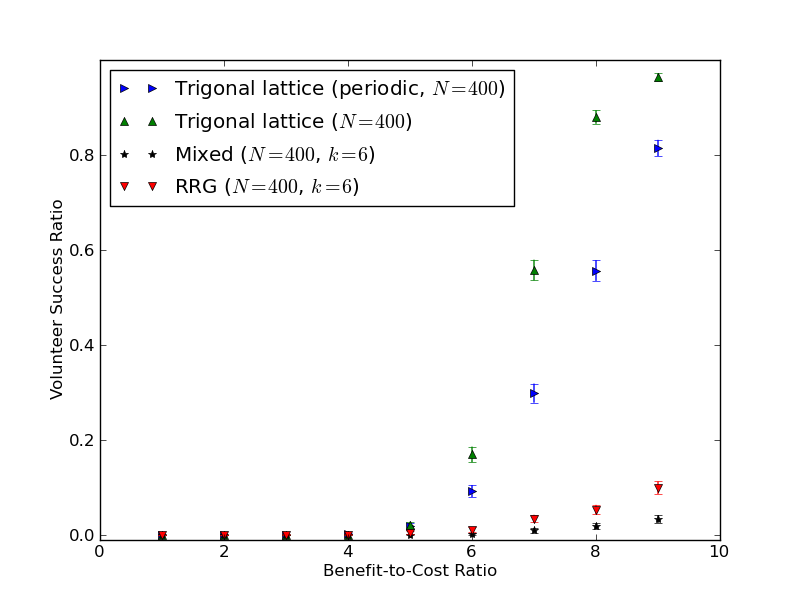}
\caption{\label{fig:Volunteer-success-vs-boverc-w0p1-2} 
  {\bf (Top)} Volunteer success rate on various topologies with degree
  $k = 4$, as a function of benefit $b$ (with cost $c = 1.0$ and
  strength of selection $w = 0.1$). {\bf (Bottom)} Volunteer success
  rate on various topologies with degree $k = 6$, as a function of
  benefit $b$ (with cost $c = 1.0$ and strength of selection $w =
  0.1$).}
\end{figure}

The results reported in this section concern what happens when a
population is initialized at random with volunteers and slackers
(heads it's one, tails it's the other).  2000 trials were performed to
obtain each data point, and the number of trials ending in ubiquitous
volunteerism was recorded.  Error bars in
Figures~\ref{fig:Volunteer-success-vs-boverc-w0p1}
and \ref{fig:Volunteer-success-vs-boverc-w0p1-2} indicate two standard
deviations.

A few general patterns emerge during the course of numerical simulations:
\begin{enumerate}
\item Boundary conditions can promote volunteerism, probably because
  nodes on an edge or in a corner have fewer neighbours, so it's
  easier for them to have a neighbourhood full of volunteers.
\item The greater the number of short loops, the larger the difference
  between the regular and random regular cases.
\item Shuffling the network at each time step always depresses the
  success rate of volunteerism.
\end{enumerate}

\subsection{Invasion}
We now test {\em invasion fitness,} by looking at how often a minority
strategy can invade a population, rather than how a balanced
population goes to one extreme or the other.  If neither strategy is
favored and the competition is neutral, then any individual in the
population is as likely as any other to become the ancestor of the
entire population.  Because the future population has to be descended
from somewhere, this implies that if an invasive strategy starts with
a single individual, the probability it will succeed and sweep to
saturation in neutral competition is $1/N$.  A strategy which suceeds
more often than this is said to be evolutionarily favored
\cite{allen2010}.  (Other definitions of evolutionary success are
possible, and the relationships among them are subtle
\cite{allen2012}.)

\begin{figure}[h]
\includegraphics[width=10cm]{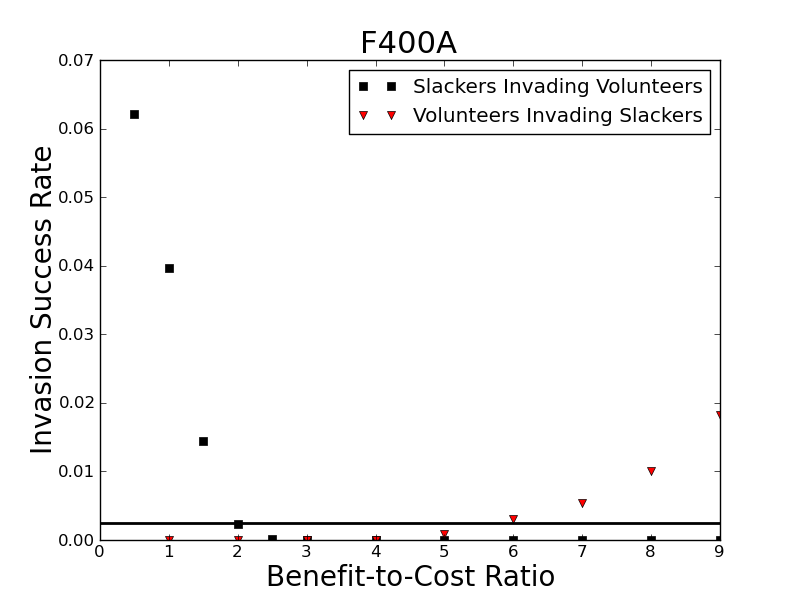}

\includegraphics[width=10cm]{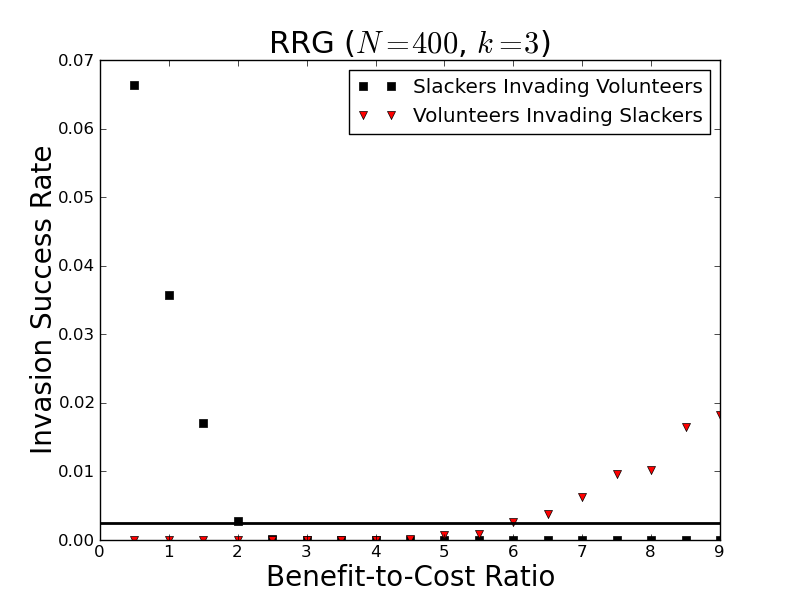}
\caption{\label{fig:invasion-success-vs-boverc-w0p1-k3} {\bf (Top)}
  Invasion success rate on the F400A topology \cite{foster2001}, a
  regular lattice with degree $k = 3$, as a function of benefit $b$
  (with cost $c = 1.0$ and strength of selection $w = 0.1$).  The
  solid horizontal line indicates the expected success rate in the
  neutral case, $1/N$. {\bf (Bottom)} Invasion success rate on a
  400-node RRG with uniform degree 3.}
\end{figure}

\begin{figure}[h]
\includegraphics[width=10cm]{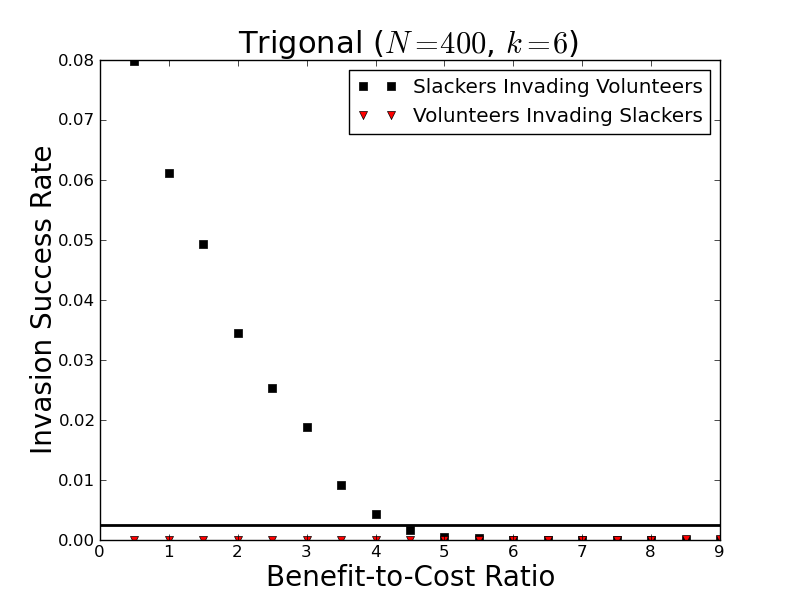}

\includegraphics[width=10cm]{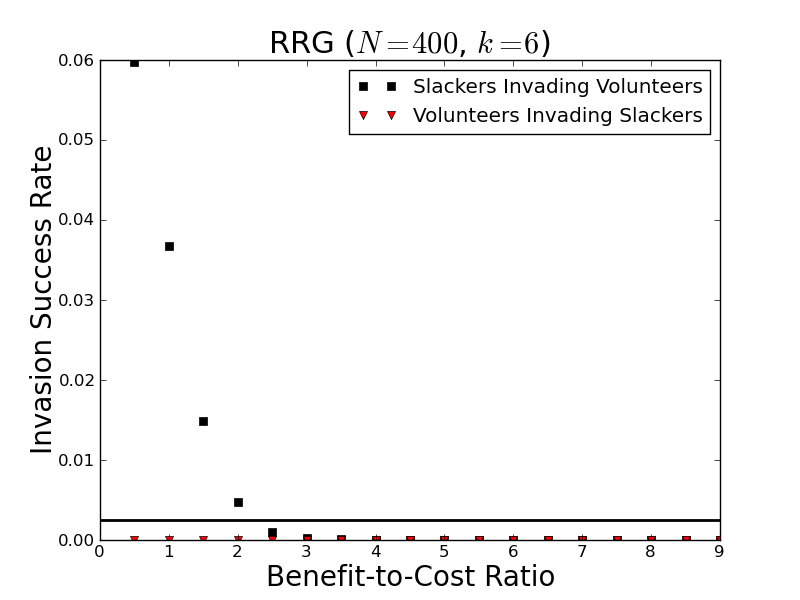}
\caption{\label{fig:invasion-success-vs-boverc-w0p1-k6} {\bf (Top)}
  Invasion success rate on a 400-node trigonal lattice with periodic
  boundary conditions, a topology in which all nodes have degree $k =
  6$, as a function of benefit $b$ (with cost $c = 1.0$ and strength
  of selection $w = 0.1$).  The solid horizontal line indicates the
  expected success rate in the neutral case, $1/N$. {\bf (Bottom)}
  Invasion success rate on a 400-node RRG with uniform degree 6.  Note
  the substantial difference in the slacker-invasion curve as compared
  to the trigonal lattice case.}
\end{figure}

Our model offers two broad categories of invasion scenarios: those in
which the resident population is composed of volunteers and those in
which it contains only slackers.  Overall, we find that the higher the
benefit-to-cost ratio $b/c$, the more favorable the situation for
volunteers invading slackers, and the less favorable for slackers
invading volunteers.  From the results reported above, one would
predict a deviation between the outcomes seen on regular lattices from
those on RRGs of the same degree if the regular lattices contain short
loops.  Comparing the results for $k = 3$,
Figure~\ref{fig:invasion-success-vs-boverc-w0p1-k3}, to those for $k =
6$, Figure~\ref{fig:invasion-success-vs-boverc-w0p1-k6}, we find this
prediction confirmed.  We also see that the $b/c$ ratio necessary to
make a volunteer population robust against invasion increases with the
network degree $k$.

\subsection{Mutation--Selection Equilibrium}

Another way to see the effect of closed loops in the underlying graph
topology is to compare the steady-state distributions for graphs of
the same vertex degrees.  If there is a nonzero probability of
mutation, then a uniform population does not have to stay uniform:
Volunteers can spontaneously appear in a lattice filled with Slackers,
and vice versa.  (We discussed this for frequency-dependent selection
in panmictic populations back in Chapter~\ref{ch:multiscale}.)
Evolutionary success means, in this context, being the more common
strategy.

\begin{figure}[h]
\includegraphics[width=12cm]{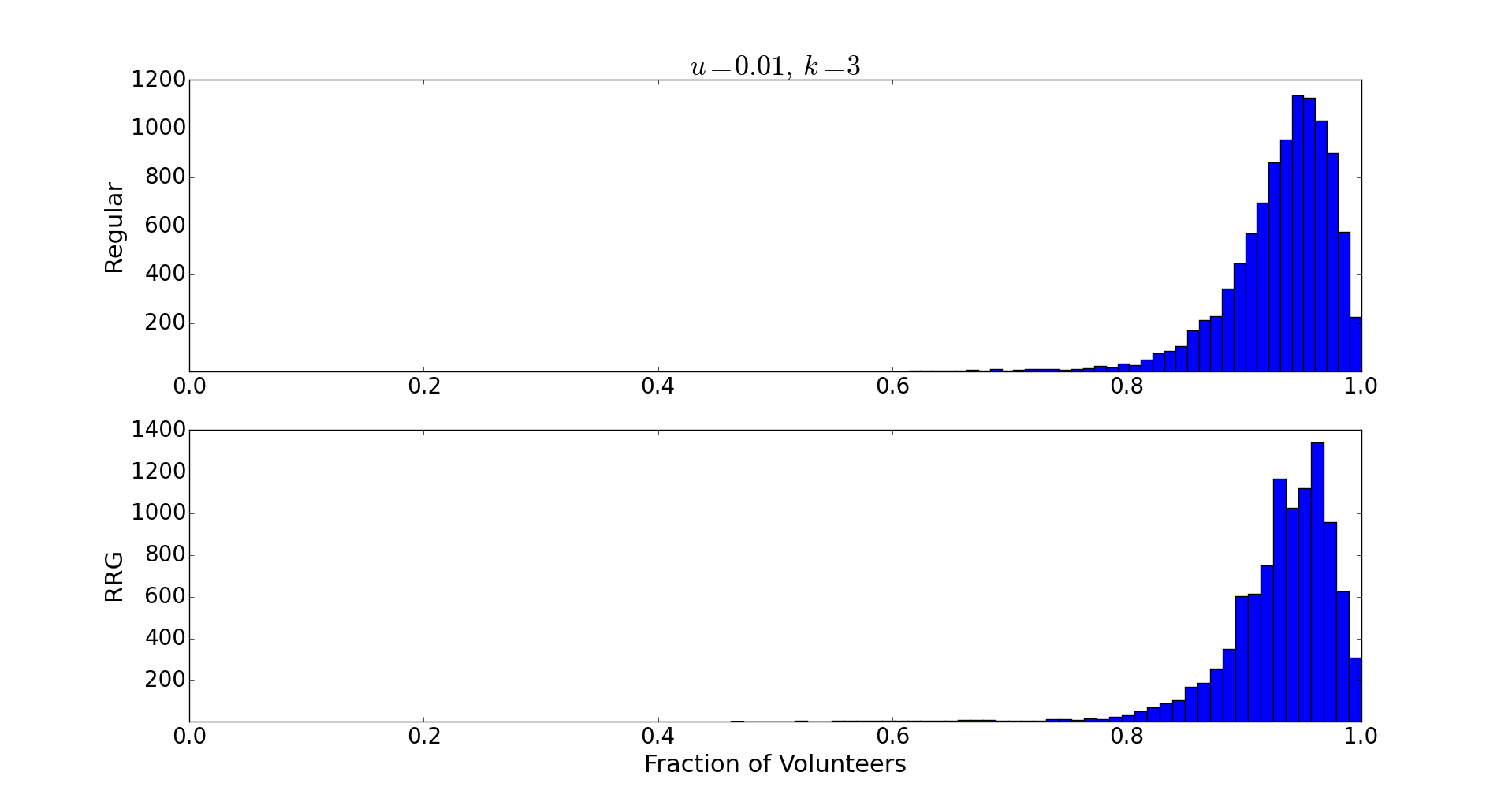}

\includegraphics[width=12cm]{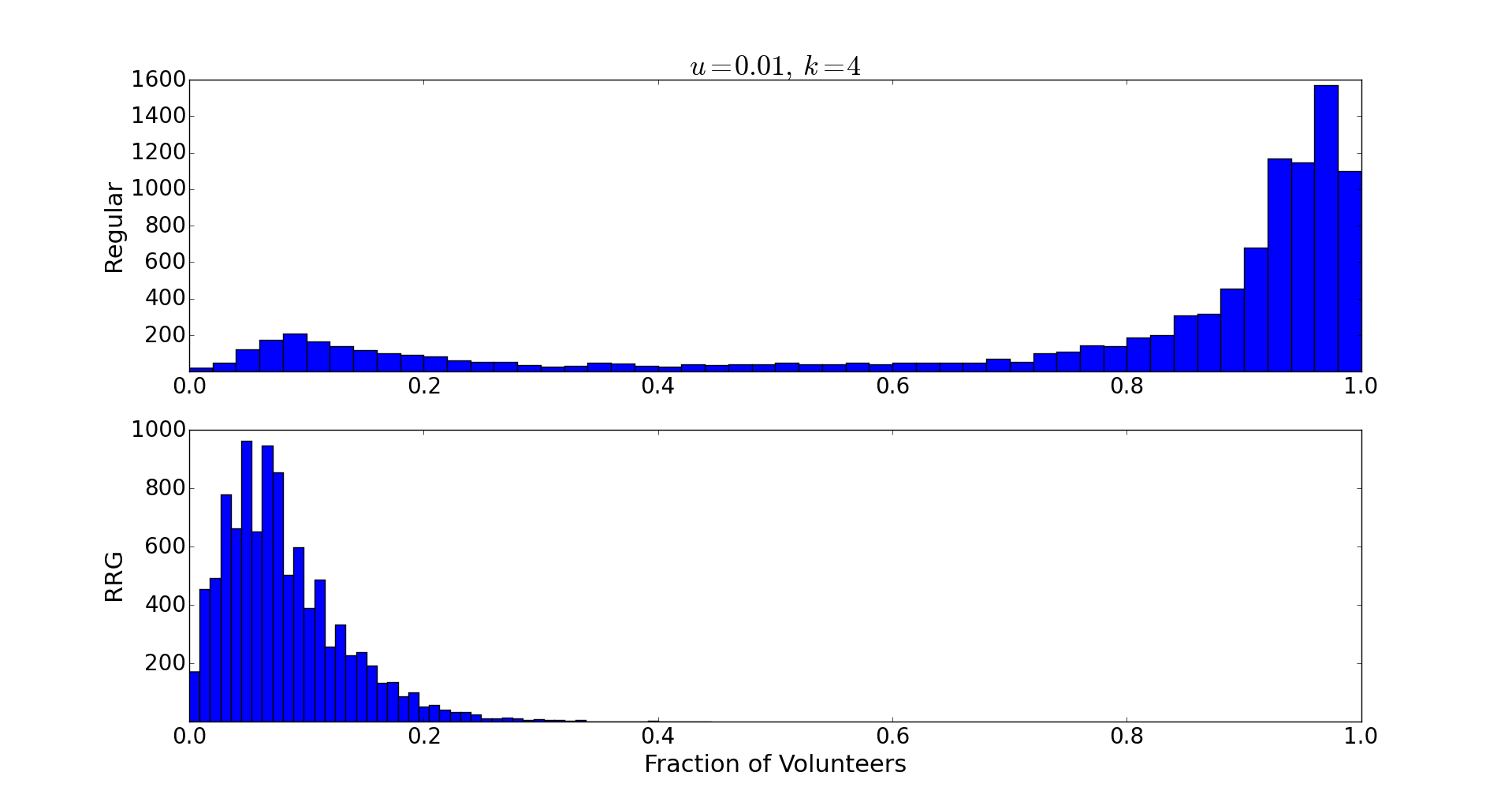}
\caption{\label{fig:comparison-k3andk4} {\bf (Top)} Histogram
indicating the steady-state distribution of the Volunteer population
density, on graphs whose vertices all have degree $k = 3$.  (Mutation
rate $u = 0.01$, with $b = 4$, $c = 1$, $w = 0.1$; computed for 10,000
Monte Carlo generations.)  The distributions for the regular lattice
and the random regular graph are quite similar.  {\bf (Bottom)} As
above, but for graphs with vertex degree $k = 4$ and benefit parameter
$b = 5$.  Here, on the square grid lattice, the distribution piles up
in the majority-Volunteer region, while on the RRG, the population is
dominated by Slackers.}
\end{figure}

Figure~\ref{fig:comparison-k3andk4} demonstrates that short closed
loops can make or break an evolutionary strategy.  This is a direct
indication of the important role which the graph topology plays, and
it is also a signal that analytical techniques which are viable for one
graph can fail for another.

\subsection{Analytical Results}
\label{sec:volunteer-analytic}
It may be possible to deduce some results analytically for the
Volunteer's Dilemma on completely filled graphs, at least in certain
limiting cases.  Our plan is to follow the logic of earlier work on
two-player games~\cite{allen2014b} and extend it to multiplayer
scenarios.  The key is to calculate the \emph{expected payoffs}
obtained by individuals at various distances from a chosen, focal
node.  By comparing these quantities at different distances, a
criterion for the success of a strategy can be determined.  The
criterion which obtains will depend, in general, on the update rule in
effect.

For the Prisoner's Dilemma with DB updating, this method indicates
that the Valiant strategy can succeed if
\begin{equation}
\frac{b}{c} > k.
\label{eq:boverc-pd}
\end{equation}
The nonlinearity of the Volunteer's Dilemma---that is, the presence of
higher-scale structure in its mapping of strategy combinations to
payoffs---makes the situation more complicated.  The numerical results
(Figures~\ref{fig:invasion-success-vs-boverc-w0p1-k3},
\ref{fig:invasion-success-vs-boverc-w0p1-k6}
and \ref{fig:comparison-k3andk4}) indicate that a success criterion
 cannot depend on the vertex degrees alone, because two graph
 structures with the same uniform degree support different
 evolutionary outcomes.  We expect that this will generally be the
 case for multiplayer games on graphs.  The reason why will become
 clear as we work through the application of the method.

Pick a node in the network, and designate it the focal node.  (We
assume that the network has sufficient symmetry that the choice of the
focal node is arbitrary, as is the case in regular lattices.)  Let the
state of the focal node be~1, that is, Volunteerism.  Next, choose a
second node somewhere else in the network.  The latter node will also
be the site of a Volunteer if it and the focal agent are related by
common ancestry without intervening mutations.  That is, an agent
other than the focus will have the same strategy as the agent at the
focal node if they are \emph{Identical By Descent} (IBD).  It is also
possible for two agents to have the same strategy even if they are not
IBD, by accident of fortuitous mutations.

When considering two-player games, it is known~\cite{allen2014b} that
evolutionary success criteria can be derived in the limit of weak
selection and low mutation rate, by considering IBD probabilities
derived for \emph{neutral drift.}  If there is no actual difference in
payoffs between the two strategies, then the dynamics are the
stochastic copying of labels without preference.  Under neutral drift,
death-birth updating reduces to the \emph{voter model,} so named
because it is an idealization of voters copying their opinions from
their neighbors.  With a nonzero mutation rate, it becomes
a \emph{noisy voter model}~\cite{granovsky1995}.  We note that the
noisy voter model closely resembles the imitation dynamic we studied
in Chapter~\ref{ch:multiscale}.

In the case of neutral drift and nonzero mutation rate, there exists a
unique and well-defined stationary probability distribution over the
possible states of the population.  We will derive our results using
expectation values evaluated with respect to this probability
distribution.  Let $\moment{s_i}$ be the expected value of site $i$,
conditional on the focal site being occupied by a Volunteer.  We will
use lower indices to label lattice sites, and superscript indices to
denote the number of steps taken in a random walk.  For example,
$\bar{f}_2$ is the expected payoff to the player at site $s_2$, while
$\bar{f}^{(2)}$ is the expected payoff for a player located at the end
of a two-step random walk which starts at the focal site.  From these
definitions, it follows that
\begin{equation}
\moment{s_0} = \moment{s^{(0)}} = 1,
\end{equation}
as well as
\begin{equation}
\bar{f}_0 = \bar{f}^{(0)}.
\end{equation}

``Evolutionary success'' can be defined in multiple ways.  For
example, we could say that the Volunteer strategy is successful if it
has more than a 50\% share of the population in mutation-selection
equilibrium.  Or, we could say that Volunteerism is successful if the
probability of its fixation when starting with a population comprising
one Volunteer and $N-1$ Slackers is greater than the Slacker
strategy's fixation probability in the reverse scenario.  In the
low-mutation-rate limit ($u \ll 1$), these conditions are
equivalent~\cite{allen2012}.  Let $b_0$ be the probability that the
focal individual reproduces, and let $d_0$ be the probability that it
is replaced by a neighbor.  These probabilities will depend on the
lattice topology, the current configuration of site values, the
benefit $b$, the cost $c$ and the strength of selection $w$.
Then~\cite{allen2014b} the Volunteer strategy is favored (for weak
selection) if
\begin{equation}
\left.\moment{\frac{\partial (b_0 - d_0)}{\partial w}}\right|_{w=0}
 > 0.
\end{equation}
For Death-Birth (DB) updating, which we considered in the numerical
work reported above, this condition reduces to
\begin{equation}
\bar{f}^{(0)} - \bar{f}^{(2)} > 0.
\label{eq:f0-f2}
\end{equation}

To see why, note that $d_0$ is constant for DB updating, so we only
need to evaluate $b_0$.  Next, we observe that $b_0$ will depend on
the payoffs $f_i$ of the players at the vertices adjacent to site~0.
Let $a_{ij} = 1$ if vertices $i$ and $j$ are adjacent, and 0
otherwise.  Then
\begin{equation}
\left.\moment{\frac{\partial b_0}{\partial w}}\right|_{w=0}
= \left.\moment{\frac{\partial}{\partial w}
                \sum_i \frac{a_{0i} e^{wf_0}}
                            {\sum_j a_{ji} e^{wf_j}}}\right|_{w=0}.
\end{equation}
Eq.~(\ref{eq:f0-f2}) follows from the quotient rule.  When we carry
out the derivative with respect to~$w$, we obtain a pair of terms:
\begin{equation}
\left.\moment{\frac{\partial b_0}{\partial w}}\right|_{w=0}
 =
\left.\moment{
\sum_i \frac{f_0 a_{0i} e^{wf_0} \sum_j a_{ji} e^{wf_j}
             - a_{0i} e^{wf_0} \sum_j a_{ji} f_j e^{wf_j}}
            {\left[\sum_j a_{ji} e^{wf_j}\right]^2}
}\right|_{w=0}.
\end{equation}
The first term gives us the $\bar{f}^{(0)}$, and the second gives us
the $\bar{f}^{(2)}$, establishing the desired relation.

Using Eq.~(\ref{eq:two-step-rw}) for a hexagonal lattice, we have that
\begin{align}
\bar{f}^{(0)} - \bar{f}^{(2)}
 &= \bar{f}^{(0)} - \frac{1}{3}\bar{f}_0 - \frac{2}{3}\bar{f}_4 \\
 &= \frac{2}{3} \left(\bar{f}_0 - \bar{f}_4\right).
\end{align}
This is positive if the expected payoff to the focal individual is
greater than that to an individual two steps away.

The expected payoff to the focal player on a hexagonal lattice is
\begin{equation}
\bar{f}_0 = -c\moment{s_0} + b\moment{s_0s_1s_2s_3}.
\end{equation}
Because the focal player is by assumption a Volunteer, this simplifies
to
\begin{equation}
\bar{f}_0 = -c + b\moment{s_1s_2s_3}.
\end{equation}
What about a player at a distance of one step from the focus?  This
player will also have three neighbors, one of whom is the focal
player.  So,
\begin{equation}
\bar{f}_1 = -c\moment{s_1} + b\moment{s_1 s_4 s_5}.
\end{equation}
And a player two steps away from the focus---for example, $s_4$---has
three neighbors, one of whom is a player who is adjacent to the focus.
\begin{equation}
\bar{f}_4 = -c\moment{s_4} + b\moment{s_1 s_4 s_6 s_7}.
\end{equation}
The site $s_1$ is one step from the focus, $s_4$ is two, and $s_6$ and
$s_7$ are at a distance of three.

A random walk which begins at the focus and has a length of two steps
can end either at the focus itself, or at a site two edges away.  The
first step always goes away from the focus, and the second will return
to it with a probability $1/k = 1/3$.  By symmetry, we can let $s_4$
stand in for all the sites at a distance of two steps from the focus.
Therefore,
\begin{equation}
\bar{f}^{(2)} = \frac{1}{3}\bar{f}_0 + \frac{2}{3}\bar{f}_4.
\label{eq:two-step-rw}
\end{equation}

The terms which are moments of a single site variable are easy to
evaluate in terms of IBD probabilities.  If the focal site and $s_i$
are IBD, then $s_i = 1$.  Otherwise, $s_i$ will take the values 0 and
1 with equal probability.
\begin{align}
\moment{s_i} &= \bar{q}_i + \frac{1}{2}(1-\bar{q}_i)
 = \frac{1 + \bar{q}_i}{2}, \\
\moment{s^{(i)}} &= \frac{1 + \bar{q}^{(i)}}{2}.
\end{align}
Evaluating the cubic and quadratic terms is more complicated.  We can
make a first stab by assuming they \emph{factor} into products of
single-variable expectation values.  This is a mean-field approximation.
It's difficult to say how drastic its effects will be without doing
the calculation, so we press forward.

This is where the fundamental difference with the two-player
Prisoner's Dilemma becomes manifest.  For that game, we have
\begin{equation}
\bar{f}^{(n)} = -c\moment{s^{(n)}} + b\moment{s^{(n+1)}},
\end{equation}
where $\moment{s^{(n)}}$ refers to the expectation value of a site
which at the end of an $n$-step random walk starting at the focus.
The only expectation values which appear are linear, and we do not
have to account for higher-order correlations.  We can, therefore,
evaluate directly:
\begin{equation}
\bar{f}^{(n)} = \frac{1}{2}
  \left(-c + b - c\bar{q}^{(n)} + b\bar{q}^{(n+1)}\right).
\end{equation}
And, in turn, the IBD probabilities $\bar{q}^{(n)}$ can be found for
symmetric graphs of arbitrary degree.  From this, one can deduce the
success criterion, Eq.~(\ref{eq:boverc-pd}).

Our primary concern, however, is the Volunteer's Dilemma on a
hexagonal lattice.  The IBD probabilities we need are given by
\begin{align}
\bar{q}^{(1)} &= 1 - u(N-1), \\
\bar{q}^{(2)} &= 1 - uN, \\
\bar{q}^{(3)} &= 1 - uN - u(N/3 - 1).
\end{align}
These can be found iteratively, through a fairly simple argument.  Let
$p^{(n)}$ denote the probability that an $n$-step random walk returns
to its starting point.  This is one way that a site at the end of a
random walk can be IBD to the focus: if that site \emph{is} the
focus.  Therefore, one contribution to~$\bar{q}^{(n)}$ will be
$p^{(n)}$.  The other way two individuals can be IBD is if one is the
offspring, without mutation, of a parent which is IBD to the other.
Let $s_i$ be a site, distinct from~$s_0$, which is separated
from~$s_0$ by an $n$-step random walk.  The parent of the individual
at~$s_i$ is an organism at a site $s_j$ which is separated from~$s_0$
by a walk which has $n + 1$ steps.  Furthermore, this walk does not
return to its starting point at the $n$\textsuperscript{th} step.  By
combining these two ways to be IBD, we arrive at the relation
\begin{equation}
\bar{q}^{(n)} = p^{(n)}
   + \left(\bar{q}^{(n+1)} - p^{(n)} \bar{q}^{(1)}\right)(1-u).
\label{eq:barq-recursive1}
\end{equation}
This somewhat heuristic argument can be made
precise~\cite{allen2014b}, with the same result.

In the limit $n \to \infty$, symmetry implies that $p^{(n)} \to 1/N$,
and $\bar{q}^{(n)}$ tends to some value which we write $\bar{q}$.
Therefore,
\begin{equation}
\bar{q} = \frac{1}{N}
 + \left(\bar{q} - \frac{\bar{q}^{(1)}}{N}\right)(1-u),
\end{equation}
which yields
\begin{equation}
(1 - u)\bar{q}^{(1)} = 1 - Nu\bar{q}.
\end{equation}
This lets us eliminate $\bar{q}^{(1)}$ from
Eq.~(\ref{eq:barq-recursive1}):
\begin{equation}
\bar{q}^{(n)} = (1-u)\bar{q}^{(n+1)} + Nu\bar{q} p^{(n)}.
\label{eq:barq-recursive2}
\end{equation}
If mutations do not occur, all individuals are IBD.  Sending $u \to 0$
in an expression for~$\bar{q}^{(n)}$ must recover this fact.  In the
limit of low mutation rate, we consequently have the useful expansion
\begin{equation}
\bar{q}^{(n)} - \bar{q}^{(n+1)}
 = u\left(N p^{(n)} - 1\right) + \mathcal{O}(u^2).
\label{eq:barq-recursive3}
\end{equation}
Dropping the higher-order terms and rearranging,
\begin{equation}
\bar{q}^{(n+1)}
 = \bar{q}^{(n)} - u\left(N p^{(n)} - 1\right).
\label{eq:barq-recursive4}
\end{equation}
The IBD probabilities we require now follow from the basic relations
\begin{equation}
p^{(0)} = \bar{q}^{(0)} = 1,
\ p^{(1)} = 0,
\ p^{(2)} = \frac{1}{k} = \frac{1}{3}.
\end{equation}
This is enough information to compute $q^{(1)}$, $q^{(2)}$ and
$q^{(3)}$ by the recursion relation, Eq.~(\ref{eq:barq-recursive4}).

Simplifying the $\bar{f}$ quantities by our mean-field approximation yields
\begin{align}
\bar{f}^{(0)} &= -c + \frac{b}{8}\left(1+\bar{q}^{(1)}\right)^3, \\
\bar{f}^{(1)} &= -\frac{c}{2}\left(1+\bar{q}^{(1)}\right)
             + \frac{b}{8}\left(1+\bar{q}^{(1)}\right)
               \left(1+\bar{q}^{(2)}\right)^2, \\
\bar{f}^{(2)} &= -\frac{c}{2}\left(1+\bar{q}^{(2)}\right)
             + \frac{b}{16}\left(1+\bar{q}^{(1)}\right)
               \left(1+\bar{q}^{(2)}\right)
               \left(1+\bar{q}^{(3)}\right)^2.
\end{align}
The next step is to substitute in the IBD probabilities.  Then, to see
whether Volunteerism can succeed with DB updating, we check the
difference $\bar{f}^{(0)} - \bar{f}^{(2)}$.  We only care
about the results to first order in the mutation rate $u$, because our
recursion relation (\ref{eq:barq-recursive4}) is only valid in that limit.

Calculating $\bar{f}^{(0)} - \bar{f}^{(2)}$ is a matter of
straightforward but tedious algebra.  The result has a form which
makes sense, but as we shall see, the numerical details are
problematic.  For the hexagonal lattice, to first order in the
mutation rate $u$,
\begin{equation}
\bar{f}^{(0)} - \bar{f}^{(2)} = Nu\left(\frac{5b}{6} - \frac{c}{2}\right).
\end{equation}
This implies that for DB updating, Volunteerism can succeed if
\begin{equation}
\frac{b}{c} > \frac{3}{5}.
\end{equation}
A criterion in terms of the ratio $b/c$ makes sense:  it is much like
the result we found in the well-mixed case, back in
Eq.~(\ref{eq:volunteer-boverc}).  However, the threshold we have found
is less than one, which is a puzzling feature.  Can the Volunteer
strategy really succeed if adopting it brings less benefit than cost?

This strongly suggests that our mean-field approximation was too severe.

Before moving on, we can make one observation.  On the hexagonal
lattice, it takes a minimum of three steps for separate paths from the
focus to converge.  At shorter distances, the hexagonal lattice looks
like a Cayley tree graph, where all the vertices have degree $k = 3$.
The logic we used to find $\bar{f}^{(0)}$, $\bar{f}^{(1)}$ and
$\bar{f}^{(2)}$ in terms of expectation values works just as well on
the $k = 3$ Cayley tree graph.  However, due to the presence of short
closed loops, this will not be the case for the square or triangular
lattices.  Even without being able to compute the expectation values,
we can say that under DB updating, Volunteer success on a hexagonal
lattice will resemble that on the $k = 3$ Cayley tree, whereas
differences will manifest between the square lattice and the $k = 4$
tree, and likewise for the triangular lattice and its tree
counterpart.  This is, indeed, what we observe in
Figures~\ref{fig:Volunteer-success-vs-boverc-w0p1}, \ref{fig:Volunteer-success-vs-boverc-w0p1-2},
\ref{fig:invasion-success-vs-boverc-w0p1-k3} and \ref{fig:comparison-k3andk4}.

One way to make progress, in the absence of a more sophisticated
calculation of higher-order expectation values, is to combine the
analytical and numerical techniques in a hybrid approach.  We can
simulate \emph{the neutral-drift process,} and use the results of
these Monte Carlo computations to assign values to~$\moment{s_4}$,
$\moment{s_1s_2s_3}$ and $\moment{s_1 s_4 s_6 s_7}$.  Then, we can
deduce a threshold benefit-to-cost ratio:
\begin{equation}
\left(\frac{b}{c}\right)_{\rm crit} = \frac{1 - \moment{s_4}}
                   {\moment{s_1s_2s_3} - \moment{s_1 s_4 s_6 s_7}}.
\label{eq:boverc-empirical-formula}
\end{equation}
In this way, we can obtain a result valid for the weak-selection
regime, using data from the case of \emph{no} selection.

Carrying out this procedure, we find
\begin{equation}
\left(\frac{b}{c}\right)_{\rm crit} = 4.05 \pm 0.37.
\label{eq:boverc-empirical}
\end{equation}
This value was obtained by simulating 10,000 generations of the
neutral-drift dynamics in equilibrium (mutation rate $u = 0.05$) and
averaging over the appropriate time series to compute values for the
moments in Eq.~(\ref{eq:boverc-empirical-formula}).  A total of 1,000
simulation runs were made, and the critical ratio computed for each
one.  Eq.~(\ref{eq:boverc-empirical}) reports the mean and standard
deviation of those results.  The average is comfortably above unity
(and, indeed, so were the ratios found in each trial).  This is a
clear improvement upon the $3/5$ threshold we found by the mean-field
argument, and it is also in the range consistent with the numerical
results for the actual Volunteer's Dilemma process
(Figures~\ref{fig:Volunteer-success-vs-boverc-w0p1}
and \ref{fig:invasion-success-vs-boverc-w0p1-k3}).

Consequently, we can say that the hybrid approach is fairly effective.
It also has the advantage that Monte Carlo results for the neutral
drift process can be applied to multiple different evolutionary games.
When we study different games, the success criterion will depend on
the fitness function and the update rule, but expectation values of
the same form will appear, and so knowledge of the neutral-drift
process will be useful in all cases.  This means that we can run Monte
Carlo simulations once and apply the results to a variety of
scenarios, potentially saving ourselves quite a bit of computer time.

One way to improve the rather na\"\i{}ve mean-field factorization of
the expectation values is to introduce two-site correlations.  We
define the pairwise correlation function
\begin{equation}
\cumulant{s_i s_j} = \moment{s_i s_j} - \moment{s_i}\moment{s_j}.
\label{eq:two-point-correlator}
\end{equation}
Recall that each of the quantities on the right are evaluated with the
condition that $s_0 = 1$.  Therefore, this condition applies to the
correlation function on the left as well.  If, given that $s_0 = 1$,
the sites $s_i$ and $s_j$ fluctuate completely independently, then
$\cumulant{s_i s_j}$ vanishes.

There is a specific, principled and well-established way to simplify a
higher-order expectation value in the approximation that no
correlations more complicated than Eq.~(\ref{eq:two-point-correlator})
are significant.  We will review this method in Chapter~\ref{ch:prob}.
For the moment, we simply avail ourselves of the results.  First, the
quantity which governs the expected benefit to the focal individual is
\begin{equation}
\moment{s_1s_2s_3} \approx \moment{s_1}\moment{s_2}\moment{s_3}
 + \moment{s_1}\cumulant{s_2 s_3}
 + \moment{s_2}\cumulant{s_1 s_3}
 + \moment{s_3}\cumulant{s_1 s_2}.
\label{eq:s1s2s3-gaussian}
\end{equation}
This expression reduces to the mean-field, factorized version if the
pairwise correlations vanish.  And, for the higher-order expectation
value we need in order to compute $f^{(2)}$, we obtain the sum
\begin{align}
\moment{s_1 s_4 s_6 s_7} \approx
 & \moment{s_1}\moment{s_4}\moment{s_6}\moment{s_7} \nonumber\\
 & + \cumulant{s_1s_4}\cumulant{s_6s_7}
   + \cumulant{s_1s_6}\cumulant{s_4s_7}
   + \cumulant{s_1s_7}\cumulant{s_4s_6} \nonumber\\
 & + \cumulant{s_1s_4}\moment{s_6}\moment{s_7}
   + \cumulant{s_1s_6}\moment{s_4}\moment{s_7} 
   + \cumulant{s_1s_7}\moment{s_4}\moment{s_6} \nonumber\\
 & + \cumulant{s_4s_6}\moment{s_1}\moment{s_7}
   + \cumulant{s_4s_7}\moment{s_1}\moment{s_6} \nonumber\\
 & + \cumulant{s_6s_7}\moment{s_1}\moment{s_4}.
\end{align}

How good is this improved approximation for neutral drift on the
hexagonal lattice?  Monte Carlo computations, under the same
conditions as before, show that this approximation for
$\moment{s_1s_2s_3}$ is accurate to within 0.3\%, while it somewhat
overestimates $\moment{s_1 s_4 s_6 s_7}$, by about 8\%.

\chapter{Techniques of Probability}

\label{ch:prob}

Probability theory is a way of managing uncertainty and making choices
when equipped with incomplete information.  In a sense, it is a theory
of knowledge and learning, one which is applicable when our attitude
about each proposition we consider, or each event we might encounter,
can be encapsulated in a real number.

\section{Basic Properties}

One way (but not the only way) to motivate the basic postulates on
which this theory relies is a parable: the story of \emph{the android
and the Ferengi bartender.}  The android walks into a bar, and the
bartender, after a quick assessment of the situation, offers a bit of
friendly sport between soon-to-be-friends.  The bartender proposes
to \emph{buy} or \emph{sell} a lottery ticket for any event the
android chooses, at any price the android desires.  For any event $E$,
the android assigns a numerical weight $p(E)$, which the android deems
to be the fair price of the lottery ticket
\begin{equation}
\ticket{Pays \$1 if the event $E$ occurs.}
\end{equation}
The android will \emph{buy} or \emph{sell} this ticket for $\$p(E)$.

The bartender offers the android the opportunity to strike a deal on
any combination of events, hoping to catch the android in an
inconsistency, forcing a \emph{sure loss.}  The goal of the android is
to avoid this eventuality.

For example, if the android assigns a price $p(E) < 0$ for some event
$E$, then the android will sell a ticket for a negative amount of
money, and the bartender wins.  Likewise if the android assigns $p(E)
> 1$: this indicates a willingness to sell a ticket for more than it
could ever be worth.

Consider two events, $E$ and $F$, which the android believes are
mutually exclusive.  The bartender proposes the following three
lottery tickets:
\begin{equation}
\ticket{Worth \$1 if ($E$ or $F$)},
\label{eq:ticket-EorF}
\end{equation}
along with
\begin{equation}
\ticket{Worth \$1 if $E$}
\label{eq:ticket-E}
\end{equation}
and finally
\begin{equation}
\ticket{Worth \$1 if $F$}.
\label{eq:ticket-F}
\end{equation}
The value of ticket (\ref{eq:ticket-EorF}) should be the same as the
total value of tickets (\ref{eq:ticket-E}) and (\ref{eq:ticket-F}).
If the android professes $p(E \hbox{ or } F) > p(E) + p(F)$, then the
bartender wins.  The bartender sells ticket (\ref{eq:ticket-EorF}) and
buys tickets (\ref{eq:ticket-E}) and (\ref{eq:ticket-F}) leaving the
android in the red, and whatever happens, the android can never recoup
the loss.  For example, if event $E$ takes place, the android
gains \$1 for the first ticket, but has to pay it back out again by
the second.

So, for mutually exclusive events,
\begin{equation}
p(E \hbox{ or } F) = p(E) + p(F).
\end{equation}
By a similar argument, the android must price the ticket
\begin{equation}
\ticket{Worth \$$\frac{m}{n}$ if $E$}
\end{equation}
at \$$\frac{m}{n}p(E)$, if $m$ and $n$ are positive integers.  A
continuity argument shows that a similar statement holds for a
positive real number $x$ in place of the rational number $m/n$.

Furthermore, if ``not $E$'' is the event that $E$ does not happen,
then we have two mutually exclusive events, and the sum of the fair
prices for their lottery tickets must be the fair price for a ticket
that is worth \$1 no matter what happens.  Therefore,
\begin{equation}
p(E) + p(\hbox{not } E) = 1.
\end{equation}
The weights of an event and its complementary event must, to be
consistent, sum up to unity.

Now, the bartender proposes a new game: \emph{conditional} lottery
tickets.  For any two events $E$ and $F$, a conditional ticket will be
worth \$1 if both $E$ and $F$ occur, but the cost of the ticket will
be returned if the event $E$ does \emph{not} occur.
\begin{equation}
\ticket{Worth \$1 if ($E$ and $F$).  But money back if (not $E$).}
\end{equation}
This is a gamble on the event ``$F$ given $E$,'' which we can denote
$F|E$.  The price of this ticket is, by definition, $ \$p(F|E) $.  So,
this ticket is the same as one written
\begin{equation}
\ticket{Worth \$1 if ($E$ and $F$).  But refund \$$p(F|E)$ if (not $E$).}
\end{equation}
To avoid a sure loss at the bartender's hands, the android must
price \emph{this} ticket as the total price of these:
\begin{equation}
\ticket{Worth \$1 if ($E$ and $F$)},
\hbox{ and } \ticket{Worth \$$p(F|E)$ if (not $E$).}
\end{equation}

The price of the final ticket must be $\$p(F|E)p(\hbox{not } E)$.  So,
consistency requires that
\begin{equation}
p(F|E) = p(E \hbox{ and } F) + p(F|E) p(\hbox{not }E).
\end{equation}
The weights of $E$ and its complement add up to 1, which implies
\begin{equation}
p(F|E) = p(E \hbox{ and } F) + p(F|E)( 1 - p(E)).
\end{equation}
We cancel the $p(F|E)$ from both sides and rearrange to find that
\begin{equation}
p(E \hbox{ and } F) = p(F|E) p(E).
\end{equation}

We saw earlier that the ticket prices for events deemed mutually
exclusive must add.  What if the android contemplates two events, $E$
and $F$, and decides that they are \emph{not} mutually exclusive?  In
this case, the events ``$E$ and (not~$F$)'' and ``not~$F$'' are still
automatically mutually exclusive in the android's judgment.
Therefore,
\begin{equation}
p\left(\hbox{[$E$ and (not~$F$)] or $F$}\right)
 = p\left(\hbox{$E$ and (not~$F$)}\right)
   + p\left( F \right).
\label{eq:not-mutually-exclusive}
\end{equation}
By the distributive rule of Boolean logic,
\begin{equation}
\hbox{[$E$ and (not~$F$)] or $F$}
 = \hbox{[$E$ or $F$] and [(not $F$) or $F$]},
\end{equation}
which simplifies to
\begin{equation}
\hbox{[$E$ and (not~$F$)] or $F$}
 = \hbox{[$E$ or $F$]}.
\end{equation}
Substituting this into the left-hand side of
Eq.~(\ref{eq:not-mutually-exclusive}) yields
\begin{equation}
p\left(\hbox{$E$ or $F$}\right)
 = p\left(\hbox{$E$ and (not~$F$)}\right)
   + p\left( F \right).
\label{eq:not-mutually-exclusive2}
\end{equation}
Using the definition of conditional lotteries, we have
\begin{equation}
p(\hbox{$E$ and (not $F$)}) = p(\hbox{not $F$}|E) p(E).
\end{equation}
By normalization, this is
\begin{equation}
p(\hbox{$E$ and (not $F$)})
 = \left(1 - p(F|E)\right) p(E).
\end{equation}
Distributing $p(E)$ over the sum and identifying $p(F|E)p(E) =
p(F\hbox{ and } E)$, we find that
\begin{equation}
p(\hbox{$E$ and (not $F$)})
 = p(E) - p(F\hbox{ and } E)
 = p(E) - p(E\hbox{ and } F).
\end{equation}
We can substitute this back into
Eq.~(\ref{eq:not-mutually-exclusive2}), yielding
\begin{equation}
p\left(\hbox{$E$ or $F$}\right)
 = p(E) + p(F) - p(E\hbox{ and } F).
\label{eq:not-mutually-exclusive3}
\end{equation}
One way to remember this relationship is in terms of areas:  if each
event is represented by a geometrical shape, then the total area of
the shape representing the event ``$E$ or $F$'' is the area of the
$E$-shape, plus the area of the $F$-shape, minus the area of the
region where they overlap, which would otherwise be double-counted.

In summary, the requirement that the android avoid a \emph{sure loss}
in \emph{any single deal} imposes a rather intricate set of
constraints on the fair-price function!  Specifically, we have shown
that prices must be bounded:
\begin{equation}
0 \leq p(E) \leq 1.
\label{eq:prob-postulate1}
\end{equation}
Also, for events believed to be mutually exclusive, the prices add:
\begin{equation}
p(E \hbox{ or } F) = p(E) + p(F).
\label{eq:prob-postulate2}
\end{equation}
For an event $C$ which the android believes is \emph{certain} to occur,
\begin{equation}
p(C) = 1.
\label{eq:prob-postulate3}
\end{equation}
If the set $\{E_i\}$ is an \emph{exhaustive} set of mutually exclusive
events, then
\begin{equation}
\sum_i p(E_i) = 1.
\end{equation}
And \emph{conditional} prices are related to \emph{joint} prices, in
accordance with
\begin{equation}
p(F|E) = \frac{p(E \hbox{ and } F)}{p(E)}.
\label{eq:prob-bayesrule}
\end{equation}
It is known~\cite{jeffrey2004, caves2005, fuchs2011} that the android
can avoid defeat at the hands of the bartender if and only if the
price assignment function satisfies Eqs.~(\ref{eq:prob-postulate1}),
(\ref{eq:prob-postulate2}) and (\ref{eq:prob-postulate3}).  An
assignment which meets these criteria is called \emph{coherent.}

Note that the bartender does not have to pick any numbers or make any
price assignments.  All the choices are made by the android; the
bartender merely exposes inconsistencies in the numerical weightings
which the android attaches to events.  Indeed, we could restate the
whole scenario as an inner challenge the android poses in the pursuit
of self-consistency.  Furthermore, note that we have not
mentioned \emph{repeated experiments} or \emph{sequences of trials} at
all.  Avoiding loss at the bartender's hands imposes consistency
conditions on numerical weightings, even for events which can only
happen once.

The chain of inferences leading to
Eqs.~(\ref{eq:prob-postulate1})--(\ref{eq:prob-postulate3}) is
typically known as a \emph{Dutch book argument.}  The players whom we
have designated as the android and the bartender are, in the standard
parlance, the bettor and the bookie, and the bettor strives to achieve
Dutch-book coherence.  The choice of terminology in this section
emphasizes the importance of these ideas for machine
learning~\cite{hodges1983, zabell1995, koller2009}, while in addition
underscoring the distinction between a mathematicized standard of
behavior and the way human beings actually conduct themselves in
gambling establishments.  Furthermore, the historical origin of the
``Dutch book'' term is rather obscure, anyway~\cite{humphreys2008,
fuchs2014b}.

The coherence conditions (\ref{eq:prob-postulate1}),
(\ref{eq:prob-postulate2}) and (\ref{eq:prob-postulate3}) are
familiar: together, they say that $p$ is a \emph{probability
distribution.}  Dutch- or Ferengi-book coherence provides an
operational meaning to probability, which as we saw earlier is
relevant even for experiments which can be performed only once.  We
could, on a purely mathematical level, have defined probability
axiomatically, declaring that a probability distribution is a set with
a particular kind of additional structure.  Such a
definition~\cite{tao2012} might read like the following:
\begin{quotation}
\noindent A probability space $(\Omega, \mathcal{B}, {\bf P})$ is a tuple
comprising a set $\Omega$, a nonempty collection $\mathcal{B}$ of
subsets of~$\Omega$ such that $\mathcal{B}$ satisfies the properties
of a $\sigma$-algebra, and a function ${\bf P}: \mathcal{B} \to [0,1]$
which sends each element of~$\mathcal{B}$ to a real number in the unit
interval.  The function ${\bf P}$ is countably additive, and ${\bf
P}(\Omega) = 1$.
\end{quotation}
In fact, a mathematician might elect to abstract further from this
definition and work with an algebra of random variables and
expectation values, eliding the basic set $\Omega$, the event set
$\mathcal{B}$ and the mapping ${\bf P}$.  This turns out to be useful
for some subjects, such as random matrix theory~\cite{tao2012}, but it
is not a perspective we will need to pursue for this chapter.

Moreover, we will not have to stress greatly over questions
of \emph{continuous} versus \emph{discrete} sets of propositions for
events.  We will casually switch between discrete probability
distributions, normalized as
\begin{equation}
\sum_i p(E_i) = 1,
\end{equation}
and continuous probability densities, which are normalized as
\begin{equation}
\int_{-\infty}^\infty dx\,p(x) = 1.
\end{equation}
The interpretation of the function $p(x)$ is as follows: for any real
number $x_0$, the quantity $p(x = x_0)dx$ is the probability of the
event that $x$ lies between $x_0$ and $x_0 + dx$.  We will not have to
be more exacting in our definitions than this.  In other words, we
will be living by the physicists' standard, which is indifferent to
mathematical subtleties until they become unavoidable, treating them
as niceties which are no more relevant than the question of how best
to construct the real numbers from the integers in the first
place~\cite{conway1976}.

All of our considerations so far in this section have concerned the
android's probability ascriptions at \emph{a single time.}  Even the
conditional probabilities, as in Eq.~(\ref{eq:prob-bayesrule}), are
statements of how the android is willing, at a particular moment, to
gamble on events, even if one of those events may chronologically
precede the other.  We have yet to address how the android might make
changes to probability assignments in a self-consistent way.

When we consider probabilities changing with time, our android's
probability ascriptions gain a time index.  Let $p_0$ be a function
from events to the interval $[0,1]$ which expresses the android's
gambling commitments at the time $t = 0$.  Similarly, $p_\tau$ denotes
the android's gambling commitments at a later time, $t = \tau$.  Our
problem is to relate $p_0$ and $p_\tau$ according to some standard of
consistency.

Some probability assignments may carry two time indices.  For example,
the bartender can suggest a wager on how many drinks the bar patrons
will order in an hour.  We can define an event $E(N,T)$ as the event
that $N$ drinks will be bought during the interval from time $T$ until
one hour later.  A probability ascription for this event has one time
label in the event itself, and another index which denotes the time at
which the ascription is made: $p_t(E(N,T))$.  Our concern here is to
relate probabilities with different subscripts, which is a problem
distinct from the question of how $p_t(E(N,T))$ relates
to~$p_t(E(N,T'))$.  Starting in section~\S\ref{sec:clt}, we will
address the latter question to a larger extent, and much of
Chapter~\ref{ch:adaptive} will be devoted to examining it for a
particular example.

Suppose that at $t = 0$, the android is willing to gamble on two
events, $E$ and $D$, in such a way that
\begin{equation}
p_0(E|D) = \frac{p_0(E \hbox{ and } D)}{p_0(D)} = q.
\end{equation}
Now, if the event $D$ occurs between $t = 0$ and $t = \tau$, what
should $p_\tau(E)$ be?  The simplest choice is to use the number we
already have on hand, and force it equal to~$q$:
\begin{equation}
p_\tau(E) = p_0(E|D) = q,
\ \hbox{if $D$ occurs.}
\label{eq:prob-update}
\end{equation}
However, we \emph{cannot deduce this} from a coherence argument like
we have used so far, because all those coherence arguments concern
probability assignments \emph{at a single time.}  They are, to use a
Greek-derived word, \emph{synchronic} statements, when what we need is
a \emph{diachronic} rule.

We will now investigate the conditions under which
Eq.~(\ref{eq:prob-update}) is a reasonable updating scheme.  Suppose
that at time $t = 0$, the android regards $\$p_0(E)$ as the fair price
for a lottery ticket worth \$1 if the event $E$ occurs.  Then, at some
later time $t = \tau$, the android's evaluation of the world has
changed, perhaps in response to new information, and the new fair
price is $\$p_\tau(E)$, where $p_\tau(E) < p_0(E)$.  There is
nothing \emph{irrational} or inconsistent about this: Being willing to
sell a ticket at a lower price is just a matter of cutting one's
losses.

However, after the android explains this, the bartender proposes a new
gamble, this time wagering on the android's own future beliefs.  The
bartender offers to buy or sell a ticket of the form
\begin{equation}
\ticket{Worth \$1 if $p_\tau(E) = q$}.
\end{equation}
Already at $t = 0$, the android can assign a fair price for this
ticket, which would be $\$p_0(p_\tau(E) = q)$.  If the android is
supremely confident that $p_\tau(E)$ will be $q$, then the fair price
of this ticket is \$1.

Now, if $q < p_0(E)$, then the android will be willing to buy a ticket
for~$\$p_0(E)$ and then sell that ticket for a lower price later.  The
android faces a sure loss, \emph{one that is already apparent at $t =
0$}, and the bartender has won.  Likewise, if $q > p_0(E)$, the
android will sell a ticket for~$\$p_0(E)$ and buy it back later at a
higher price, ensuring a sure loss again.

What can the android do to avoid this eventuality?  Defeating the
bartender in this scenario requires adhering to the condition
\begin{equation}
p_0(E|p_\tau(E) = q) = q.
\label{eq:prob-reflection}
\end{equation}
Like all the consistency conditions derived so far, this is a
requirement placed upon the current gambling commitments, though in
this case, the space of events include the android's future
declarations of belief.  This condition is an example of
the \emph{reflection principle,} an idea due to van
Fraassen~\cite{vanfraassen1984, fuchs2011}.  It can be derived from a
more relaxed assumption: instead of $p_0(p_\tau(E) = q) = 1$, we can
deduce Eq.~(\ref{eq:prob-reflection}) if
\begin{equation}
p_0(p_\tau(E) = q) > 0.
\end{equation}

We are now in a position to apply the reflection principle to the task
of choosing a probability-updating scheme.  Let $E$ be an event, and
let $Q_q$ be the event that at time $t = \tau$, we will have $p_\tau(E)
= q$.  Next, suppose that there exists a set $\{D_q\}$ of
possible \emph{data-acquisition events,} such that there is a
bijection between $\{D_q\}$ and the possible values for~$p_\tau(E)$.
Then
\begin{equation}
p_0(E|D_q) = p_0(E|D_q, Q_q) = p_0(E|Q_q).
\end{equation}
The reflection principle states that
\begin{equation}
p_0(E|Q_q) = p_0(E|p_\tau(E) = q) = q.
\end{equation}
Therefore,
\begin{equation}
p_0(E|D_q) = q.
\end{equation}
This is a statement of a gambling commitment made at time $t = 0$: the
android is, at $t = 0$, willing to pay $\$q$ for the lottery ticket
\begin{equation}
\ticket{Worth \$1 if ($E$ and $D_q$). But money back if (not $D_q$).}
\end{equation}
By definition, $q$ is $p_\tau(E)$ if the event $Q_q$ occurs.
Consequently, if the android goes ahead and does what the android had
been confident about doing, then
\begin{equation}
p_\tau(E) = p_0(E|D_q).
\end{equation}
This is just the rule we guessed in the first place,
Eq.~(\ref{eq:prob-update}).  It is known as the \emph{Bayes rule.}

We can, again, arrive at this point from a more relaxed assumption.
The essential requirement is that the android be able to identify at
$t = 0$ an event which, the android expects, can determine the future
gambling commitments.  That is, there must be some event $D$ such that
\begin{equation}
p_0(p_\tau(E) = q|D) = 1.
\label{eq:prob-needed-for-bayes}
\end{equation}

What if no such event $D$ exists, and
Eq.~(\ref{eq:prob-needed-for-bayes}) does not hold?  For example, the
android may expect that whatever happens between $t = 0$ and $t
= \tau$, the data will be too ambiguous to warrant upgrading the
confidence in any proposition to 100\%.  Instead, the android expects
that some uncertainty will necessarily remain, which can be
represented by a probability distribution over data-acquisition
events:
\begin{equation}
\sum_D p_\tau(D) = 1.
\end{equation}
In this case, we can use a more general probability-updating scheme,
the \emph{Jeffrey rule}:
\begin{equation}
p_\tau(E) = \sum_{D'} p_0(E|D') p_\tau(D').
\label{eq:prob-jeffrey-rule}
\end{equation}
This reduces to the Bayes rule, Eq.~(\ref{eq:prob-update}), if
\begin{equation}
p_\tau(D') = \delta_{D,D'},
\end{equation}
for some event $D$.  And, like Eq.~(\ref{eq:prob-update}), the Jeffrey
rule (\ref{eq:prob-jeffrey-rule}) can be justified using the
reflection principle~\cite{skyrms1987}.

Diaconis and Zabell~\cite{diaconis1982} give an example where this
more general scheme of updating would be applicable:
\begin{quotation}
\noindent suppose we are about to hear one of two recordings of
Shakespeare on the radio, to be read by either Olivier or
Gielgud, but are uncertain as to which, and have a prior
with mass $\frac{1}{2}$ on Olivier and $\frac{1}{2}$ on Gielgud.
After hearing the recording, one might judge it fairly likely, but by
no means certain, to be by Olivier.  The change in belief takes place
by direct recognition of the voice; all the integration of sensory
stimuli has already taken place at a subconscious level.  To demand a
list of objective vocal features that we condition on in order to
affect the change would be a logician's parody of a complex
psychological process.
\end{quotation}

Furthermore~\cite{diaconis1986},
\begin{quotation}
\noindent  If the \emph{only} impact of hearing the recording is to
change the odds on Olivier and Gielgud, in the sense that for any $A$,
$P_\tau(A|O) = P_0(A|O)$ and $P_\tau(A|G) = P_0(A|G)$, then after
assessing $P_\tau(O)$ we may proceed to apply Jeffrey's rule.  (Of
course, the former might well \emph{not} be the case; for example the
quality of the recording might convey additional information as to its
date or manufacture.)\footnote{I have adjusted the notation slightly
in this passage to be consistent with the rest of this section.}
\end{quotation}

In general, we may say that the updating of probabilities is a subtle
subject.  The next section will relate this question, which seems to
live in the realm of machine learning, to mathematical biology.  When
we establish this connection, the idea of mesoscale environmental
structure will make an appearance.

\section{An Analogy with Evolutionary Dynamics}

Let $\{H_i\}$ be a set of $n$ mutually exclusive and exhaustive
hypotheses, so that at any time $t$,
\begin{equation}
\sum_{i=1}^n p_t(H_i) = 1.
\end{equation}
In the previous section, we showed that at least under some
circumstances, if an event $E$ happens in between times $t = 0$ and $t
= \tau$, we are justified in updating the probabilities according to
the following rule:
\begin{equation}
p_\tau(H_i) = \frac{p_0(E|H_i) p_0(H_i)}{p_0(E)}.
\label{eq:prob-condit}
\end{equation}
This follows from the simple method for updating probabilities, the
Bayes rule that we wrote down in Eq.~(\ref{eq:prob-update}).

Now, we jump sideways and consider a simple model of evolutionary
dynamics in a panmictic population~\cite{page2002}.  We suppose there
are $n$ types of organism.  These could be different species,
different genotypes in the same species, or in principle, genetically
identical individuals who adhere to different social behaviors.  We
represent the configuration of the population by an $n$-tuple of
nonnegative real numbers:
\begin{equation}
x = (x_1,x_2,\ldots,x_n).
\end{equation}

By assuming panmixia, we deliberately blur over all spatial
organization or other kinds of population structure.  We neglect
stochasticity, and we assume that at each instant, the population
configuration is definitely specified.  In other words, for this
problem we consider evolution as a \emph{deterministic} dynamical
system, one which we will take to operate in discrete time.  We
establish the dynamics by writing an update rule, which yields a new
tuple $x'$ when given $x$.

That a model which neglects all stochasticity should relate to
probability theory may be surprising, but we shall soon see that it is
the case, and the relationship is quite direct.

To implement the idea of natural selection in this context, we
introduce a \emph{fitness function} which maps population tuples to
real numbers.  Each of the $n$ types has its own fitness function:
\begin{equation}
f_i = f_i(x_1,x_2,\ldots,x_n).
\end{equation}
Types which are more fit should be represented more strongly in the
next generation.  Our update rule for~$x_i$ should, consequently, have
the form
\begin{equation}
x_i' \propto x_i f_i(x).
\end{equation}
It is convenient to keep the overall population size constant, and in
that case, we might as well use the $x_i$ to represent proportions:
\begin{equation}
\sum_{i=1}^n x_i = 1.
\end{equation}
To ensure that this normalization is maintained, we introduce the
average fitness
\begin{equation}
\bar{f}(x) = \sum_{i=1}^n x_i f_i(x).
\end{equation}

Our dynamical update law, the \emph{replicator equation,} is
\begin{equation}
x_i' = \frac{x_i f_i(x)}{\bar{f}(x)}.
\end{equation}
This is formally analogous~\cite{harper2009} to the rule for updating
probabilities by conditionalization, Eq.~(\ref{eq:prob-condit}).  To
see the relationship, we make the following substitutions:
\begin{align}
\{p_0(H_i)\} &\to x = (x_1,\ldots,x_n), \\
p_0(E|H_i) &\to f_i(x), \\
p_0(E) &\to \bar{f}(x), \\
p_\tau(H_i) &\to x' = (x'_1,\ldots,x'_n).
\end{align}
Natural selection, in the situation modeled by the replicator
equation, can be thought of as a learning process, in which the
population gains information about the fitness landscape.

What about the more flexible Jeffrey rule?  Adapting
Eq.~(\ref{eq:prob-jeffrey-rule}) to this context, we find that
\begin{equation}
p_\tau(H_i) = \sum_{E'} p_0(H_i|E') p_\tau(E')
 = \sum_{E'} \frac{p_0(H_i) p_0(E'|H_i)}{p_0(E')} p_\tau(E').
\label{eq:prob-condit-jeffrey}
\end{equation}
This reduces to the simpler updating scheme,
Eq.~(\ref{eq:prob-condit}), if $p_\tau(E') = \delta_{E,E'}$.  If
$p_\tau(E')$ is not a delta function, then multiple potential events
$E'$ are relevant.  The evolutionary analogue of this is the
possibility of \emph{multiple fitness landscapes.}\footnote{To my
knowledge, the question of an evolutionary analogue of the Jeffrey
rule hasn't been raised before.  Harper~\cite{harper2009} and
Baez~\cite{baez2012b}, for example, stop with the Bayes rule,
Eq.~(\ref{eq:prob-condit}).}

Imagine that we have a test tube filled with various types of
bacteria.  We pipette a fraction $w_j$ of its contents into each
of~$m$ new test tubes.  The conditions in these tubes can differ from
one another:  perhaps they are illuminated under different frequencies
of light, or they contain varying nutrient mixtures.  The bacteria
interact, and their population proportions change due to natural
selection.  We then pour the contents of the $m$ tubes all back
together again.  The total population size remains constant throughout
the whole process.

Let $f_i^{(j)}$ be the fitness function for type $i$ in tube $j$.  As
before, the fitness depends on the proportions of the different types
present in the environment.  If all of the populations are scaled down
by the same factor $w_j$, the relative proportions remain the same.
Therefore, the initial proportions in each of the new tubes are the
same as those in the original source, and we can write the fitnesses
as~$f_i^{(j)}(x)$.  The mean fitness in tube number $j$ is
\begin{equation}
\bar{f}^{(j)}(x) = \sum_i x_i f^{(j)}_i(x),
\end{equation}
and this is true no matter what the value of~$w_j$.

Because a fraction $w_j$ of each $x_i$ experiences the environment in
the $j$\textsuperscript{th} tube, the new value $x_i'$ will be
\begin{equation}
x_i' = x_i \sum_{j=1}^m \frac{f^{(j)}_i(x)}{\bar{f}^{(j)}(x)} w_j,
\hbox{ with } \sum_{j=1}^m w_j = 1.
\end{equation}
This is the analogue of the Jeffrey rule,
Eq.~(\ref{eq:prob-condit-jeffrey}).

\section{Biased Coin-Flips and Urn Models}

Consider the prototypical probability problem of repeatedly flipping a
coin.  Let $p$ be the probability ascribed to the outcome that the
coin comes up heads.  What probability should we ascribe to the event
$E_m$ of seeing the coin land on heads exactly $m$ times out of~$N$?
For this event to transpire, we must see heads $m$ times and tails $N
- m$ times.  The probability of first seeing $m$ instances of heads,
followed by $N - m$ instances of tails, is $p^m (1-p)^{N-m}$.
However, this is not the only way the event $E_m$ can happen:  Our
specification of the event $E_m$ coarse-grains over the details of the
order in which the desired outcomes arrive.  So, the probability
$p(E_m)$ is the value we computed before, multiplied by the number of
ways we can distribute the $m$ heads throughout the total of~$N$
flips:
\begin{equation}
p(E_m) = \binom{N}{m} p^m (1-p)^{N-m},
\label{eq:prob-binomial}
\end{equation}
This defines the \emph{binomial distribution,} in which we have used
the binomial coefficients~\cite{moriarty1867} more explicitly given as
\begin{equation}
\binom{N}{m} = \frac{N!}{m!(N-m)!}.
\label{eq:binom-coeff}
\end{equation}

We can think of this scenario in a slightly different way, which leads
to some useful modifications.  Consider an urn filled with a total
of~$N_B$ balls, and let $R = N_B p$, where $p$ is the parameter we
used before.  Exactly $R$ of the balls in the urn are red, the other
$G = N_B - R$ being green.  We draw a ball from the urn at random,
check its color and drop it back into the urn.  The probability that
the ball we pick will be red is just $p$.  If we repeat the
draw-and-replace operation $N$ times, the probability that we see red
in exactly $m$ trials is given by Eq.~(\ref{eq:prob-binomial}).

What happens if we do \emph{not} replace a ball after we withdraw it?
Then, the population of the urn changes from one trial to the next.
Call $E_m$ the event of seeing red in exactly $m$ trials out of~$N$.
The probability of this event is
\begin{equation}
p(E_m) = \frac{(\hbox{\# of ways to get $m$ red balls})
               \times(\hbox{\# of ways to get $N-m$ green balls})}
              {(\hbox{total \# of ways to make a selection})}.
\end{equation}
Each factor in this expression can be found using combinatorics.  In
fact, each quantity which enters this formula is a binomial
coefficient:
\begin{align}
(\hbox{\# of ways to get $m$ red balls})
 &= \binom{R}{m}, \nonumber\\
(\hbox{\# of ways to get $N-m$ green balls})
 &= \binom{G}{N-m}, \\
(\hbox{total \# of ways to make a selection})
 &= \binom{R+G}{N}. \nonumber
\end{align}
Putting these together, we have
\begin{equation}
p(E_m) = \frac{\binom{R}{m} \binom{G}{N-m}}
              {\binom{R+G}{N}}.
\label{eq:prob-hypergeometric}
\end{equation}
This defines the \emph{hypergeometric distribution.}  Expanding out
the binomial coefficients per Eq.~(\ref{eq:binom-coeff}),
\begin{equation}
p(E_m) = \frac{R! G!}{m!(R-m)!(N-m)!(G-N+m)!}
         \frac{N! (R+G-N)!}{(R+G)!}.
\end{equation}

The third variation of the urn problem is the following: Every time we
draw out a ball, we check its color, and we return \emph{two} balls of
that color to the urn.  This is the \emph{P\'olya urn
model}~\cite{eggenberger1923, collevecchio2013}, and in this case, if
we begin with $R$ red balls and $G$ green balls,
\begin{equation}
p(E_m) = \frac{N!}{m!(N-m)!}
         \frac{(m+R-1)!(N-m+G-1)!}{(N+R+G-1)!}
         \frac{(R+G-1)!}{(R-1)!(G-1)!}.
\end{equation}
In terms of the gamma function,
\begin{equation}
p(E_m) = \frac{\Gamma(N+1)}{\Gamma(m+1)\Gamma(N-m+1)}
 \frac{\Gamma(m+R)\Gamma(N-m+G)}{\Gamma(N+R+G)}
 \frac{\Gamma(R+G)}{\Gamma(R)\Gamma(G)}.
\label{eq:prob-beta-binomial}
\end{equation}
This is the beta-binomial distribution we encountered in
Chapter~\ref{ch:multiscale}, as the steady-state solution to imitation
dynamics on a complete graph, Eq.~(\ref{eq:network-imitation-q}).

These probability distributions are related in another way, other than
their common appearance in variations of the urn
scenario~\cite{helfand2013}.  We can find this relationship by
revisiting the conditionalization rules that we studied in the
previous sections.  Take the Bayes rule (\ref{eq:prob-condit}), which
effects a map from one probability distribution to another:
\begin{equation}
p(\theta) \to \frac{p(y|\theta) p(\theta)}{p(y)}.
\end{equation}
When confronted with a transformation, we typically like to know what
it leaves unchanged.  For example, the eigenvectors of a matrix are
the vectors which the transformation represented by that matrix
changes only by a scaling factor.  Likewise, it is useful to identify
probability distributions which, under the conditionalization map,
keep the same form.  The input to the map is known as
the \emph{prior,} and its output is the \emph{posterior.}  If the
prior and posterior have the same functional form, then they
are \emph{conjugate.}  This relation is dependent on the likelihood
function $p(y|\theta)$ used in the mapping, but that can often be
taken as fixed on other grounds.

Suppose that
\begin{equation}
p(y|\theta) = \binom{N}{y} \theta^y (1-\theta)^{N-y},
\end{equation}
for a positive integer $N$ and all integers $y$ satisfying $0 \leq
y \leq N$.  Then, the conjugate prior of the binomial distribution
(\ref{eq:prob-binomial}) is the Beta distribution, 
\begin{equation}
p(\theta) = \frac{\Gamma(\alpha+\beta)}{\Gamma(\alpha)\Gamma(\beta)}
 \theta^{\alpha-1}(1-\theta)^{\beta-1},
\end{equation}
and the conjugate prior of the hypergeometric distribution
(\ref{eq:prob-hypergeometric}) is the beta-binomial,
Eq.~(\ref{eq:prob-beta-binomial}).

We now have the tools to dig a little more deeply.  Consider a
scenario in which we wish to \emph{repeat} an experiment.  That is, we
have the budget to carry out a long experiment, made up of many
successive trials~\cite{caves2002}.  We can represent the outcome of
each trial by a random variable $x_j$, and we can assign a joint
probability distribution $p(x_1,x_2,\ldots,x_N)$ over the possible
outcomes of an $N$-trial experiment.  Motivated by common experiences
in the workaday life of a scientist, we now impose two conditions on
this joint probability distribution.  These conditions will help
dramatically in narrowing down the possible forms of the joint
distribution $p(x_1,x_2,\ldots,x_N)$.  First, we require that it
be \emph{finitely exchangeable}: its value is invariant under
permutations of its arguments.  If $\pi$ is any permutation of the
numbers $\{1,\ldots,N\}$, then
\begin{equation}
p(x_1,\ldots,x_N) = p(x_{\pi(1)},\ldots,x_{\pi(N)}).
\end{equation}
This property recalls the exchange symmetry we invoked in
Chapter~\ref{ch:multiscale} to simplify the complexity profile.

Second, we require that $p(x_1,\ldots,x_N)$ be extensible, in the
following manner.  For any integer $M > 0$, there is a finitely
exchangeable distribution with more arguments, $p_{N+M}$, such that
\begin{equation}
p(x_1,\ldots,x_N)
 = \sum_{x_{N+1},\ldots,x_{N+M}}
    p_{N+M}(x_1,\ldots,x_N,x_{N+1},\ldots,x_{N+M}).
\end{equation}
These two requirements make precise the idea that our probability
assignment $p$ derives from an arbitrarily long sequence of random
variables, the order of which is inconsequential.  We say that a $p$
which satisfies both conditions, finite exchangeability and
extensibility, is \emph{exchangeable.}

Let us say we have an exchangeable probability assignment for $M$
binary random variables $x_1,\ldots,x_M$.  Define $p(n,N)$ to be the
probability of obtaining $n$ 1's in $N$ trials.  Because the $n$
appearances of the outcome 1 can arrive in any order, this is
\begin{equation}
p(n,N) = \binom{N}{n} p(x_1=1,\ldots,x_n=1,
                        x_{n+1}=0,\ldots,x_N=0).
\end{equation}
This is equal to
\begin{equation}
p(n,N) = \binom{N}{n} \sum_{m=0}^M
 p(x_1=1,\ldots,x_n=1,x_{n+1}=0,\ldots,x_M=0|m,M) p(m,M).
\end{equation}

Exchangeability means that all sequences with $m$ occurrences of ``1''
in $M$ trials are equiprobable.  So, we have again an urn problem,
specifically, the one that led us to the hypergeometric distribution.
If we stock an urn with~$M$ balls, $m$ of which are red, then the
hypergeometric distribution tells us how likely we are to find $n$ red
balls in~$N$ draws made without replacement.
\begin{equation}
p(x_1=1,\ldots,x_n=1,x_{n+1}=0,\ldots,x_M=0|m,M) 
 = \frac{\binom{m}{n}\binom{M-m}{N-n}}
        {\binom{M}{N}}.
\end{equation}
Defining
\begin{equation}
(r)_q = \prod_{j=0}^{q-1} (r-j) = \frac{r!}{(r-q)!},
\end{equation}
we have after a spot of algebra that
\begin{equation}
p(n,N) = \binom{N}{n} \sum_{m=0}^M \frac{(m)_n(M-m)_{N-n}}
                                       {(M)_N} p(m,M).
\end{equation}
We can rewrite this sum as an integral:
\begin{equation}
p(n,N) = \binom{N}{n} \int_0^1 dz\,
 \frac{(zM)_n [(1-z)M]_{N-n}}{(M)_N} P_M(z),
\end{equation}
where we have defined
\begin{equation}
P_M(z) = \sum_{m=0}^M p(zM,M) \delta(z - m/M).
\end{equation}

Thanks to the extendibility property, we can take the limit
$M\to\infty$.  In this limit, we find that $P_M(z)$ approaches a
continuous curve, $P_\infty(z)$, and the rest of the integrand goes to
$z^n(1-z)^{N-n}$.
\begin{equation}
p(n,N) = \binom{N}{n} \int_0^1 dz\, z^n(1-z)^{N-n} P_\infty(z).
\label{eq:de-finetti-binary}
\end{equation}
Interestingly, the result has the form of an integral over what we
might call a ``meta-probability.''  Note that $z^n(1-z)^{N-n}$, times
the binomial coefficient out front, has the form of a binomial
distribution with probability equal to~$z$.  The curve $P_\infty(z)$
has, by construction, the normalization property of a probability
density:
\begin{equation}
\int_0^1 dz\,P_\infty(z) = 1.
\end{equation}
It is as if we can write the function $p(n,N)$ in terms of ``the
meta-probability $P_\infty(z)$ that the probability is $z$.''

Eq.~(\ref{eq:de-finetti-binary}) can be generalized readily enough
beyond the case of binary random variables.  The result is \emph{de
Finetti's theorem.}  Let $\Delta_k$ denote the space of valid
probability assignments over $k$ outcomes:
\begin{equation}
\Delta_k = \left\{ \vec{p}: p_j \geq 0 \hbox{ for all } j
 \hbox{ and } \sum_j p_j = 1 \right\}.
\end{equation}
Then, exchangeability implies that
\begin{equation}
\boxed{p(x_1,\ldots,x_N)
 = \int_{\Delta_k} d\vec{p}\, P(\vec{p}\,)\, p_{x_1}\cdots p_{x_N}
 = \int_{\Delta_k} d\vec{p}\, P(\vec{p}\,)\, p_1^{n_1}\cdots p_k^{n_k},}
\label{eq:de-finetti}
\end{equation}
where $P(\vec{p})$ is properly normalized over $\Delta_k$:
\begin{equation}
\int_{\Delta_k} d\vec{p}\, P(\vec{p}\,) = 1.
\end{equation}

When we used the parable of the Ferengi bartender to motivate the
basic rules of probability, the story left no room for an ``unknown
probability,'' per se.  The android's probability assignments at any
time are known to the android, and the idea of an ``unknown known''
sounds like a contradiction in terms.  However, de Finetti's theorem
gives meaning to it.  The locution ``unknown probability'' is
a \emph{shorthand for a scenario in which the android is gambling on a
sequence of trials that the android judges to be exchangeable.}

\section{Shannon Information}

Intuitively speaking, \emph{information} is that which removes our
uncertainty.  We can quantify an amount of information by specifying
how many questions are necessary to overcome the uncertainty we have
about something.  For example, suppose we have an experiment which
yields one of~$M$ different outcomes, stochastically.  If we repeat
this experiment $N$ times, the number of different possible sequences
of outcomes is $M$ to the $N$\textsuperscript{th} power.  Each of
these sequences can be represented as a string of symbols, drawn from
an alphabet of size $M$.  The number of yes-or-no questions we would
need to narrow down this set of possibilities to a single result is
$\log_2 M^N = N\log_2 M$. However, we may have \emph{expectations}
about the experiment, meaning that we will find some strings of
outcomes less surprising than others.  Suppose that we ascribe a
probability $p_i$ to each of the $M$ possible outcomes of an
individual trial in the sequence.  We have a set of nonnegative
numbers which together satisfy
\begin{equation}
\sum_{i=1}^M p_i = 1.
\end{equation}
This set defines a \emph{random variable.}  We do not know in advance
exactly how many times the $i$\textsuperscript{th} outcome will occur,
but the \emph{number of such occurrences we will find least
surprising} is easily computed:
\begin{equation}
N_i = Np_i.
\end{equation}
How many of the $M^N$ possible strings are, in terms of our
probability assignment, unsurprising?  A minimally surprising string
is one where each of the $M$ symbols in our alphabet occurs $N_i$
times.  The number of such strings having total length $N$ is, by
elementary combinatorics,
\begin{equation}
K = \frac{N!}{\prod_{i=1}^M N_i!}
  = \frac{N!}{\prod_{i=1}^M (Np_i)!}.
\end{equation}
How many bits do we need to specify one message out of a set
of \emph{this} size?  Applying the algebraic properties of logarithms,
this is
\begin{equation}
\log_2 K = \log_2 N! - \sum_{i=1}^M \log_2(Np_i)!.
\end{equation}

Here, it is useful to invoke Stirling's approximation,
\begin{equation}
\log N! \approx N\log N - N.
\end{equation}
With this, we can evaluate the natural log of $K$ as
\begin{equation}
\log K \approx N\log N - N - \sum_i \left[Np_i \log (Np_i) - Np_i\right].
\end{equation}
We have assumed here that each of the $N_i$ is large enough for
Stirling's approximation to be viable.  Now, we simplify.  By
normalization, the $p_i$ must sum up to one, so we can cancel the
second and final terms, leaving
\begin{equation}
\log K \approx N\log N - N\sum_i p_i \left[\log N + \log p_i\right].
\end{equation}
Invoking normalization again, we see that we are left with
\begin{equation}
\log K \approx -N\sum_i p_i \log p_i,
\end{equation}
which we can easily convert to a base-2 logarithm by division.  Then,
if we divide by~$N$, we can obtain a result in terms of bits per symbol.

This result provides us with a \emph{measure of information}
associated with the probability distribution $\{p_i\}$.  If the
probabilities $p_i$ express our expectations of the stochastic
experiment, then the number of yes-or-no questions which we should
expect to require in order to remove our uncertainty about each
iteration of that experiment is
\begin{equation}
\boxed{ H[\{p_i\}] = -\sum_i p_i \log_2 p_i. }
\end{equation}
This is the \emph{Shannon information} of the probability
distribution, also known variously as the Shannon entropy and the
Shannon index.  The base of the logarithm is a convention which
depends on the subfield of science that one is currently studying.
The quantity $H[\{p_i\}]$ vanishes if $p_i = \delta_{ij}$ for some
$j$, and it is maximized by the uniform probability distribution over
all $i$.  This expresses the fact that if our expectations favor all
possibilities equally, we cannot expect to gain any advantage by
asking questions cleverly (as we could, for example, with English
text: ``Is the next letter an \emph{E}?'').  Contrariwise, if we are
supremely confident that a particular outcome will obtain, we expect
that we will require zero questions to ascertain the result.

Back in Chapter~\ref{ch:multiscale}, we used a general idea of an
information function to quantify the concept of multiscale structure.
We can apply Shannon information in that context if we consider not
just one random variable, but combinations of them.  If we have two
random variables $X$ and $Y$ and a joint probability distribution
$p(x,y)$, the total joint information of~$X$ and $Y$ considered
together is, by the Shannon formula,
\begin{equation}
H(X,Y) = - \sum_{x,y} p(x,y) \log p(x,y).
\end{equation}
If we only care about one of the two random variables, we can sum over
the other:
\begin{equation}
p_X(x) = \sum_y p(x,y),\ p_Y(y) = \sum_x p(x,y).
\end{equation}
In turn, we can use these probability distributions in the Shannon
formula just as well.

When we developed our formalism in Chapter~\ref{ch:multiscale}, we
considered a measure of \emph{shared information} which satisfied the
relation
\begin{equation}
\boxed{I(X;Y) = H(X) + H(X) - H(X,Y).}
\end{equation}
Is there a formula for $I(X;Y)$ in terms of Shannon indices?  We start
by re-expressing the Shannon indices of the individual variables in
terms of their probability distributions:
\begin{equation}
I(X;Y) = -H(X,Y) - \sum_x p_X(x) \log p_X(x) - \sum_y p_Y(y) \log p_Y(y).
\end{equation}
This, in turn, is the same as
\begin{align}
I(X;Y) &= -H(X,Y) - \sum_{x,y} p(x,y) \log p_X(x) 
              - \sum_{x,y} p(x,y) \log p_Y(y) \nonumber\\
 &= -H(X,Y) + \sum_{x,y} p(x,y) [-\log p_X(x) - \log p_Y(y)].
\end{align}
Using the definition of $H(X,Y)$ and the familiar properties of
logarithms, we arrive at the \emph{mutual information} between $X$ and
$Y$:
\begin{equation}
I(X;Y) = \sum_{x,y} p(x,y) \log\left(\frac{p(x,y)}{p_X(x) p_Y(y)}
                                    \right).
\end{equation}
Note that $I(X;Y)$ vanishes if $p(x,y)$ factors neatly into $p_X(x)$
and $p_Y(y)$.  In this case, knowing the value of~$x$ does not reduce
the number of questions we require to ascertain the value of~$y$, and
vice versa.

Consider a set $A$ of random variables, described by a joint
probability distribution over all possible outcomes of all the
variables in the set.  The Shannon index can be shown to satisfy the
following two properties.
\begin{itemize}
\item \emph{Monotonicity:} The Shannon index of a subset $U \subset A$ that is
  contained in a subset $V \subset A$ cannot be larger than that
  of~$V$.  That is, if $U \subset V$, then $H(U) \leq H(V)$.
\item \emph{Strong subadditivity:} Given two subsets, the Shannon
  index of their union cannot exceed the Shannon index of each
  separately minus the index of their intersection:
\begin{equation}
H(U \cup V) \leq H(U) + H(V) - H(U \cap V).
\end{equation}
\end{itemize}
Therefore, Shannon's formula is an information function which can be
used in the multiscale structure formalism of Chapter~\ref{ch:multiscale}.

Moreover, the Shannon index is a \emph{natural} information function
in the context of probabilities and random variables, because it is
the unique functional which satisfies certain basic and fairly
intuitive desiderata~\cite{aczel1974}.  First, the value of
information function applied to a probability distribution should not
change if we expand the set of outcomes with new elements which are
assigned probability zero.  Second, an information function should be
invariant under permutations of the probabilities: we are just as
uncertain if our expectations are
\begin{equation}
p_0 = x,\ p_1 = 1 - x,
\end{equation}
as we are if
\begin{equation}
p_0 = 1 - x,\ p_1 = x.
\end{equation}
Third, for two experiments represented by random variables $X$ and
$Y$, an information function should always satisfy
\begin{equation}
H(X \cup Y) \leq H(X) + H(Y).
\end{equation}
This follows from strong subadditivity, if we take $U = \{X\}$ and $V
= \{Y\}$.  Fourth, if the random variables $X$ and $Y$
are \emph{independent,} we should have equality in the previous
relation.

If we impose these four desiderata, then the only possible information
functions are linear combinations of the Shannon index and the
quantity
\begin{equation}
K[\{p_i\}] = \log_2 \left|\{p_i | p_i \neq 0 \}\right|,
\end{equation}
which counts the number of outcomes assigned greater than zero
probability.  This is known as the \emph{Hartley entropy.}  We can
rule out a contribution of this form if we require in addition that
for a random variable which has two possible outcomes, the information
is small if the probability of one outcome is near zero.

Many other derivations of the Shannon index from sets of desiderata
have been made.  For a recent example, see Baez, Fritz and
Leinster~\cite{baez2011}.  The characterization described here has the
advantage that it relates fairly clearly to the axioms of the
multiscale structure formalism.

\section{Moments and Cumulants}
\label{sec:diag-moments}
The mean of a probability distribution $p(x)$ is
\begin{displaymath}
\moment{x} \equiv \sum_{x = \infty}^\infty p(x) x,
\end{displaymath}
or for a continuous distribution,
\begin{displaymath}
\moment{x} \equiv \int dx\, p(x) x.
\end{displaymath}
This expression readily generalizes to that for the
\emph{expectation value} of a function $f$,
\begin{equation}
\moment{f(x)} \equiv \int dx\, f(x) p(x).
\end{equation}
Expectation values of powers of $x$ are called \emph{moments} of
$x$:
\begin{equation}
\moment{x^n} = \int dx\, p(x) x^n.
\end{equation}
Moments are interesting things to know about a probability
distribution, but they are not always the most convenient values for
characterizing how such distributions work.  For example, suppose we
have two random variables,\index{random variable} $X$ and $Y$, which
independently of one another take values according to probability
distributions $p_X(x)$ and $p_Y(y)$.  What can we say about a third
random variable $Z \equiv X + Y$?  And, is there some characterization
of~$p_X$ and $p_Y$ such that we can just add properties of~$p_X$ and
$p_Y$ to obtain the corresponding property of~$p_Z$?

If $Z$ takes the value $z$ and $X$ takes the value $x$, then $Y$ must
have taken the value $z - x$.  Following the basic rules of
probability, we deduce that because $X$ and $Y$ are independent, the
probability of $X$ taking the value $x$ and $Y$ the complementary
value $z - x$ is $p_X(x) p_Y(z - x)$.  But this is only one way for
a measurement of $Z$ to give the value $z$; the sum $x + y$ can work
out to $z$ for any value of $x$.  So, adding up the probabilities, we
find
\begin{equation}
p_Z(z) = \int dx\, p_X(x) p_Y(z - x).
\end{equation}
The probability distribution for the sum of two random variables is
the \emph{convolution} of the distributions for the random
variables being added.  This indicates that the higher moments
of~$p_X$ and $p_Y$ will not combine neatly when we construct $p_Z$.

A useful result, the \emph{convolution theorem}, says that the Fourier
transform of a convolution is the product of the Fourier transforms of
the functions convolved, up to a constant depending on how the
transformation was normalized.  (The proof is both standard and fairly
direct; see a good text on mathematical methods~\cite{gbur2011}.)
This suggests a way to proceed.

The \emph{characteristic function} of a probability distribution is
just its Fourier transform, which we can write as an expectation
value:
\begin{equation}
\tilde{p}(k) \equiv \moment{e^{-ikx}} = \int dx\, p(x) e^{-ikx}.
\end{equation}
Thus,
\begin{equation}
p(x) = \frac{1}{2\pi} \int dk\, \tilde{p}(k) e^{ikx}.
\end{equation}
When we add random variables, their characteristic functions multiply,
which means that the {\em logarithms} of their characteristic
functions {\em add.}  So, if we want to define quantities which are
like moments but which add conveniently when we combine random
variables, we should look at logarithms of characteristic functions.

Recall that
\begin{equation}
e^z = 1 + z + \frac{z^2}{2!} + \frac{z^3}{3!} + \cdots,
\end{equation}
so we can expand the characteristic function as
\begin{equation}
\tilde{p}(k) = \moment{\sum_{n=0}^\infty \frac{(-ik)^n}{n!} x^n}
  = \sum_{n=0}^\infty \frac{(-ik)^n}{n!} \moment{x^n}.
\label{eq:characteristic-expand}
\end{equation}
This relates the characteristic function to the moments.

As indicated, we're interested in logarithms of characteristic
functions, because characteristic functions multiply under
convolution, and logarithms turn products into sums.  Expanding the
logarithm of $\tilde{p}(k)$ as we expanded $\tilde{p}(k)$ itself, we
define the \emph{cumulants} of $x$, written with the notation
$\cumulant{x^n}$.
\begin{equation}
\log \tilde{p}(k) \equiv \sum_{n = 1}^\infty \frac{(-ik)^n}{n!}
                         \cumulant{x^n}.
\label{eq:def-cumulant}
\end{equation}
When manipulating logarithms, it is sometimes useful to recall the
Taylor expansion of the logarithm function for arguments near~1:
\begin{equation}
\log(1 + \epsilon) = \sum_{n = 1}^\infty (-1)^{n+1} \frac{\epsilon^n}{n}.
\end{equation}

How do the cumulants relate to the moments?\footnote{This discussion
is based on Kardar~\cite{kardar2007, kardar2007b}, but with more
intermediate steps worked out.}  Can we find a simple formula for one
set of numbers in terms of the other?  We have two formulas involving
the characteristic function $\tilde{p}(k)$, one using moments and the
other using cumulants.  If we equate the expressions for
$\tilde{p}(k)$ in Eqs.~(\ref{eq:characteristic-expand}) and
(\ref{eq:def-cumulant}), we can derive the relationship between the
moments $\{\moment{x^n}\}$ and the cumulants $\{\cumulant{x^n}\}$.
From Eqs.~(\ref{eq:characteristic-expand}) and
(\ref{eq:def-cumulant}), we see that
\begin{align}
\sum_{m = 0}^\infty \frac{(-ik)^m}{m!} \moment{x^m} 
 &= \exp\left[\sum_{l=1}^\infty \frac{(-ik)^l}{l!}
  \cumulant{x^l}\right] \nonumber\\
 &= \prod_{l=1}^\infty \exp\left[\frac{(-ik)^l}{l!}
  \cumulant{x^l}\right] \nonumber\\
 &= \prod_{l=1}^\infty \sum_{n_l = 0}^\infty
  \left[\frac{(-ik)^{ln_l}}{n_l!}
    \left(\frac{\cumulant{x^l}}{l!}\right)^{n_l} \right].
\end{align}

The powers of $(-ik)^m$ on both sides have to be equal, so
\begin{equation}
\frac{\moment{x^m}}{m!} = \sum_{\{n_l\}}^\prime 
   \prod_l \frac{\cumulant{x^l}^{n_l}}{n_l! (l!)^{n_l}},
\end{equation}
in which the prime on the $\Sigma$ means we are restricting the sum
such that $\sum l n_l = m$.  Moving the $m!$ to the right-hand side,
we find
\begin{equation}
\moment{x^m} = \sum_{\{n_l\}}^\prime m!
    \prod_l \frac{1}{n_l! (l!)^{n_l}} \cumulant{x^l}^{n_l}.
\label{eq:moment-cumulant-1}
\end{equation}
The numerical factor multiplying $\cumulant{x^l}^{n_l}$ turns out to
have a fairly simple combinatorial interpretation: it is the number of
ways of breaking $m$ points into clusters, satisfying the condition
that for each value of $l$, the assignment has $n_l$ clusters of $l$
points.

This is easiest to see if we work out a simple example first.  Suppose
we have three points, which we wish to regroup into a set of two
points and a third point left to itself.  We can do this in three
distinct ways, grouping together 1 and 2, 1 and 3 or 2 and 3.  In
other words, there exist three ways to assign a total of three
particles to a 1-cluster and a 2-cluster.

\includegraphics[width=12cm]{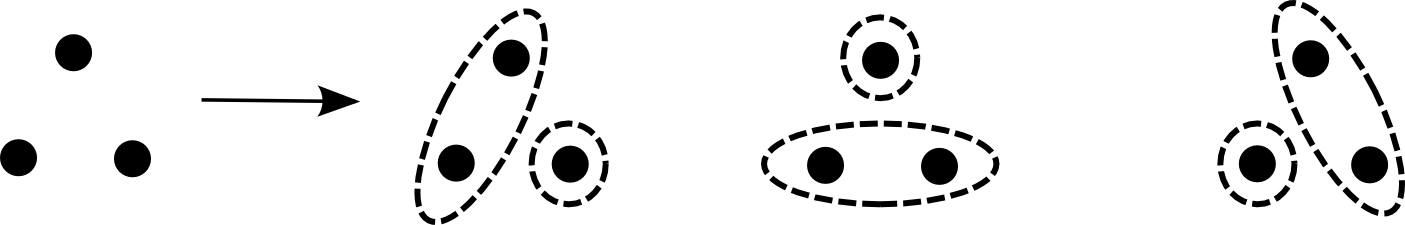}

If we write $n_1$ for the number of 1-clusters and $n_2$ for the
number of 2-clusters in our decomposition, we can say that the number
of ways $W$ to organize our 3-point set is
\begin{equation}
W(n_1 = 1, n_2 = 1) = 3.
\label{eq:W-3-cluster}
\end{equation}

We now explore the behavior of $W$ for a general set of $m$ points.
First, because every point has to be placed within one cluster or
another, the cluster sizes times the number of clusters used have to
add up to the total size of the set:
\begin{equation}
\sum_l l n_l = m.
\end{equation}
We know that we can order an $m$-element set in $m!$ ways.  We can
find a general formula for $W$ if we divide these $m!$ total
permutations by the number of equivalent ones, that is, by the number
of permutations which give indistinguishable results.  Inside each
subset of $l$ elements, we can permute the labels in $l!$ different
ways and still get an equivalent grouping.  Furthermore, we can
rearrange $n_l$ subsets in $n_l!$ ways.  The former gives us a factor
of $(l!)^{n_l}$, and the latter a factor of $n_l!$.  Thus,
\begin{equation}
W(\{n_l\}) = \frac{m!}{\prod_l n_l! (l!)^{n_l}}.
\label{eq:W-formula}
\end{equation}
To sanity-check this formula, note that for $n_1 = n_2 = 1$,
\begin{equation}
W(1, 1) = \frac{3!}{(1! 1!^1)(1! 2!^1)} = \frac{3\cdot 2}{2} = 3.
\end{equation}
Putting everything together, we use this combinatorial notation to
rewrite Eq.~(\ref{eq:moment-cumulant-1}):
\begin{equation}
\boxed{\moment{x^m} = \sum_{\{n_l\}}^\prime W(\{n_l\})
  \prod_l \cumulant{x^l}^{n_l},
  \hbox{ where } \sum_l ln_l = m.}
\label{eq:moment-cumulant-2}
\end{equation}
Given any sequence $\{\cumulant{x^l}\}$ of cumulants, we can compute
the moments.  This means that if we have two independent random
variables $X$ and $Y$, described by the probability assignments $p_X$
and $p_Y$, we can compute the moments of the new random variable $X +
Y$.  We will apply this to very good effect in section~\S\ref{sec:clt}.
 
It is a straightforward matter to work out the first few moments in
terms of the first few cumulants.  (And these relations are convenient
to have on hand for reference.)  First, because there's only one way
to make a linked cluster out of one point, the first moment is the
same as the first cumulant:
\begin{equation}
\moment{x} = \cumulant{x}.
\label{eq:first-cumulant}
\end{equation}
The second moment, $\moment{x^2}$, will have two terms, because we can
make a diagram with two disconnected points (that is, two one-point
clusters) and a diagram with two connected points (a 2-cluster).
\begin{equation}
\moment{x^2} = \cumulant{x^2} + \cumulant{x}^2
\end{equation}
The third moment $\moment{x^3}$ is slightly more complicated, since we
can chop up the three-vertex diagram into a 2-cluster and a 1-cluster
in three ways.  (We worked this out explicitly earlier.)  This means
that our formula will have a symmetry factor.

In diagram form,
\begin{equation}
\includegraphics[width=10cm]{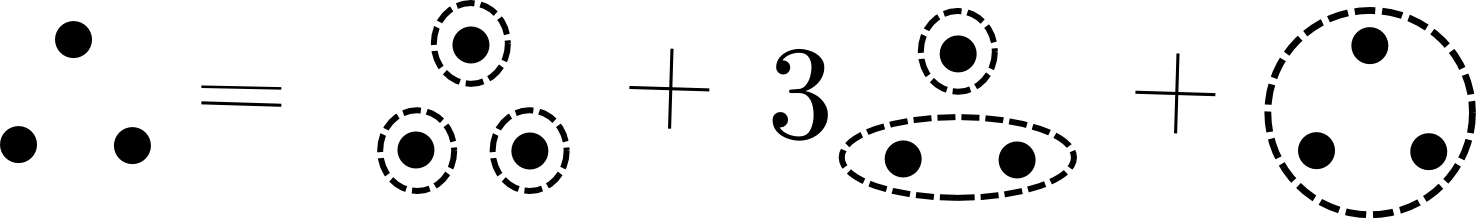}
\end{equation}
An alternate way to draw this picture is as a forest of trees:
\begin{equation}
\includegraphics[width=10cm]{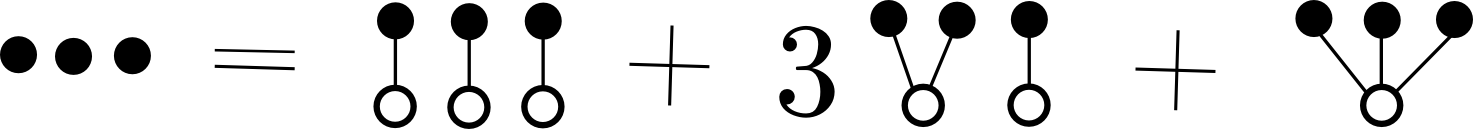}
\end{equation}
Here, the filled circles on the upper row indicate factors of~$x$, and
they belong to the same group if they are linked to the same circle on
the bottom row.  Each group, which stands for a cumulant, is a rooted
tree, and each little forest is a term in the expansion
of~$\moment{x^3}$.  Either way, the equivalent in algebraic symbols is
\begin{equation}
\moment{x^3} = \cumulant{x^3} + 3\cumulant{x^2}\cumulant{x}
   + \cumulant{x}^3.
\end{equation}

The expectation value of $x^4$ works analogously.  Expanded out as
products of cumulants, $\moment{x^4}$ is a sum of five terms:
\begin{equation}
\moment{x^4} = \cumulant{x^4} + 4\cumulant{x^3}\cumulant{x}
   + 3 \cumulant{x^2}^2 + 6\cumulant{x^2}\cumulant{x}^2
   + \cumulant{x}^4.
\label{eq:cumulant-to-moment}
\end{equation}
Note that the total degree of each term on the right-hand side is
equal to~4.  If $x$ is a quantity which carries units, this is
necessary for dimensional consistency.  The coefficients are
combinatorial symmetry factors that come from the number of ways a set
of four points can be partitioned.  There are four ways to pull out a
single point, three ways to divide the set into two pairs, and six
ways to group together a pair of points while leaving the other two
points out.  As before, we have multiple equivalent ways to derive
this: by counting ways to circle points, by counting our options to
make forests out of trees, or by plugging into our algebraic formula
for~$W(\{n_l\})$.  In fact, we can also build up these formulas
recursively, by considering the operation of merging a new tree into a
forest in all possible ways.

We have derived the first four moments in terms of the first four
cumulants, using Eq.~(\ref{eq:moment-cumulant-2}).  Inverting these
relationships yields the first four cumulants in terms of the first
four moments:
\begin{align}
\cumulant{x} &= \moment{x}\nonumber\\
\cumulant{x^2} &= \moment{x^2} - \moment{x}^2 \nonumber\\
\cumulant{x^3} &= \moment{x^3} - 3\moment{x^2}\moment{x}
   + 2\moment{x}^3 \nonumber\\
\cumulant{x^4} &= \moment{x^4} - 4\moment{x^3}\moment{x}
   -3\moment{x^2}^2 + 12\moment{x^2}\moment{x}^2
   -6\moment{x}^4.
\label{eq:moment-to-cumulant}
\end{align}
The first cumulant is just the mean.  The second cumulant,
the \emph{variance,} is a measure of spread about the mean.  Together,
they are the cumulants referred to and made use of most regularly.
Soon, in \S\ref{sec:clt}, we will see why this is the case.  The
relations in Eq.~(\ref{eq:moment-to-cumulant}) are, in practice, about
as high-order as one needs to calculate.  Cumulants $\cumulant{x^n}$
with $n > 4$ aren't invoked as often.

We are now in a position to understand the approximations we made
in \S\ref{sec:volunteer-analytic}.  The only extra wrinkle is that in
those calculations, we used \emph{cross-cumulants} of multiple
different random variables (for example, $s_1$ and $s_2$).  This
breaks the symmetry which yields the numerical coefficients, but the
logic is otherwise the same:
\begin{equation}
\moment{xyz} = \cumulant{xyz}
   + \cumulant{xy}\cumulant{z}
   + \cumulant{yz}\cumulant{x}
   + \cumulant{zx}\cumulant{y}
   + \cumulant{x}\cumulant{y}\cumulant{z}.
\end{equation}
We obtain the result we used in the previous chapter,
Eq.~(\ref{eq:s1s2s3-gaussian}), by imposing the approximation that the
third-order cross-cumulant $\cumulant{xyz}$ vanishes.  If the
second-order cross-cumulants vanish as well, then the expectation
value of the product is just the product of the means.

The coefficients in Eq.~(\ref{eq:cumulant-to-moment}) show up
somewhere else, too.  Consider two functions, $f(x)$ and $g(x)$.
Composing them yields a new function, $f(g(x))$, and the chain rule
tells us its derivative:
\begin{equation}
\frac{d}{dx}\left[f(g(x))\right]
 = \frac{df}{dg}\frac{dg}{dx}.
\end{equation}
Now, let's differentiate again.  Using the product rule,
\begin{equation}
\frac{d^2}{dx^2}\left[f(g(x))\right]
 = \frac{d^2f}{dg^2}\left(\frac{dg}{dx}\right)^2
   + \frac{df}{dg}\frac{d^2g}{dx^2}.
\end{equation}
We can repeat this process, cranking through the algebra:
\begin{equation}
\frac{d^3}{dx^3}\left[f(g(x))\right]
 = 3\left(\frac{dg}{dx}\right)\frac{d^2f}{dg^2}
    \frac{d^2g}{dx^2}
   + \left(\frac{dg}{dx}\right)^3
     \frac{d^3 f}{dg^3}
   + \frac{df}{dg}
     \frac{d^3g}{dx^3}.
\end{equation}
And, taking it one more step,
\begin{align}
\frac{d^3}{dx^3}\left[f(g(x))\right]
 = &\ \frac{d^4f}{dg^4}\left(\frac{dg}{dx}\right)^4 \nonumber\\
   & + 6\frac{d^3f}{dg^3}\frac{d^2g}{dx^2}
        \left(\frac{dg}{dx}\right)^2 \nonumber\\
   & + 3\frac{d^2f}{dg^2}
        \left(\frac{d^2g}{dx^2}\right)^2 \nonumber\\
   & + 4\frac{d^2f}{dg^2}\frac{d^3g}{dx^3}
        \frac{dg}{dx} \nonumber\\
   & + \frac{df}{dg}\frac{d^4g}{dx^4}.
\end{align}
This is sufficient to see a pattern take shape.

The coefficients in each sum are the same numbers, $W(\{n_l\})$, which
we encountered when converting between moments and cumulants.  In the
ordinary chain rule, we have one term, and its coefficient is 1.  For
the second derivative, we get two terms, each of which has a
coefficient of~1.  In the third derivative, one term has a coefficient
of~3, and when we go to the fourth derivative, the coefficients are
the same as those in the expression for the moment $\moment{x^4}$ in
terms of the first four cumulants.

We can make the constructions exactly analogous if we identify
a \emph{connected cluster of $k$ points} with the $k^{\rm th}$
derivative of~$g(x)$.  Each term is a derivative of~$f$, times one or
more factors which are derivatives of~$g$.  The diagram we used to
demonstrate Eq.~(\ref{eq:W-3-cluster}) also tells us the coefficients
in the third derivative of~$f(g(x))$.

Why should this be the case?  We shall see in the next section.

\section{Generating Functions}
\label{sec:generating-functions}
Our diagrammatic methods assign a {\em weight} to a graph, by
interpreting that graph as an integral over a probability
distribution.  By construction, these graph weightings satisfy two
properties.  First, the weight of any graph is the product of the
weights of its pieces, and second, the total weight of any collection
of graphs is the sum of the weights of the individual graphs.  These
properties combine to yield an interesting and quite general result.

The moment $\moment{x^n}$ is the summed weight of all graphs having
$n$ vertices.  To understand this series better, we ``hang it on a
clothesline''~\cite{wilf1994} by writing its \emph{generating
function}:
\begin{equation}
\cQ(z) = \sum_{n=0}^\infty \frac{z^n}{n!} \moment{x^n}.
\end{equation}
Using Eq.~(\ref{eq:moment-cumulant-2}), we can rewrite this generating
function in terms of the cumulants $\cumulant{x^n}$:
\begin{equation}
\cQ(z) = \sum_{n=0}^\infty \frac{z^n}{n!}
\sum_{\{n_l\}}^\prime W(\{n_l\})
  \prod_l \cumulant{x^l}^{n_l},
  \hbox{ where } \sum_l ln_l = n.
\end{equation}
The $l$-th cumulant $\cumulant{x^l}$ is the weight of an $l$-vertex
connected graph.

Because we are summing over all $n$, we can remove the restriction on
the summation over $\{n_l\}$.  First, we recall the definition of
$W(\{n_l\})$.
\begin{equation}
\cQ(z) = \sum_{n = 0}^\infty \frac{z^n}{n!} 
     \sum_{\{n_l\}}^\prime 
     \frac{n!}{\prod_l n_l! (l!)^{n_l}}
     \prod_l \cumulant{x^l}^{n_l}.
\end{equation}
One way to think of this is as follows: having ``primed'' the
summation symbol means that we've inserted an implicit delta
function.  We can really sum over all sets $\{n_l\}$, as long as it's
understood that each term is multiplied by a Kronecker delta which is
zero whenever $\sum_l ln_l$ is not equal to $n$.  But the outer
summation symbol means that we have terms for all values of $n$,
meaning that we can really take $\{n_l\}$ to be anything we like, as
long as in that term, $n$ is set equal to $\sum_l ln_l$.

In frighteningly abstracted notation, we are saying that
\begin{equation}
\sum_{n=0}^\infty \sum_{\{n_l\}} \delta_{\sum_l ln_l, n}
 = \sum_{\{n_l\}},
\end{equation}
so we can rewrite the generating function as
\begin{align}
\cQ(z) &= \sum_{\{n_l\}} z^{\sum_l ln_l}
     \prod_l \frac{\cumulant{x^l}^{n_l}}{n_l! (l!)^{n_l}} \nonumber\\
  &= \sum_{\{n_l\}} \prod_l \frac{1}{n_l!}
     \left(\frac{z^l \cumulant{x^l}}{l!}\right)^{n_l}.
\end{align}
Interchanging the product and summation symbols and employing the
definition of the exponential,
\begin{equation}
\cQ(z) = \prod_l \exp\left(z^l \frac{\cumulant{x^l}}{l!}\right),
\end{equation}
or,
\begin{equation}
\boxed{\cQ(z) = \exp\left(\sum_{l=1}^\infty
                     \frac{\cumulant{x^l}}{l!} z^l\right).}
\label{eq:exp-of-generating}
\end{equation}
The generating function over all graphs turns out to be the
exponential of the generating function for \emph{connected} graphs.
This is known as the \emph{linked-cluster theorem.}

We used only very general properties of graph weights in order to
derive Eq.~(\ref{eq:exp-of-generating}).  As long as those basic
conditions are met, the same relationship will hold, no matter what
interpretation we give to the vertices and edges of our graphs.

Another application occurs in statistical physics.  If the vertices of
our graphs represent particles and the graph weighting is determined
by pairwise interactions between particles, then the generating
function $\cQ(z)$ has the form of the grand partition function, with
the formal variable $z$ playing the role of the fugacity
$e^{\beta\mu}/\lambda^3$, which represents the difficulty of changing
particle number by thermal fluctuations~\cite{kardar2007}. The grand
partition function includes all ``Mayer graphs'', connected or not;
Eq.~(\ref{eq:exp-of-generating}) states that the sum of all graphs is
given by the exponential of the sum over connected graphs.

Now, we can see why the pattern of coefficients we saw in the iterated
chain rule relates to moments and cumulants.  If $f(z)$ is a
generating function over some formal variable $z$,
\begin{equation}
f(z) = \sum_{n=0}^\infty \frac{z^n}{n!} f_n,
\end{equation}
then we can extract the coefficient $f_n$ by differentiating $n$ times
and evaluating the result at zero:
\begin{equation}
f_n = \left.\frac{d^n}{dz^n}f(z)\right|_{z=0}.
\end{equation}
The differentiations remove all terms of lower order than $n$, and
setting $z=0$ removes all terms of higher order, leaving only $f_n$.

Consider the generating function defined by
\begin{equation}
f(z) = \exp\left[g(z)\right],
\end{equation}
where $g(z)$ is itself a generating function,
\begin{equation}
g(z) = \sum_{n=0}^\infty \frac{z^n}{n!}g_n.
\end{equation}
If we associate the expansion coefficients $g_n$ with connected
diagrams, then we are back to the linked-cluster theorem again.  The
values of~$f_n$ depend on the $g_n$.  But we know that $f(z)$ is
the \emph{composition} of~$g(z)$ with the exponential function, and we
know that the $f_n$ relate to the derivatives of~$f(z)$, so we can
compute the $f_n$ by iterating the chain rule.  The pictorial
interpretation is that the operation of merging a new tree into a
forest in all possible ways represents the product rule, which says
that a derivative acts on a product to create a sum of new products,
each of which has one factor modified.  In probability theory, we also
merge a new tree into a forest, when we go from the cumulant expansion
of~$\moment{x^n}$ to that of~$\moment{x^{n+1}}$.

\section{Central Limit Theorem}
\label{sec:clt}
What happens when we add together many independent random variables?
The same logic we used in \S\ref{sec:diag-moments} still applies:
cumulants of their distributions add.  If the random variable $Y$ is
given by
\begin{equation}
Y = \sum_{i=1}^N X_i,
\end{equation}
where $X_i$ are independent random variables described by
distributions $p(x_i)$, then the $n$th cumulant of $Y$,
$\cumulant{y^n}$, is simply the sum
\begin{equation}
\cumulant{y^n} = \sum_{i=1}^N \cumulant{x_i^n}.
\end{equation}
To take the simplest case, suppose that the distributions of all the
$X_i$ are identical.  Then $\cumulant{y^n} = N \cumulant{x^n}$,
meaning that the new random variable
\begin{equation}
Z \equiv \frac{Y - N\cumulant{x}}{\sqrt{N}}
\end{equation}
has a mean of zero, and higher cumulants which scale as
$\cumulant{z^n} \propto N^{1 - n/2}$.  For $N$ sufficiently big, all
cumulants higher than the second die off, and $Z$ becomes \emph{Gaussian.}
\begin{equation}
\boxed{\lim_{N\rightarrow\infty} 
    p\left(z = \frac{\sum_i x_i - N\cumulant{x}}{\sqrt{N}}\right)
  = \frac{1}{2\pi\cumulant{x^2}}
    \exp\left(-\frac{z^2}{2\cumulant{x^2}}\right).}
\end{equation} 

We have arrived at the \emph{Central Limit Theorem}: the sum of many
uncorrelated random variables is itself a random variable, whose
probability distribution is approximately Gaussian.  A Gaussian
distribution is specified by two parameters, which we denote here
as~$\mu$ and $\sigma$:
\begin{equation}
p(x) = \frac{1}{\sqrt{2\pi\sigma^2}}
       \exp\left(-\frac{(x - \mu)^2}{2\sigma^2}\right).
\label{eq:gaussian-def}
\end{equation}
The Fourier transform of a Gaussian curve is itself a Gaussian:
\begin{equation}
\tilde{p}(k) = \int_{-\infty}^\infty \frac{dx}{\sqrt{2\pi\sigma^2}}
  e^{-ikx} \exp\left(-\frac{(x - \mu)^2}{2\sigma^2}\right)
  = \exp\left(-ik\mu - \frac{k^2\sigma^2}{2}\right).
\end{equation}
We can read off the cumulants:
\begin{equation}
\cumulant{x} = \mu,\ \cumulant{x^2} = \sigma^2,
\end{equation}
and $\cumulant{x^n} = 0$ for all $n > 2$.  The Gaussian distribution
is characterized entirely by its first two cumulants, namely the mean
$\mu$ and the variance $\sigma^2$.

A neat thing happens when we take Eqs.~(\ref{eq:first-cumulant})
through (\ref{eq:cumulant-to-moment}) and set all cumulants of higher
order than the variance to zero.  Carrying this out, we get,
\begin{align}
\moment{x} &= \mu \nonumber\\
\moment{x^2} &= \sigma^2 + \mu^2\nonumber\\
\moment{x^3} &= 3\sigma^2\mu + \mu^3 \nonumber\\
\moment{x^4} &= 3\sigma^4 + 6\sigma^2\mu^2 + \mu^4.
\end{align}
In diagram language, our graphs are allowed to have clusters of one
point and clusters of two points, but clusters of three or more points
cause the graph value to vanish.  Alternatively, in the forest
picture, trees with more than two leaf nodes evaluate to zero.  The
numerical value of a forest is the product of the values of its
constitutent trees, so any forest containing a tree with too many leaf
nodes will, likewise, contribute nothing to the sum total.

Recall the approximation we made in \S\ref{sec:volunteer-analytic}, in
order to understand random neutral drift (and, thus, weak-selection
evolutionary dynamics) on the hexagonal lattice.  As explained above,
that approximation involves setting cross-cumulants of third and
higher order to zero.  We can, therefore, consider it a \emph{Gaussian
closure.}

One of the many applications of the Central Limit Theorem is
to \emph{random walks.}  Consider a walker that starts at $x = 0$
takes steps whose direction and length are chosen at random, according
to a probability distribution $p_1(l)$.  We can choose,
conventionally, that negative increments are motion toward the left,
and positive increments are steps toward the right.  What is the
probability to be at position $x$ after $N$ steps?  How do we
characterize this probability distribution, and how does it change
over time, as the walker takes more steps?

The position after $N$ steps is
\begin{equation}
x = l_1 + l_2 + \cdots + l_N.
\end{equation}
Because the $l_i$ come from identical independent distributions, the
average displacement is
\begin{equation}
\moment{x} = \moment{l_1} + \cdots + \moment{l_N}
  = N \moment{l}.
\end{equation}
However, the final displacement of the walker can be greater or
smaller than this value.  The probability distribution for the net
displacement will have some variance,
\begin{align}
\cumulant{x^2} &= \sum_{i,j=1}^N
  ( \moment{l_il_j} - \moment{l_i} \moment{l_j})\\
  &= \sum_{i,j=1}^n \cumulant{l^2} \delta_{ij}\\
  &= N \cumulant{l^2}.
\end{align}

All the cumulants follow this general pattern: after $N$ steps, their
value is multiplied by $N$.  This is a specific example of the general
property which we established for cumulants of independent random
variables added together.  By the Central Limit Theorem, we can
neglect the higher-order cumulants for~$N\gg 1$.  Thus, the final
probability distribution for large $N$ will be a Gaussian, which we
can write
\begin{equation}
p(x,N) = \exp\left[-\frac{(x - N\moment{l})^2}{2N\cumulant{l^2}}\right]
\frac{1}{\sqrt{2\pi N\cumulant{l^2}}}.
\end{equation}

Because $p(x,N)$ is an exponential function, taking its derivative
yields back $p(x,N)$ again.  We work out the first two derivatives
with respect to position, and the first derivative with respect to~$N$:
\begin{align}
\pdf{p}{x} &= -\frac{x - N\moment{l}}{N\cumulant{l^2}} p,\\
\pdfh{p}{x}{2} &= \frac{(x - N\moment{l})^2}
  {N^2 \cumulant{l^2}^2}p - \frac{1}{N\cumulant{l^2}} p,
\end{align}
and, formally,
\begin{equation}
\frac{\partial p}{\partial N} = \left[
\frac{\moment{l} (x - N\moment{l})}{N\cumulant{l^2}}
  + \frac{(x - N\moment{l})^2}{2N^2\cumulant{l^2}} - \frac{1}{2N}
\right]p.
\end{equation}
We can use the first two formulae to simplify the third:
\begin{equation}
\frac{\partial p}{\partial N} = -\moment{l} \frac{\partial
  p}{\partial x} + \frac{\cumulant{l^2}}{2} \frac{\partial^2
  p}{\partial x^2}.
\end{equation}
Instead of expressing how $p$ changes with the number of steps $N$, we
can say how $p$ changes over time, if we introduce a timestep $\tau$.  Define
\begin{equation}
v = \frac{\moment{l}}{\tau},\ D = \frac{\cumulant{l^2}}{\tau}.
\end{equation}
Then, we obtain a \emph{diffusion equation,} which relates the
spatial and temporal partial derivatives of the probability density:
\begin{equation}
\frac{\partial p(x,t)}{\partial t} = -v \frac{\partial p}{\partial x}
+ \frac{D}{2} \frac{\partial^2 p}{\partial x^2},
\end{equation}
If the random walk is \emph{unbiased,} then $v = 0$, and the diffusion
equation simplifies to
\begin{equation}
\boxed{\frac{\partial p(x,t)}{\partial t} =
 \frac{D}{2} \frac{\partial^2 p}{\partial x^2}.}
\label{eq:diffusion}
\end{equation}

We derived this diffusion equation in terms of the probability that a
single random walker will arrive at position $x$.  The equation
codifies how our expectations for the walker at different times relate
to each other.  If we have many walkers moving along the same axis but
not interacting with each other, then the fraction of those walkers
which we expect to fall within a given interval scales with the
integral of the probability density $p(x,t)$ over that interval.  That
is, the probability density governs our expectation for the number density.

This picture would become significantly more involved if the different
random walkers in the population interacted with one another.  In that
case, we would have no reason to assume that the joint probability
distribution over all walker positions should factor nicely into a
product of single-walker probability densities.

The left-hand side of the diffusion equation (\ref{eq:diffusion}) is a
time derivative, so its units are those of~$p$ divided by time.  On
the right-hand side, we have a second derivative with respect to
position, which introduces units of length$^{-2}$.  In order for the
dimensions on both sides to match, the constant $D$ must have units of
length$^2$ per time.  This is one way to remember the intuition that
for a diffusive process, variance grows linearly with time, or to say
it another way, a characteristic length scale increases with the
square root of the time elapsed.

\section{Variations on Diffusion}
\label{sec:variations-on-diffusion}
Diffusion becomes more complicated when multiple types of particle are
moving together through the same space, and the overall amount of
space available to move through is limited.  By considering the effect
of volume limitations at the rate-equation level, we can derive in a
simpler way some interesting modifications to the ordinary diffusion
equation, alterations which have been seen to emerge from a highly
involved analysis~\cite{lugo2008, biancalani2010}.

We consider three sites, with occupation numbers $a_1$, $a_2$ and
$a_3$.  Particles jump away from a site at an overall rate $k$, and
jumps are equally likely in either direction.  We make the
approximation that the \emph{actual} flow is the same as
the \emph{most probable} flow, so we can quantify everything in terms
of average rates.  The change of the population at site~2 is a result
of flows into site~2 from sites~1 and 3, and the flows out of site~2
into its neighbors:
\begin{equation}
\frac{da_2}{dt} = \frac{k}{2} a_1 - ka_2 + \frac{k}{2}a_3.
\end{equation}
We rearrange this as follows:
\begin{align}
\frac{da_2}{dt} &= \frac{k}{2}(a_1 - 2a_2 + a_3) \\
 &= \frac{k}{2}(a_1 - a_2 + a_3 - a_2) \\
 &= \frac{k}{2}[(a_3 - a_2) - (a_2 - a_1)].
\end{align}
The time derivative of $a_2$ is the difference of two differences,
which we can express as
\begin{equation}
\frac{da_2}{dt} = \frac{k}{2}[(\Delta a)_2 - (\Delta a)_1],
\end{equation}
or in terms of the \emph{discrete Laplacian operator} as
\begin{equation}
\boxed{\frac{da_2}{dt} = \frac{k}{2} (\Delta^2 a)_2.}
\end{equation}
This is exactly the discrete analogue of the diffusion equation for a
continuous line, Eq.~(\ref{eq:diffusion}).

A more intricate case is the diffusion of two substances through a
common space, with a limit imposed on how much total substance can be
at any point:
\begin{equation}
a_i + b_i \leq N,\ \forall\ i.
\end{equation}
Writing $e_i$ for the amount of empty space at location $i$, we can
turn this inequality into an equality:
\begin{equation}
a_i + b_i + e_i = N.
\end{equation}
Because flow can only happen into empty space, the population of type
$a$ at site~2 changes as
\begin{equation}
\frac{da_2}{dt} = -k(e_1+e_3)a_2 + \frac{k}{2}a_1e_2
 + \frac{k}{2}a_3e_2.
\end{equation}
Solving for $e_i$ in terms of $a_i$ and $b_i$, this becomes
\begin{equation}
\frac{da_2}{dt} = \frac{kN}{2}(a_1 - 2a_2 + a_3)
 + \frac{k}{2} \left[(b_1 + b_3)a_2 - b_2(a_1 + a_3)\right].
\end{equation}
The first term, which carries a factor of $N$, has the same functional
dependence on the $\{a_i\}$ that we saw before, with the ordinary
diffusion process.  The second term is more complicated, having an
interdependence between the $\{a_i\}$ and the $\{b_i\}$.  A bit of
algebra reveals a convenient way to state that interdependence:
\begin{align}
(b_1 + b_3)a_2 - b_2(a_1 + a_3)
 &= a_2(b_1 + b_3) - 2a_2b_2 - b_2(a_1 + a_3) + 2b_2a_2 \\
 &= a_2(b_1 - 2b_2 + b_3) - b_2(a_1 - 2a_2 + a_3).
\end{align}
This has the structure of ``self times the Laplacian of the other,
minus the other times the Laplacian of the self.''  The same holds
true for the time evolution of~$b_i$, with the roles of the $a$- and
$b$-variables interchanged.

Now, we express the $\{a_i\}$ and the $\{b_i\}$ as \emph{fractions} of
the volume at each site, $N$:
\begin{equation}
a_i = Nf_i,\ b_i = Ng_i.
\end{equation}
The time derivative of $a_2$ tells us the time derivative of~$f_2$,
which in terms of the $f$- and $g$-variables works out to be
\begin{equation}
\frac{df_2}{dt} = \frac{kN}{2}(f_1 - 2f_2 + f_3)
 + \frac{kN}{2}\left[(g_1+g_3)f_2 - g_2(f_1+f_3)\right].
\end{equation}
Using the discrete Laplacian operator,
\begin{equation}
\boxed{\frac{df_i}{dt} = \frac{kN}{2} (\Delta^2 f)_i
 + \frac{kN}{2} \left[f_i (\Delta^2 g)_i - g_i (\Delta^2 f)_i\right].}
\label{eq:anomalous-cross-diffusion}
\end{equation}
The differential equation we had before is modified
by \emph{cross-diffusion} terms, which couple the time evolutions of
the two fields.  These new terms arise from the fact that limiting the
total available volume at each site curtails the maximum possible flow
rate.

What if spreading happens by \emph{reproduction} instead of by
hopping?  This is, for example, characteristic of the spatial
host--consumer model we studied at length in
Chapter~\ref{ch:redgreen}, and the evolutionary game-based lattice
model of Chapter~\ref{ch:volunteer}.  As before, we require empty
space to move into, so the growth rate from site~$i$ into site~$j$
should go like
\begin{equation}
ka_i e_j = ka_i(N - a_j - b_j).
\end{equation}

There is no hopping out of a site, only the possibility of budding
into it.  So, the rate of change at site $a_2$ must be given by the
rate at which budding events can happen, which depends upon the number
of particles present at the neighboring sites, modulated by the amount
of empty space availabe at~$a_2$.  Implementing this idea in equations,
\begin{align}
\frac{da_2}{dt} &= ka_1(N-a_2-b_2) + ka_3(N-a_2-b_2) \\
 &= k(N-a_2-b_2)(a_1+a_3) \\
 &= k(N-a_2-b_2)\left[(\Delta^2a)_2 + 2a_2\right].
\end{align}
Examining this result, we see an \emph{effective diffusion} term
appear: as before, the time derivative of~$a_2$ can be written using
$(\Delta^2 a)_2$.

Cross-diffusion terms like those in
Eq.~(\ref{eq:anomalous-cross-diffusion}) become important in the study
of \emph{reaction-diffusion} systems.  Suppose we have two substances,
whose densities are given by the functions $a(x,t)$ and $b(x,t)$.
The particles of both substances can spread diffusively, but when they
bump into each other, reactions can take place, which introduce the
possibility of creating or destroying particles.  How many such
reactions happen depends on the concentration of particles present.
So, we write a system of equations to express how the densities should
change:
\begin{align}
\frac{\partial a(x,t)}{\partial t}
 &= D_1 \frac{\partial^2 a}{\partial x^2} + f_1(a, b), \\
\frac{\partial b(x,t)}{\partial t}
 &= D_2 \frac{\partial^2 b}{\partial x^2} + f_2(a, b).
\end{align}
It is a common practice to start with a model defined with a single
location, writing functions $f_1$ and $f_2$ to represent the local
dynamics, and then promote this construction to a spatial model by
making $a$ and $b$ position-dependent and adding the diffusion terms.
However, if the effective available volume at each position is
limited, then this is not correct~\cite{lugo2008, biancalani2010}.  We
have in that case to include the cross-diffusion terms:
\begin{align}
\frac{\partial a(x,t)}{\partial t}
 &= D_1 \left[\frac{\partial^2 a}{\partial x^2}
              + a\frac{\partial^2 b}{\partial x^2}
              - b\frac{\partial^2 a}{\partial x^2}
        \right] + f_1(a, b), \\
\frac{\partial b(x,t)}{\partial t}
 &= D_2 \left[\frac{\partial^2 b}{\partial x^2}
              + b\frac{\partial^2 a}{\partial x^2}
              - a\frac{\partial^2 b}{\partial x^2}
        \right] + f_2(a, b).
\end{align}

The cross-diffusion terms introduce an extra interdependence between
the $a(x,t)$ and $b(x,t)$, above and beyond that which might be
defined in the reaction terms $f_1(a,b)$ and $f_2(a,b)$.

\chapter{Stochastic Adaptive Dynamics}

\label{ch:adaptive}
In Chapter~\ref{ch:volunteer}, we examined the dynamics of
evolutionary games with a discrete set of strategies, and we wrote
coupled differential equations to define dynamical systems.  The
competing populations in our dynamical systems were continuously
variable in size.  Now, we consider a complementary type of scenario,
in which the game-players' strategies themselves are continously
variable.  For example, the amount to which an individual is willing
to contribute to a group-level effort to attain some social good could
be a continuous quantity.  

To simplify our analysis, we shall follow a precedent set in the
literature and focus on cases where the total population size is
constant.  This is another way in which the current chapter is
complementary to Chapter~\ref{ch:volunteer}.  The models we study in
this chapter belong to an area of evolutionary theory known
as \emph{adaptive dynamics.}  Our goal will be to extend recent
results in adaptive dynamics beyond the deterministic limit and into a
stochastic regime.

\section{Justifying the Fokker--Planck Equation}

Consider a random walk over a set of sites, in which the walker moves
from site~$i$ to site~$j$ at a rate~$\Pi_{ji}$.  That is, the
probability to jump in a short time~$dt$ is $\Pi_{ji}dt$.  How do the
occupation probabilites $\{P_i(t)\}$ change over time?  A given
occupation probability can change because the walker is likely to jump
{\em to} that site or because a walker at that site is likely to jump
{\em away.}  In mathematics,
\begin{equation}
\frac{dP_i}{dt} = - \sum_j \Pi_{ji} P_i + \sum_j \Pi_{ij} P_j.
\label{eq:master}
\end{equation}
This relation is known as the {\em master equation.}   By taking the
  continu\"um limit, in which the sites are very closely squished, we
  can arrive at another useful relation.  We essentially make the
  transformations
\begin{equation}
i \rightarrow x,\ P_i \rightarrow P(x,t),\ \Pi_{ji} \rightarrow
\Pi(x'-x,x).
\end{equation}
Note that we have re-parametrized the jump probability to use the
difference of positions, rather than the positions themselves.  (The
two parametrizations are, of course, entirely equivalent.)  The result
of transforming Eq.~(\ref{eq:master}) in this fashion is
\begin{equation}
\frac{\partial}{\partial t} P(x,t) =
 - \int dx'\, \Pi(x'-x,x)P(x,t) 
 + \int dx'\, \Pi(x-x',x') P(x',t).
\label{eq:master-transformed}
\end{equation}
The transition rates depend on the separation between the old position
and the new position, $y = x' - x$.  A small change in the new
position~$x'$ implies a small change in the separation: $dx' = dy$.
Using the new variable~$y$, Eq.~(\ref{eq:master-transformed}) becomes
\begin{equation}
\frac{\partial}{\partial t} P(x,t) =
 - \int dy\, \Pi(y,x)P(x,t) + \int dy\, \Pi(y,x-y)P(x-y,t).
\label{eq:master-substituted}
\end{equation}
For most physical applications, we expect typical changes to be local,
{\em i.e.,} that $\Pi$ is dominated by small $y$.  We can then expand
the second integral in Eq.~(\ref{eq:master-substituted}) as a Taylor
series.  The first term in this expansion cancels with the first
integral in Eq.~(\ref{eq:master-substituted}).  The next two terms
yield
\begin{equation}
\frac{\partial}{\partial t} P(x,t) =
 - \int dy\, y \frac{\partial}{\partial x}(\Pi(y,x) P(x,t))
 + \frac{1}{2} \int dy\, y^2 \frac{\partial^2}{\partial x^2}
   (\Pi(y,x) P(x,t)).
\end{equation}
We can take the derivatives outside of the integrals, like so:
\begin{equation}
\frac{\partial}{\partial t} P(x,t) =
-\frac{\partial}{\partial x}
\left[P(x)\int dy\, y \Pi(y,x)\right] +
 \frac{1}{2} \frac{\partial^2}{\partial x^2}
\left[\int dy\, y^2 (\Pi(y,x) P(x,t))\right].
\end{equation}
This becomes a close analogue of the diffusion equation, which we
derived in Eq.~(\ref{eq:diffusion}).  Each term involves a factor
which is essentially a moment of $\Pi(y,x)$ with respect to~$y$.
Define
\begin{equation}
F(x) = \int dy\, y \Pi(y,x),
\end{equation}
and
\begin{equation}
D(x) = \int dy\, y^2 \Pi(y,x).
\end{equation}
Then
\begin{equation}
\boxed{\frac{\partial P(x,t)}{\partial t} =
- \frac{\partial}{\partial x}\left[
  F(x) P(x,t)\right]
+ \frac{1}{2}\frac{\partial^2}{\partial x^2}\left[
  D(x) P(x,t)\right].}
\label{eq:FP}
\end{equation}
Eq.~(\ref{eq:FP}) is known as the {\em Fokker--Planck equation.}  For
many applications, $D(x)$ can be taken as a constant, and so,
\begin{equation}
\frac{\partial P(x,t)}{\partial t} =
 - \frac{\partial}{\partial x}\left[F(x) P(x,t)\right]
 + \frac{D}{2} \frac{\partial^2 P(x,t)}{\partial x^2}.
\label{eq:FP2-a}
\end{equation}
An appropriate change of variables can in fact transform away the
position dependence of~$D(x)$, so this special case can be used even
when the ease of diffusion is position-dependent~\cite{risken1996}.

If we apply the product rule, we can expand the derivatives in
Eq.~(\ref{eq:FP}), obtaining
\begin{equation}
\frac{\partial P(x,t)}{\partial t}
 = -\frac{D}{2}P'' + (D'-F)P'
   + \left(\frac{D''}{2} - F'\right)P.
\end{equation}
Note that this expression includes both the first derivative of~$F(x)$
and the second derivative of~$D(x)$.  This will be important soon.

To find the steady-state solution, $P^*(x)$, for the Fokker--Planck
equation with constant diffusion, set the derivative with respect to
time equal to zero.  This implies
\begin{equation}
\frac{\partial}{\partial x} \left[F(x) P^*(x)\right]
= \frac{D}{2} \frac{\partial^2 P^*(x)}{\partial x^2},
\end{equation}
or
\begin{equation}
\frac{\partial^2 P^*(x)}{\partial x^2}
 = \frac{2}{D} \frac{\partial}{\partial x} \left[F(x) P^*(x)\right].
\end{equation}
Integrating once over $x$,
\begin{equation}
\frac{\partial P^*(x)}{\partial x}
 = \frac{2F(x)}{D} P^*(x).
\end{equation}
When the derivative of a function gives back the function itself, the
function is an exponential.  The question is, an exponential of what?
The derivative of its argument with respect to $x$ must be $2F(x)/D$.
Recall that the derivative of an integral with respect to its upper
limit is the integrand evaluated at that limit.  So,
\begin{equation}
P^*(x) \propto
     \exp\left(\frac{2}{D}
               \int^x dx'\,F(x')\right).
\label{eq:FP-solution}
\end{equation}
The choice of the lower limit and the constant of integration glossed
over earlier can be absorbed into the prefactor which establishes the
proper normalization,
\begin{equation}
\int dx\,P^*(x) = 1.
\end{equation}

If the diffusion function $D(x)$ is not constant, then we can find the
steady-state distribution in much the same way.  The result is the
slightly modified formula
\begin{equation}
P^*(x) \propto \frac{1}{D(x)}
 \exp\left(2 \int^x dx' \frac{F(x')}{D(x')}\right).
\label{eq:FP-equilibrium}
\end{equation}

In what follows, we will use the Fokker--Planck equation in the
following way.  The variable $x$ will denote the \emph{value of a
  genetic trait in a population.}  We will assume that the population
is uniform: each time we survey it, all the organisms have the same
genotype.  However, in between observations, mutations can occur,
creating individuals with slightly different genotypes.  If the
offspring of a mutant take over the population, then the value of~$x$
will change by a small amount.  If the mutants fail to supplant the
native population, then $x$ stays the same.  The timescale of this
ecological competition will, in our analysis, be much shorter than the
timescale over which we follow the variable $x$.

Even though the population is genetically uniform at each observation,
we might not be \emph{certain} about its genetic composition.  For
example, in a practical context we might only be able to carry out a
few observations, and the measurements we take might be confounded by
environmental factors, meaning that multiple values of~$x$ could be
consistent with the data we gather.  We therefore summarize our
knowledge of the population by a probability distribution over~$x$.
If we know how mutations can happen, then we can say something about
what $x$ might be at a later time.  However, because the outcomes of
mutation events and competition are not certain, our statements about
what $x$ might be in the future must also be probabilistic.  The
function $P(x,t)$ expresses what we can deduce about what $x$ might be
at time~$t$, given the information currently at hand.  When the
time~$t$ actually rolls around, we might go out and gather new data,
allowing us to refine our information about what genotypes could be
present.  (This is the kind of problem for which we derived
probability-update rules in Chapter~\ref{ch:prob}.)  For the most
part, we will not be concerned here with that side of the problem; we
will in this chapter focus on the computation of expectations in the
absence of new information.

\section{The Deterministic Limit}

We define an evolutionary process in the following way: at any time, a
mutant can arise in an otherwise uniform population.  The mutation
rate can in general depend upon the current trait value, and the
mutated trait value is near that of the resident population, but
displaced by a random small amount.  Whether the mutant variety takes
over the population or not depends upon how the two varieties
interact, which we codify as a game.  The payoff for a mutant
individual with trait value $x_M$ playing against another individual
drawn from a population having trait value $x_R$ is $A(x_M;x_R)$.

An article by Allen, Nowak and Dieckmann~\cite{allen2013d} studies
this scenario in considerable detail, and for the remainder of this
section, we follow their notation fairly closely.

As time progresses, the population's trait value can change.  Let
\begin{equation}
a_{ij} = A(x_i;x_j),\ i,j \in \{M,R\}.
\end{equation}
We write the payoff matrix for mutant- and resident-type individuals
interacting via two-player games as
\begin{equation}
G = \left(\begin{array}{cc}
          a_{MM} & a_{MR} \\
          a_{RM} & a_{RR}
          \end{array}\right).
\end{equation}
The probability that a variety with trait value $x'$ takes over a
population with trait value $x$ is $\rho(x';x)$, which is some
function of the matrix $G$.

Because we are considering two-player games with a set of two
strategies, we can make use of a considerable amount of theory that
has been developed for that case.  It can be proved that, in a
two-player game with two strategies, if the effect of selection is not
too strong, then a strategy $R$ is favored over another strategy $M$
if the following inequality is satisfied:
\begin{equation}
\sigma(a_{RR} - a_{MM}) + a_{RM} - a_{MR} > 0.
\end{equation}
Here, $\sigma$ denotes the Tarnita \emph{structure coefficient,} a
number that depends on the population structure and the update
rule, but is independent of the payoff matrix~\cite{tarnita2009}.  If
this inequality is satisfied, then the fixation probability of~$R$ is
greater than that of~$M$:  a single $R$-type individual in a
population of type $M$ is more likely to take over the ecosystem than
a single $M$-type individual in the reverse scenario.  Structure
coefficients have been calculated for several different cases of
interest~\cite{tarnita2009, allen2013d, allen2015b}.

Denote the mutation rate per individual by~$u(x)$, and let the
mutation step size be described by a probability density $\mu(z)$,
which we take to be centered at zero.  We assume that $\mu(z)$ is a
narrow distribution, such that the expectation value of~$|z|^2$ is
much greater than that of~$|z|^3$.  Because many of our expressions
will involve the variance of~$\mu(z)$, we'll abbreviate it as~$\nu$:
\begin{equation}
\nu = \int dz\,z^2 \mu(z).
\end{equation}

Suppose the current trait value is established to be $x$.  By how much
can we expect the trait value to change?  We find this by integrating
over the step size $z$:
\begin{equation}
\moment{\Delta x} = \int dz\, z
                         Nu(x)\mu(z)\rho(x+z;x)
 = Nu(x) \int dz\, z \mu(z)\rho(x+z;x).
\end{equation}
If we expand the probability $\rho(x+z;x)$ to first order in the jump
distance $z$,
\begin{equation}
\rho(x+z;x) = \rho(x;x) + z\left.\frac{\partial \rho(x';x)}
                                      {\partial x'}\right|_{x'=x}
              + \mathcal{O}(|z|^2),
\end{equation}
then our expectation value $\moment{\Delta x}$ becomes
\begin{align}
\moment{\Delta x} =&\ Nu(x)\rho(x;x)\int dz\,z\mu(z) \nonumber\\
 & + Nu(x) \int dz\,z^2\mu(z)\left.\frac{\partial \rho(x';x)}
                                       {\partial x'}\right|_{x'=x}
   \\
 & + \mathcal{O}\left[\int dz\,\mu(z) |z|^3\right]\nonumber.
\end{align}
The first term vanishes by symmetry, and the third is negligible by
assumption.  Therefore,
\begin{equation}
\moment{\Delta x}
 = Nu(x) \int dz\,z^2\mu(z)\left.\frac{\partial \rho(x';x)}
                                     {\partial x'}\right|_{x'=x}
 = Nu(x) \nu \left.\frac{\partial \rho(x';x)}
                       {\partial x'}\right|_{x'=x}.
\label{eq:first-moment-delta-x}
\end{equation}
This relates the expected change in~$x$ to the derivative of the
fixation probability $\rho(x';x)$.

The simplest way to turn this into a dynamical system is to impose the
condition that the \emph{actual} change in~$x$ is given by
this \emph{expected} change in~$x$.  Doing so, we arrive at
the \emph{canonical equation} for the time evolution of the trait
value:
\begin{equation}
\boxed{\frac{dx}{dt} = Nu(x)\nu \left.\frac{\partial \rho(x';x)}
                                    {\partial x'}\right|_{x'=x}.}
\label{eq:ad-canonical}
\end{equation}
The derivative on the right-hand side can be written using the chain
rule as
\begin{equation}
\left.\frac{\partial \rho(x';x)}
           {\partial x'}\right|_{x'=x}
 = \frac{1}{A(x;x)}
   \sum_{j,k} \left.\frac{\partial\rho}{\partial a_{jk}}\right|_{G=J}
   \left.\frac{\partial a_{jk}}{\partial x'}\right|_{x'=x}.
\end{equation}

\section{A Fokker--Planck Equation for Adaptive Dynamics}

Having derived Eq.~(\ref{eq:ad-canonical}) for the deterministic
limit, we now push our understanding into new territory by including
the effects of stochastic fluctuations.  Instead of a single
coordinate value which depends on time, $x(t)$, we will consider a
time-varying probability density over the possible population states,
$P(x,t)$.

In the previous section, we equated expected change in $x$ with actual
change, thereby establishing a deterministic dynamic.  Now, we relax
that requirement.  Our calculation of the expected change,
Eq.~(\ref{eq:first-moment-delta-x}), is still valid, but it no longer
tells the full story.  Looking back over our derivation of the
Fokker--Planck equation, we see that need a quantity which represents
how much the probability density $P(x,t)$ spreads out over time.  If
we define
\begin{equation}
\moment{(\Delta x)^2} 
 = \int dz\,z^2 u(x) \mu(z)\rho(x+z;x),
\end{equation}
then we find that
\begin{equation}
\moment{(\Delta x)^2} = u(x)\nu\rho(x;x).
\label{eq:second-moment-delta-x}
\end{equation}

We can use Eqs.~(\ref{eq:first-moment-delta-x}) and
(\ref{eq:second-moment-delta-x}) to construct a Fokker--Planck
equation for the time evolution of~$P(x,t)$.  The result is
\begin{equation}
\boxed{\frac{\partial P(x,t)}{\partial t} =
- \frac{\partial}{\partial x}\left[
  u(x)\nu \left.\frac{\partial \rho(x';x)}
                     {\partial x'}\right|_{x'=x}
  P(x,t)\right]
+ \frac{1}{2}\frac{\partial^2}{\partial x^2}\left[
  u(x)\nu\rho(x;x) P(x,t)\right].}
\label{eq:FP-adaptive}
\end{equation}
Note that the time evolution of the probability density depends
on \emph{second} derivatives of~$\rho$.  In contrast, the
deterministic system defined by the canonical equation,
Eq.~(\ref{eq:ad-canonical}), depends only on \emph{first} derivatives
of~$\rho$.

We said that the game which governs the invasion dynamics is a
two-player interaction.  What if it is a multiplayer game instead?
Recently, Allen, Nowak and Dieckmann~\cite{allen2013d} proved that for
the deterministic system based on Eq.~(\ref{eq:ad-canonical}), a
multiplayer interaction is actually equivalent to a two-player game.
If the payoff function for the multiplayer game is
$B(x;y_1,y_2,\ldots,y_{n-1})$, and the game is symmetric under
permutations of the $y$-arguments, then the dynamics are equivalent to
those of a model defined in terms of the pairwise game
\begin{equation}
A(x;y) = B(x;y,y,\ldots,y).
\end{equation}
However, their proof depends crucially on the fact that the
time-evolution equation uses only first derivatives of~$\rho$.  This
means that it does not apply in the general stochastic context.

The Fokker--Planck equation (\ref{eq:FP-adaptive}), which we can think
of as a ``canonical diffusion equation for adaptive dynamics,'' is
equivalent to the stochastic differential equation derived by
Champagnat and Lambert~\cite{champagnat2007}.  The route we have taken
is more in line with a physicist's approach to stochastic processes.
A similar derivation was recently given by Van
Cleve~\cite{vancleve2014}.

From the steady-state solution to the Fokker--Planck equation,
(\ref{eq:FP-equilibrium}), we can derive the mutation-selection
equilibrium for stochastic adaptive dynamics:
\begin{equation}
P^*(x) \propto \frac{1}{u(x)\nu\rho(x;x)}
 \exp\left(2 \int^x dx' \frac{1}{\rho(x';x')}
                         \left.\frac{\partial \rho(y;x')}
                                    {\partial y}\right|_{y=x'}
     \right).
\end{equation}
The integrand in the exponential includes a first derivative
of~$\rho$, but not any higher derivatives.  We might be able to rescue
the theorem of Allen, Nowak and Dieckmann!  Specifically, we might be
able to relate the mutation-selection equilibrium distribution for a
multiplayer game to that of a pairwise game, even if
the \emph{approach} to equilibrium does not match.  However, before we
can do that, we have to understand $\rho(x;x)$, the probability that a
mutant organism with the same genotype as the resident population can
take over the ecosystem by genetic drift.  This depends on the
population structure and the update rules.  So, in order to reduce
multilateral interactions to pairwise ones, we have to construct an
appropriate structure of pairwise relationships and rules of
succession.

That is a very general problem.  For the moment, we focus instead on a
more specific one, zeroing our attention in on a particular two-player
game.  In Chapters~\ref{ch:multiscale} and \ref{ch:volunteer}, we
examined the Prisoner's Dilemma, implementing the basic idea of it in
two different settings.  Can we do the same here?  Within the world of
adaptive dynamics, strategies are continuously variable, rather than
binary.  And indeed, simplifying multiplayer games to dyadic ones
hinges upon that smooth variation.

We can write the payoff function for a continuous Prisoner's Dilemma
as
\begin{equation}
A(x;y) = -C(x) + B(y).
\end{equation}
The value $C(x)$ is the cost which an individual having genotype $x$
pays in order to benefit another, and the amount of benefit which the
other obtains is given by the function $B$.  We set
\begin{equation}
C(0) = B(0) = 0,
\end{equation}
and we require that both $C(x)$ and $B(x)$ are strictly increasing,
with the inequality $C(x) < B(x)$ satisfied for~$x > 0$.  For the
mathematics to work smoothly, we also posit that $C(x)$ and $B(x)$ are
twice differentiable.\footnote{We note here a difference between our
approach and that of Van Cleve~\cite{vancleve2014}.  In this chapter,
the continuously variable trait $x$ is a \emph{strength of
cooperation,} while in Van Cleve's study, the evolvable trait is
the \emph{fraction of time} in which an agent cooperates in
a \emph{discrete} game.}

Assuming that the neutral-drift takeover probability can be given in
terms of an effective population size $N_e$, we find that
\begin{equation}
P^*(x) \propto \frac{N_e}{u(x)\nu}
 \exp\left(2 N_e\int^x_0 dx'
                         \left.\frac{\partial \rho(y;x')}
                                    {\partial y}\right|_{y=x'}
     \right).
\end{equation}
Plugging in the Prisoner's Dilemma, this becomes
\begin{equation}
P^*(x) \propto \frac{N_e}{u(x)\nu}
 \exp\left(2 N_e\int^x_0 \frac{dx'}{A(x';x')}
                \left[-C'(x') + \frac{\sigma-1}{\sigma+1}B'(x')\right]
     \right),
\end{equation}
where $\sigma$ is the Tarnita structure coefficient, which as we said
earlier depends on the population structure and on the update
rule~\cite{tarnita2009}.  To simplify matters, take the mutation rate
$u(x)$ to be constant.  Carrying out the integral in the exponential
is difficult; however, we can understand a great deal about the
solution by using the fact that the exponential will be peaked where
its argument is maximized.  To find this extremum, we can
differentiate the argument with respect to~$x$, and by the Fundamental
Theorem of Calculus, this just extracts the integrand, evaluated at
$x' = x$.

If $\sigma \leq 1$, then the argument of the exponential is always
nonpositive, so the probability density $P^*(x)$ piles up at~$x = 0$.
On the other hand, if $\sigma > 1$, then $P^*(x)$ is peaked at the
value of~$x$ which satisfies
\begin{equation}
\boxed{\frac{B'(x)}{C'(x)} = \frac{\sigma+1}{\sigma-1}.}
\label{eq:boxed-PD-peak}
\end{equation}
Ratios of benefits to costs are commonplace in the study of how social
behaviors evolve.  Recall that in Chapter~\ref{ch:volunteer}, we found
the stability criterion for the nonspatial Volunteer's Dilemma in
terms of just such a ratio.  We will see similar ratios again in
Chapters~\ref{ch:invasion-fitness} and \ref{ch:price-mls}.
Eq.~(\ref{eq:boxed-PD-peak}) is an interesting variant, in that it is
a comparison of \emph{marginal} quantities, \emph{i.e.,} of the
derivatives of the benefit and cost functions.  It tells us that
the peak of~$P^*(x)$ depends on the population structure and the
details of the organisms' life cycles, through the structure
coefficient $\sigma$.

Note that if the mutation rate $u$ is not constant but instead varies
with~$x$, then the shape of the mutation-selection equilibrium curve
$P^*(x)$ will change.  This is another example of a theme noted by
Allen and Tarnita~\cite{allen2012}: standards of evolutionary success
depend upon mutation rates.  We saw this issue arise in
Chapter~\ref{ch:multiscale}, where the mutation rate had a strong
effect upon the shape of the mutation-selection equilibrium curve, and
in turn, upon the complexity profile.  It is a general, but
underappreciated, phenomenon~\cite{tarnita2014}.

Another scenario of evolutionary interest is the \emph{Snowdrift
game.}  Doebeli \emph{et al.}~\cite{doebeli2004} provide a good
description:
\begin{quotation}
\noindent [T]wo drivers are trapped on either side of a snowdrift and
have the option of staying in their cars or removing the snowdrift.
Letting the opponent do all the work is the best option, but if both
players refuse to shovel they can't get home.  The essential feature
of the snowdrift game is that defection is better than cooperation if
the opponent cooperates but worse if the opponent defects.
\end{quotation}
If the amount of cooperation is a continuously variable quantity, then
this fits into the framework we have developed in this chapter.  The
payoff to an agent who cooperates an amount $x$ when met with another
agent who cooperates to the extent $y$ is
\begin{equation}
A(x;y) = -C(x) + B(x + y).
\end{equation}
Note that if we take a partial derivative with respect to~$x$ or with
respect to~$y$, we pick up a $B'$ either way.

The mutation-selection equilibrium distribution for the Snowdrift game
is
\begin{equation}
P^*(x) \propto  \frac{N_e}{u(x)\nu}
 \exp\left(2 N_e\int^x_0 \frac{dx'}{A(x';x')}
                \left[-C'(x')
                      + \left(1+\frac{\sigma-1}{\sigma+1}\right)B'(2x')\right]
     \right).
\end{equation}
Again, simplifying the integral is not straightforward, but we can
find the position where $P^*(x)$ is peaked.

Defining the convenient abbreviation
\begin{equation}
r = \frac{\sigma - 1}{\sigma + 1},
\end{equation}
the argument of the exponential is extremized where
\begin{equation}
\frac{B'(2x)}{C'(x)} = \frac{1}{1+r}.
\label{eq:snowdrift-extremum}
\end{equation}

In the deterministic limit, the Snowdrift game is known to exhibit
interesting behavior with the quadratic cost and benefit functions
\begin{equation}
C(x) = c_2x^2 + c_1x,\ B(x) = b_2x^2 + b_1x.
\end{equation}
Given these forms for the cost and benefit curves, then
Eq.~(\ref{eq:snowdrift-extremum}) is satisfied when
\begin{equation}
x = \frac{c_1 - (1+r)b_1}{4(1+r)b_2 - 2c_2}.
\end{equation}
At $\sigma = 1$, we have $r = 0$, and our expression for $x$ reduces to
\begin{equation}
x = \frac{c_1 - b_1}{4b_2 - 2c_2},
\end{equation}
which is the equilibrium value that Doebeli \emph{et al.}\ derive for
deterministic dynamics in a panmictic population~\cite{doebeli2004}.

\begin{figure}[h]
\includegraphics[width=12cm]{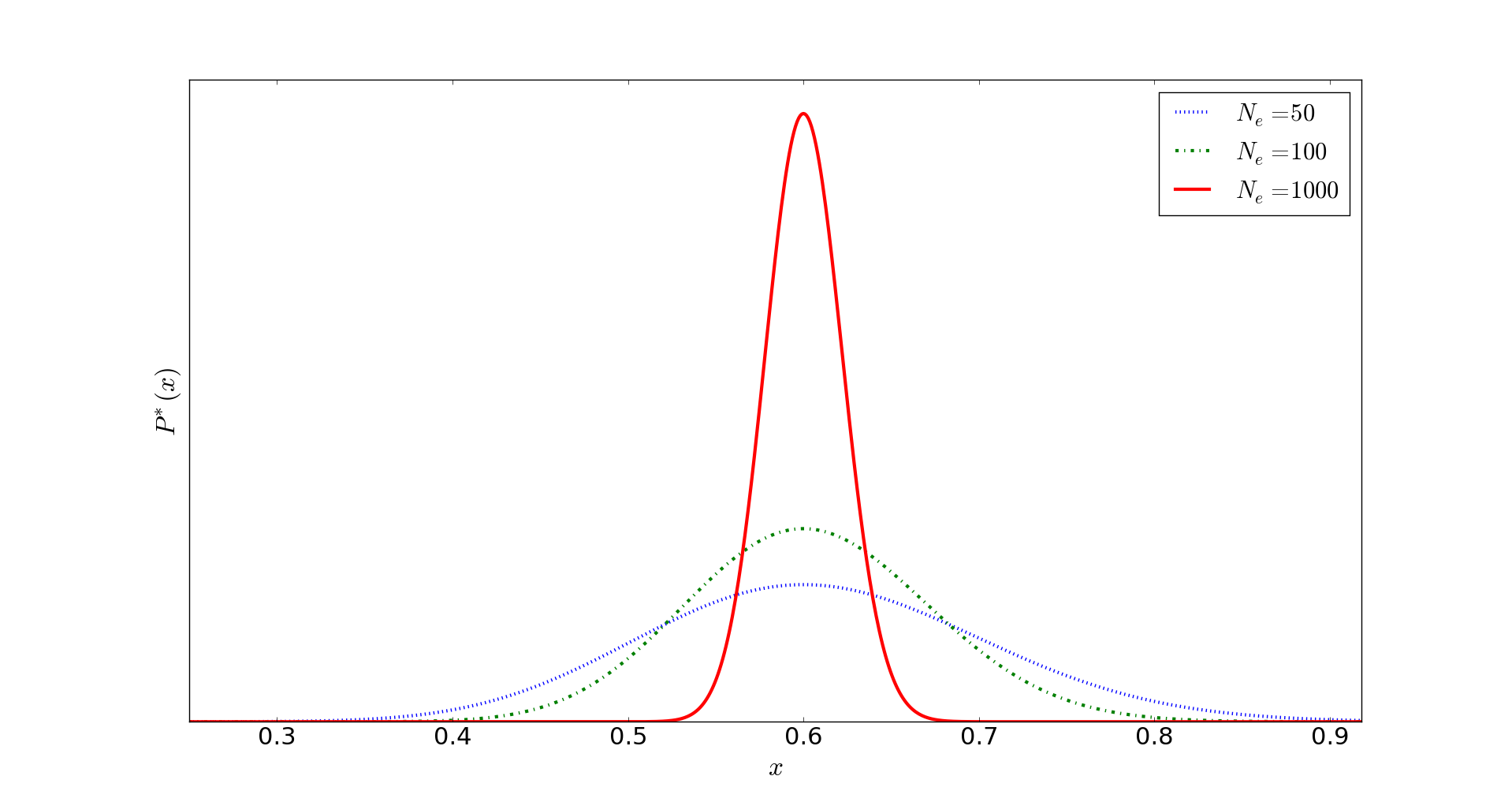}
\caption{\label{fig:snowdrift-effect-of-N} Steady-state probability
distributions for the Snowdrift game, computed with three different
choices of the effective population size $N_e$.  As we increase $N_e$
from~1 to~10 and then 100, the distribution becomes narrower.  The
peak of the distribution is, in each case, located at the ESS value of
the deterministic dynamics, $x = 0.6$ ($b_1 = 7$, $b_2 = -1.5$, $c_1 =
4.6$, $c_2 = -1.0$, $\sigma = 1.0$).}
\end{figure}

The values of $b_1$, $b_2$, $c_1$ and $c_2$ govern whether or not the
fixed point of the deterministic dynamics is stable.  By adjusting
these parameters appropriately, we can have fixed points at the same
value of~$x$ but opposite stabilities.  For example,
\begin{equation}
b_1 = 7,\ b_2 = -1.5,\ c_1 = 4.6,\ c_2 = -1.0
\end{equation}
yields a fixed point at $x = 0.6$, as does
\begin{equation}
b_1 = 3.4,\ b_2 = -0.5,\ c_1 = 4.0,\ c_2 = -1.5.
\end{equation}
However, in the former case, $x = 0.6$ is an Evolutionary Stable
Strategy, while in the latter, it is a repellor
point~\cite{doebeli2004}.  The analogue of this in the stochastic case
is the fact that two probability distributions can have extrema in the
same positions, one having a minimum and the other a maximum in that
location.  Now, we investigate this issue in more detail.  In
particular, we'd like to know how varying the structure coefficient
$\sigma$ affects stability.

Whether an extremum of $P^*(x)$ is a minimum or a maximum depends on
the second derivative of the integral inside the exponential, which is
the first derivative of the integrand.  Applying the quotient rule to
the integrand yields a ratio whose denominator is $A(x;x)^2$.  This is
always positive, so the sign can only depend on the sign of the
numerator.  Furthermore, the numerator itself simplifies at an
extremum, leaving us with the condition
\begin{equation}
2(1+r)B''(2x) - C''(x) < 0.
\end{equation}
For quadratic cost and benefit functions, this becomes
\begin{equation}
2(1+r)b_2 - c_2 < 0.
\end{equation}
If this inequality is satisfied, then the extremum $x$ is a maximum,
and thus is evolutionarily stable.

\begin{figure}[h]
\includegraphics[width=12cm]{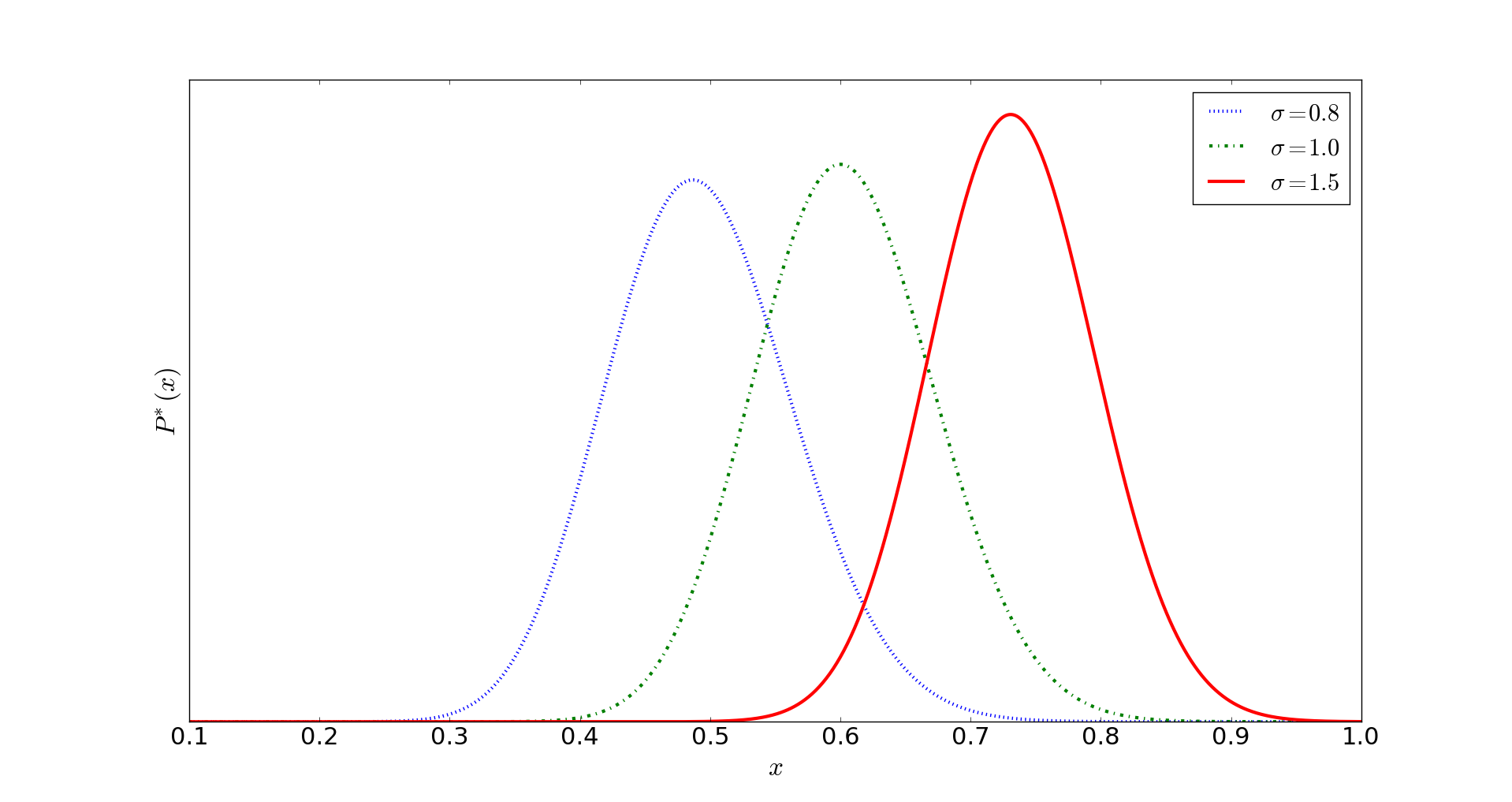}
\caption{\label{fig:snowdrift-effect-of-sigma} Steady-state probability
distributions for the Snowdrift game, computed with three different
choices of the Tarnita structure coefficient $\sigma$.  As we increase
$\sigma$, the peak of the steady-state distribution moves to higher
$x$, indicating an increased disposition to cooperation ($b_1 = 7$, $b_2 =
-1.5$, $c_1 = 4.6$, $c_2 = -1.0$, $N_e = 100$).}
\end{figure}

Incorporating stochastic effects by way of our Fokker--Planck equation
goes beyond prior work on the continuous Snowdrift game.  We can go
further by varying the structure coefficient $\sigma$, which
corresponds to implementing the Snowdrift game with different
short-term ecological dynamics.  We plot typical results in
Figure~\ref{fig:snowdrift-effect-of-sigma}.  Generally, as we increase
$\sigma$, the peak of the steady-state distribution moves to higher
$x$, indicating an increased disposition to cooperation.

\section{Concurrent Mutations, Discreteness and Multi-strategy Games}

The message of adaptive dynamics, whether deterministic or stochastic,
is that calculations become easier when traits vary continuously and
mutations are both small and rare.  In essence, adaptive dynamics
illustrates the principle that when we can use Taylor expansion, we
can simplify.  Now, we consider factors that can spoil that clean
conceptual simplicity.

Suppose that each player in a population can choose from the same set
of $n$ strategies, where $n$ is possibly larger than two.  Let's
assume that the game interactions still occur on a pairwise basis.
That is, we can write the payoffs as a matrix $a_{ij}$, where $i$ and
$j$ both range from~1 to~$n$.  We can define four different meaningful
averages:
\begin{align}
\bar{a}_{**} &= \frac{1}{n}\sum_{i=1}^n a_{ii}, \\
\bar{a}_{r*} &= \frac{1}{n}\sum_{i=1}^n a_{ri}, \\
\bar{a}_{*r} &= \frac{1}{n}\sum_{i=1}^n a_{ir}, \\
\bar{a} &= \frac{1}{n^2}\sum_{i,j=1}^n a_{ij}.
\end{align}
If the effect of selection is not too strong, then the condition for
strategy $r$ to be evolutionarily favored is
\begin{equation}
\sigma_1(a_{rr} - \bar{a}_{**})
 + \sigma_2(\bar{a}_{r*} - \bar{a}_{*r})
 + \sigma_3(\bar{a}_{r*} - \bar{a}) > 0.
\label{eq:tarnita-3sigma}
\end{equation}
The coefficients $\sigma_1$, $\sigma_2$ and $\sigma_3$ depend upon the
population structure and the update rule, just as $\sigma$ did for
two-strategy games, but they are independent of the payoff
function~\cite{tarnita2011, mcavoy2015}.  By dividing through by a
nonzero structure coefficient, we can convert
Eq.~(\ref{eq:tarnita-3sigma}) into a two-parameter condition.

This becomes relevant to adaptive dynamics if multiple mutations can
be present in the population at the same time, even if only for brief
intervals.  This is, in fact, a biologically important
scenario~\cite{maddamsetti2015}.  If an individual in the population
can pick from among more than two strategies, then we will in general
require at least two structure coefficients in order to decide which
strategy will end up dominant.  This rather spoils the convenience of
having one $\sigma$ that controls the outcome.

What if the trait value of interest is not continuously variable, but
instead discretized?  One problem which motivates an investigation
into discretized adaptive dynamics is the evolution of drug
resistance.  Take \emph{Plasmodium falciparum,} a parasitic protozoan
that is responsible for malaria.  The standard treatment for it was a
drug called chloroquine, but over the years, \emph{P.\ falciparum}
evolved the ability to resist the medication~\cite{hastings2002,
escalante2009}.  Chloroquine resistance (CQR) arose independently
multiple times during the twentieth century---quite probably, more
times than we know about, because any mutation that arises in the wild
must proliferate to an extent before researchers can detect it.  A
recent study~\cite{summers2014} examined the evolution of CQR in
detail and found that there exist multiple paths of mutations leading
from low resistance to high.  Intermediate stages provide partial
resistance, the levels of which can be quantified.

Drug resistance can be treated in terms of evolutionary games,
particularly when some organisms produce compounds that are harmful to
others~\cite{kirkup2004}.  Furthermore, we expect the same general
pattern---a spread of intermediates between extreme trait values,
discretized fundamentally by the genetic code---in other microbial
social behaviors.

This suggests the following idealized scenario:

Most of the time, the population is genetically uniform.
Occasionally, a mutation arises, and the new variety challenges the
established resident type for dominance.  We again make the
approximation that at most one mutant type is ever present at any
given time, but unlike before, we do not treat the trait value or the
mutation step size as continuously variable.  Mutations are rare, but
they are no longer small.

In a two-player game with two strategies, a resident strategy $x$ is
favored over a mutant strategy $x'$ provided that
\begin{equation}
\sigma[A(x;x) - A(x';x')] + A(x;x') - A(x';x) > 0.
\label{eq:sigma-simplest}
\end{equation}
Let the separation between the possible values of $x$ be $\Delta$.
This defines the mutation step size.  If a strategy $x$ is favored
over both $x+\Delta$ and $x-\Delta$, then we must have
\begin{align}
\sigma[A(x;x) - A(x+\Delta;x+\Delta)]
 + A(x;x+\Delta) - A(x+\Delta;x) &> 0, \nonumber\\
\sigma[A(x;x) - A(x-\Delta;x-\Delta)]
 + A(x;x-\Delta) - A(x-\Delta;x) &> 0.
\end{align}
A value of $x$ where these conditions are met is a local equilibrium.

Because the mutational steps are discrete, we cannot trust the
decomposition of a multiplayer game into dyadic ones.  Therefore, if
the basic game is a multiplayer interaction, then we need a criterion
analogous to Eq.~(\ref{eq:sigma-simplest}) for $K$-player games with
two strategies.

Denote the strategies by $A$ and $B$.  Let $a_j$ and $b_j$ be the
payoffs to a focal $A$- or $B$-type player, respectively, given that
they interact with $j$ neighbors who play strategy $A$.  Assuming weak
selection, strategy $A$ is favored if
\begin{equation}
\sum_{j=0}^{K-1} \sigma_j(a_j - b_{K-1-j}) > 0.
\end{equation}
That is, for two-strategy games, the number of structure coefficients
grows linearly with the number of players~\cite{wu2013, mcavoy2015}.
This means that when the trait value $x$ is discretized, we will need
more than one structure coefficient to locate the equilibrium point,
even if mutations are rare and the population is monomorphic in
between competition events.

A more general and potentially more realistic modeling method is to
picture the allowed trait values not as points on a line, but as
vertices in a graph.  Each vertex represents a possible genotype, for
example, one of the variations of the \emph{PfCRT} gene responsible
for chloroquine resistance.  We connect two vertices with an edge if a
mutation, like a substitution of nucleotides, can convert between
those two forms of the gene.  If we treat the competition between two
types as an evolutionary game, we can use the payoff function and the
structure coefficients to assign a direction to each edge of the
graph.  Local equilibria are sinks of the flow, \emph{i.e.,} vertices
that have all their edges pointing inwards.  (More specifically, such
points should ``collect probability,'' even though it is
not \emph{guaranteed} that the system must always move along the
direction of the edges.  The $\sigma$-rule analysis tells us which
fixation probability is greater, not the probabilities themselves.
Moving against an arrow is possible, but we bet against its happening.
This is another place where the discrete case is more complicated than
the continuous.  In the adaptive dynamics of continuous quantities,
the smoothness of all the functions involved means that we only need
to know which fixation probability is greater, not their specific
values~\cite{allen2013d}.)  Assuming that none of the mutations
fundamentally change the population structure and life cycle, then we
can use the same structure coefficients for all the edges.  This may
not always be a reasonable assumption.  For example, a mutation in a
virus could allow it to jump into a new host species.  In that event,
using the same $\sigma$ values before and after the mutation would
clearly not be reasonable.  However, for mutations of lesser
consequence, it seems a viable starting point.

Up until this point, we have treated the mutation rate $u(x)$ as
independent of the trait value $x$.  But we can argue that this does
not have to be the case.  We saw back in Chapter~\ref{ch:multiscale}
that many different nucleotide sequences can map to the same protein.
Because these sequences are equivalent as far as their products are
concerned, within the space of all possible sequences there exist
regions inside of which movement is unaffected by selection pressure.

Mutations happen stochastically with a certain probability per base
pair.  With more genetic material overall, we expect more mutations.
The rate at which we see changes in the trait value $x$ depends upon
how easily a mutation can push a genetic sequence from the space that
maps to one phenotype into a space that maps to another.  The volumes
and surface areas of these spaces do not have to be constant over~$x$!
Consequently, some values of~$x$ can be more \emph{robust} than
others.  We can define the robustness of a nucleotide sequence
quantitatively: we consider the set of all sequences that can be
produced from the original by mutations, and we find the fraction of
that set for which the phenotype is unchanged.  Then, we can average
this quantity over all sequences that map to the same phenotype in
order to define a robustness for that phenotype~\cite{wagner2008}.
For systems that are small enough to study with simulations---the
folding of short RNA sequences is a popular choice---values of the
robustness can vary over an order of magnitude from one phenotype to
another~\cite{partha2014}.  In order to apply adaptive dynamics, we
originally assumed that the trait value $x$ could vary smoothly.  A
heritable trait with an effectively continuous spectrum of possible
values will almost certainly be the product of many individual genes
and regulatory elements, acting in concert.  The number of nucleotide
sequences that will map to the same value of~$x$ will, therefore, be
vast.  Nevertheless, the same considerations should apply.  If the
robustness varies with $x$, then so too will the effective mutation
rate $u(x)$.  The more rapid the variations in~$u(x)$, the more the
effects will be seen in the mutation--selection equilibrium
distribution $P^*(x)$, particularly for smaller population sizes.

We have discussed the cases of discrete and of continuous $x$.
Somewhat of a conceptual intermediary is the case where $x$ is
continuous, but \emph{mutation step sizes can be large.}  When we
derived the Fokker--Planck equation (\ref{eq:FP}), we assumed
explicitly that changes are dominated by small increments, and larger
fluctuations are all negligible.  Then we carried this assumption over
into adaptive dynamics, where we said that the mutation step size
distribution $\mu(z)$ was sufficiently narrow that we only needed to
know its variance.  However, considerations of \emph{systems biology}
suggest that in some situations, we ought to allow $\mu(z)$ to be much
wider.  The interaction networks of intracellular components often
have long-tailed degree distributions: most components have few
interaction partners, but a few have many, and there is a continuum of
variation in between the extremes~\cite{janjic2014}.  This hints that
many mutations will have limited effects, while a small number---those
that directly impact the components of highest network degree---will
sway the trait value $x$ much more strongly.

The distribution of mutation effect sizes is currently a research
topic with a great many question marks, both empirically and
theoretically~\cite{orr2010}.  Furthermore, it is not clear how
applicable the extant theoretical results are to our problem; their
basic assumptions were not formulated with adaptive dynamics in mind.
We therefore have a certain freedom to go off in new directions.  One
interesting possibility comes from the study of \emph{L\'evy flights,}
random walks in which the distribution of step lengths is
long-tailed.  Investigating this type of stochastic process reveals
that the Fokker--Planck equation must be replaced with an analogue
that is written using \emph{fractional derivatives.}

Again, this is a situation in which we should not expect that we could
decompose a many-player game into dyads.

\section{Interspecies Interactions}

We now return to the continuous Prisoner's Dilemma.  Pleasingly, the
peak value of~$x$ computed by Eq.~(\ref{eq:boxed-PD-peak}) is also the
equilibrium value for the deterministic time-evolution equation
derived by Allen, Nowak and Dieckmann~\cite{allen2013d}.  Up to a
constant prefactor of little interest,
\begin{equation}
\frac{dx}{dt} \propto -C'(x) + \frac{\sigma-1}{\sigma+1}B'(x).
\end{equation}
What must be satisfied for an equilibrium point of this dynamical rule
to be stable?  Let $dx/dt = 0$ at~$x = \bar{x}$, and define the
separation from equilibrium by $\xi = x - \bar{x}$.  Rescale time as
convenient, so that the proportionality becomes an equality.  Then
\begin{align}
\frac{d\xi}{dt} &= \frac{d}{dt}(x - \bar{x}) \\
 &= -C'(\bar{x} + \xi) + \frac{\sigma-1}{\sigma+1}B'(\bar{x}+\xi) \\
 &\approx -\left[C'(\bar{x}) + \xi C''(\bar{x})\right]
    + \frac{\sigma-1}{\sigma+1}
      \left[B'(\bar{x}) + \xi B''(\bar{x})\right] \\
 &=\xi \left[-C''(\bar{x})
            + \frac{\sigma-1}{\sigma+1}B''(\bar{x})\right].
\end{align}
This is a linear response equation for the displacement $\xi$:
\begin{equation}
\frac{d\xi}{dt} = -\xi \left[C''(\bar{x})
            - \frac{\sigma-1}{\sigma+1}B''(\bar{x})\right].
\label{eq:PD-linear-response}
\end{equation}
Since we required that both $B(x)$ and $C(x)$ be twice differentiable,
this is fine.  The condition that the equilibrium $\bar{x}$ be stable
is that the quantity in square brackets is positive.  Therefore,
stability requires
\begin{equation}
C''(\bar{x}) > \frac{\sigma-1}{\sigma+1}B''(\bar{x}).
\end{equation}

Now, let us consider what happens if we have \emph{multiple} species,
each evolving in accord with the deterministic dynamics specified by a
cost function and a benefit function.  For $i = 1,2,\ldots,M$, we
have, if the species do not interact,
\begin{equation}
\frac{dx_i}{dt} \propto -C_i'(x_i) + \frac{\sigma_i-1}{\sigma_i+1}B_i'(x_i).
\end{equation}
We assume that a stable equilibrium $\bar{x}_i$ exists for each
species.  So, in terms of the displacements $\xi_i = x_i - \bar{x}_i$,
\begin{equation}
\frac{d\xi_i}{dt} = -\xi_i \left[C_i''(\bar{x}_i)
            - \frac{\sigma_i-1}{\sigma_i+1}B_i''(\bar{x}_i)\right].
\end{equation}
What happens if the species in this ecosystem start to interact with
one another?  The most straightforward way to incorporate interactions
into this dynamical system is to include cross terms:
\begin{equation}
\frac{d\xi_i}{dt}
 = -\xi_i \left[C_i''(\bar{x}_i)
                - \frac{\sigma_i-1}{\sigma_i+1}B_i''(\bar{x}_i)\right]
   + \alpha \sum_{j=1}^M J_{ij} \xi_j.
\end{equation}
The parameter $\alpha$ controls the strength of the interactions, and
the matrix element $J_{ij}$ indicates how much species $j$ influences
species $i$.

We lose nothing of real interest if we assume that the factors
multiplying the $\xi_i$ are equal for all~$i$, and so we can set them
to unity.  This is a further simplification, but a reasonable one,
since we expect that the timescales with which each species in
isolation settles to its own equilibrium are roughly equal.

The equilibrium we had without interactions was stable.  It will
remain stable with the interactions turned on if the eigenvalues
$\lambda_a$ of the coupling matrix $J$ all satisfy the inequality
\begin{equation}
\alpha\lambda_a - 1 \leq 0,
\end{equation}
which is the same thing as saying that the largest eigenvalue is less
than or equal to~$1/\alpha$.  Consequently, if the elements of the
matrix $J$ are chosen at \emph{random,} then the \emph{probability}
that the equilibrium will remain stable is the probability that the
largest eigenvalue of~$J$ is bounded by~$1/\alpha$.

This is exactly the kind of problem studied in random matrix theory.
In fact, we have recovered (with a new meaning for the symbols) an old
problem of May~\cite{may1972}, which has deep connections to the
statistical physics of coupled random variables~\cite{majumdar2014}.

\chapter{Spatial Stochastic Mechanics}

\label{ch:doi}
The goal of this chapter is to provide, by analytical means,
quantitative predictions for the values of some observations made in
Chapters~\ref{ch:redgreen} and \ref{ch:volunteer}.  We will do this
not by solving the specific models we studied there, but rather by
understanding related models, and seeing how results gleaned from
those solutions can be translated over to our original problems of
interest.  We touched upon this method of exploration in
Chapter~\ref{ch:redgreen}, when we introduced the concept of a
universality class (\S\ref{sec:percolation}).  We learn about one
system in a universality class by studing another and finding the
features which apply across the class in its entirety.  Now, it is
time to apply this method in detail.  We will begin by grounding
ourselves in the essential ideas of \emph{renormalization} theory,
which underpins the study of universality classes.  Then, we will
build the infrastructure necessary to express the stochastic spatial
models of Chapters~\ref{ch:redgreen} and \ref{ch:volunteer}
analytically.  With that development complete, we will be able to
justify the theoretical predictions to which we compared the
phase-transition phenomena of those models.

Our subject for the next few sections will be ``the renormalization
group.''  This term is a little like the Holy Roman Empire, in that
the latter, as Voltaire observed, is neither holy, Roman nor an
empire.  The word \emph{group} in this context is a technical term,
but mathematically speaking, it is not really the \emph{right}
technical term.  \emph{Renormalization} is a bit of jargon that hails
from quantum electrodynamics, and it dates to a time many years before
people figured out what they were doing and why their techniques
actually worked.  Consequently, it is not very indicative.
Even \emph{the} isn't quite right: it carries the implication that
``\emph{the} renormalization group'' is a single, specific
construction, like ``the quadratic formula.''

Since the full name of the subject is a misnomer, it is convenient to
elide that name as much as possible, and so we will use the common
abbreviation \emph{RG.}  We begin, therefore, our discussion
of \emph{RG theory} and \emph{RG transformations.}

\section{The Central Limit Theorem by RG Transformations}

To ease ourselves into RG theory, we will revisit a topic we addressed
before, this time approaching it from a different, albeit related
perspective.  Our first example of RG theory is the central limit
theorem, which we first proved in~\S\ref{sec:clt}.  A loose but useful
statement of this result is that the sum total of many uncorrelated
small influences is statistically distributed following a Gaussian
curve.

When we introduced complex systems, back in
Chapter~\ref{ch:multiscale}, we defined systems made of components,
and we quantified relationships among those components using
information theory.  Let us bring the central limit theorem into this
context.  Consider a system with many components, each of which is a
random variable.  We describe the system probabilistically by
ascribing a joint distribution to the set of random variables.  In
this case, we postulate that all higher-order complexities vanish,
meaning that the probability distribution factors neatly into
single-variable distributions:
\begin{equation}
p(X_1 = x_1, X_2 = x_2, X_3 = x_3,\ldots)
 = p(X_1 = x_1)p(X_2 = x_2)p(X_3 = x_3)\cdots
\end{equation}

To start with, let the components of our system be coins, which we say
are independent of each other and equally likely to come up either
heads ($+1$) or tails ($-1$).  A snapshot of this system will be, we
expect, a speckling of pluses and minuses in roughly equal proportions
(we might get a large fluctuation one way or the other, but we're not
gambling on that).  What happens if we blur our view, grouping
together pairs of components?

The total value of a pair of coins can be $-1$, $0$ or $+1$.  For a
single coin, we have
\begin{equation}
p_1(-1) = \frac{1}{2},\ p_1(+1) = \frac{1}{2}.
\end{equation}
There are two ways the total could sum to~$0$, and so
\begin{equation}
p_2(-2) = \frac{1}{4},
\ p_2(0) = \frac{1}{2},
\ p_2(+2) = \frac{1}{4}.
\end{equation}
We see that the distribution stays symmetrical and becomes peaked in
the middle.

\begin{figure}[h]
\includegraphics[width=8cm]{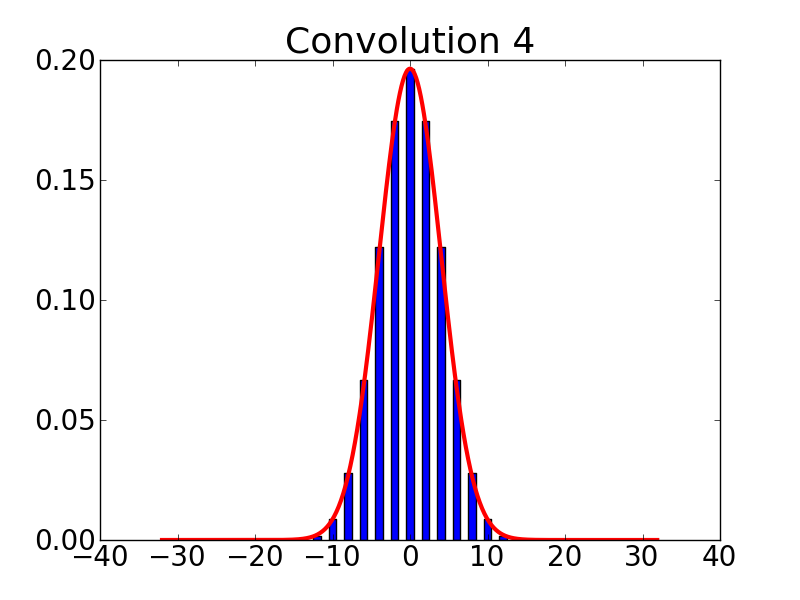}
\caption{\label{fig:coinflips-curve} 
Discrete probability distribution obtained by coarse-graining
uncorrelated fair coins four successive times.  The solid curve is a
Gaussian with mean zero and variance $\sigma^2 = 16$.}
\end{figure}

Following the same logic, we can repeat the blurring procedure.  The
arithmetic becomes more involved, but the concept is straightforward.
After just a few repetitions, the symmetrical, peaked distribution
comes to resemble a Gaussian curve, as shown in
Figure~\ref{fig:coinflips-curve}.

\begin{figure}[h]
\includegraphics[width=8cm]{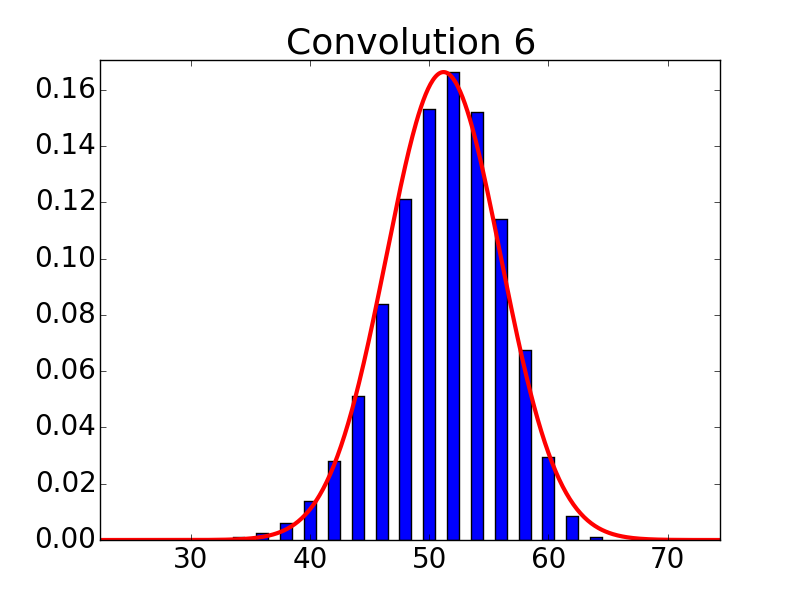}
\caption{\label{fig:biased-coinflips-curve} 
Discrete probability distribution obtained by coarse-graining
uncorrelated loaded coins six successive times.  The solid curve is a
Gaussian with mean 51.2 and variance $\sigma^2 = 23.0$.}
\end{figure}

What happens if we start our gambling on the assumption that the coins
are biased?  Say, for example, that we had chosen
\begin{equation}
p_1(-1) = 0.1,\ p_1(+1) = 0.9.
\end{equation}
Then we would have
\begin{equation}
p_2(-2) = 0.01,
\ p_2(0) = 0.18,
\ p_2(+2) = 0.81.
\end{equation}
This $p_2$ is not symmetrical at all!  However, as we repeat the
blurring procedure, the asymmetry partially washes out.  Examinining
Figure~\ref{fig:biased-coinflips-curve}, we see that the statistics
for the coarse-grained view are still approximately Gaussian, but the
mean of the Gaussian curve has shifted in the direction of the
original tilt.

Let's explore the effects of successive coarse-graining in more
generality.  Take $p(x)$ to be the probability distribution for each
of our original variables, and suppose for convenience that its mean
is zero.  When we convolve $p(x)$ with itself, we double all its
cumulants.  Now, after we coarse-grain, let us \emph{rescale the axis}
by a factor $1/\sqrt{2}$.  This restores the variance to its original
value, because the variance involves two powers of~$x$.  We repeat the
operations of coarse-graining and rescaling $K$ times in succession.
The mean and the variance remain unchanged, but what about the higher
cumulants?

The cumulants of the original distribution are $\cumulant{x^n}$.  Each
coarse-graining doubles $\cumulant{x^n}$, and each rescaling multiplies
$\cumulant{x^n}$ by a factor $(1/\sqrt{2})^n$.  So, if
$\cumulant{z_K^n}$ denotes the result of~$K$ iterations, then
\begin{equation}
\cumulant{z_K^n} = \cumulant{x^n} \left(\frac{2}{2^{n/2}}\right)^K
 = \cumulant{x^n} \left(2^{1-n/2}\right)^K.
\end{equation}
For $n > 2$, repeating the operations of coarse-graining and rescaling
will send $\cumulant{z_K^n}$ to zero.  This is just the conclusion we
drew in \S\ref{sec:clt}, phrased in terms of an iterative procedure.
We say that the higher cumulants are \emph{irrelevant,} because they
shrink under the iteration.

If we hadn't invested time in developing our understanding of
cumulants, we could have arrived at the same conclusion by going back
to the basics of Fourier transforms.  (This is the approach taken, for
example, in Sethna's textbook~\cite{sethna2006}.)  We define the
Fourier transform of a probability density function $p(x)$ as
\begin{equation}
\cF[p](k) = \int_{-\infty}^\infty dx\,e^{-ikx} p(x) = \tilde{p}(k).
\end{equation}
The inverse transformation is
\begin{equation}
p(x) = \frac{1}{2\pi}\int_{-\infty}^\infty dk\,e^{ikx} \tilde{p}(k).
\end{equation}

We coarse-grain by convolution:
\begin{equation}
(p\star p)(x) = \int_{-\infty}^\infty dy\,p(x-y)p(y).
\end{equation}
By the convolution theorem, the Fourier transform of a convolution is
the product of the Fourier transforms of the functions convolved.  In
this case, convolving a curve with itself,
\begin{equation}
\cF[p\star p](k) = \left(\tilde{p}(k)\right)^2.
\end{equation}
The next step is to re-scale the curve, under the assumption that the
mean is zero.  This has the effect of creating a new function,
\begin{equation}
\cS_{\sqrt{2}}[p](x) = \sqrt{2}p(\sqrt{2}x).
\end{equation}
The prefactor of $\sqrt{2}$ is necessary to preserve normalization, as
we can see by integrating:
\begin{align}
\int_{-\infty}^\infty dx\,\cS_{\sqrt{2}}[p](x)
 &= \sqrt{2}\int_{-\infty}^\infty dx\,p(\sqrt{2}x) \nonumber\\
 &= \sqrt{2}\int_{-\infty}^\infty
    \frac{dy}{\sqrt{2}} p(y) \nonumber\\
 &= 1.
\end{align}
The change of length scale is equivalent to an inverse change of
frequency:
\begin{align}
\cF\left[\cS_{\sqrt{2}}[p]\right](k)
 &= \int_{-\infty}^\infty dx\,e^{-ikx} 
    \sqrt{2}p(\sqrt{2}x) \nonumber\\
 &= \sqrt{2} \int_{-\infty}^\infty \frac{dy}{\sqrt{2}}
    e^{-iky/\sqrt{2}} p(y) \nonumber\\
 &= \tilde{p}(k/\sqrt{2}).
\end{align}

It follows that the Gaussian curve
\begin{equation}
p^*(x) = \frac{1}{\sqrt{2\pi\sigma^2}}
         \exp\left(-\frac{x^2}{2\sigma^2}\right)
\end{equation}
is a fixed point of the coarse-grain-and-rescale transformation.  We
shall now prove this explicitly.

For brevity, we denote this composite transformation by~$\cT$.  The
$\cT$ operator effects a convolution and a rescaling:
\begin{equation}
\cT[p](x) = \cS_{\sqrt{2}}[p\star p](x)
 = \sqrt{2}\int_{-\infty}^\infty dy\,p(\sqrt{2}x-y)p(y).
\end{equation}
Explicitly,
\begin{equation}
\cT[p^*](x) = \sqrt{2}\int_{-\infty}^\infty dy\,\frac{1}{2\pi\sigma^2}
 \exp\left(-\frac{(\sqrt{2}x-y)^2}{2\sigma^2}\right)
 \exp\left(-\frac{y^2}{2\sigma^2}\right).
\end{equation}
Expanding out and combining the arguments of the exponentials,
\begin{equation}
\cT[p^*](x) = \frac{\sqrt{2}}{2\pi\sigma^2}\int_{-\infty}^\infty dy\,
 \exp\left(-\frac{2x^2-2\sqrt{2}xy+2y^2}{2\sigma^2}\right).
\end{equation}

With a bit of algebra,
\begin{align}
2x^2-2\sqrt{2}xy+2y^2
 &= 2y^2-2(\sqrt{2}y)x + 2x^2 \nonumber\\
 &= 2y^2 -2(\sqrt{2}y)x + x^2 + x^2 \nonumber\\
 &= (\sqrt{2}y-x)^2 + x^2,
\end{align}
we can make the integral over $y$ a Gaussian integral, which we know
how to evaluate.  This eliminates the $y$ dependence, leaving us with
\begin{equation}
\cT[p^*](x) = \frac{1}{\sqrt{2\pi\sigma^2}}
 \exp\left(-\frac{x^2}{2\sigma^2}\right) = p^*(x),
\end{equation}
as desired.

The Fourier transform of $p^*$ is
\begin{equation}
\tilde{p}^*(k) = \cF[p^*]
 = \int_{-\infty}^\infty \frac{dx}{\sqrt{2\pi\sigma^2}}
   e^{-ikx - x^2/(2\sigma^2)}
 = \exp\left(-\frac{k^2\sigma^2}{2}\right).
\end{equation}
Generally, the Fourier transform of a $\cT$-transformed function is
\begin{equation}
\tilde{\cT}[\tilde{p}](k) = \cF\left[\cT[p]\right](k)
 = \left(\tilde{p}(k/\sqrt{2})\right)^2,
\label{eq:fourier-ct}
\end{equation}
with which it is easy to verify that
\begin{equation}
\tilde{\cT}[\tilde{p^*}](k) = \cF\left[\cT[p^*]\right](k) = \tilde{p}(k).
\end{equation}
That is, $p^*$ is a fixed point in the space of probability
distributions, and $\tilde{p}^*$ is a fixed point of the analogous
transformation in the space of Fourier representations.

Let $p$ be a distribution which is close to $p^*$:
\begin{equation}
p(x) = p^*(x) + \epsilon f(x).
\end{equation}
Then, because the Fourier transform is linear,
\begin{equation}
\tilde{p}(k) = \tilde{p}^*(k) + \epsilon\tilde{f}(k).
\end{equation}
What are the eigenfunctions and eigenvalues of $\cT$?
\begin{align}
\cT[p^* + \epsilon f_n]
 &= p^* + \lambda_n \epsilon f_n + \cO(\epsilon^2), \\
\tilde{\cT}[\tilde{p}^* + \epsilon \tilde{f}_n]
 &= \tilde{p}^* + \lambda_n \epsilon \tilde{f}_n + \cO(\epsilon^2).
\end{align}
Using Eq.~(\ref{eq:fourier-ct}),
\begin{align}
\tilde{\cT}[\tilde{p}^*+\epsilon\tilde{f}_n]
 &= \left[\tilde{p}^*(k/\sqrt{2})
          + \epsilon\tilde{f}_n(k/\sqrt{2})\right]^2 \\
 &= \left[\tilde{p}^*(k/\sqrt{2})\right]^2
    + 2\epsilon\tilde{p}^*(k/\sqrt{2})\tilde{f}_n(k/\sqrt{2})
    + \cO(\epsilon^2) \\
 &= \tilde{p}^*(k) 
    + 2\epsilon\tilde{p}^*(k/\sqrt{2})\tilde{f}_n(k/\sqrt{2})
    + \cO(\epsilon^2).
\end{align}
Therefore,
\begin{equation}
\tilde{f}_n(k) = \frac{2}{\lambda_n}
 \tilde{p}^*\left(\frac{k}{\sqrt{2}}\right)
 \tilde{f}_n\left(\frac{k}{\sqrt{2}}\right).
\end{equation}

Proposal: let
\begin{equation}
\tilde{f}_n(k) = (ik)^n \tilde{p}^*(k).
\end{equation}
With this ansatz,
\begin{align}
\tilde{p}^*\left(\frac{k}{\sqrt{2}}\right)
\tilde{f}_n\left(\frac{k}{\sqrt{2}}\right)
 &= \left(\frac{ik}{\sqrt{2}}\right)^n
    \tilde{p}^*\left(\frac{k}{\sqrt{2}}\right) \\
 &= \frac{(ik)^n}{(\sqrt{2})^n}
    \tilde{p}^*(k) \\
 &= \frac{1}{(\sqrt{2})^n} \tilde{f}_n(k).
\end{align}
Consequently,
\begin{equation}
\tilde{f}_n(k) = (\sqrt{2})^n
 \tilde{p}^*\left(\frac{k}{\sqrt{2}}\right)
 \tilde{f}_n\left(\frac{k}{\sqrt{2}}\right),
\end{equation}
meaning that the eigenvalue $\lambda_n$ is given by
\begin{equation}
\frac{2}{\lambda_n} = (\sqrt{2})^n = 2^{n/2}
\ \Rightarrow\ \lambda_n = 2^{1-n/2}.
\end{equation}

The first eigenmode, associated with the eigenvalue $\lambda_0 = 2$,
is
\begin{equation}
\tilde{f}_0(k) = \tilde{p}^*(k).
\end{equation}
When applied as a perturbation to $\tilde{p}^*$, this yields
\begin{equation}
\tilde{p}(k) = \tilde{p}^*(k) + \epsilon\tilde{p}^*(k),
\end{equation}
meaning that
\begin{equation}
p(x) = (1+\epsilon)p^*(x).
\end{equation}
This is not a normalized probability distribution function, unless
$\epsilon = 0$.  Therefore, we can neglect the $\lambda_0$ mode as
meaningless.

The next eigenfunction is
\begin{equation}
\tilde{f}_1(k) = ik\tilde{p}^*(k).
\end{equation}
Applying this perturbation to $\tilde{p}^*$ yields a probability
distribution
\begin{equation}
p(x) = \frac{1}{2\pi} \int dk\,(1+\epsilon ik)e^{ikx}
       \tilde{p}^*(k).
\end{equation}
Because $\epsilon$ is a small number, we can use a Taylor
approximation:
\begin{equation}
p(x) = \frac{1}{2\pi} \int dk\,e^{ik\epsilon} e^{ikx}
       \tilde{p}^*(k)
 =  \frac{1}{2\pi} \int dk\,e^{ik(\epsilon + x)}
       \tilde{p}^*(k).
\end{equation}
Therefore,
\begin{equation}
p(x) = p^*(x+\epsilon).
\end{equation}
That is, perturbing in the $\lambda_1$ eigenmode is equivalent to
shifting the mean of~$p^*(x)$.

The operation of coarse-graining and then rescaling is our first
example of an RG transform, and the cumulants provide our first
encounter with the concept that parameters can be relevant or
irrelevant under an RG transform.  We think of a \emph{flow} in the
space of parameters:  in this case, the flow induced by the RG
transform is the change in the cumulant values.  The flow takes us to
a \emph{fixed point} which is, in this example, a Gaussian curve.
Generally, a quantity is \emph{relevant} if repeated applications of
the RG transform cause it to grow, and a quantity is \emph{irrelevant}
if it shrinks as we approach the RG fixed point.  (Quantities can also
be \emph{marginal,} if a linear stability analysis of an RG fixed
point fails to indicate whether they will increase or decrease.)

This example also illustrates how fixed points of RG flows relate
to \emph{universality.}  Gaussian curves show up throughout the
sciences, because they arise whenever one considers the sum total
effect due to a large number of uncorrelated small influences.
Furthermore, once you understand one Gaussian curve, you've pretty
much understood them all.  This is the essence of a universality
class, as we defined that concept in Chapter~\ref{ch:redgreen}.  It is
the fact that so many details are irrelevant, in the RG sense, that
makes the Central Limit Theorem so powerful.

Using the calculus of variations, one can show that Gaussian curves
maximize the Shannon index (also known as the Shannon entropy) for a
given variance.  That is, if we fix the variance to be $\sigma^2$,
then the largest value of the Shannon index compatible with this
constraint is
\begin{equation}
S_G(\sigma^2) = \frac{1}{2}\log_2 (2\pi e \sigma^2).
\end{equation}
As we repeat the RG transform, the variance remains constant, and the
probability distribution looks more and more Gaussian, meaning that
the Shannon index will approach $S_G$ from below.  The Shannon index
is an example of a quantity which increases along with the RG flow,
attaining a maximum at the flow's fixed point.  When RG ideas are
applied in the realm of field theory, this idea becomes
the $c$-theorem of Zamolodchikov~\cite{zamolodchikov1986}.

Finally, we saw that \emph{exponents} turn out to be important
quantities: in this example, they are the eigenvalues $\lambda_n =
2^{1-n/2}$.  The importance of exponents that govern scaling behavior
holds true across RG theory.

\section{Isotropic Percolation}

In the previous section, we did not put any spatial structure on the
set of system components.  When we decided to coarse-grain two
variables together, any two variables were as good as any other pair.
Now, we will move on to a problem where we can apply RG theory to a
system that has spatial structure.  Our next example of an RG
analysis, which will move us closer to eco-evolutionary models, is the
topic of \emph{isotropic percolation.}

In Chapter~\ref{ch:redgreen}, we introduced percolation as the study
of flow through randomized media.  Our focus was on \emph{directed}
percolation, in which there is a preferred direction (\emph{e.g.,}
downhill).  It turns out to be somewhat easier to obtain quantitative
results by relaxing this restriction and considering \emph{isotropic}
percolation, in which no spatial orientation is picked out as special.

Typically, the first step in defining an isotropic percolation problem
is to construct a regular lattice.  Each point, or vertex, in the
lattice is connected to the same number of other vertices.  One way to
proceed is to imagine that each edge can be colored with one of two
possible pigments, \emph{e.g.,} white and black.  The color of each
edge is picked at random, independently of the choices for all other
edges.  For example, we can say that the probability an edge is marked
with black is $p$, and so an edge chosen at random will be white with
probability $1-p$.  We can then ask, for any value of~$p$, what the
sizes of the clusters formed by black edges will be.  This defines
a \emph{bond percolation} problem.

Alternatively, we can make the lattice points our variables of
interest.  For example, we can say that each lattice point is
either \emph{filled} or \emph{empty,} and we fill sites at random with
probability $p$.  Depending on the value of~$p$, we will have
different statistical distributions for the sizes of the clusters
formed by adjacent filled points.  This variant is known as \emph{site
percolation.}

What constitutes an RG transform in this context?  It is easy to
imagine what coarse-graining a lattice might mean: we can simply blur
out the fine-scale details.  This operation will create a new lattice
based on the configuration of the original.  One way to do this for
site percolation is to consider each lattice point together with its
immediate neighbors.  We ``collapse'' this set of points down to a
single site in the new lattice, and we choose whether the new site is
filled or not based on how many of the original sites are filled.
Applying this to all the sites in the original lattice creates a new
graph in which the small-scale information has been blurred away.  It
is then convenient to ``step back from the picture,'' rescaling the
lengths of the lattice edges so that neighboring points in the derived
lattice are separated by the same distance as those in the original.
This combination of coarse-graining followed by rescaling is an RG
transform.

Suppose that $p$ is small.  Then the lattice sites will mostly be
empty, though we will have some small islands of filled sites amidst
the sea of emptiness.  Applying the RG transform will tend to make
each of these islands smaller, because filled sites on the edges of
the islands will have too many empty neighbors for their
coarse-grained images to be filled.  So, heuristically speaking, the
RG transform will create a new lattice configuration that looks like
what we would find \emph{for a smaller value of~$p$.}

At the opposite extreme, suppose that $p$ is close to 1.  Then most of
the sites will be filled, with some low density of empty regions.  The
RG transform will blur away small details, making the smallest empty
regions vanish.  Empty regions which aren't quite small enough to
vanish entirely will be shrunk in the transformed image.  So, the
result will be a lattice configuration that looks like what we would
typically find \emph{for a larger initial choice of~$p$.}

If small values of the filling probability flow to even smaller ones,
and large values flow to even larger ones, then there must be a
``continental divide'' somewhere in the middle, where $p$ is mapped
to \emph{itself.}  At this point, we expect to find clusters in a
continuous range of sizes: the RG transform erases the smallest
clusters, so there must be slightly larger ones to replace them, and
so forth.  When $p$ equals this \emph{critical} value $p_c$, the
distribution of cluster sizes will be \emph{scale-invariant.}  Because
the lattice will contain clumps of all sizes, when we study the
overall properties of the system, the details of how the connections
between vertices look at the smallest scale should not matter much.
This is the root of why we can group these systems into universality
classes.

To make these qualitative considerations concrete, we will work
through an example of site percolation.  We take a triangular lattice
and fill in the lattice sites at random.  Let $p$ denote the probability
that a lattice site chosen at random is filled.  We will see how a
simple RG transform can send one value of~$p$ to another.

The coarse-graining operation should preserve the connectedness of
clusters.  This is rather tricky to enforce with a simple mapping, but
one way of coming close is to use a \emph{majority rule.}  We map a
triplet to a filled site if at least two of the original three sites
are filled.  This turns out to provide a simple approximation that
gives remarkably good agreement with the best known results for the
triangular lattice.

How does the majority rule create a mapping between values of~$p$?
With probability $p^3$, all three vertices of a triangle will be
filled, and when this configuration occurs, we coarse-grain it to a
single occupied site:
\begin{equation}
\includegraphics[width=6cm]{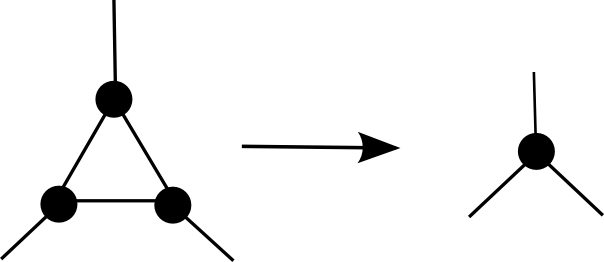}
\end{equation}

Likewise, if two out of three vertices are occupied, the corresponding
site in the coarse-grained lattice is filled:
\begin{equation}
\includegraphics[width=6cm]{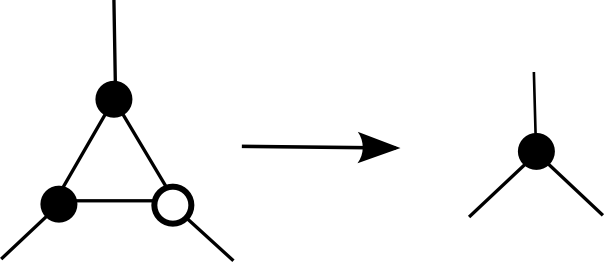}
\end{equation}
Because there are three ways to choose which vertex is unoccupied,
this situation carries a symmetry factor, occuring with probability
$3p^2(1-p)$.

We obtain an empty vertex in the coarse-grained lattice if two sites
are open:
\begin{equation}
\includegraphics[width=6cm]{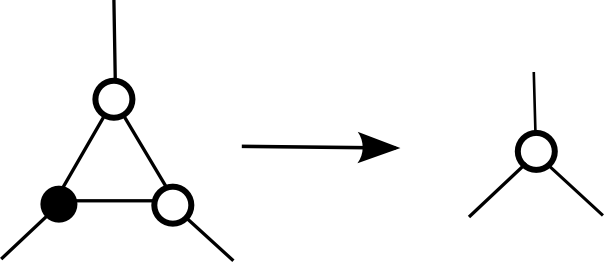}
\end{equation}
Or if all three vertices in the original triangle are unfilled:
\begin{equation}
\includegraphics[width=6cm]{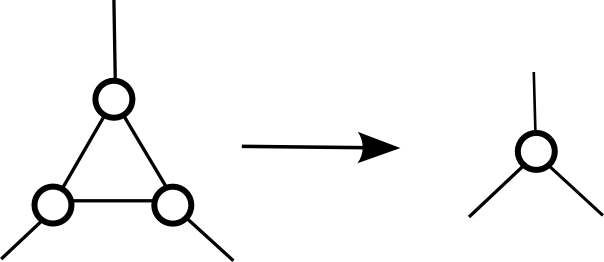}
\end{equation}

Let $p'$ denote the probability that a randomly chosen site in the
coarse-grained lattice is filled.  From the relations above, we see
that
\begin{equation}
p' = p^3 + 3p^2(1 - p).
\label{eq:rg-flow-trigon}
\end{equation}
The fixed points of this mapping are given by
\begin{equation}
p_c = p_c^3 + 3p_c^2(1 - p_c).
\end{equation}
Trivially, this equation is solved by 0 and by 1, and it has a less
obvious solution at
\begin{equation}
\boxed{p_c = \frac{1}{2}.}
\end{equation}

We will call this the \emph{critical} value of the site-filling
probability.  What happens if $p$ is almost but not quite equal to the
critical value?  Let us define
\begin{equation}
p = p_c + \delta p,\ p' = p_c + \delta p'.
\end{equation}
We can relate $p$ and $p'$ by the RG flow equation (\ref{eq:rg-flow-trigon}):
\begin{equation}
p_c + \delta p' = (p_c + \delta p)^3
 + 3(p_c + \delta p)^2(1 - p_c - \delta p).
\end{equation}
Expanding out the binomials, multiplying and simplifying, we arrive at
\begin{equation}
p_c + \delta p' = p_c + 6p_c(1-p_c)\delta p.
\end{equation}
Therefore,
\begin{equation}
\boxed{\delta p' = \frac{3}{2}\delta p.}
\label{eq:rg-trigon-boxed1}
\end{equation}
We see that a small deviation from the critical value becomes a larger
one.

Are there exponents for the percolation problem as there were in the
probability scenario we studied in the previous section?  In fact,
there are.  The most important in practice is $\nu$, which relates the
characteristic length scale---the typical size of connected
clumps---to the distance from the critical threshold:
\begin{equation}
\xi \propto |p-p_c|^{-\nu}.
\end{equation}
Here, we are measuring length in units of the lattice spacing.  That
the characteristic length $\xi$ varies in this fashion is perhaps
most directly appreciated by numerical simulation.  We expect that the
characteristic length will diverge at the critical threshold,
thanks to the scale-invariance we described above.

Coarse-graining a lattice will produce a new grid in which the
vertices are further apart.  It is traditional to denote the factor by
which the vertex separation is increased by the letter $b$.  If
$\xi(p)$ is the characteristic cluster size for filling probability
$p$, then $\xi(p')$ is the characteristic cluster size on the lattice
produced by coarse-graining, measured in units of the coarse-grained
lattice spacing.  But this must be related to the original average
cluster size by the scaling factor:
\begin{equation}
\xi(p) = b\xi(p').
\end{equation}
For example, if we projected our original lattice on a wall and
measured the typical cluster size to be four centimeters, then
coarse-graining will produce a new image, containing fewer pixels per
unit area, in which the typical cluster size must still be four
centimeters.

If we rewrite our previous expression in terms of the distance from
the threshold,
\begin{equation}
\xi(p_c + \delta p) = b\xi(p_c + \delta p'),
\end{equation}
substituting in the scaling form involving the exponent $\nu$ yields
\begin{equation}
|\delta p|^{-\nu} = b|\delta p'|^{-\nu}.
\end{equation}
We know from Eq.~(\ref{eq:rg-trigon-boxed1}) how $\delta p$ and
$\delta p'$ are related, and so we can say that
\begin{equation}
|\delta p|^{-\nu} = b\left(\frac{3}{2}\right)^{-\nu}|\delta p|^{-\nu}.
\end{equation}
Canceling the common factor and solving for $\nu$,
\begin{equation}
\nu = \frac{\log b}{\log (3/2)}.
\end{equation}
We can show geometrically that for this coarse-graining operation on
the triangular lattice, $b = \sqrt{3}$.  Therefore,
\begin{equation}
\boxed{\nu = \frac{\log \sqrt{3}}{\log (3/2)} \approx 1.355.}
\label{eq:rg-trigon-boxed2}
\end{equation}

The \emph{critical exponents} like $\nu$ are constant across all
members of a universality class.  However, other quantities, like the
locations of the critical thresholds, are typically model-specific.
This means that $\nu$ will be the same on a square lattice, for
example, while $p_c$ will be different (in fact, it is roughly 0.593).

In order to illustrate the RG concepts, this section has worked
through an example where the theory is fairly straightforward to
apply.  It turns out that had we chosen to start with a square lattice
instead of a triangular one, our task would have been significantly
more difficult, and we'd have to sweat a bit more even to get
an \emph{approximation} for~$\nu$ and for~$p_c$.

We have seen RG theory at work in two rather different contexts.  It
can be deployed in many more, and I am not aware of a comprehensive
reference that covers them all, or even the ones that are by now
well-established.  Introductory texts include those by Creswick,
Farach and Poole~\cite{creswick1992} and by McComb~\cite{mccomb2003}.
The textbooks by Kardar~\cite{kardar2007b} and by Zee~\cite{zee2003}
cover some aspects of RG in field theory.  Bar-\!Yam~\cite{baryam2003}
provides a succinct first course on RG that includes its classic use
in chaos theory.  The admirably slim volume by Cardy~\cite{cardy1996}
includes a study of directed percolation, which is relevant to our
concerns here.

\section{Doi Formalism}

The next step is to write stochastic dynamical processes in such a way
  that RG theory is applicable.  We can employ the operator tools
  pioneered by Masao Doi \cite{doi1976, doi1976b, mattis1998,
  janssen2005, dodd2009, buice2009, baez2012} to do so.  After some
  preliminaries, we will be able to use this formalism to describe
  evolutionary ecological systems.

We can take our system to consist of a series of components indexed by
$i$, where $i$ might range over individuals, genes or sites of a
lattice or other network, as appropriate.  The state of the $i$-th
component will be denoted $x_i$; we can then represent the
configuration of the entire system as $\underline{x}$.  The
probability of being in the state $\underline{x}$ can increase if the system
can transition into it, and it can decrease if transitions can take
the system out of it.  We thus write the master equation for the
system dynamics:
\begin{equation}
\frac{dp(\underline{x})}{dt} = 
  \sum_{\underline{y}} \pi\left(\underline{y}\rightarrow\underline{x}\right) 
                        p\left(\underline{y}\right)
- \sum_{\underline{y}} \pi\left(\underline{x}\rightarrow\underline{y}\right) 
                        p\left(\underline{x}\right).
\label{eq:ME-0}
\end{equation}
The form of the transition rates
$\pi\left(\underline{x}\rightarrow\underline{y}\right)$ depends upon
the detailed interactions of the system---how hosts reproduce, how
infections are transmitted and so forth.  We can make the master
equation Eq.~(\ref{eq:ME-0}) tractable if we make a simplifying
approximation; of course, any such specialization carries its own cost
in biological realism.

The mean-field simplification of Eq.~(\ref{eq:ME-0}) is
\begin{equation}
\frac{dp(x)}{dt} =
  \sum_{y} \pi(y\rightarrow x) p(y)
 -\sum_{y} \pi(x\rightarrow y) p(x).
\label{eq:ME-MFA}
\end{equation}
In Eq.~(\ref{eq:ME-0}), the ``vector'' $\underline{x}$ labeled a
microstate; in Eq.~(\ref{eq:ME-MFA}), $x$ without the arrow labels a
macrostate to which many distinct microstates map under a
coarse-graining transformation.

To take a concrete example, consider a Lotka--Volterra model of hosts
(prey) and consumers (predators).  The number of hosts, $H$, changes
through birth and predation, while the number $C$ of consumers
increases through predation and decreases due to death.  The
differential equation for the probability $p(C,H)$ will have a
predation term proportional to $p(C-1,H+1)$ and to the consumer
voraciousness $\lambda$, since each consumption act must increase $C$
by 1 and decrement $H$ by the same amount.  By the same reasoning, the
consumer death term must be proportional to $p(C+1,H)$ and to the
death rate $\mu$; likewise, prey reproduction generates a term
involving $p(C,H-1)$ and the growth rate $\sigma$.  Finally, we must
have a negative term reflecting how $p(C,H)$ decreases due to death,
birth and predation ``taking probability away'' from the $(C,H)$
macrostate.  We can thus write the mean-field master equation
\cite{mobilia2006b}
\begin{align}
\frac{dp(C,H)}{dt} = &\ \lambda(C-1)(H+1) p(C-1,H+1) \nonumber\\
  &+ \mu(C+1)p(C+1,H) \nonumber\\
  &+ \sigma(H-1) p(C,H-1) \nonumber\\
  &- (\mu C + \sigma H + \lambda CH)p(C,H).
\label{eq:lotka-volterra}
\end{align}
To rewrite this equation in a form more accessible to the tools we
have from our physics education, we introduce creation and
annihilation operators.  In this case, we require one pair of
operators for each species; more generally, the operators will be
indexed by some variable $i$.  Annihilation operators $a_i$ and
creation operators $a_j^\dag$ satisfy the commutation relation
\begin{equation}
\comm{a_i}{a_j^\dag} = \delta_{ij}.
\label{eq:basic-commutator}
\end{equation}
A person excessively steeped in category theory would say that this is
just the natural thing to do when dealing with a set of objects whose
size can be incremented or decremented \cite{baez2000}.  The reason is
that we are really considering the {\em number of ways to perform some
  manipulation} on a collection of objects: there are $n$ ways to draw
a toy out of a box containing $n$ of them, but only one way to drop a
new toy in, and so the operations of ``removing a toy'' and ``adding a
toy'' fail to commute by one unit.  Thanks to the product rule,
differentiation and multiplication by a variable $x$ satisfy the
commutator in Eq.~(\ref{eq:basic-commutator}).  This is ultimately why
we can discuss probability distributions using generating functions.

We build up states by acting on the vacuum $\ket{0}$ with creation
operators.  The state so built will be labeled by the occupation
numbers, {\em i.e.,} by how many times we acted with $a_i^\dag$, for
all allowed values of $i$.
\begin{equation}
\ket{\underline{n}} = \left(\prod_i \left(a_i^\dag\right)^{n_i} \right)
                \ket{0}.
\end{equation}
The normalization we have chosen implies that the action of $a_i$ is
to lower the label of a state and produce a prefactor.  In the
simplest case, where we have only one type of object,
\begin{equation}
a\ket{n} = a(a^\dag)^n \ket{0} = n(a^\dag)^{n-1} \ket{0}
         = n\ket{n-1}.
\end{equation}
From this, we can conclude
\begin{equation}
a^\dag a\ket{n} = n\ket{n}.
\end{equation}
We turn occupation probabilities into state vectors by
\begin{align}
\ket{\phi(t)} &= \sum_{\underline{n}} P(\underline{n},t)
                 \prod_i \left(a_i^\dag\right)^{n_i} \ket{0} \nonumber\\
              &= \sum_{\underline{n}} P(\underline{n},t) \ket{\underline{n}}.
\end{align}
Of special importance will be the coherent states, defined via
\begin{equation}
\cket{\underline{\eta}} = \exp\left(\sum_i \eta_i a_i^\dag\right) \ket{0}.
\label{eq:coherent}
\end{equation}
These will play an important role in the path integrals studied later.
One particular coherent state,
\begin{equation}
\cket{1} = e^{\sum_i a_i^\dag} \ket{0},
\end{equation}
is an essential part of calculating probabilities.  We know that the
scheme for turning state vectors into probabilities cannot be the same
as that we use in quantum mechanics; physically, we're working with a
wholly classical system, while mathematically, a Dirac bracket
$\braketop{\phi(t)}{\cO}{\phi(t)}$ would be bilinear in the
occupation probabilities instead of linear.

If we have only one type of particle, then we only need one number to
label a coherent state:
\begin{equation}
\cket{\eta} = e^{\eta a^\dag} \ket{0}
 = \sum_{n=0}^\infty \frac{\eta^n (a^\dag)^n}{n!} \ket{0}
 = \sum_{n=0}^\infty \frac{\eta^n}{n!} \ket{n}.
\end{equation}
The \emph{Poisson distribution} is a probability distribution over the
nonnegative integers, defined by
\begin{equation}
p_n = \frac{\eta^n e^{-\eta}}{n!}.
\end{equation}
The mean of this distribution is equal to~$\eta$.  (In
fact, \emph{all} of the cumulants are equal to~$\eta$.)  So, up to
normalization, a coherent state is ``an over-educated way of talking
about a Poisson distribution''~\cite{baez2012}.  This will be
important later, because we will often want to initialize a stochastic
dynamical system with Poissonian starting conditions.

The coherent states defined by Eq.~(\ref{eq:coherent}) are
eigenstates of the annihilation operators $a_i$ with eigenvalue
$\eta_i$; the proof of this goes through just as that for the quantum
harmonic oscillator.  Consequently, the adjoint coherent states
\begin{equation}
\cbra{\underline{\eta}} = \bra{0} \exp\left(\sum_i \eta_i a_i\right)
\end{equation}
are {\em left} eigenstates of the {\em creation} operators
$a_i^\dag$.  In particular, because $ \bra{0}e^{\sum_i a_i} $ is a
left eigenstate of $a_i^\dag$ with eigenvalue unity, we can write the
expectation value of an operator $\cO$ as
\begin{equation}
\moment{\cO(t)} = \projectop{1}{\cO}{\phi(t)}.
\label{eq:expectation-O}
\end{equation}

It is useful to know the inner product of two coherent states:
\begin{equation}
\cbraket{\phi_1}{\phi_2}
 = \exp\left(-\frac{1}{2}\left|\phi_1\right|^2
             -\frac{1}{2}\left|\phi_2\right|^2
             + \phi_1^*\phi_2\right).
\label{eq:coherent-overlap}
\end{equation}

Using these tools, we can rewrite the Lotka--Volterra system of
Eq.~(\ref{eq:lotka-volterra}) in operator language.  To do so, we
introduce two sets of creation and annihilation operators, which for
brevity we denote with $c$ and $h$, satisfying
\begin{equation}
\comm{c}{c^\dag} = \comm{h}{h^\dag} = 1,\ \comm{c}{h} = \comm{c}{h^\dag} = 0.
\end{equation}
In terms of these operators, Eq.~(\ref{eq:lotka-volterra}) takes the
form of an imaginary-time Schr\"odinger equation:
\begin{equation}
\partial_t \ket{\phi(t)} = -\hamil\ket{\phi(t)},
\label{eq:time-evolution}
\end{equation}
where
\begin{align}
\hamil = &\ \lambda c^\dag c h^\dag h + \mu c^\dag c + \sigma h^\dag h
         \nonumber\\
         &-\lambda c^\dag c^\dag c h^\dag h - \mu c + \sigma h^\dag h^\dag h,
\end{align}
which we can reorganize into
\begin{equation}
\hamil = \lambda(1 - c^\dag) c^\dag c h^\dag h
         + \mu(c^\dag - 1) c
         + \sigma(1 - h^\dag) h^\dag h.
\label{eq:lotka-volterra-hamiltonian}
\end{equation}

This example illustrates the general way of
constructing \emph{stochastic Hamiltonian operators,} which we can
summarize as follows~\cite{taeuber2005}:
\begin{quotation}
   \noindent [F]or each new particle species additional occupation
   numbers, second-quantized operators, and fields are to be
   introduced. The details of the reaction are coded into the master
   equation, though after some practice, it is actually easier to
   directly start with the Doi time evolution operator, as it is a
   more efficient representation. The general result is as follows:
   For a given reaction, two terms appear in the quasi-Hamiltonian (as
   in the original master equation). The first contribution, which is
   positive, contains both an annihilation and creation operator for
   each reactant, normal-ordered. For example, for the
   $A+A \rightarrow 0$ and $A+A \rightarrow A$ reactions this term
   reads $\hat{a}^{\dag 2} \hat{a}^2$, whereas one obtains for the $A
   + B \rightarrow 0$ reaction
   $\hat{a}^\dag \hat{b}^\dag \hat{a}\hat{b}$. These contributions
   indicate that the respective second-order processes contain the
   particle density products $a^2$ and $ab$ in the corresponding
   classical rate equations. The second term in the quasi-Hamiltonian,
   which is negative, entails an annihilation operator for every
   reactant and a creation operator for every product,
   normal-ordered. For example, in $A + A \rightarrow 0$ this term
   would be $\hat{a}^2$, whereas for $A + A \rightarrow A$ it becomes
   $\hat{a}^\dag \hat{a}^2$, and for $A + B + C \rightarrow A + B$ it
   would read $\hat{a}^\dag \hat{b}^\dag \hat{a}\hat{b}\hat{c}$. These
   terms thus directly reflect the occurring annihilation and creation
   processes in second-quantized language.
\end{quotation}

As mentioned, it becomes fairly easy with practice to write the
time-evolution operator directly from the description of the
reactions.  We will do this for some examples in the next section.
Before moving on, however, we note one subtlety which is relevant on
occasion when reading papers in this field.

Calculating probabilities using Eq.~(\ref{eq:expectation-O})
introduces a new wrinkle as we move into more advanced computations,
because the standard machinery of field theory expects operators to be
{\em normal-ordered,} that is, to have annihilation operators to the
right of creation operators in all products thereof.  Since the
coherent state $\cbra{1}$ involves an exponential of annihilation
operators $a_i$, expressions involving $\cbra{1}$ and an operator like
the Hamiltonian of Eq.~(\ref{eq:lotka-volterra-hamiltonian}) will not
be normal-ordered.  To resolve this, we can commute $e^{\sum_i a_i}$
through the Hamiltonian.  Because
\begin{align}
e^a a^\dag &= \sum_{n=0}^\infty \frac{a^n}{n!} a^\dag\nonumber\\
   &= \sum_{n=0}^\infty \frac{a^\dag a^n + na^{n-1}}{n!} \nonumber\\
   &= \left(1 + a^\dag\right) e^a,\nonumber
\end{align}
the effect of commuting the exponential through the operator being
bracketed is to shift $a_i^\dag \rightarrow a_i^\dag + 1$.

\section{Examples}

First, let's apply the procedure described above to the reaction
$A \to 0$, which we shall say happens at rate $\lambda$.  The
Hamiltonian operator is
\begin{equation}
\hamil = \lambda(a^\dag a - a) = \lambda(N - a).
\end{equation}
We would like to know how the expected number of particles present,
$\moment{N}$, will change with time.  From
Eqs.~(\ref{eq:expectation-O}) and (\ref{eq:time-evolution}), we have
\begin{align}
\frac{d}{dt}\moment{N} &= -\projectop{1}{N\hamil}{\phi(t)} \\
 &= -\lambda \projectop{1}{N(N-a)}{\phi(t)}.
\end{align}

At this juncture, it is useful to note a couple relations that follow
from the actions of the $a$ and $a^\dag$ operators on the vectors
$\ket{n}$:
\begin{align}
\projectop{1}{a^\dag}{\phi} &= \project{1}{\phi}, \\
\projectop{1}{a}{\phi} &= \projectop{1}{N}{\phi} = \moment{N}.
\end{align}
In addition, the fundamental commutator between $a$ and $a^\dag$
implies that
\begin{equation}
\comm{a}{N} = a.
\end{equation}
Now, we compute:
\begin{align}
\lambda\projectop{1}{N(a-N)}{\phi}
 &= \lambda\projectop{1}{(Na - N^2)}{\phi} \\
 &= \lambda\projectop{1}{(aN - N - N^2)}{\phi} \\
 &= \lambda\projectop{1}{(N^2 - N - N^2)}{\phi} \\
 &= -\lambda\projectop{1}{N}{\phi}.
\end{align}
Therefore,
\begin{equation}
\boxed{\frac{d}{dt}\moment{N} = -\lambda\moment{N}.}
\end{equation}
We have shown that the reaction $A \to 0$ implies \emph{exponential
decay.}

Likewise, we can see that the reproduction reaction $A \to A + A$
produces exponential growth.  The stochastic Hamiltonian for this
reaction is
\begin{equation}
\hamil = \lambda\left[a^\dag a - (a^\dag)^2 a\right].
\end{equation}
As before, we find the time derivative of the expectation value
$\moment{N}$:
\begin{equation}
\frac{d}{dt}\projectop{1}{N}{\phi}
 = -\lambda\projectop{1}{N\left[a^\dag a - (a^\dag)^2 a\right]}
                     {\phi}.
\end{equation}
The expression in the middle is equivalent to
\begin{equation}
N\left[N - (a^\dag)^2a\right]
 = N^2 - Na^\dag N.
\end{equation}
We can commute the $a^\dag$ through the number operator $N$, producing
\begin{equation}
N\left[N - (a^\dag)^2a\right]
 = N^2 - a^\dag (N+1) N.
\end{equation}
Therefore,
\begin{align}
-\lambda\projectop{1}{\left[N^2 - a^\dag (N+1) N\right]}{\phi}
 &= \lambda\projectop{1}{\left[(N+1)N - N^2\right]}{\phi} \\
 &= \lambda\projectop{1}{N}{\phi}.
\end{align}
And we see that indeed
\begin{equation}
\frac{d}{dt}\moment{N} = \lambda\moment{N}.
\end{equation}

We conclude this section with an example in which there are multiple
locations at which particles can be present.  Suppose that we have two
boxes, which we call box~1 and box~2, and particles can hop
stochastically from one site to the other.  We can treat this as a
particle being annihilated in box~1 and a replacement being created
within box~2.  If this takes place at a rate $\lambda$, then
\begin{equation}
\hamil_{1\to2} = \lambda(a_1^\dag - a_2^\dag)a_1.
\end{equation}
For the reverse process,
\begin{equation}
\hamil_{2\to1} = \lambda(a_2^\dag - a_1^\dag)a_2.
\end{equation}
If we let transitions happen in both directions,
\begin{equation}
\hamil = \hamil_{1\to2} + \hamil_{2\to1}
 = \lambda(a_1^\dag a_1 - a_2^\dag a_1
           + a_2^\dag a_2 - a_1^\dag a_2),
\end{equation}
which we can factor as
\begin{equation}
\hamil = \lambda(a_1^\dag - a_2^\dag)(a_1 - a_2).
\end{equation}

Generalizing this to a large set of boxes, we can write a stochastic
Hamiltonian for hopping between adjacent locations, in terms of a sum
over nearest-neighbor sites:
\begin{equation}
\hamil = \lambda
 \sum_{\langle kj \rangle} (a_k^\dag - a_j^\dag)(a_k - a_j).
\label{eq:stochastic-H-diffusion}
\end{equation}
This models a \emph{diffusion} process, if we take the rate $\lambda$
as given by a diffusion coefficient $D$, divided by the square of an
inter-site spacing $\Delta x$.

\section{Coherent-State Path Integrals}

Path integrals in field theory are much like partition functions in
statistical mechanics.  In both cases, we take a sum over terms, each
of which is an exponential of a formula encoding the system dynamics,
the sum running over all possible ways the system can do something.
When we calculated a partition function in elementary
statistical mechanics, we summed over all the {\em states} of a
system; when we compute a path integral, we sum over all {\em paths}
which the system can take from one state to another.

Start with some initial state $\ket{\phi(0)}$, which we produce by
acting on the vacuum $\ket{0}$ with some combination of raising
operators.  This state evolves in time according to
Eq.~(\ref{eq:time-evolution}); at some later time $t$, we compute the
expectation value of some operator $\cO$:
\begin{equation}
\moment{\cO(t)} = \projectop{1}{\cO e^{-\cH t}}{\phi(0)}.
\end{equation}
We could find the same expectation value by time-advancing the system
in~$M$ repeated small increments of size $\Delta t = t/M$.
\begin{equation}
\exp(-\cH t) = \lim_{M\rightarrow\infty} (1 - \cH\Delta t)^M.
\end{equation}
Various ways of developing the path integral all come from different
choices for writing the options available to the system at each
timestep.  This amounts to a choice of basis.  We have a long string
of factors of~$(1-\cH\Delta t)$, each one denoting the advance
from~$t_i$ to~$t_{i+1}$, where $t_i = i\Delta t$.  At time $t_i$, a
system in any one state has some probability of transitioning into any
other state; a {\em path} is a set of~$M$ such transitions.  The
normalization of our basis states gives us a ``resolution of unity'':
between every two successive factors of~$(1-\cH\Delta t)$, we insert a
sum over expressions of the form $\ket{x}\bra{x}$ which works out to
1.  In our case, we employ the coherent-state resolution of unity,
\begin{equation}
\int \frac{d\tilde{\underline{\eta}} d\underline{\eta}}{\pi^N}
     \cket{\underline{\eta}} \cbra{\underline{\eta}} = 1.
\end{equation}
The factors of $\pi$ are introduced to cancel those which arise from
Gaussian integrals over the state labels.  (Recall the normalization
of the basic Gaussian curve in Eq.~(\ref{eq:gaussian-def}).)

The {\em coherent-state representation} of an operator $\cO$ is
\begin{equation}
\cO[\tilde{\psi}(t), \phi(t)] \equiv
 \braketop{\psi(t)}{\cO(t)}{\phi(t)}.
\end{equation}

When we insert the coherent-state resolution of the identity into our
formal solution for~$\moment{\cO(t)}$, we end up with an
integral of the form
\begin{equation}
\int \cdots \cket{\phi_{t+\Delta t}}\cbra{\phi_{t+\Delta t}}
 e^{-\cH\Delta t} \cket{\phi_{t}}\cbra{\phi_{t}}
 e^{-\cH\Delta t} \cket{\phi_{t-\Delta t}}\cbra{\phi_{t-\Delta t}}
 \cdots,
\end{equation}
where we are integrating over $t/\Delta t$ different variables.  Each
factor in the long integrand is of the form
\begin{equation}
\cbra{\phi_t} e^{-\cH\Delta t}\cket{\phi_{t-\Delta t}}
 = e^{-\cH(\phi_t^*, \phi_{t-\Delta t}) \Delta t}
   \cbraket{\phi_t}{\phi_{t-\Delta t}}.
\end{equation}
Using the overlap between coherent states,
Eq.~(\ref{eq:coherent-overlap}), this becomes
\begin{equation}
\cbra{\phi_t} e^{-\cH\Delta t}\cket{\phi_{t-\Delta t}}
 = e^{-\cH(\phi_t^*, \phi_{t-\Delta t}) \Delta t}
   \exp\left(-\frac{1}{2}\left|\phi_t\right|^2
             -\frac{1}{2}\left|\phi_{t-\Delta t}\right|^2
             + \phi_t^*\phi_{t-\Delta t}\right).
\end{equation}
We can approximate this by
\begin{equation}
\cbra{\phi_t} e^{-\cH\Delta t}\cket{\phi_{t-\Delta t}}
 \approx e^{-\cH(\phi_t^*, \phi_t) \Delta t}
   \exp\left(-\frac{1}{2}\left|\phi_t\right|^2
             -\frac{1}{2}\left|\phi_{t-\Delta t}\right|^2
             + \phi_t^*\phi_{t-\Delta t}\right).
\end{equation}
And, by considering the terms produced by adjacent factors in the big
integrand, we can simplify this to
\begin{equation}
\cbra{\phi_t} e^{-\cH\Delta t}\cket{\phi_{t-\Delta t}}
 \approx e^{-\cH(\phi_t^*, \phi_t) \Delta t}
   \exp\left(-\phi_t^* \partial_t \phi_t \Delta t\right).
\end{equation}

Piling up many time slices and taking the limit $\Delta t \to 0$, we
get that $e^{-Ht}$ becomes
\begin{equation}
\int \mathcal{D}\phi^* \mathcal{D}\phi \exp\left(-\int_0^t dt'\,
 [\phi^* \partial_{t'}\phi + H(\phi^*,\phi)]\right).
\end{equation}
In this integral, the product of the integration volumes
$d^2\phi_j/\pi$ has become $\mathcal{D}\phi^* \mathcal{D}\phi$.

The \emph{action} which determines the weighting of each trajectory in
the path integral is
\begin{equation}
S[\phi^*, \phi] =
\int_0^t dt'\,
 [\phi^* \partial_{t'}\phi + H(\phi^*,\phi)].
\end{equation}
We can include the initial condition $\ket{\phi(0)}$ and the
projection state $\cbra{1}$ by adding a couple terms
to the action.  For example, if we start with a Poisson distribution
with mean $n_0$, then
\begin{equation}
S[\phi^*, \phi] =
\int_0^t dt'\,
 [\phi^* \partial_{t'}\phi + H(\phi^*,\phi)]
- \phi(t) - n_0\phi^*(0).
\end{equation}

Now, the expectation value $\moment{\cO(t)}$ is
\begin{equation}
\moment{\cO(t)} = \frac{1}{\mathcal{N}}
\int D\phi_j^* D\phi_j
 \cO\left(\phi(t)\right) e^{-S[\phi^*,\phi]}.
\end{equation}
The prefactor is a normalization constant that won't concern us.

We can eliminate the term that was brought in by the projection state
$\cbra{1}$, by defining a new variable:
\begin{equation}
\phi^* \to 1 + \tilde{\phi}.
\end{equation}
The time-derivative term inside the action integral becomes
\begin{equation}
\int_0^t dt'\,(1 + \tilde{\phi}) \partial_{t'} \phi
 = \phi(t) - \phi(0) + \int_0^t dt'\, \tilde{\phi}
                       \partial_{t'} \phi.
\end{equation}
This shift also transforms $H$, taking it to $H(1 + \tilde{\phi}, \phi)$.

If we have multiple lattice sites, then the action is a function of
all the $\phi_k$ and the $\tilde{\phi}_k$:
\begin{equation}
S[\{\tilde{\phi}_k\}, \{\phi_k\}] =
\sum_k \left[\int_0^t dt'\,
             \left[\tilde{\phi}_k \partial_{t'} \phi_k
                   + H(\{\tilde{\phi}_k\}, \{\phi_k\})
             \right]
            - n_0 \tilde{\phi}_k(0)
 \right].
\label{eq:stochastic-action}
\end{equation}

Our notation in this section largely follows the lectures by
Vollmayr-Lee~\cite{vollmayrlee2009}, which provide some additional
details.

\section{Spatial Dependence}

Incorporating spatial extent into a model means promoting our
creation and annihilation operators to sets thereof, indexed by
spatial position, and adding appropriate movement terms to the
stochastic Hamiltonian.  This procedure creates an expression that can
be studied by means of RG theory.

Most of the work done on this topic assumes that particles move about
by diffusion and react in some way if they happen to bump into one
another.  Therefore, we start with a mathematical model of a diffusion
process.  The diffusion action, assuming Poisson initial conditions,
is just the diffusion Hamiltonian (\ref{eq:stochastic-H-diffusion}) in
our new coherent-state language:
\begin{equation}
S_D = \int dt \left[\sum_i \tilde{\phi}_i \partial_{t'} \phi_i
 + \frac{D}{\Delta x^2}
   \sum_{\langle ij \rangle} (\tilde{\phi}_i - \tilde{\phi}_j)
                         (\phi_i - \phi_j)\right]
 - \sum_i n_0 \tilde{\phi}_i(0).
\end{equation}
If we take the continuum limit, $\phi$ and $\tilde{\phi}$ become
fields, and the action becomes
\begin{equation}
S_D = \int dt\,d^dx \left[ \tilde{\phi} \partial_{t'}\phi
 + D \nabla \tilde{\phi} \cdot \nabla{\phi}
 - n_0 \tilde{\phi}\delta(t)\right].
\end{equation}
Integrating by parts, we find
\begin{equation}
S_D = \int dt\,d^dx
 \left[ \tilde{\phi} (\partial_{t'} - D\nabla^2)\phi
  - n_0 \tilde{\phi}\delta(t)\right].
\end{equation}
This action is linear in the response field $\tilde{\phi}$.
Therefore, finding its extremum is easy:
\begin{equation}
\frac{\delta S_D}{\delta \tilde{\phi}}
 = \partial_t \phi - D\nabla^2\phi - n_0\delta(t) = 0.
\end{equation}
This is satisfied when
\begin{equation}
\partial_t \phi = D\nabla^2\phi + n_0\delta(t),
\end{equation}
which is the familiar diffusion equation, with initial conditions
specified at time $t = 0$.

Mobilia \emph{et al.}\ look in particular at the sustainability
transition, where the carnivores are consuming just enough to keep up
their numbers and not go extinct~\cite{mobilia2006, mobilia2006b}.
After a lot of redefinitions and rescalings, they get an action which
is an integral of a Lagrangian density of the form
\begin{equation}
\mathcal{L}[\tilde{\psi}, \psi] = \tilde{\psi}(\partial_t + D_c(r_c -
\nabla^2))\psi - u\tilde{\psi}(\tilde{\psi} - \psi)\psi +
\tau\tilde{\psi}^2 \psi^2.
\end{equation}
The field $\psi$ is a shifted and rescaled carnivore density, and
$\tilde{\psi}$ is its conjugate ``response field''. The various
constants are mixed-together combinations of the parameters we started
with.\footnote{This is, incidentally, the same form of Lagrangian
which defines the ``Reggeon field theory'' used back in the '70s for
high-energy scattering physics, with the last term providing what I
think they'd call a quadruple-Pomeron vertex~\cite{abarbanel1974,
grassberger1978, cardy1980}.} There are a fair many assumptions in the
derivation of this Lagrangian which ought to be unpacked (such as the
decision to expand perturbatively around diffusive motion), and of
course, once we've squeezed the dynamics we want to study into this
formalism, we'd like to get answers \emph{out} of it, and \emph{that}
appears to involve a whole lot of elaborate RG machinery.

\section{Rate Equations from Tree-Level Calculations}

In this field-theoretic formalism, rate equations arise as tree-level
results, \emph{i.e.,} the results obtained by using Feynman diagrams
containing no closed loops.  This is easiest to appreciate in a simple
model of diffusion-limited annihilation.  In this model, particles
move about by Brownian motion, and if they come into contact, they
annihilate with some probability.  Feynman diagrams for this model
have a natural interpretation in terms of ways a particle can survive
over an interval of time.

Suppose we release a particle at time 0 and find it again at some later
time $t$.  What could have happened to the particle in between, given
that it survived until~$t$?  The simplest possibility is that it
encountered no other particles and so had no opportunity to
annihilate.  The diagram for this process is just a straight line.
Alternatively, the particle could have bumped into one or more other
particles but failed to annihilate each time.  Diagrams for these
processes have vertices where two lines come in and one or two lines
come out.  The total probability of surviving until time $t$ is the
sum over all the ways the particle could survive, which means a sum
over all possible diagrams.

\begin{figure}[h]
\includegraphics[width=10cm]{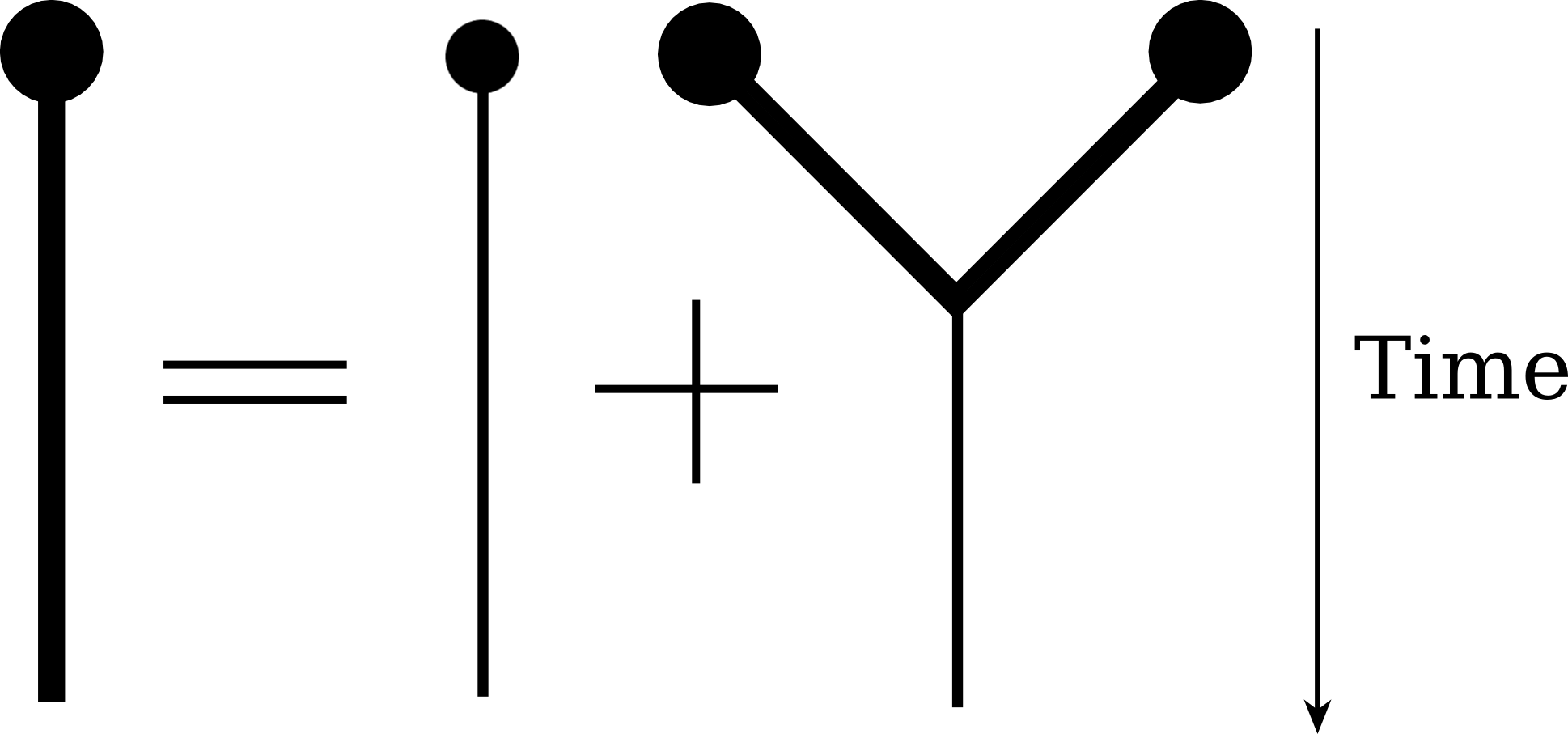}
\caption{\label{fig:dla-self-consistency} Self-consistency calculation
  for the diffusion-limited annihilation model's tree-level
  propagator.}
\end{figure}

Neglecting correlations means neglecting the possibility that two
particles coming together at a certain time have done so already in
the past.  This means that neglecting correlations allows us to omit
all diagrams containing closed loops from our sum.  This calculation
is significantly easier than the complete version.
Figure~\ref{fig:dla-self-consistency} shows the reason in diagrammatic
form:  the heavy line, representing all possible tree-shaped diagrams,
obeys a self-consistency condition.

The \emph{propagator} tells us how particles get from one point to
another if nothing happens in between. We're saying that these
particles move by diffusion, so the propagator in this case is the
Green's function for the diffusion equation.
\begin{equation}
G_D(k, \omega) = \frac{1}{-i\omega + D k^2}.
\end{equation}
Transforming from frequency back into the time domain,
\begin{equation}
G_D(k, t) = \int \frac{d \omega}{2\pi} \frac{e^{-i \omega t}}{-i \omega + D k^2}.
\end{equation}
For $t > 0$, this is
\begin{equation}
G_D (k, t) = \exp(-D k^2 t).
\end{equation}
In position space,
\begin{equation}
G_D(x, t > 0) = \frac{e^{-x^2 / (4 D t)}}{(4 \pi D t)^{d / 2}}.
\end{equation}
We can think of this as saying that the response to a delta-function
spike at $t = 0$ is a Gaussian curve which spreads out as time passes,
its standard deviation growing as the square root of the elapsed time.

To each trivalent vertex, we associate a factor $-2\lambda_0$, and
each initial vertex gets a $n_0$. Wave-vector (or ``momentum'')
conservation applies at each vertex. We can read off the
self-consistency condition for the tree-level contributions directly
from the diagrams:
\begin{equation}
a_{tree}(t) = n_0 + \int_0^t d t_1 G_D(0, t - t_1) (-2 \lambda_0) a_{tree}(t_1)^2.
\end{equation}
The propagator with $k = 0$ is just 1. Differentiating both sides of
the self-consistency equation yields that the time derivative
of~$a_{tree}$ is the integrand evaluated at~$t$.
\begin{equation}
\frac{d a_{tree}}{d t} = -2\lambda_0 a_{tree}^2
\end{equation}
This is just a rate equation for $a_{tree}$. With the initial
condition $a_{tree}(0) = n_0,$ this has the solution
\begin{equation}
\boxed{a_{tree}(t) = \frac{n_0}{1 + 2\lambda_0 n_0 t}.}
\end{equation}

Going beyond tree-level calculations requires RG theory.  The
computations are, at least for diffusion-limited annihilation, not all
that difficult on a technical level, once the basic concepts are
grasped.  One fact which enables considerable simplifications is that
the reactions in this model cannot increase the number of particles
present.  Applying RG to models in which this condition does not hold
is more complicated.

\begin{quotation}
\noindent \textbf{Question:} \emph{Since we require the concepts of
  RG at such an early stage of our analysis, would a more
  sophisticated mathematical treatment of that theory be useful?  It
  appears that the Hopf-algebraic study of RG~\cite{connes2000} would
  be just as applicable in this stochastic setting as it is in QFT.
  Perhaps diffusion-limited annihilation might even provide a simpler
  setting for it.}
\end{quotation}

\section{Directed Percolation}

Back in Chapter~\ref{ch:redgreen}, we introduced directed percolation
as an idealized model of fluid flow through a porous medium.  We began
with the mental image of a regular lattice of channels, some of whose
junction points were blocked.  If the fraction of blocked points was
too large, fluid flowing through the lattice from top to bottom would
always be stymied, but if the blockages were sparse, the fluid could
percolate downwards through the channels.  We can treat this model in
any number of dimensions; the crucial point is that there is a
preferred direction.  And because the fluid only spreads downhill,
there is another picture available.  We can also think of the tip of
each stream as a corpuscle executing a \emph{random walk.}  These
walkers can merge, if two fluid streams come together at a common
point, and they can reproduce, which happens when a fluid stream
splits and its offshoots progress along two different channels.  The
walkers can also perish:  this corresponds to a stream which
encounters a point where no further propagation is possible.

In short, directed percolation in a $(d+1)$-dimensional lattice, with
one dimension singled out as the axis of gravity, can equally well be
thought of as \emph{merging, reproducing and perishing random walks}
in a $d$-dimensional space.  This latter conception of the problem is
amenable to the techniques we have developed in this chapter.

Consider the spatial stochastic process defined by the combination of
particle decay ($A \to 0$), reproduction ($A \to A + A$) and merging
($A + A \to A$) with diffusive motion.  Call the decay or mortality
rate $\mu$, the reproduction rate $\sigma$ and the merging rate $\lambda$.

The stochastic action is
\begin{align}
S = \int d^dx \left[-\phi(t)
  + \int_0^t dt'\left(\vphantom{\nabla^2}\right.\right. &
 \phi^*(\partial_{t'} - D\nabla^2)\phi
 - \mu(1 - \phi^*)\phi
 + \sigma(1 - \phi^*)\phi^* \phi \nonumber\\
 & \left.\vphantom{\int}\left. - \lambda(1 - \phi^*)\phi^* \phi^2 \right)
   - n_0 \phi^*(0)\right].
\end{align}
Performing the field shift $\phi^* = 1 + \tilde{\phi}$, and writing $r
= (\mu - \sigma)/D$, we obtain
\begin{equation}
S = \int d^dx\,dt'
 \left[\tilde{\phi}\left(\partial_{t'} + D(r - \nabla^2)\right)\phi
       - \sigma\tilde{\phi}^2\phi
       + \lambda \tilde{\phi} \phi^2
       + \lambda \tilde{\phi}^2 \phi^2
 \right].
\end{equation}
We have deliberately dropped the initial-conditions term, because
particle decay and generation will scramble the original
configuration.  Note that $\delta S/\delta \phi = 0$ is satisfied
by~$\tilde{\phi} = 0$, and combining this with the other variation
$\delta S/\delta \tilde{\phi} = 0$ yields the equation of motion
\begin{equation}
\partial_t \phi = -D(r - \nabla^2)\phi - \lambda \phi^2.
\end{equation}

If particles can \emph{annihilate} as well as merge ($A + A \to 0$),
then the action is modified:
\begin{equation}
S = \int d^dx\,dt'
 \left[\tilde{\phi}\left(\partial_{t'} + D(r - \nabla^2)\right)\phi
       - \sigma\tilde{\phi}^2\phi
       + (\lambda + 2\lambda')\tilde{\phi} \phi^2
       + (\lambda + \lambda')\tilde{\phi}^2 \phi^2
 \right].
\end{equation}
And the rate equation becomes
\begin{equation}
\partial_t \phi = -D(r - \nabla^2)\phi - (\lambda + 2\lambda')\phi^2.
\end{equation}

It is interesting to follow the history of the directed-percolation
concept.  It was first proposed in 1957~\cite{broadbent1957}. The
mathematics necessary to treat it cleverly was invented (or, rather,
adapted from a different area of physics) in the 1970s, and then
forgotten, and then rediscovered by somebody else~\cite{mattis1998}.
Connections with other subjects were made.  Experiments were carried
out on systems which \emph{almost} behaved like the idealization, but
always turned out to differ in some way\ldots\ until 2007, when the
behavior was finally caught in the wild~\cite{takeuchi2009}. And this
experiment, which at last observed a DP-class phase transition with
quantitative exactness, used a liquid crystal substance
(\emph{N}-(4-Methoxybenzylidene)-4-butylaniline) which wasn't
synthesized until 1969~\cite{kelker1969}.

This rather puts the lie to the notion that a scientific hypothesis
must be amenable to immediate falsification by experiment.  The model
here is not an esoteric proposal for quantum gravity, but an
idealization of water flowing through coffee grounds.  And not only
did it take half a century to go from a mathematician's mind to a
laboratory bench, but that journey depended on tools which did not
exist when the model was first conceived.

\section{Prior Relevant Results and Difficulties}

One can find in the literature various applications of the Doi
field-theoretic formalism to systems which resemble our host--consumer
model.  Field-theory tools allow one to calculate, for example, the
critical exponents which describe the dynamics of a predator
population near its extinction threshold~\cite{taeuber2005,
  mobilia2006b}.  They also appear to be fairly successful at locating
where the critical point will occur in an epidemic model, in at least
some regions of parameter space~\cite{peruani2013}.

One significant problem blocks the path to applying these techniques
to the spatial host--consumer model studied in this report.  All the
prior work assumed \emph{diffusive motion,} as we did in our brief
encounter with annihilating particles above.  The host--consumer model
has no such feature: all motion is due to \emph{reproduction} into
adjacent lattice sites (empty sites for hosts, host-occupied sites for
consumers).  The validity of expanding around a diffusion propagator
is, therefore, questionable.  It may be a viable approximation in some
cases, however, thanks to universality.  In a later section, we will
make an argument to this effect in more depth.

Simulations such as those depicted in Figure~\ref{fig:animation}
suggest that treating the advances of host and consumer populations as
\emph{surface growth} may be a useful approach.  The field-theoretic
study of the Kardar--Parisi--Zhang model may, therefore, be an area to
draw from~\cite{zee2003, kardar2007b}.  (Visually, a consumer wave
eating its way through a field of hosts does resemble burning paper,
which has been studied as a possible instance of KPZ-class dynamics.)

\section{Carrying Capacity}

Realistic ecosystem models typically have some notion of a
\emph{carrying capacity}: a given environment can only sustain a
living population of limited size.  In our spatial host--consumer
model, population density is limited by the fact that each lattice
site can only hold one H-type or one C-type individual.  (As we
discussed in Chapter~\ref{ch:redgreen}, each H-type or C-type agent in
the model can be thought of as an idealization of a homogeneous
subpopulation.)  To make progress, we need to incorporate carrying
capacity into the field-theoretic formalism developed above.  One way
to do this is to introduce rate-limiting terms into the Lagrangian,
which ensure that the probability of the population density growing
too large at any point in space is negligible.  Another way is to
curtail the state space of the theory itself.  This latter approach
relates to some interesting topics in mathematical physics, so we'll
explore it at greater length.

In the Schwinger oscillator model, we have two sets of harmonic
oscillator creation and annihilation operators,
\begin{equation}
[a_-,a_-^\dag] = 1,\quad [a_+,a_+^\dag] = 1,
\end{equation}
such that operators pertaining to one oscillator commute with those
for the other:
\begin{equation}
[a_+,a_-^\dag] = [a_-,a_+^\dag] = 0.
\end{equation}
Excited states---that is, states with nonzero particle number---are
built by acting repeatedly with the creation operators:
\begin{equation}
\ket{n_+ n_-} = \frac{(a_+^\dag)^{n_+}(a_-^\dag)^{n_-}}
                     {\sqrt{n_+!}\sqrt{n_-!}}
                \ket{0,0}.
\end{equation}
The number operators $N_+ = a_+^\dag a_+$ and $N_- = a_-^\dag a_-$
are eigenvalues of~$\ket{n_+ n_-}$ with eigenvalues $n_+$ and $n_-$,
respectively.  Under the substitution
\begin{equation}
n_+ \to j + m,\quad n_- \to j - m,
\end{equation}
we can write
\begin{equation}
\ket{j,m} = \frac{(a_+^\dag)^{j+m}(a_-^\dag)^{j-m}}
                 {\sqrt{(j+m)!}\sqrt{(j-m)!}}
                \ket{0,0}.
\end{equation}
Defining
\begin{equation}
J_+ = \hbar a_+^\dag a_-,
\quad J_- = \hbar a_-^\dag a_+,
\quad J_z = \frac{\hbar}{2}(N_+ - N_-),
\end{equation}
the operators $J_\pm$ and $J_z$ satisfy the angular momentum algebra
$\mathfrak{su}(2)$:
\begin{align}
[J_z,J_\pm] &= \pm J_\pm, \\
[J_+,J_-] &= 2J_z.
\end{align}

To relate Schwinger's work and stochastic mechanics, take a system
composed of two particle species, which we can call type $a$ and type
$b$.  The different normalization does not affect the operator
algebra.  We can still define $J_+ = a^\dag b$, $J_- = b^\dag a$ and
$J_z = (a^\dag a - b^\dag b)/2$.  Acting on a ket labeled by particle
number, we still have
\begin{equation}
J_z\ket{n_a,n_b} = \frac{1}{2}(n_a - n_b)\ket{n_a,n_b},
\end{equation}
and the commutators among our operators are again
\begin{align}
[J_z,J_\pm] &= \pm J_\pm, \\
[J_+,J_-] &= 2J_z.
\end{align}
Defining
\begin{equation}
J^2 = J_z^2 + \frac{1}{2}(J_+J_- + J_-J_+),
\end{equation}
we find that
\begin{equation}
J^2\ket{n_a,n_b} =
\frac{N}{2}\left(\frac{N}{2} + 1\right)\ket{n_a,n_b},
\qquad \hbox{where} \qquad
N = n_a + n_b.
\end{equation}
This means that we can label basis kets just as well by~$j$ and $m$ as
we could by~$n_a$ and $n_b$.

If the total number of individuals present is $N$, then we are working
with the $N$-dimensional representation of~$\mathfrak{su}(2)$.

In the host--consumer model, each lattice site can be in one of three
states.  If we write $H_i$, $C_i$ and $E_i$ for the number of hosts,
consumers and empty slots at position $i$, then
\begin{equation}
H_i + C_i + E_i = 1.
\end{equation}
Introducing the number operator
\begin{equation}
N_i = h_i^\dag h_i + c_i^\dag c_i + e_i^\dag e_i,
\end{equation}
we have that $N_i$ takes the value 1 on all admissible states.  $N_i$
commutes with all nine operators which annihilate an individual and
then create one: $h_i^\dag h_i$, $h_i^\dag c_i$, $c_i^\dag h_i$ and so
on.  So, $N_i$ commutes with all linear combinations of operators of
this form.  We have, then, an eight-dimensional space of operators,
which is the Lie algebra $\mathfrak{su}(3)$, in Schwinger
form.\footnote{Actually, it's $\mathfrak{sl}(3)$, the complexification
  of $\mathfrak{su}(3)$.  See~\cite{baez2013}.}

\section{A Common Denominator}

One stochastic model recurs in many of the settings we have explored
in different chapters of this thesis.  This model is a stochastic
process which includes \emph{reproduction by budding into empty space}
and \emph{death leaving empty space behind,} implemented on a lattice
or other network.  By understanding this stochastic process, we can
obtain quantitative predictions for the systems we studied in
Chapters~\ref{ch:redgreen} and \ref{ch:volunteer}.

We can identify the basic reactions in this model as follows.  Let
$A_i$ denote the presence of a particle at site~$i$, and let $E_i$
denote the fact of site~$i$ being empty.  Writing $i$ and $j$ for site
indices, and $\partial(i)$ for the neighborhood of site~$i$, we have
\begin{align}
A_i + E_j &\to A_i + A_j,\ j \in \partial(i); \\
A_i &\to E_i.
\end{align}
The first reaction is reproduction; and the second, death.  Each of
these processes happens with some probability whenever the conditions
are right to allow it, \emph{i.e.,} whenever the appropriate reactants
are present in the proper juxtaposition.  Together, these reactions
define a kind of \emph{contact process.}  

The eco-evolutionary systems that have motivated us to write this
simplified model typically include higher-order complications.  The
rates at which organisms can die and empty sites can be filled will,
in general, be affected by the surroundings.  Therefore, we include one
generalization beyond the standard contact process: The probability of
the reaction taking place, when it is allowed, can depend on the local
environment.  Specifically, we are concerned with cases where the
probability of a particle at site $i$ reproducing to make a new
particle at site $j$ depends on all the sites which are adjacent
to~$i$.  This means that the probability of site $j$ becoming occupied
can depend on the \emph{next-to-nearest} neighbors of~$j$, which is
unlike the basic contact process.

This model includes the single-species cases of the Volunteer's
Dilemma, which we studied in \S \ref{sec:vs-lattice}.  When the
lattice contains only Volunteers or only Slackers, the only difference
between the two species' basic dynamical rules are how the presence of
neighboring individuals affect the rate of the budding process.
Furthermore, this model can serve to approximate the spatial
host--consumer model of Chapter~\ref{ch:redgreen}, in a scenario where
the ecosystem is filled with hosts, and only a small number of
consumers are present in a localized area.  When the consumer
transmissibility is just barely large enough to keep the consumer
population from fizzling, and the lattice contains no empty space, we
can gain some understanding from a simplified model which has only two
possible states for each lattice site.  In essence, we can to a first
approximation neglect the possibility of empty sites, as long as the
host growth rate is nonzero, because any sites that are emptied will
be refilled quickly enough not to be of concern.

The empty lattice is an absorbing state for this model: fluctuations
can take us into it, but never out.  Experience in nonequilibrium
statistical mechanics, codified in a conjecture of Grassberger and
Janssen~\cite{janssen1981, grassberger1982}, suggests that the phase
transition between the active and absorbing regimes will belong to the
directed percolation universality class.  Indeed, if our system had
diffusive particle motion and no occupation restrictions, we could
make that identification immediately.  Likewise, the transition in the
standard contact process is known to be a DP-class
phenomenon~\cite{hinrichsen2000}.

The question now arises: does the modification we made to the contact
process affect the universality class to which the model belongs?  If
we can argue that it does \emph{not,} then we can use off-the-shelf
results about DP-class transitions for the spatial host--consumer
model and for the Volunteer's Dilemma.  This would represent a
substantial advance:  even if the computation of DP critical exponents
is a laborious calculation, it only has to be done for one system.

We now develop two ways of writing equations for stochastic contact
processes.  This will help tie the model we have defined here into the
prior literature.  First, we will apply as directly as possible the
tools of raising and lowering operators we developed in earlier
sections.

If we were not concerned about the need for empty space, we could
write the following stochastic Hamiltonians to encapsulate the
reactions we described above:
\begin{align}
H_d &= \sum_i \lambda_i(a^\dag_i a_i - a_i), \\
H_r &= \sum_{\langle i, j\rangle} 
  \sigma_i(a^\dag_i a_i - a^\dag_i a^\dag_j a_i).
\end{align}
Here, we have introduced the potentially site-dependent transition
probabilities $\sigma_i$ and $\lambda_i$.  In the simplest case, these
can be taken as constant across all sites, but more generally, they
will depend on the configuration of other sites in the vicinity
of~$i$.  That is, whatever higher-order interactions among organisms
might arise, we can roll them into the $\sigma_i$ and $\lambda_i$.

However, because each site can hold at most one particle at a time,
events which violate that constraint must be disallowed.  We can
represent this formally by introducing delta functions which make a
contribution to the Hamiltonian vanish if conditions are not right.
For example, the event of a particle at site~$i$ budding to produce an
offspring at site~$j$ can only take place if exactly one particle
exists at~$i$ and exactly zero exist at~$j$.  So, we write the
modified stochastic Hamiltonians,
\begin{align}
H_r &= \sum_{\langle i, j\rangle} \sigma_i(a^\dag_i a_i - a^\dag_i a^\dag_j a_i)
              \delta_{n_i,1}\delta_{n_j,0}, \\
H_d &= \sum_i \lambda_i(a^\dag_i a_i - a_i)
       \delta_{n_i,1}.
\end{align}
As before, both $\sigma_i$ and $\lambda_i$ can be functions of the
local environment around site~$i$.

When we pass to the coherent-state representation and field theory,
the delta functions in our stochastic Hamiltonians will become
exponential factors~\cite{vanwijland2001, stollenwerk2011}.  This is a
consequence of the Fourier representation of the Kronecker delta:
\begin{equation}
\delta_{n,m} = \frac{1}{2\pi}\int dx\, e^{ix(n-k)}.
\end{equation}

Now for the second approach, in which we incorporate the
site-occupation restrictions directly into our definitions of the
creation and annihilation operators.  Grassberger and de la
Torre~\cite{grassberger1979} define the following
\emph{simplified Gribov process,} which is known to have a DP-class
phase transition.  Construct a regular lattice, whose
points will be labeled by the index $i$.  Each lattice point can be
occupied by at most one particle, \emph{i.e.,} we restrict the
occupation number $\nu_i$ to be 0 or 1.  Each particle can
spontaneously decay with rate $\kappa$, and each particle can produce
another in an adjacent empty lattice site with rate $\kappa'$.

The creation operator for site $n$ acts on a basis state by the rule
\begin{equation}
c_i^\dag \ket{\cdots \nu_i \cdots} =
(1-\nu_i)\ket{\cdots \nu_i+1 \cdots},
\end{equation}
and the annihilation operator on site $n$ acts as
\begin{equation}
c_i \ket{\cdots \nu_i \cdots} =
\nu_i\ket{\cdots \nu_i-1 \cdots}.
\end{equation}
Note that acting twice with either operator destroys the vector.  If
we apply the combination $c_i^\dag c_i$ to a state,
\begin{equation}
c_i^\dag c_i \ket{\cdots \nu_i \cdots}
 = \nu_i(2-\nu_i)\ket{\cdots \nu_i \cdots},
\end{equation}
we see that the result vanishes if $\nu_i = 0$, which is consistent.
This justifies using $c_i^\dag c_i$ as the number operator for
site~$i$.  Likewise,
\begin{equation}
c_i c_i^\dag \ket{\cdots \nu_i \cdots}
 = \nu_i(1-\nu_i)(1+\nu_i)\ket{\cdots \nu_i \cdots},
\end{equation}
which vanishes if $\nu_i = 1$.

Define $\ket{0}$ to be the state annihilated by all $c_i$, and let
\begin{equation}
V = \prod_i (1 + c_i).
\end{equation}
Then the probability to find particles at locations $n_1,\ldots,n_k$
irrespective of what is happening elsewhere on the lattice is
\begin{equation}
\rho^k(n_1,\ldots,n_k | \Phi) = \bra{0} c_{n_1}\cdots
c_{n_k} \ket{\Phi}.
\end{equation}
We can check this for a system composed of a single site.  Let the
state $\ket{\Phi}$ be
\begin{equation}
\ket{\Phi} = (1-p)\ket{0} + p\ket{1}.
\end{equation}
Then we have that
\begin{align}
\bra{0} c_1(1-c_1) \ket{\Phi}
 &= \bra{0} c_1(1-c_1) \ket{\Phi} \nonumber \\
 &= \bra{0} c_1(1-p)\ket{0}
    + \bra{0} c_1 p\ket{1}
    - \bra{0} c_1^2(1-p)\ket{0}
    = \bra{0} c_1^2p\ket{1} \nonumber \\
 &= p\braket{0}{0} \nonumber \\
 &= p.
\end{align}
And we see indeed that the probability of finding a particle in the
single site of our system is $p$, as it should be.

For a properly normalized state $\ket{\Phi}$, we have
\begin{equation}
\bra{0} V \ket{\Phi} = 1.
\end{equation}
The time evolution of $\ket{\Phi}$ is given by
\begin{equation}
\frac{d}{dt} \ket{\Phi} = -L\ket{\Phi},
\end{equation}
with the operator $L$ defined as
\begin{equation}
L = \kappa\sum_i \left(c_i^\dag - 1\right)c_i
    + \frac{\kappa'}{z} \sum_{\langle i,j \rangle}
      (c_i - 1) c_i^\dag c_j^\dag c_j.
\label{eq:stochastic-hamiltonian-c}
\end{equation}
The first sum is over all lattice sites $i$, and the second is over
all pairs of nearest neighbors $i$ and $j$.  In the prefactor of the
second sum, we have used $z$ to denote the coordination number of the
lattice.

Returning to the first way we set up the problem, let us suppose now
that the transition probabilities $\sigma_i$ and $\lambda_i$ are
independent of position.  In the absence of site-occupation
restrictions, the stochastic Hamiltonian is
\begin{equation}
H = \lambda \sum_i (a^\dag_i a_i - a_i)
    + \sigma \sum_{\langle i,j \rangle}
    (a^\dag_i a_i - a^\dag_i a^\dag_j a_i)
\end{equation}
Incorporating the restriction that a lattice site can contain at most
one particle, the stochastic Hamiltonian becomes
\begin{equation}
H = \lambda \sum_i (a^\dag_i a_i - a_i) \delta_{n_i,1}
    + \sigma \sum_{\langle i,j \rangle}
    (a^\dag_i a_i - a^\dag_i a^\dag_j a_i)
    \delta_{n_i,1}\delta_{n_j,0}.
\label{eq:stochastic-hamiltonian-a}
\end{equation}
Note that we have defined the basic interactions of our model in terms
of pairs; higher-order complications would emerge should we try
developing the model in perturbation theory to make numerical computations.

Is there a relationship between $L$ and $H$?  The annihilation
operators $a_i$ are analogous to the $c_i$, and likewise for
$a_i^\dag$ and $c_i^\dag$.  We saw above that the operator $c_i^\dag
c_i$ acts like a Kronecker delta function, comparing the occupancy of
site~$i$ to~1.  Consequently, $(1 - c_i^\dag c_i)$ can be thought of
as a delta function that compares the population size at site~$i$
to~0.

If we replace $\delta_{n_i,1}$ with $c_i^\dag c_i$, $\delta_{n_i,0}$
with $(1 - c_i^\dag c_i)$, $a_i$ with $c_i$ and $a_i^\dag$ with
$c_i^\dag$, then we turn our operator $H$ into the operator we
constructed earlier, $L$.  That is, we have established that
Eq.~(\ref{eq:stochastic-hamiltonian-c}) and
Eq.~(\ref{eq:stochastic-hamiltonian-a}) are equivalent ways of
representing the dynamics of the contact process.

What happens if $\sigma$ and $\lambda$ are no longer constant across
the system?  Does this bump the model out of the DP universality
class?  We can argue that in fact it does not, for the following
reason.  Although for brevity we wrote the transition rates as
depending on the site index $i$, we defined them as depending upon the
configuration of particles near $i$.  The value of $\sigma_i$ can
change as the population in the vicinity of site $i$ fluctuates.  We
could instead have defined a model in which $\sigma_i$ is chosen at
random for each $i$ following some probability distribution, and then
all the $\sigma_i$ remain constant over time.  This latter approach is
known in the language of statistical physics as \emph{quenched
disorder.}  And it is quenched disorder that pushes models out of the
DP universality class~\cite{moreira1996, janssen1997,
hooyberghs2004}.  In contrast, because the fluctuations due to
higher-order interactions among organisms are not frozen, they are
irrelevant, in the RG sense of the term.

Therefore, we expect that the DP critical exponents will be applicable
to the phase transitions in our spatial host--consumer model and the
lattice implementation of the Volunteer's Dilemma.  Referring to
Figures~\ref{fig:percolation}, \ref{fig:survival-prob-slackers}
and \ref{fig:survival-prob-volunteers}, it is satisfying to report
that the simulations agree.

\chapter{Invasion Fitness by Moment Closure Approximations}

\label{ch:invasion-fitness}
\section{Introduction and Overview}

\textbf{Moment closures} are a way of forgetting information about a system in a controlled fashion, in the hope that an incomplete, fairly heavily ``coarse-grained'' picture of the system will still be useful in figuring out what will happen to it. Sometimes, this is a justifiable hope, but in other cases, we are right to wonder whether all the algebra it generates actually leads us to any insights. Here, we'll be concerned with a particular application of this technology: studying the vulnerability of an ecosystem to invasion. We shall find expressions for \textbf{invasion fitness,} the expected relative growth rate of an initially-rare species or variety.

Consider a lattice, each site of which can occupied by an individual of ``resident'' type ($R$), occupied by a mutant ($M$), or empty ($0$). The difference between the mutant-type and resident-type individuals is encoded in the choice of transition rules representing death, birth and migration. We can get an aggregate measure of the situation by finding the probability that a randomly chosen site will be in state $a$, where $a$ can take values in the set $\{R, M, 0\}$. A finer degree of distinction is provided by the conditional probabilities $q_{a|b}$, where, for example, $q_{R|M}$ denotes the probability that a randomly chosen neighbor site to a randomly chosen mutant is of resident type. Note that if a mutant is injected into a native resident population and its offspring form a geographical cluster, $q_{M|M}$ can be much larger than $p_M$: few individuals are mutants overall, but the probability of a mutant life-form interacting with another mutant is high.

The \emph{pair dynamics} of the system involves the time evolution of the probabilities $p_{a b}$, that is, the probability that a randomly selected lattice edge will have $a$ on one end and $b$ on the other. The differential equation for $d p_{R M}/d t$, for example, will have terms reflecting the processes which can form and destroy $R M$ pairs: $R M\rightarrow R R$ is one possibility, and $R M\rightarrow M M$ is another. Death, which comes for organisms and leaves empty spaces behind, introduces processes like $R M \rightarrow R 0$, $R M \rightarrow 0 M$ and $R M \rightarrow 0 0$. Reproduction can lead to formerly empty spaces becoming occupied: $R 0 \rightarrow R R$ and $M 0 \rightarrow M M$. We've moved beyond just creating and annihilating residents and mutants, and now we're dynamically changing the number of ``resident--resident'' and ``resident--mutant'' \emph{pairs.}

Each term in our differential equations will have a transition rate dependent upon a conditional probability of the form $q_{a|b c}$, denoting the probability that a $b$ of a $b c$ pair will have a neighbor of type $a$. The differential equations for the pair probabilities $p_{a b}$ thus depend on triplet probabilities $p_{a b c}$, which depend upon quadruplet probabilities and so forth. To make progress, we truncate this hierarchy, brutally cutting off higher-order correlations by declaring that
\begin{equation}
q_{a|b c} \approx q_{a|b}.
\end{equation}
This imposition, a \emph{pair approximation,} destroys information
about spatial structure and thereby introduces bias which in an ideal
world ought to be accounted for. In theoretical ecology, this maneuver
dates back at least to Matsuda \emph{et al.}\ in
1992~\cite{matsuda1992}, though it has antecedents in statistical
physics, going back to the kinetic theory work of Bogoliubov, Born,
Green, Kirkwood and Yvon, for whom the ``BBGKY hierarchy'' is named.

Invasion fitness is judged in the following manner. We start with a lattice devoid of mutants ($p_{M a} = 0$) and find the equilibrium densities $p_{R R}^*$ and $p_{R 0}^*$ by setting
\begin{equation}
\frac{d p_{R0}}{d t} = \frac{d p_{R R}}{d t} = 0.
\end{equation}
The exact form of $p_{R R}^*$ and $p_{R0}^*$ will depend upon interaction details which we won't worry about just yet. We then inject a mutant strain into this situation; as the mutants are initially rare, we can say they do not affect the large-scale dynamics of the resident population. Summarizing the pair probabilities $p_{M a}$ with the shorthand $\underline{p}$, we write the differential equation in matrix form
\begin{equation}
\frac{d\underline{p}}{d t} = T(q_{a|b c})\underline{p},
\end{equation}
where the matrix $T(q_{a|b c})$ encapsulates the details of our chosen dynamics. The pair approximation, in which we discard correlations of third and higher order, lets us simplify this to
\begin{equation}
\frac{d\underline{p}}{d t} = T(q_{a|b})\underline{p}.
\end{equation}
When people started doing simulations of lattice models like these, they found that the conditional probabilities $q_{a|M}$ equilibrate. That is to say, even if the global density of mutants $p_M$ changes, the local statistical structure of a mutant cluster remains constant. This is the key statement which allows us to linearize the dynamics and write the behavior of $\underline{p}$ in terms of eigenvectors and eigenvalues:
\begin{equation}
\underline{p}(t) = C\underline{e}_A \exp(\lambda t).
\end{equation}
The dominant eigenvalue $\lambda$ of $T$ is the ``invasion exponent'' which characterizes whether an invasion will fail ($\lambda < 0$) or succeed ($\lambda > 0$). The eigenvector $\underline{e}_A$ associated with $\lambda$ describes the vehicle of selection for the mutants' particular genetic variation, by summarizing the structure of their geographical cluster.

All of this, of course, is only as good as our linearization! If
something interesting happens further away from the fixed point, just
looking at the eigenvalues we got from our matrix $T$ won't tell us
about it. In addition, if the actual dynamics of the system tend to
form patterns which can't be represented very well by pairwise
correlations, then pair approximation will run into a wall. As long
ago as 1994, Tainaka~\cite{tainaka1994} pointed out that, in a
rock-paper-scissors system,
\begin{quotation}
\noindent The failure of the mean-field theory and PA model implies that the long-range correlation is essentially important for the pattern formation.
\end{quotation}
 Minus van Baalen puts the issue in the following way:
\begin{quotation}
\noindent The extent to which pair-dynamics models are satisfactory depends on the goal of the modeler. As we have seen, these models do not capture all of the phenomena that can be observed in simulations of fully spatial probabilistic cellular automata. Basically, the approximation fails whenever spatial structures arise that are difficult to ``describe'' using pairs alone. More technically, the method fails whenever significant higher-order correlations arise -- that is, whenever the frequency of parituclar triplets (or triangles, squares, or all sorts of star-like configurations) starts to diverge from what one would expect on the basis of pair densities. Thus, pair-dynamics models satisfactorily describe probabilistic cellular automata in which only ``small-scale'' patterns arise. Larger, ``meso-scale'' patterns such as spirals are difficult to capture using this method.

---in {Dieckmann \emph{et al.} (2000)}, chapter 19~\cite{vanbaalen2000}.
\end{quotation}
 
It's also pretty easy for the algebra involved in a pair-approximation calculation to blow up far beyond the point of being useful. For example, Dobrinevski \emph{et al.}~\cite{dobrinevski2014} study a four-species system, where the pair approximation turns out to require 256 coupled differential equations. The only way to tackle that problem is to give it back to the computer and solve those equations numerically---and when they do that, it doesn't even work all that well!

\section{Example 1: Birth, Death, Movement}

In Chapter~\ref{ch:redgreen}, and again in the previous section, we
introduced the idea of \emph{pair approximation,} by which we try to
understand a system by tracking the joint probability distributions
for pairs of its pieces. Now, we'll look at this machinery in more
detail by focusing on a specific example. The ecosystem which we shall
study will contain one species living on a regular lattice, and the
individual organisms of that species can move about, give birth and
die. That is, our pair dynamics will include three processes, each
occurring stochastically with its own characteristic
rate: \emph{movement} or \emph{migration,} \emph{birth}
and \emph{death.} We follow the notation of van
Baalen~\cite{vanbaalen2000}.

We write $z$ for the ``coordination number'' of the lattice. That is, each lattice site will have $z$ neighbors.  We can represent the birth process as follows:
\begin{equation}
R + 0 \rightarrow R + R,
\end{equation}
and we say this takes place with rate $b/z$.

Similarly, the death process can be represented as
\begin{equation}
R + a \rightarrow 0 + a,
\end{equation}
for any site type $a$.  This takes place with rate $d/z$.

Movement or migration from one site to a neighboring location is the reaction
\begin{equation}
R + 0 \rightarrow 0 + R.
\end{equation}
This reaction occurs with rate $m/z$.

Given these reactions, we can write differential equations for the time derivatives of the pairwise densities.  The density of~$R0$ pairs changes as
\begin{equation}
\begin{array}{rcl}
\frac{d p_{R 0}}{d t} & = & -p_{R 0}[b/z + d + (z - 1)q_{0|R0}m/z + (z - 1)q_{R|0R}(b + m)/z] \\
& & + p_{00} (z - 1) q_{R|00} (b + m)/z \\
& & + p_{R R} [d + (z - 1) q_{0|R R} m/z].
\end{array}
\end{equation}
Likewise,
\begin{equation}
\frac{d p_{0 0}}{d t} = -p_{00} 2(z - 1) q_{R|00} (b + m)/z + p_{R 0} 2 [d + (z - 1) q_{0|R 0} m/z].
\end{equation}
Finally,
\begin{equation}
\frac{d p_{R R}}{d t} = p_{R 0} 2[b/z + (z - 1)q_{R|0 R} (b + m)/z] - p_{R R} 2[d + (z - 1) q_{0|R R} m/z].
\end{equation}

Summing over pairwise densities recovers overall densities:
\begin{equation}
p_i = \sum_j p_{i j}.
\end{equation}
From this, we deduce that
\begin{equation}
\frac{d p_R}{d t} = (b q_{0|R} - d) p_R.
\end{equation}
If we ignore spatial structure altogether, we can say that
\begin{equation}
q_{0|R} = p_0,
\end{equation}
which by normalization of probability means
\begin{equation}
q_{0|R} = 1 - p_R.
\end{equation}
So,
\begin{equation}
\frac{d p_R}{d t} = (b(1 - p_R) - d) p_R.
\end{equation}
This should look familiar: it's a logistic equation for population growth, with growth rate $b - d$ and equilibrium population $1 - d/b$.

It's worth pausing a moment here and using this result to touch on a more general concern. Often, a logistic-growth model is presented with the growth rate and the equilibrium population size as its parameters. When we see the model in that form, we naturally start thinking of those parameters as \emph{independently variable} quantities. We imagine that a mutation or a change in the environmental conditions could change one without affecting the other. However, if the growth rate and the equilibrium population size are both functions of other parameters taken together, then the changes \emph{which are biologically reasonable to consider} will likely affect both of them. To understand which quantities we should treat as independent, we need to spend time looking at how the numbers which apply to population-scale phenomena arise from the smaller-scale physiological and ecological goings-on~\cite{fox2011}.

\section{Example 2: Epidemic in an Adaptive Network}

We can juice things up a little by considering another example, one which is topical for these jittery times. Let's look at \emph{the spread of an epidemic.} There's a classic genre of models for this, which we can call after a prototypical representative, the Susceptible--Infected--Recovered or SIR model. In the SIR model, we imagine a population of individuals through which a disease can spread. Each individual is either \textbf{S}usceptible to the disease, \textbf{I}nfected with it or \textbf{R}ecovered from it. 

Contact with an Infected individual can turn a Susceptible one into an Infected, so an $S$ plus an $I$ becomes an $I$ plus an $I$. If the disease runs its course in an individual, they gain the status of $R$ and are thereafter immune to further infection. (Perfect immunity is, mathematically, the same as death, but we'll be optimistic with our labels today.) We can complicate the model in many ways, for example by making the immune response \emph{imperfect,} so that individuals who have recovered can be re-infected later. This could happen by the immunity fading over time, so that $R$ individuals transition back to $S$, or the immunity might only be partial, so that we have transitions from $R$ directly back to $I$. We can also add \emph{population structure}: interesting things happen when the individuals are not all in direct contact with one another. This complication is obviously something we'll have to address if we want to do epidemiology with real-world diseases! Speaking from a more mathematical perspective, we find neat phase-transition effects when we put these epidemic models on a lattice; see Chapter~\ref{ch:redgreen} and references therein.

The complicating factor we'll consider in this section is the
following: the spread of a disease through a social network can itself
change the way people contact each other. This makes epidemiology a
candidate subject for the study of adaptive networks: graph-structured
systems in which the states associated with the vertices and the
topology of the edges can change on the same timescales, feeding back
on one another~\cite{gross2006, gross2008, gross2009, do2009,
shaw2009, shaw2010, vansegbroeck2010, sunclipse2011, demirel2012,
sayama2015}.

To the SIR problem we described earlier, we add two extra wrinkles: first, the individuals are arranged in a network, and infection spreads only along the links within that network. Second, if a susceptible individual is in contact with an infected one, that link can be broken, and a new link established to another susceptible individual, which we pick at random from the pool of eligible susceptibles (that is, those who aren't already neighbours---we are disallowing multiple links). This \emph{rewiring rule} makes this scenario an adaptive-network problem.

For the moment, let's say that once they're infected, these organisms don't recover. We have only $S$ and $I$ in the population at any time. Therefore, the probability that an organism chosen at random has status $S$ and the probability that an organism chosen at random has status $I$ sum to 1:
\begin{equation}
p_S + p_I = 1.
\end{equation}
This simplifies the SIR model to an SIS model.  A node can start in the $S$ state, become infected and enter the $I$ state, and then potentially recover and return to the $S$ state again.  We neglect the possibility of immunity: nodes which have been infected and recovered are just as susceptible to the disease as those which have never been in the $I$ state.

Following the literature~\cite{gross2006, do2009}, we choose to normalize our pairwise densities in the following way:
\begin{equation}
p_{S S} + p_{S I} + p_{I I} = \langle k \rangle,
\end{equation}
where $\langle k \rangle$ is the average degree of the nodes in the network.

The number of infected individuals decreases as organisms recover from the disease, while it increases as the contagion spreads over links from those already infected. We say how quickly recovery happens using the parameter $r$, and we encode the ease with which the disease travels along network links with the transmissibility $\tau$. With these definitions, we can write the following rate equation for the density of infected individuals, $p_I$:
\begin{equation}
\frac{d}{d t} p_I = \tau p_{S I} - r p_{I}.
\end{equation}
In the original SIR example, where everyone was in contact with everyone else, we could say $p_{S I} \approx \langle k \rangle p_S p_I$. But we ought to be wary of using this approximation here, for two reasons. First, any time we have a system which has a chance of developing heterogeneity, of forming lumps in one region which don't directly affect lumps in another, then averaging over the whole system becomes a risky business. Second, more specifically, this approximation can't capture rewiring. Links are breaking and re-forming all the time in our model, but the product $p_S p_I$ stays the same when we move links around. So, to write an equation for how $p_I$ changes, we need to write down how the density of $S I$ pairs will change, but in order to do that, we have to include the rewiring effect.

Let's introduce a third parameter, $w$, to indicate how much rewiring is going on.  The $w$ parameter will be a rate, having units of inverse time, just like $r$ and $\tau$.  The density of $S S$ pairs will go \emph{down} as the disease spreads to them from infected nodes, but it will go \emph{up} as nodes recover \emph{and} as susceptible nodes rewire their links.  Consequently, the time derivative of~$p_{SS}$ must include a positive term which depends on~$r$ and $w$, and a negative term which depends upon~$\tau$.
\begin{equation}
\frac{d}{d t} p_{S S} = (r + w) p_{S I} - \tau p_{S S I}.
\end{equation}
Similarly, $p_{II}$ is increased by the disease being transmitted, and decreased by the rewiring effect:
\begin{equation}
\frac{d}{d t} p_{I I} = \tau(p_{S I} + p_{I S I}) - 2r p_{I I}.
\end{equation}
Now comes the moment-closure step.  We enforce a pair approximation, by writing three-way probabilities in terms of lower-order ones:
\begin{equation}
p_{I S I} = \frac{\langle q \rangle}{\langle k \rangle} \cdot \frac{p_{S I}^2}{p_S}.
\end{equation}
The ``mean excess degree'' $\langle q \rangle$ is the number of additional links we expect to find after we follow a random link. It turns out~\cite{do2009} that $\langle q \rangle = \langle k \rangle$ is a reasonable approximation.

Our dynamical system is defined by three equations. First, we have the rule we wrote before,
\begin{equation}
\frac{d}{d t} p_I = \tau p_{S I} - r p_I,
\end{equation}
and then the rules we deduced using pair approximation:
\begin{equation}
\frac{d}{d t} p_{S S} = (r + w)p_{S I} - 2 \tau p_{S I} \frac{p_{S S}}{p_S},
\end{equation}
and
\begin{equation}
\frac{d}{d t} p_{I I} = \tau p_{S I} \left(1 + \frac{p_{S I}}{p_S}\right) - 2r p_{I I}.
\end{equation}
Having written these equations, we can compare the behaviour of the dynamical system they define to that of a simulation of the original model. They actually agree pretty well~\cite{gross2006, do2009}.

Note that we have made an important simplifying assumption, not in the analytical treatment of our model but in the original definition of it: we chose a rewiring rule which lets nodes form new links to \emph{S}-type nodes anywhere else in the system. That is, the rewiring rule lacks any notion of \emph{proximity.} Depending on the phenomenon we're trying to model, it might be reasonable to allow rewiring only to neighbours of a node's current neighbours, for example. This could be the case if information about available opportunities for building new connections had to spread through existing links. Or, if the nodes' spatial positions are significant, perhaps links which span longer geographical distances should be disfavoured. These are the sorts of complications which can make pair approximations inapplicable.

The study of dynamical processes on and of graph structures brings together the subjects of ``network theory'', as the term has been used in stochastic process research, and ``complex networks'', to invoke a term which has been trending elsewhere.

\section{Example 3: Evolution of Altruism}
\label{sec:pairapprox-altruism}
One application of this machinery has been to understand how evolution
can produce \emph{altruistic behaviour}~\cite{vanbaalen1998}. A
behaviouristic definition of ``altruism'' would go something like,
``Acting to increase the reproductive success of another individual at
the expense of one's own''. This can be written out using game theory;
one defines a parameter $b$ to stand for benefit, $c$ to stand for
cost and a payoff function which depends on $b$, $c$ and the strategy
employed by an organism.

We shall consider a lattice of sites, each one of which can be in one of three states: \emph{empty,} denoted by 0; occupied by a \emph{selfish organism,} denoted by $S$; and occupied by an \emph{altruist,} denoted by $A$.

In the reproduction process, an organism spawns an offspring into an adjacent empty lattice site. This turns a pair of type $S 0$ into a pair of type $S S$. At what rate should the transition $S 0 \rightarrow S S$ occur? If we presume some baseline reproductive rate, call it $b_0$, then the presence of altruistic neighbours should \emph{augment} that rate. We'll say that if the number of nearby altruists is $n_A$, then selfish individuals will reproduce at a rate $b_0 + B n_A / n$, where the parameter $B$ specifies how helpful altruists are. The reproduction process for altruists, which we can write $A 0 \rightarrow A A$, occurs at a rate $b_0 + B n_A / n - C$. Here, the parameter $C$ is the \emph{cost} of altruism: it's how much an altruist gives up to help others.

In the differential equation for $d p_{S 0}/d t$, the $S 0 \rightarrow S S$ transition contributes a term proportional to the density of $S 0$ pairs:
\begin{equation}
-[(1 - \phi) b_S + (b_S + m) \phi q_{S|0 S}] p_{S 0},
\end{equation}
where we have written $\phi$ for $1 - 1/z$, to save a little ink. All things told, the rate of change of $p_{S 0}$ is given by
\begin{equation}
\begin{array}{rcl}
\frac{d p_{S 0}}{d t} &= &(b_s + m) \phi q_{S|0 0} p_{0 0}\\
& & - [d_S + m \phi q_{0|S 0}] p_{S 0}\\
& & - [(1 - \phi) b_s + (b_S + m) \phi q_{S|0 S}] p_{S 0}\\
& & + [d_S + m \phi q_{0|S S}] p_{S S}\\
& & - (b_A + m) \phi q_{A|0 S} p_{S 0}\\
& & + [d_A + m \phi q_{0|A S}] p_{S A}.
\end{array}
\end{equation}
Yuck! After writing a few equations like that, it's easy to wonder if
maybe we should look for \emph{new mathematical ideas} which could
help us better organise our thinking.  But, for the nonce, we will
simply press on with the algebra.

The next steps follow the general plan we laid out above. We write differential equations for the pairwise probabilities $p_{a b}$, which depend on triplet quantities $q_{a|b c}$. Then, we impose a pair approximation, declaring that $q_{a|b c} = q_{a|b}$, which gives us a closed system of equations. Next, we find the fixed point with $p_A = 0$, and we perturb around that fixed point to see what happens when a strain of altruists is introduced into a selfish population. The dominant eigenvalue $\lambda$ of the time-evolution matrix $T$ tells us, in this approximation, whether the altruistic strain will invade the lattice or wither away. The condition $\lambda > 0$ can be written in the form
\begin{equation}
B \phi \tilde{q}_{A | A} - C > 0.
\label{eq:hamilton-invasion-condition}
\end{equation}
Here, we've written $\tilde{q}_{A | A}$ for the conditional probability of altruists contacting altruists which obtains as the local densities equilibrate. That is to say, an attempted invasion by altruists will succeed if a measure of benefit, $B$, multiplied by an indicator of ``assortment'' among genetically similar individuals, is greater than the cost of altruistic behaviour, $C$.

After all our mucking with eigenvalues, we have found a condition which is strongly reminiscent of a classic and influential idea from mid-twentieth-century evolutionary theory. In biology, the inequality
\begin{equation}
\hbox{(benefit)} \times \hbox{(relatedness)} > \hbox{(cost)}
\end{equation}
is known as \emph{Hamilton's rule}~\cite{vandyken2011}. This is a rule-of-thumb for when natural selection can favour altruistic behaviour: altruists can prosper when the inequality is satisfied. Hamilton's rule was originally derived for unstructured populations, with no network topology or spatial arrangement to them. We can understand Hamiltion's rule in this context in the following way:

How well an organism fares in the great contest of life depends on the environment it experiences. During the course of its life, an individual member of a species will interact with a set of others, which we could call its ``social circle''. The composition of that social circle affects how well an individual will propagate its genetic information to the next generation---its \emph{fitness.} In an unstructured population, we can think of such circles being formed by taking random samples of the population. An altruist, by our definition, sacrifices some of its own potential so that offspring of other individuals can prosper. A social circle of altruists can fare better than a social circle of selfish individuals, increasing the chances that social circles which form in the next generation will contain altruists~\cite{vandyken2011}.

It's common to treat ``benefit'' and ``cost'' as parameters of the system. We could potentially derive them from more fundamental dynamics, if we looked more closely at the interactions within a particular ecosystem, but right now, they're just knobs we can turn. What about the remaining quantity in Hamilton's rule: what does ``relatedness'' mean? Excellent question! We can get a feel for where the term came from by taking a gene's-eye view: copies of many of my particular genetic variants will be sitting inside the cells of my close relatives. Consequently, as far as my genes are concerned, if my relatives survive, that's almost as good as my surviving. When reckoning the benefit of altruism against its cost, then, the aid one organism brings to another ought to be weighted by how ``related'' they are.

So, we can say that we have ``recovered Hamilton's rule as an emergent property of the spatial dynamics''---\emph{if} we are willing to draw a circle around the middle of our formula and declare those terms to be the ``relatedness''.

Knowing where our invasion condition came from, we can appreciate some of the caveats which scientists have raised in connection with Hamilton's rule.

Simon \emph{et al.}\ address this specifically~\cite{sfd2012}:
\begin{quotation}
\noindent In particular, $r$ is often taken to be the average relatedness of interacting individuals, as compared to the average relatedness in the population, in which case inequality (1) $[r B > C]$ is referred to as Hamilton's rule. It is important to note that inequality (1) is only a description of whether the current level of assortment as subsumed in the parameter $r$ is sufficient to favour cooperation, but not a description of the mechanisms that would lead to such assortment. It has been suggested repeatedly that the problem of cooperation can be understood entirely based on Hamilton's rules of the form (1). Even though often taken as gospel, this claim is wrong in general, for two reasons.

First, and foremost, even if a rule of the form (1) predicts the direction of selection for cooperation at a given point in time, the long-term evolution of cooperation cannot be understood without having a dynamic equation for the quantity $r$, i.e., without understanding the temporal dynamics of assortment. The dynamics of $r$ in turn cannot be understood based solely on the current level of cooperation, and hence expressions of type (1) are in general insufficient to describe the evolutionary dynamics of cooperation. Second, the quantity $r$, which measures the average relatedness among interacting individuals, is insufficient to construct Hamilton's rule in models that account for variable individual-level death rates and/or group-level events.
\end{quotation}

Damore and Gore~\cite{damore2011} have more to say on this point:
\begin{quotation}
\noindent Contrary to the popular use of the word, ``relatedness'' describes a population of interacting individuals, where $r$ refers to how assorted similar individuals are in the population.
\end{quotation}
And in further detail:
\begin{quotation}
[E]very definition of relatedness must take into account the population. Therefore, relatedness is not the percent of genome shared, genetic distance, or any extent of similarity between two isolated individuals in a larger population. Also, because horizontal gene transfer is commonplace between microbes and selection is strong, phylogenetic distance or any other indirect genetic measure is likely to be inaccurate. Many of these false definitions live on partly because ambiguous heuristics like ``$\frac{1}{2}$ for brothers, $\frac{1}{8}$ for cousins,'' which require very specific assumptions, are repeated in the primary literature. Also, most non-theoretical papers simply define relatedness as ``a measure of genetic similarity'' and do not elaborate or instead leave the precise definition to the supplemental information $[\ldots]$ Unfortunately, scientists can easily misinterpret this ``measure of genetic similarity'' to be anything that is empirically convenient such as genetic distance or percent of genome shared. Largely because of this confusion, we support the more widespread use of the term ``assortment,'' which is harder to misinterpret $[\ldots]$ For similar reasons of reader understanding, we also encourage authors to make calculations more explicit, either in the main or supplemental text, and to avoid repeating previous results without giving the assumptions that went into deriving them.
\end{quotation}

It is for this reason that we called $\phi \tilde{q}_{A | A}$ a measure of ``assortment'' earlier. Of course, even with this careful choice of terminology, the limitations of our Hamilton-esque rule still apply: we know that because we derived it from the condition that the dominant eigenvalue be positive, it will miss any effects which a fixed-point eigenvalue analysis is not sensitive to.

Stepping back for a moment, notice that although the terms and
coefficients started to proliferate on us, we haven't introduced any
remarkably ``advanced'' or ``esoteric'' mathematics. Derivatives,
matrices, eigenvalues---this is \emph{undergraduate} stuff! The amount
of algebra we've been able to stir up without really even trying is,
however, a little worrying. We can \emph{invent} a mathematical model
for some particular biological scenario, and we might even be able to
solve it, or at least tell how it'll behave in certain interesting
circumstances. But what if we want \emph{general} results which extend
across models, or ideas which will help us identify the common
features and the key disparities among a host of examples?  We will
return to this question in Chapter~\ref{ch:speculations}.

We conclude this section with a more detailed derivation of the
invasion condition (\ref{eq:hamilton-invasion-condition}).  First, for
convenience's sake, we define the abbreviations
\begin{align}
\alpha_j &= (b_j + m)\phi = (b_j + m)(1 - 1/z), \\
\beta_j &= (1 - \phi)b_j = \frac{b_j}{z}, \\
\gamma_{jk} &= d_j + m\phi q_{0|jk}.
\end{align}
These denote, respectively, the effective rates of arrivals from
neighboring sites, birth events within pairs of sites, and the
vacation of sites by migration or death.  If there are no altruists in
the ecosystem, then the $S0$ and $SS$ pair densities evolve according
to the coupled equations
\begin{align}
\frac{dp_{S0}}{dt} &= \alpha_S q_{S|00} p_{00}
 - (\beta_S + \alpha_S q_{S|0S} + \gamma_{S0}) p_{S0}
 + \gamma_{SS} p_{SS}, \\
\frac{dp_{SS}}{dt} &= 2(\beta_S + \alpha_S q_{S|0S}) p_{S0}
 - 2 \gamma_{SS} p_{SS}.
\end{align}

Setting these rates to zero determines the equilibrium point for the
altruist-less ecosystem.  At this equilibrium,
\begin{equation}
\alpha_S \bar{q}_{S|0S} \bar{p}_{00} - \gamma_{S0} \bar{p}_{S0} = 0.
\end{equation}
Invoking the pair approximation, we drop some indices:
\begin{equation}
\alpha_S \bar{q}_{S|0} \bar{p}_{00} - \gamma_{S} \bar{p}_{S0} = 0,
\ \hbox{where}\ \gamma_s = d_S + m\phi q_{0|S}.
\end{equation}
Recalling that
\begin{equation}
\bar{q}_{S|0} = \frac{\bar{p}_{S0}}{\bar{p}_0},
\ \bar{q}_{0|0} = 1 - \bar{q}_{S|0} = \frac{\bar{p}_{00}}{\bar{p}_0},
\end{equation}
we conclude that
\begin{equation}
\alpha_S \bar{q}_{0|0} - \gamma_S = 0.
\label{eq:bar-q-00}
\end{equation}
This tells us the density of adjacent empty sites in an equilibrated
$S$-type population.

Next, we consider a resident population of selfish individuals that
has come to an equilibrium and is invaded by a strain of altruists.
Will the invasion be successful or not?  We make the approximation
that initially, the altruists are sufficiently rare that they do not
affect the resident population, so the demographic statistics we
derived above are still applicable.

The pairwise densities involving altruists then evolve according to
the following three coupled differential equations:
\begin{align}
\frac{dp_{A0}}{dt} &= \alpha_A q_{A|00} \bar{p}_{00}
 - (\beta_A + \alpha_A q_{A|0A})p_{A0}
 - \alpha_S q_{S|0A} p_{A0}
 + \gamma_{SA} p_{AS}
 - \gamma_{A0} p_{A0}, \\
\frac{dp_{A0}}{dt} &= \alpha_A q_{A|0S} \bar{p}_{0S}
 + \alpha_S q_{S|0A} p_{A0}
 - (\gamma_{AS} + \gamma_{SA}) p_{AS}, \\
\frac{dp_{A0}}{dt} &= 2(\beta_A + \alpha_A q_{A|0A}) p_{A0}
 - 2\gamma_{AA} p_{AA}.
\end{align}
We are treating the barred quantities as constant, so the dynamical
variables are $p_{A0}$, $p_{AS}$ and $p_{AA}$.  Collecting these into
a three-element vector $(p_{A0}, p_{AS}, p_{AA})^{\rm T}$, we can
express this dynamical system as
\begin{equation}
\frac{d\vec{p}}{dt} = M(q_{j|kl}) \vec{p},
\end{equation}
where the matrix $M$ is a function of the triplet densities:
\begin{equation}
M(q_{j|kl}) = \left(\begin{array}{ccc}
 \alpha_A q_{A|00} - (\beta_A + \alpha_A q_{A|0A}) - \gamma_{A0}
 - \alpha_S q_{S|0A}
 & \gamma_{AS} & \gamma_{AA} \\
 (\alpha_S + \alpha_A) q_{S|0A}
 & -\gamma_{AS} - \gamma_{SA}
 & 0 \\
 2(\beta_A + \alpha_A q_{A|0A})
 & 0 & -2 \gamma_{AA}
 \end{array}\right).
\end{equation}
Here, we have used the fact that
\begin{equation}
q_{A|0S} \bar{p}_{0S} = q_{S|0A} p_{0A} = p_{A0S}
\end{equation}
to simplify the middle element in the left column.

Again calling upon the pair approximation, we simplify the matrix $M$
to
\begin{equation}
M(q_{j|k}) = \left(\begin{array}{ccc}
 \alpha_A - \beta_A - (\alpha_A + \alpha_S) \bar{q}_{S|0}
 & \gamma_S & \gamma_A \\
 (\alpha_S + \alpha_A) \bar{q}_{S|0} & - \gamma_A - \gamma_S & 0 \\
 2\beta_A & 0 & -2\gamma_A
\end{array}\right).
\end{equation}

We are now in the familiar terrain of linear stability analysis.  The
altruist invasion will be successful, according to all the
approximations we have made so far, if the all-$S$ fixed point is
unstable.  The transition between stability and instability occurs
when the dominant eigenvalue of~$M$ changes from negative to
positive.  The product of the eigenvalues of~$M$ is the determinant
of~$M$, so we seek the location in parameter space where the
determinant of~$M$ vanishes.  This yields the condition
\begin{equation}
(\alpha_A - \gamma_A)(\gamma_A + \gamma_S)
 - (\alpha_A + \alpha_S) \gamma_A \bar{q}_{S|0} = 0.
\end{equation}
Having derived $\bar{q}_{0|0}$ in Eq.~(\ref{eq:bar-q-00}), we have a
value we can use for~$\bar{q}_{S|0}$.  Making this substitution,
\begin{equation}
(\alpha_S + \gamma_A)(\alpha_A\gamma_S - \alpha_S\gamma_A) = 0,
\end{equation}
meaning that the transition lies at
\begin{equation}
\frac{\alpha_A}{\gamma_A} = \frac{\alpha_S}{\gamma_S}.
\end{equation}
The all-$S$ fixed point is unstable when the ratio $\alpha_A
/ \gamma_A$ exceeds $\alpha_S / \gamma_S$.  This provides our
condition for invasion success:
\begin{equation}
\frac{(b_A + m)\phi}{d_A + m\phi\tilde{q}_{0|A}}
>
\frac{(b_S + m)\phi}{d_S + m\phi\tilde{q}_{0|S}},
\end{equation}
where $\tilde{q}_{0|j}$ are given by the eigenvector of~$M$
corresponding to the dominant eigenvalue.

To simplify still further, let's see what happens when the migration
rate goes to zero, and the death rates $d_A$ and $d_S$ are equal.
Then the invasion condition reduces to $b_A > b_S$, or
\begin{equation}
\boxed{B\phi \tilde{q}_{A|A} - C > 0.}
\end{equation}
This has the form we promised: the altruists invade if the benefit
parameter, multiplied by an assortment factor that depends on
population structure, is greater than the cost parameter.  Note that
the assortment factor will generally depend upon the population
dynamics, through quantities like migration rates!

\section{Host--Consumer Dynamics}

Let us consider a spatial host--consumer system containing a single
type of consumer.  In such an ecosystem, each lattice site can be in
one of three possible states: $0$, $H$ and $C$.  Consequently, there
are six different site-site correlations.  Because
\begin{equation}
\sum_j P(ij) = P(i),
\end{equation}
only three of the six variables are independent.  Following de
Aguiar \emph{et al.}, we choose the independent correlations to be
\begin{equation}
u = P(H0),\ r = P(HC)\ \hbox{and}\ w = P(0 C).
\end{equation}
That is, our independent variables will be $x$, $y$, $u$, $r$ and
$w$.  This choice implies that if we write $z = 1 - x - y$ for the fraction
of empty space, the other three correlations are given by 
\begin{equation}
q = P(00) = z-u-w,\ p = P(HH)=x-r-u \hbox{ and } s =
P(CC)=y-r-w.
\end{equation}

A given host can only be consumed once.  Therefore, the simplest way
to find the probability of its consumption is to calculate
one \emph{minus} the probability that it is \emph{not} consumed.  If
a host has $m$ consumer neighbors, then this probability is $1 -
(1-\tau)^m$.  Analogous considerations apply to empty sites being
colonized by hosts.

Let $\zeta$ be the number of nearest-neighbor sites (so, on the
default square lattice, $\zeta = 4$).  We define
\begin{equation} 
\label{eq:aux} 
h_{\zeta}(\alpha) =  1-(1-\alpha)^{\zeta}.
\end{equation} 

The host population fraction $x$ increases if hosts can reproduce into
empty space, which depends upon the contact rate between host sites
and empty sites.  Therefore, $dx/dt$ will have a positive contribution
that depends on~$u$.  It will also have a negative contribution that
depends upon the probability that consumers will be adjacent to hosts
and thus able to consume them.  Together, these two processes combine
to yield
\begin{equation} 
\label{eq:pair4a} 
\frac{dx}{dt} = z \, h_{\zeta}(g u/z) -  x \, h_{\zeta}(\tau r/x).
 \end{equation} 
The rate of change in $y$ can be written similarly.  However,
consumers do not require the presence of any particular entities in
their neighborhood to die off, so the negative term will only depend
on~$y$ itself:
\begin{equation} 
\label{eq:pair4b} 
\frac{dy}{dt} = x \, h_{\zeta}(\tau r/x) - v y .
\end{equation} 
If we make the approximation $r = xy$, this becomes
\begin{equation}
\frac{dy}{dt} = x \, h_{\zeta}(\tau y) - vy,
\end{equation}
which is the mean-field equation we would write for consumer dynamics.

On much the same principles, we can write time-evolution equations for
the pairwise correlation variables.  The appropriate equations turn
out to be
\begin{align}
\frac{du}{dt} &= (q-u) \, h_{\zeta-1}(g u/z) + v r -   
  u \, h_{\zeta-1}(\tau r/x) \nonumber \\
  &\ -\, g u [1-h_{\zeta-1}(g u/z)],
\label{eq:pair4c} 
\end{align} 
and
\begin{align}
\displaystyle{\frac{dr}{dt}} &= (p-r) \, h_{\zeta-1}(\tau r/x) - v r +  
  w \, h_{\zeta-1}(g u/z) \nonumber \\
  &\ - \, \tau r [1-h_{\zeta-1}(\tau r/x)] 
\label{eq:pair4d} 
\end{align}
and 
\begin{equation} 
\label{eq:pair4e} 
\frac{dw}{dt} = u \, h_{\zeta-1}(\tau r/x) + v (s-w) -  
w \, h_{\zeta-1}(g u/z)  \; . 
\end{equation} 

This is the system of differential equations we solved numerically in
order to find the pair-approximation results in
Chapter~\ref{ch:redgreen}.  They can be extended to include two types
of consumer with differing $\tau$ values.  However, investigating that
augmented system of equations reveals a problem~\cite{aguiar2003b,
aguiar2004, aguiar2003b-errata}: higher $\tau$ almost always wins out
over lower $\tau$.  This means that, as we saw in
Chapter~\ref{ch:redgreen}, whatever we learn from pair approximation
must be augmented by knowledge from another source if we wish to
understand evolutionary processes.

\section{Modified Mean-Field}
\label{sec:mmf}
We should at this point mention a proposal by Pascual \emph{et
al.}~\cite{pascual2011} to incorporate spatial effects into mean-field
models by altering the functional dependence on mean-field quantities
in differential equations, rather than by moment closures.
Pascual \emph{et al.}\ begin with a simple Lotka--Volterra model for
a plant species, whose population density they denote $p$, and an
herbivore species, whose density is denoted $h$:
\begin{align}
\dot{p} &= \alpha pe - \beta ph, \\
\dot{h} &= \beta ph - \delta h.
\end{align}
Here, $e$ stands for the density of empty space, $1 - (p+h)$.  By
rescaling the units of time, we can eliminate one parameter, which we
pick to be the consumption rate $\beta$:
\begin{align}
\dot{p} &= \alpha pe - ph, \\
\dot{h} &= ph - \delta h.
\end{align}
This system of equations has a stable equilibrium point at
\begin{equation}
p^* = \delta,\ h^* = \frac{\alpha(1-\delta)}{1+\alpha}.
\end{equation}

Pascual \emph{et al.}\ suggest modifying the dependence of~$\dot{p}$
and $\dot{h}$ on the variables $p$, $e$ and $h$, so that each variable
no longer enters linearly:
\begin{align}
\dot{p} &= \alpha c_1 p^{a_1}e^{b_1} - c_2 p^{a_2}h^{b_2}, \\
\dot{h} &= c_2 p^{a_2}h^{b_2} - \delta h.
\end{align}
This is the ``modified mean-field'' (MMF) model.  The new parameters
$a_i$ and $b_i$ are intended to represent the deviations in the
effective population density from mean-field expectations which an
individual organism experiences due to spatial fluctuations.  The
other new parameters, $c_1$ and $c_2$, incorporate linear effects of
spatial pattern formation which ``can decrease or increase (when $0 <
c_i < 1$ or $c_i > 1$ respectively) the overall rate but cannot not
modify its functional form''~\cite{pascual2011}.  The MMF may appear
to be a rather ad-hoc alteration of the original Lotka--Volterra
system, but the equations are simpler than those produced by pair
approximation, and by fitting the new parameters, the trajectories
of~$p$ and $h$ seen in a lattice system can be reproduced.

For our purposes, however, we need more.  Can the MMF say anything
about the \emph{competition} between two types of herbivore with
different consumption rates?  This is the minimum we require when we
advance from ecological to eco-evolutionary dynamics.  To investigate
this, we consider an augmented MMF system with population densities
$p$, $h$ and $k$.  Now, the amount of empty space is given by
\begin{equation}
e = 1 - (p+h+k).
\end{equation}
The dynamical equations are
\begin{align}
\dot{p} &= \alpha c_1 p^{a_1}e^{b_1}
           - c_2 p^{a_2} h^{b_2}
           - c_3 p^{a_2} k^{b_2}, \\
\dot{h} &= c_2 p^{a_2} h^{b_2} - \delta h, \\
\dot{k} &= c_3 p^{a_2} k^{b_2} - \delta k.
\end{align}
The new parameter $c_3$ indicates the consumption rate of the
additional herbivore species, which we can think of as a mutant
variety.  Without loss of generality, we take $c_3 > c_2$.  If $k =
0$, we recover the two-species MMF system.  Moreover, the only
difference between the two types of herbivore is the rate at which
they consume plants, so we have no reason to think that the exponents
should be different (and in fact Pascual \emph{et al.}\ find no
significant dependence of the fitted exponents $a_2$ and $b_2$ on the
rate parameters).

From the difference
\begin{equation}
\frac{d}{dt}(h-k) = \dot{h}-\dot{k}
 = p^{a_2} (c_2 h^{b_2} - c_3 k^{b_2}) - \delta(h-k),
\end{equation}
we see that if $h = k$, then the right-hand side is negative, and so
$\dot{h} < \dot{k}$.  Consequently, if the resident and mutant
herbivore population densities are ever equal, then the mutant
population is growing faster, or is diminishing less rapidly.  This
implies that the resident population density cannot cross from below
the mutant density to above.  Invasion by a strain with lower
voraciousness is, in the MMF, impossible.  However, we see it
happening easily in the spatial host--consumer model.

\begin{figure}[h]
\includegraphics[width=12cm]{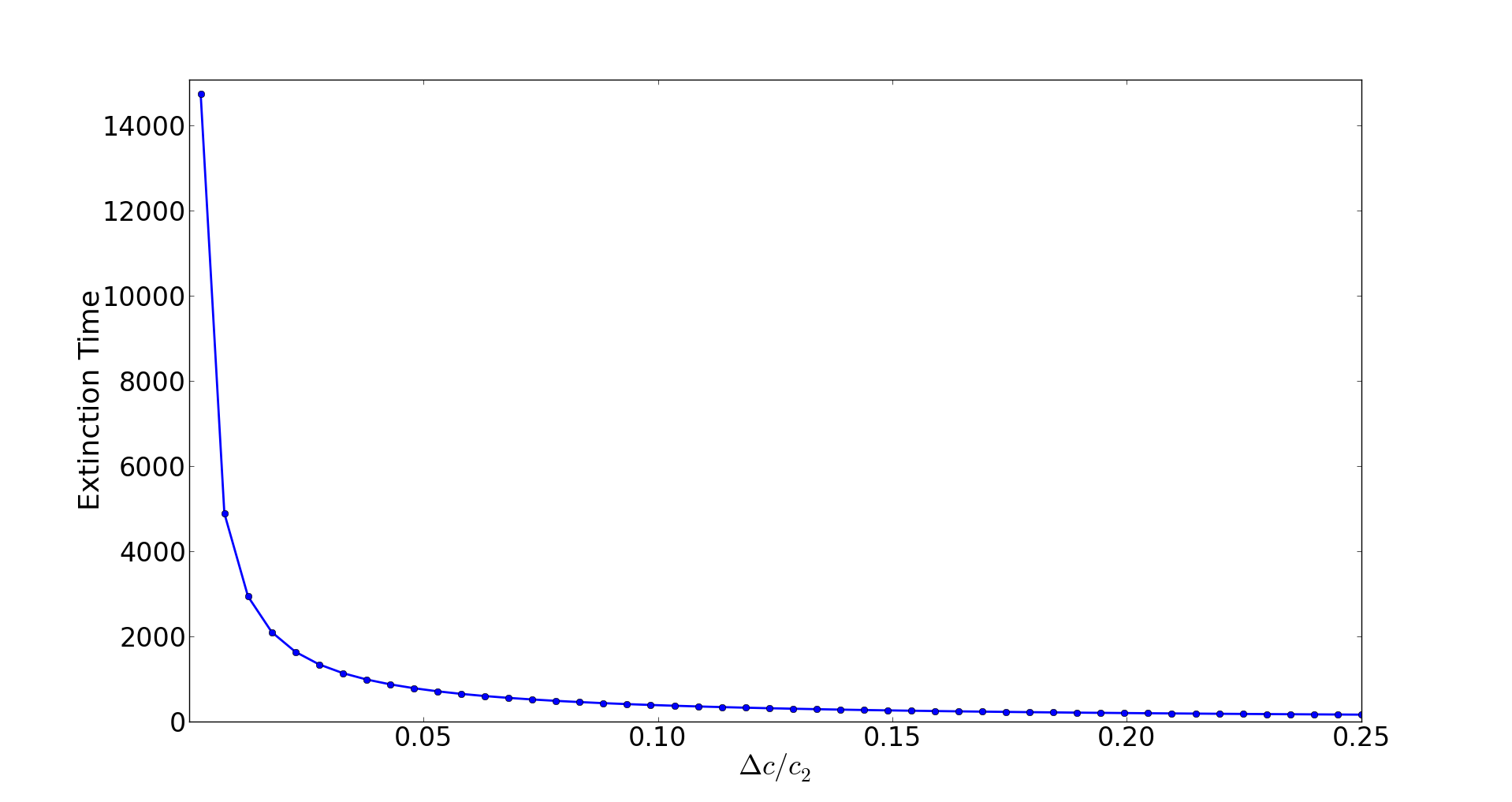}
\caption{\label{fig:extinction-times-modmeanfield} Time required for a
mutant herbivore population to take over an ecosystem, as a function
of the difference in consumption rates, $\Delta c = c_3 - c_2$.  The
mutant population density is initialized to~$1/100$ of the resident,
and the dynamics are simulated until the resident population density
has diminished to~$1/100$ of its initial value.  To facilitate
comparison with the original paper~\cite{pascual2011}, parameter
values for these computations were set at $a_1 = 0.2$, $b_1 = 0.59$,
$c_1 = 0.19$, $a_2 = 0.59$, $b_2 = 1.0$, $c_2 = 0.4$, $\alpha = 0.6$,
$\delta = 0.25$.}
\end{figure}

Numerical analysis (Figure~\ref{fig:extinction-times-modmeanfield})
shows that a small amount of a mutant herbivore strain with higher
voracity can take over a two-species system which has settled into its
equilibrium.  In other words, the MMF approach isn't any better at
handling coexistence or ``the resident strikes back'' effects than
moment closure is~\cite{aguiar2003b, aguiar2003b-errata}.  This is
only to be expected, since neither approach captures the
descendent-shading we examined in Chapter~\ref{ch:redgreen}.  One can
hack the exponents until the MMF model matches the two-species
ecological dynamics, but it is inadequate for \emph{evolutionary}
studies.

\chapter{The Varietes of Multilevel Selection}

\label{ch:price-mls}

\section{Introduction}

Ever since Darwin, biologists have tried to explain how social
behaviors---cooperation, even self-sacrifice---could emerge from
natural selection, which seems to embody a dog-eat-dog, dog-eat-cat
world.  One explanation for how evolution could yield altruism,
popular up through the 1950s, went something like this: imagine a
species divided into several smaller sub-populations, like flocks,
tribes or wandering herds. Some of the herds might be full of selfish
individuals who act without regard to others in their vicinity, but
other sub-populations might be full of altruists, whose cooperative
behavior allows those sub-populations to produce more sub-populations,
also full of altruists. If something in the environment kills off
sub-populations, then the clusters of altruists will produce more
offspring clusters than those comprised of selfish individuals, and
this biased survival of stochastically varying replicators---in this
case, the replicating units being taken as clusters---will lead to a
species dominated by altruists.

In fact, this idea goes back to Darwin~\cite{darwin1872}, who wrote,
\begin{quotation}
\noindent Natural selection will modify the structure of the young in 
relation to the parent, and of the parent in relation to the young. In
social animals it will adapt the structure of each individual for the
benefit of the whole community; if the community profits by the
selected change.
\end{quotation}
And, in more detail~\cite{darwin1871}:
\begin{quotation}
\noindent When two tribes of primeval man, living in the same country, 
came into competition, if the one tribe included (other circumstances
being equal) a greater number of courageous, sympathetic, and faithful
members, who were always ready to warn each other of danger, to aid
and defend each other, this tribe would without doubt succeed best and
conquer the other. [\ldots\!] Selfish and contentious people will not
cohere, and without coherence nothing can be effected. A tribe
possessing the above qualities in a high degree would spread and be
victorious over other tribes; but in the course of time it would,
judging from all past history, be in its turn overcome by some other
and still more highly endowed tribe. Thus the social and moral
qualities would tend slowly to advance and be diffused throughout the
world.
\end{quotation}

However, as Darwin recognized, this process might be vulnerable to
subversion from within:
\begin{quotation}
\noindent It is extremely doubtful whether the offspring of the more 
sympathetic and benevolent parents, or of those which were the most
faithful to their comrades, would be reared in greater number than the
children of selfish and treacherous parents of the same tribe.
\end{quotation}
Understanding this problem more deeply requires a better knowledge of
the mechanisms of heredity than was available in Darwin's time.
Indeed, genetics did not become a serious part of evolutionary
theorizing until the early twentieth century~\cite{stoltzfus2014}.
Now we know that each human being has far more cells in their brain
than genes in their genome, so we must expect that genes code
for \emph{patterns of development.}  In turn, the structures formed by
those developmental patterns then display ranges of behavior, in ways
which depend on the environment.  And, as for our species, so for the
others.  A justifiably popular aphorism observes, ``Evolution is the
control of development by ecology''~\cite{myers2012}.  When we perform
mathematical modeling, we often elide these intermediate steps and
pretend that genetic information can specify behaviors directly.  How
much we lose by doing this is difficult to say in advance---but even
the bower birds we modeled back in Chapter~\ref{ch:multiscale} turn
out to have culture~\cite{madden2004}.

We can refer to the idea that natural selection acts upon groups of
organisms as \emph{group selection.}  The general notion of natural
selection effecting change at different scales within a population
is \emph{multilevel selection.}  This has been a remarkably
controversial idea~\cite{vanveelen2014, pruitt2014, goodnight2015c},
with a significant fraction of the controversy arising from people
arguing words instead of mathematics.  And, when mathematics was
employed, it turned out to be too much statistics and too little
dynamics~\cite{allen2013b, allen2013c}.

In order to sort this out, we have to be careful and delineate the
various different ideas which have been paraded under the ``multilevel
selection'' banner.  All too often, the terminology of the subject has
brought the appearance of precision, more than the actuality.

A useful parallel can be drawn between the different types of
multilevel selection and the hierarchy of approximations used in
spatial ecology.  This relationship clarifies which modeling methods
are equivalent, and it points the way to future extensions of
multilevel selection theory.  The key question is one of
\emph{description}: how much and what kind of information does a
modeling method use to represent the state of being of an ecosystem?
The different types of multilevel selection (MLS) answer this in
different ways and thus relate to different kinds of ecological model.
We shall see that ``MLS-A'' stands on a level with mean field theory,
and ``MLS-B'' is akin to pair approximation.

Emphasizing the idea that context matters takes us beyond the mean
field approximation.  This extension becomes necessary if our model
includes explicit group-level processes like budding and fusion, or if
the geographical extent of our ecosystem fosters spatial
heterogeneity. The much-discussed equivalence of multilevel selection
and inclusive fitness (IF) is easy to establish for MLS-A and can be
proven mathematically in the mean field, but real differences arise
when we move from MLS-A to MLS-B and beyond.

The mathematical study of biological evolution, as it stands today,
includes a wide variety of particular models but lacks an overall
theory to organize those models and clarify their interrelationships
\cite{allen2012}.  This provides an opportunity for interdisciplinary
collaboration.  Mathematicians and physicists can bring not just new
techniques but also new \emph{attitudes,} potentially creating a novel
way of thinking about generalizations, specializations, equivalences
and other ways mathematical models are interrelated.  The challenge
for the physicist is to know enough biology that one does not waste
time on an obsolete or irrelevant problem, while also avoiding
absorption in the internal politics and factionalism of long-running
biologist debates.  Interdisciplinary research must steer a course
between the Scylla of dilettantism and the Charybdis of tedium.

Simon \emph{et al.}\ \cite{sfd2012} characterize two types of MLS
models, which we shall designate \emph{MLS-A} and \emph{MLS-B}.  Their
description of MLS-A states,
\begin{quotation}
\noindent Key features typically include that all groups form at the
same time from a randomly mixed (or mated) global population and
contain two types (e.g. cooperators and defectors), the number of
groups is constant or infinite, groups start out all the same size,
groups vary only in the reproduction rate of the individuals, and all
groups cease to exist at the same time.  Groups then contribute their
individuals to the random global mixing (or mating) phase from which
new groups are again formed.
\end{quotation}
In MLS-A, the ``groups'' are social circles of organisms which form at
random, typically once each generation.  This random formation and
reformation implies that the only information used to predict the
environment experienced by any organism are \emph{aggregate} measures
taken over the entire population.  Any single organism interacts with
its local environment, \emph{i.e.}, the randomly-formed group of which
it is a part, but that local environment is modeled using a global
average.  This coarse-graining of the ecological picture is
a \emph{mean-field approximation} \cite{baryam1999}. (We recall from
Chapter~\ref{ch:redgreen} that this term, used in spatial
ecology \cite{tgoei2000, aguiar2004} and evolutionary dynamics
\cite{gross2009}, derives originally from statistical physics
\cite{kardar2007}.)  Additional complications, such as nonlinear
fitness functions, can make the mathematical analysis more cumbersome,
but as long as the model uses only global averages to stand in for the
ecosystem's configuration, it remains within the mean field.

By contrast, MLS-B models ``contain more explicit group-level events''
such as group fissioning, group merging, mass dispersal and so forth
\cite{sfd2012}.  An MLS-B model does not treat all configurations with
the same total numbers of cooperators and defectors as equivalent.
The allocation of individuals into groups matters, and most
importantly, the composition of each group is not set by global
averages \cite{sfd2012, nielsen2012}.  An organism's local environment
is no longer assumed to be representative of the aggregate properties
of the whole population, and vice versa.

This advance is, in concept if not in algebraic detail, essentially
the same as the move in spatial ecology from mean-field theory to
higher-order \emph{moment closures} such as \emph{pair approximation.}
We can illustrate this parallel by example: let us consider how one
would treat a population of cooperators and defectors, arranged in a
spatial or a network structure, by a pair approximation method.
Because the population is structured, rather than being well-mixed,
having a clump of cooperators in one small region is a significantly
different circumstance than having the same number of cooperators
spread uniformly throughout the whole ecosystem.  To attempt to
capture this distinction, we augment the bare-bones, coarse-grained
description of the ecosystem, moving beyond overall averages and
including some measure of how local environments can differ.  Consider
the probability that an organism chosen at random from the population
is a cooperator.  This is an overall, global average; we can denote it
$p_C$.  Its complement, the probability that a randomly-picked
organism is a defector, is $p_D = 1 - p_C$.  (For simplicity, we
assume that the region we are studying is entirely filled up, with no
empty spaces.)  If all the cooperators are clustered together in a
patch, then the \emph{conditional} probability that the \emph{neighbor
  of a cooperator} is also a cooperator will be significantly
different from the overall density of cooperators in the population at
large.  If we pick an organism at random and find that it is a
cooperator, we should be willing to bet more heavily on the chance
that picking one of its neighbors at random will find another
cooperator.  In short, the conditional probability $q_{C|C}$ can
differ from $p_C$.  A pair approximation for this model assumes that
these pairwise conditional probabilities capture enough of the
ecosystem's heterogeneity to make all the predictions in which we are
interested.

The move from mean-field theory to a pair approximation, like the move
from MLS-A to MLS-B, brings in the idea that variation among local
environments matters.  Furthermore, both MLS-B and pair approximation
share an important simplifying assumption of their own: variation
among locales is important, but \emph{the set of all locales has no
  structure.}  Consider, for example, the MLS-B model of Simon
\emph{et al.}~\cite{sfd2012}.  This model represents the state of the
entire population at time $t$ by a list of numbers, namely, the number
of groups of type $\vec{x}$ at that time $t$, for all possible
group-types $\vec{x}$.  There is no sense in which some groups are
geographically closer to each other than others.  A group of type
$\vec{x}$ in one spot is just as good as a group of type $\vec{x}$ in
any other spot.  Likewise, when we develop a pair approximation, a
pair of a certain type (for example, cooperator--defector) occurring
in one location counts the same as a pair of that type occurring
anywhere else.

As we mentioned earlier, the parallel between MLS-B and pair
approximation is at a conceptual level, not necessarily an exact
mathematical one.  (We shall see plenty of exact mathematics soon
enough.)  The essential fact is that both extend our thinking beyond
the mean field, while both are themselves limited in the same way:
context matters, but the relationships among contexts are neglected.

One can extend the basic notion of pair approximation to higher
orders.  Instead of reducing everything to pairs, we can describe an
ecosystem using triples, for example.  Mathematically, if we write a
probability distribution over all possible states an ecosystem can be
in, then imposing a mean-field approximation means replacing that with
a product of many independent single-variable probability
distributions.  In other words, mean-field theory depends on the
assumption that the joint probability distribution can be
\emph{factored} in a drastic way without losing important information.
A pair approximation is the imposition of a slightly less drastic
factorization, namely one into probabilities for pairs of variables,
instead of single variables.  Factorizations into higher-order
probability distributions yield higher-order moment closures.  (The
\emph{exact} analogue of the Simon \emph{et al.}~MLS-B model falls
within this picture, but is not necessarily the pair approximation
which we used to illustrate the basic concept.)

We note that the terms ``type 1'' and ``type 2 multilevel selection''
are also employed in the literature.  However, their usage is not
quite consistent~\cite{goodnight2015}, and since the distinction which
Smith \emph{et al.}\ make is the most convenient for our purposes, we
will employ the MLS-A and MLS-B designations for the categories they
describe.  (We will, in this way, bypass talk of ``the level at which
fitness is assigned''~\cite{goodnight2015c}.)  In this case, we are
confronted with a tradeoff, and we choose attempting to reduce
ambiguity over maintaining continuity with older writings.

\section{Fisher's Fundamental Theorem}
\label{sec:fisher}
We have in this subject an interesting situation.  The mathematics
that has been employed to date has not been very elaborate.  It
involves a lot of algebra, some of it linear, much of it high-school.
The challenge lies in the \emph{stories told about the mathematics,}
stories which have historically been muddled, overzealous in their
claims, communicated at cross-purposes and, dare I say it,
politicized.  Later, we will try to sort some of that out.  But before
we reach that point, we will get the important equations in place.

Consider two populations, each composed of individuals of different
types in various amounts.  Let the proportion of type $i$ in the first
population be $p_i$, and denote the proportion of type $i$ in the
second population by $p_i'$.  We can let the index $i$ range from~1 to
however large it needs to be to enumerate all types present in both
populations.  The basic statements of normalization are
\begin{equation}
\sum_i p_i = 1,\ \sum_i p_i' = 1.
\end{equation}

Now, introduce a third set of quantities to express the relationship
between the two populations, thusly:
\begin{equation}
p_i \widetilde{w}_i = p_i'.
\end{equation}
This defines $\widetilde{w}_i$ in terms of the given information about the
population demographics.  We suppose that this is always
possible, \emph{i.e.,} that there is no $i$ for which $p_i = 0$ but
$p_i' \neq 0$.

If we take the mean of $\widetilde{w}_i$ across all types present in the
first population, weighted by their abundances, we find that
\begin{align}
\overline{\widetilde{w}} &= \sum_i p_i \widetilde{w}_i \nonumber \\
 &= \sum_i p_i' \nonumber \\
 &= 1.
\end{align}
If we instead take the mean weighted by the proportions in
the \emph{second} population, we obtain
\begin{align}
\overline{\widetilde{w}}' &= \sum_i p_i' \widetilde{w}_i \nonumber \\
 &= \sum_i p_i \widetilde{w}_i^2.
\end{align}
The difference between the two averages can be written
\begin{equation}
\overline{\widetilde{w}}' - \overline{\widetilde{w}}
 = \sum_i p_i \widetilde{w}_i^2 - 1
 = \sum_i p_i \widetilde{w}_i^2 - \left(\sum_j p_j \widetilde{w}_j\right)^2.
\label{eq:fisher}
\end{equation}
The right-hand side has the form of a variance, as we defined in
Chapter~\ref{ch:prob}: the second moment minus the square of the
first.  It is often called as such.  And, if we designate
$\widetilde{w}_i$ the \emph{relative fitness} of type~$i$, then we can
read Eq.~(\ref{eq:fisher}) as saying that the change in fitness from
one population to the other is the variance of fitness in the first
population.

When we say it in words, this statement sounds like the definition of
a dynamical system, or perhaps the beginnings of such a definition.
However, it really isn't: we took both populations as given, and we
declared that the $\{w_i\}$ have the right values to relate them.
Eq.~(\ref{eq:fisher}) holds by fiat.  It is neither a predictive
statement---given some $\{p_i\}$ and $\{p_i'\}$ obtained
experimentally, it tells us nothing we did not already know---nor an
update rule for a dynamical system that we can iterate and thereby
explore mathematically.

The word ``fitness'' makes Eq.~(\ref{eq:fisher}) sound like biology,
and in that context, it is known as \emph{Fisher's fundamental
theorem}~\cite{goodnight2015b}.  But as an arithmetic identity,
Eq.~(\ref{eq:fisher}) has nothing necessarily to do with genetics or
evolution.  To illustrate, let the first population be a set of books,
and let the second be a set of movie and TV adaptations.  Many books
never make the journey to the small or large screens.  Some books are
adapted once, and some are adapted multiple times.  For example,
Dashiell Hammett's novel \emph{The Maltese Falcon} (1929) was made
into a movie by that name in 1931, another film titled \emph{Satan Met
a Lady} in 1936 and, most famously, the version starring Humphrey
Bogart opposite Mary Astor in 1941.  We could say that the novel's
fitness in Hollywood is $w_1 = 3$, if its index on our list is $i =
1$.  Assigning ``fitnesses'' $w_i$ in this way for all values of~$i$,
we can then define
\begin{equation}
\widetilde{w}_i = \frac{w_i}{\sum_j p_j w_j},
\end{equation}
and the sum of the $\widetilde{w}_i$, weighted by~$p_i$, will be unity.
Fisher's fundamental theorem tells us that the change in fitness due
to moving from the written word to the moving picture is the fitness
variance among the original novels.

This should make clear the difference between an arithmetic identity
and a predictive model.  The former can be ``always true'' or
``universally applicable''; nevertheless, without additional
assumptions, it doesn't go anywhere.  In turn, those additional
assumptions can make the application of the identity invalid.  For
example, suppose one wishes to compare two configurations of an
ecosystem separated by a long interval of time.  The ecosystem
dynamics are defined in terms of short-term ecological or game-based
interactions (as we did in Chapters~\ref{ch:redgreen}
and \ref{ch:volunteer}).  The fitnesses which arise as functions of
game payoffs generally will not be the ones necessary to fill the role
of the $\{w_i\}$ relating the two population configurations.

\section{The Price Equation: Motivation and Shortcomings}

Biological evolution is the change over generations of the genetic
composition of populations due to natural factors, typically including
significant randomness. Describing this mathematically, and developing
quantitative tools to predict what might evolve under which
conditions, is a great challenge. One place to begin is by describing,
in a nice way, a population's change in genetic character from one
generation to the next. By ``a nice way'', we mean that we'd like to
be able to attribute changes to the appropriate influences. What
changes are due to random mutations creating new variations, for
example, and what changes are due to natural selection winnowing out
varieties which cannot survive in their environment?

We can make a crude measure of a population's genetic composition by
counting up how many organisms in the population have a certain gene
of interest. We can express this amount as a percentage of the total
population, saying, for example, ``The frequency of gene $A$ in this
population is 0.22.''  This, of course, is a mean-field statement.  We
know that such statements can be insufficient for making viable
predictions about \emph{dynamics,} but in this section, we will assume
a more modest aim, and try only to manipulate the \emph{description,}
as we did in deriving Fisher's theorem.

In this section, we use the notation of van
Veelen~\cite{vanveelen2005}.

We consider two populations, $S_1$ and $S_2$. All the offspring of
organisms in $S_1$ belong to $S_2$, and all the parents of organisms
in $S_2$ are in $S_1$. We write $N$ for the size of population
$S_1$. For an individual $i \in S_1$, the frequency of gene $A$ is
\begin{equation}
q_i = \frac{g_i}{l_z},
\end{equation}
where $l_z$ is the zygotic ploidy, \emph{i.e.,} the number of copies
of each chromosome carried by a fertilized egg. The frequency of gene
$A$ in population $S_1$ is
\begin{equation}
Q_1 = \frac{\sum_{i\in S_1} g_i}{l_z N} = \frac{\sum_i q_i}{N}.
\end{equation}
We want to relate $Q_2$ and $Q_1$. One simple way to do so is to take
their difference:
\begin{equation}
\Delta Q = Q_2 - Q_1.
\end{equation}
We can write $Q_2$ as
\begin{equation}
Q_2 = \frac{\sum_i g_i'}{l_g\sum_i z_i},
\end{equation}
where $l_g$ is the gametic ploidy, $z_i$ is the number of successful gametes from individual $i$, and $g_i'$ is the number of $A$-type genes in the set of all successful gametes from individual $i$. The proportion of $A$-type genes in that set is
\begin{equation}
q_i' = \frac{g_i'}{z_i l_g}.
\end{equation}
From this,
\begin{equation}
Q_2 = \frac{\sum_i z_i l_g q_i'}{l_g \sum_i z_i} = \frac{\sum_i z_i q_i'}{\sum_i z_i}.
\end{equation}
Therefore,
\begin{equation}
\Delta Q = \frac{\sum_i z_i q_i'}{\sum_i z_i} - \frac{1}{N}\sum_i q_i.
\end{equation}
We'd like our expression for the change in $Q$ to be written in terms of the changes in the individual $q_i$, so we subtract and add a sum over $q_i$:
\begin{equation}
\Delta Q = \frac{\sum_i z_i (q_i' - q_i)}{\sum_i z_i} + \frac{\sum_i z_i q_i}{\sum_i z_i} - \frac{1}{N}\sum_i q_i.
\end{equation}
Next, we gather the last two terms over a common denominator:
\begin{equation}
\Delta Q = \frac{\sum_i z_i (q_i' - q_i)}{\sum_i z_i} + \frac{\sum_i z_i q_i - \frac{1}{N}\sum_i q_i\sum_j z_j}{\sum_i z_i}.
\end{equation}
Now, we factor an $N$ out of the latter term.
\begin{equation}
\Delta Q = \frac{\sum_i z_i (q_i' - q_i)}{\sum_i z_i} + \frac{N}{\sum_i z_i} \left[\frac{1}{N}\sum_i z_i q_i - \frac{1}{N^2}\sum_i q_i\sum_j z_j\right].
\end{equation}
We rearrange this just a bit to yield the \textbf{Price equation}:
\begin{equation}
\boxed{\Delta Q = \frac{N}{\sum_i z_i} \left[\frac{1}{N}\sum_i z_i q_i - \left(\frac{1}{N}\sum_i q_i\right)\left(\frac{1}{N}\sum_j z_j\right)\right]
+ \frac{\sum_i z_i (q_i' - q_i)}{\sum_i z_i}.}
\label{eq:price-boxed}
\end{equation}
This is just an algebraic identity: we took the compositions of the
two populations as given, and we wrote a fancy expression for the
change of gene frequency between them. \emph{We have not said anything
about dynamics from which this change could be derived, nor have we
made any claims about what changes are more probable than others.}
Eq.~(\ref{eq:price-boxed}) is a rearrangement of the given
information, not an update rule for a dynamical system.  It is not
even a statement about probabilities, although the expression in
brackets formally resembles the covariance between two random
variables.  Nothing in the derivation of Eq.~(\ref{eq:price-boxed})
was a random variable; despite this, the Price equation is typically
written in ``covariance'' notation.  This poor tradition of notation
has contributed more than a little confusion to the subject.

Van Veelen \emph{et al.}~\cite{vanveelen2012} make the point in the
following way:
\begin{quotation}
\noindent[W]hat is most important is that we realize that the numerical input of the Price equation is a list of numbers. It is a list that concerns two generations, and which tracks who is whose offspring. But whatever it reflects, it is crucial to realize that the point of departure is nothing but a list of numbers. This list of numbers is used twice. First we use it to compute the frequencies of the gene under consideration in generations 1 and 2, respectively, and subtract the latter from the former. This amounts to the change in gene frequency. Then we use the same list to compute a few other, slightly more complex quantities. The essence of the Price equation is that these quantities also add up to the change in gene frequency. One way of computing the change in frequency therefore can be rewritten as the other and vice versa. What they are, therefore, is nothing but two equivalent ways to compute the change in gene frequency, given a list of numbers concerning genes in two subsequent generations \ldots\ Whether this particular second generation is likely to follow the first or not, the two ways of computing the change in frequency return the same number. 
\end{quotation}
To make a physics analogy, what we have done is like starting with Newton's second law, $\vec{F} = m\vec{a}$, and writing it as
\begin{equation}
m\vec{a} = m\vec{a}.
\end{equation}
We could then rewrite the $\vec{a}$ vectors in some elaborate way. For example, we could write one side of the equation in Cartesian coordinates and the other in spherical coordinates, giving some complicated formulas involving trigonometric functions all over the place. These formulas would be \emph{true,} in the sense that Euclidean geometry is \emph{true,} but they would contain no \emph{physics.} In some circumstances, they might be useful, but we could not wring value out of them without some extra assumptions about the dynamics at work.

We now make a series of assumptions geared towards turning the Price
equation (\ref{eq:price-boxed}) into something more like an update
rule for a dynamical system.

First, we specialize our considerations to a scenario in which
individuals interact by donating effort or assistance to one another.
Donors of effort increase the number of successful gametes produced by
the recipient, at the expense of their own. We parameterize this in
the following way: denote by $c$ a donor's decrease in successful
gametes of its own, and denote by $b$ the increase in successful
gametes of the recipient. We idealize interactions as pairwise events,
and so we keep track of them using matrices. The first index, $i$,
denotes an individual in population $S_1$. The second index, $\alpha$,
ranges over the occasions on which interactions can take place.
\begin{equation}
\Delta Q = \frac{N}{\sum_i z_i} \left[\frac{1}{N}\sum_i z_i q_i - \left(\frac{1}{N}\sum_i q_i\right)\left(\frac{1}{N}\sum_j z_j\right)\right]
+ \frac{\sum_i z_i (q_i' - q_i)}{\sum_i z_i}.
\end{equation}
We can be slightly more general and allow each individual to have their own ploidy, $l_i$. So, instead of using the population size $N$, we use $\sum_i l_i$. Following the literature, we calculate the number of successful gametes per haploid set, $w_i = z_i / l_i$.
\begin{equation}
\Delta Q = \frac{\sum_i l_i}{\sum_i l_i w_i} \left[\frac{\sum_i l_i w_i q_i}{\sum_i l_i} - \left(\frac{\sum_i l_i w_i}{\sum_i l_i}\right)\left(\frac{\sum_i l_i q_i}{\sum l_i}\right)\right]
+ \frac{\sum_i l_i w_i (q_i' - q_i)}{\sum_i l_i w_i}.
\end{equation}
We now make two additional assumptions:
\begin{enumerate}%
\item The second term in this form of the Price identity is negligible.
\item The fitnesses $z_i$ can be written
\begin{equation}
z_i = l_i w_i = f_i + b \sum_\alpha S_{i\alpha} - c\sum_\alpha Q_{i\alpha}.
\end{equation}
Here, $\sum_\alpha S_{i\alpha}$ is the total number of times individual $i$ received a benefit, and $\sum_\alpha Q_{i\alpha}$ is the number of times individual $i$ incurred a cost.
\end{enumerate}
We introduce the abbreviation
\begin{equation}
\overline{q} = \frac{\sum_i l_i q_i}{\sum_i l_i}.
\end{equation}
Dropping the last term of $\Delta Q$ and substituting in our chosen form for $l_i w_i$, we arrive after some algebra at the following:
\begin{equation}
\Delta Q = \left(\frac{\sum_{i,\alpha} Q_{i\alpha}(q_i
- \overline{q})}{\sum_i l_i
w_i}\right) \left[\left(\frac{\sum_{i,\alpha} S_{i\alpha}(q_i
- \overline{q})}{\sum_{i,\alpha} Q_{i\alpha}(q_i
- \overline{q})}\right) b - c \right].
\label{eq:price-hamiltonish1}
\end{equation}
The quantity in square brackets has the form of Hamilton's condition, if we identify the quotient multiplying $b$ as a measure of assortment:
\begin{equation}
r = \frac{\sum_{i,\alpha} S_{i\alpha}(q_i
- \overline{q})}{\sum_{i,\alpha} Q_{i\alpha}(q_i - \overline{q})}.
\label{eq:price-hamiltonish2}
\end{equation}

\section{Interconverting MLS-A and Inclusive Fitness}
\label{sec:mls-and-inclusive}

Hamilton~\cite{hamilton1964} defines inclusive fitness in the
following manner.
\begin{quotation}
\noindent Inclusive fitness may be imagined as the personal fitness 
which an individual actually expresses in its production of adult
offspring as it becomes after it has been first stripped and then
augmented in a certain way. It is stripped of all components which can
be considered as due to the individual's social environment, leaving
the fitness which he would express if not exposed to any of the harms
or benefits of that environment. This quantity is then augmented by
certain fractions of the quantities of harm and benefit which the
individual himself causes to the fitnesses of his neighbours. The
fractions in question are simply the coefficients of relationship
appropriate to the neighbours whom he affects: unity for clonal
individuals, one-half for sibs, one-quarter for half-sibs, one-eighth
for cousins, [\ldots\!] and finally zero for all neighbours whose
relationship can be considered negligibly small.
\end{quotation}
Allen~\cite{allen2013c} comments,
\begin{quotation}
At this point you may be asking, ``Wait, does it really make sense to
divide offspring into those  produced on one's own versus those
produced by help from others?''  This is exactly the problem!  Aside
from the obvious point that no one reproduces without help in sexual
species, nature is full of synergistic and nonlinear interactions, so
that making clean divisions like this is impossible in most
situations.  Thus the idea of inclusive fitness theory only works in
simplified toy models of reality.
\end{quotation}

Inclusive fitness has variously been claimed to supersede multilevel
selection, to be mathematically equivalent to MLS or to be a subset of
it.  In order to clarify the relationship between MLS and inclusive
fitness (IF) models, it helps to have a specific example in hand.
Fortunately, Bijma and Wade~\cite{bijma2008} have provided an
explication which, made slightly more general, is quite helpful.

The remainder of this section will be specific to MLS-A.  That is,
while we will sometimes speak of ``groups'' within the population,
these groups will be formed at random from the pool of available
individuals, rather than being entities which have their own explicit
dynamics included in the update rules.  Furthermore, we will consider
only short-term changes, comparing one generation to the next, instead
of entire evolutionary trajectories.  This is just the kind of
comparison where we can apply the Price equation which we derived in
the previous section.

We shall consider a quantitative genetic model in which the trait
value of an individual organism affects its fitness as well as the
fitness of those with which it interacts.  The trait value of
individual $i$, which we denote $P_i$, depends on a genetic component
and a nonheritable, environmental component:
\begin{equation}
P_i = G_i + E_i.
\end{equation}
The personal fitness of individual $i$ depends on that individual's
trait value, $P_i$, as well as the trait values of those in its social
group.  For simplicity, we imagine that all groups have size 2; that
is, each individual $i$ interacts with a partner $j$.  In addition, we
assume that the effect of trait value on fitness is linear, and that
the combination of the self-value and partner-value effects is also
linearly additive.  Including residual effects (due, say, to exogenous
environmental variation), we can write
\begin{equation}
W_i = \hbox{const.} + \beta_{i,i} P_i + \beta_{i,j} P_j + e_i.
\label{eq:W-inclusive}
\end{equation}
It could be that, for the evolutionary ecosystem one is trying to
model, the relationship between trait value and fitness is not linear.
One could still run a linear regression on experimental data obtained
from that system, but the output from the statistical software package
would not be meaningful \cite{damore2011}.  We shall neglect this
complication for the moment and return to it in a later section.

The next step in model-building would be to define a rule which gives
the following generation in terms of the current one and the set of
fitness values $\{W_i\}$.  Typically, one uses the Price equation to
do so, saying that ``the change in gene frequency is given by the
covariance of genes and fitness.''  This amounts to pretending that a
purely descriptive equation has predictive value.  In other words,
taking this step requires importing additional assumptions.  (The
language of the invocation, with its talk of ``covariance,'' also
propagates the confusion about the Price equation and its status,
which we touched upon in the previous section.)

Furthermore, the update rule one invents by this scheme does not even
really define a dynamical system: it requires more information about
the current generation than it can yield about the next generation, so
it cannot be iterated.  We can say that this update rule is not
``dynamically sufficient.''

This ``inclusive'' or ``neighbor-modulated'' fitness calculation,
Eq.~(\ref{eq:W-inclusive}), represents the state of the population by
the trait values $\{P_i\}$.  We can equally well use a group selection
calculation---that is, an MLS-A model---wherein we say that the
personal fitness of individual $i$ depends on the mean trait value of
its social circle,
\begin{equation}
\overline{P}_g = \frac{1}{2}(P_i + P_j),
\end{equation}
and on how far $P_i$ deviates from that average,
\begin{equation}
\Delta P_i = P_i - \overline{P}_g
 = \frac{1}{2}(P_i - P_j).
\end{equation}
If we once again assume linear relationships, we can write the
personal fitness $W_i$ in terms of a between-group component and a
within-group component:
\begin{equation}
 W_i = \hbox{const.} + e_i + \beta'_{i,g} \overline{P}_g
       + \beta'_{i,\Delta} \Delta P_i.
\label{eq:W-group}
\end{equation}
If we substitute our definitions of $\overline{P}_g$ and $\Delta P_i$ into
this equation, we find
\begin{align}
W_i &= \hbox{const.} + e_i
       + \beta'_{i,g} \frac{P_i + P_j}{2}
       + \beta'_{i,\Delta} \frac{P_i - P_j}{2} \nonumber\\
    &= \hbox{const.} + e_i
       + \frac{1}{2}(\beta'_{i,g} + \beta'_{i,\Delta}) P_i
       + \frac{1}{2}(\beta'_{i,g} - \beta'_{i,\Delta}) P_j.
\label{eq:W-transformed}
\end{align}
Comparing this expression with the ``inclusive'' or
``neighbor-modulated'' fitness expression in
Eq.~(\ref{eq:W-inclusive}), we see that
\begin{equation}
\beta_{i,i} = \frac{1}{2}(\beta'_{i,g} + \beta'_{i,\Delta}),
\label{eq:beta-transform-1}
\end{equation}
and
\begin{equation}
\beta_{i,j} = \frac{1}{2}(\beta'_{i,g} - \beta'_{i,\Delta}).
\label{eq:beta-transform-2}
\end{equation}

On the one hand, we have the formula based on inclusive fitness,
Eq.~(\ref{eq:W-inclusive}), and on the other, we have the
levels-of-selection formula, Eq.~(\ref{eq:W-group}).  The relation
between the two is a \emph{linear transformation of coordinates}: we
change from one way of tallying trait values to another.  When we
transform our coordinate system, the parameters of our model get mixed
up with one another, as seen in Eqs.~(\ref{eq:beta-transform-1}) and
(\ref{eq:beta-transform-2}).

A mechanical analogy is illustrative: if we wish to study the
collision of two billiard balls on a table, we can look at the
collision in countless different ways.  We could describe what happens
from the perspective of an observer standing at rest with respect to
the pool table.  Or, alternatively, we could view the situation from
the perspective of an observer who is at rest with respect to the
center of mass of the two billiard balls.  We can even switch
perspectives in the middle of a calculation: starting with information
given in terms of the table rest frame, we transform into the
center-of-mass rest frame to see what Newton's laws imply, and then we
transform back into the table rest frame to predict what the observer
standing beside the table will see.  The biological situation is
analogous: we have transformed the trait values from a laboratory
reference frame into a ``center of group'' frame.

We can express this transformation more generally using matrix
notation.  If we write the trait values as a vector, then the
group-based trait values are given by the vector of individual trait
values multiplied by a matrix:
\begin{equation}
\left(\begin{array}{c} \overline{P}_g \\ \Delta P_i \end{array}\right)
 = 
\frac{1}{2}
\left(\begin{array}{cc} 1 & 1 \\ 1 & -1 \end{array}\right)
\left(\begin{array}{c}  P_i \\ P_j \end{array}\right).
\end{equation}
Writing ${\bf \Lambda}$ for the transformation matrix,
\begin{equation}
{\bf \Lambda} = 
\frac{1}{2}
\left(\begin{array}{cc} 1 & 1 \\ 1 & -1 \end{array}\right),
\end{equation}
we are saying that the MLS-A model's trait values, ${\bf P}'$, are
related to the inclusive-fitness trait values, ${\bf P}$, by the
equation
\begin{equation}
{\bf P}' = {\bf \Lambda} {\bf P}.
\end{equation}
We know that the results of the two calculations must agree.  In
matrix form, this requirement means
\begin{equation}
{\bf \beta}' {\bf P}'
 = {\bf \beta} {\bf P},
\end{equation}
where the matrix ${\bf \beta}$ is
\begin{equation}
{\bf \beta}
= \left(\begin{array}{cc} \beta_{i,i} & \beta_{i,j} \\
                          \beta_{j,i} & \beta_{j,j}
        \end{array}\right).
\end{equation}
By substituting in the transformation rule for ${\bf P}$, we see that
\begin{equation}
{\bf \beta}' ({\bf \Lambda}{\bf P})
 = {\bf \beta} {\bf P},
\end{equation}
which in turn means that
\begin{equation}
{\bf \beta} = {\bf \beta}' {\bf \Lambda}.
\end{equation}
This is the matrix version of Eqs.~(\ref{eq:beta-transform-1}) and
(\ref{eq:beta-transform-2}).

The matrix-algebra statement of the relationship between IF and MLS-A,
although somewhat more abstract than the original equations, brings
forth the essential point: when we change the way we describe trait
values, we must make a corresponding change in the parameters which
control the evolutionary dynamics.  If we go from ${\bf P}$ to ${\bf
  P}'$ by the transformation ${\bf \Lambda}$, then we change our
coefficients ${\bf \beta}$ by the inverse of ${\bf \Lambda}$:
\begin{equation}
{\bf \beta}' = {\bf \beta} {\bf \Lambda}^{-1}.
\end{equation}

Because our IF model and our MLS-A model are different perspectives on
the same mathematics, problems with one will be problems with the
other, only seen from a different angle.  For example, to predict what
strategy will be evolutionarily favored, the IF model relies on
Hamilton's concept of ``relatedness,'' and in practical scenarios,
``relatedness'' is problematic.  But if ``relatedness'' is ill-defined
or inapplicable, mixing it with another parameter by means of a
coordinate transformation will do no good, either.

\section{An Example of MLS-B}

In order to understand how MLS-B goes beyond MLS-A, we now turn to an
example of an MLS-B dynamical system, constructed by van
Veelen \emph{et al.}~\cite{vanveelen2014}.  Per the definition of
MLS-B, the population is organized into groups, and the dynamical
update rules make explicit reference to the group level of structure.
In this particular model, all groups are taken to have the same size,
and this size is constant over time.  This is accomplished by
balancing reproduction events with death events.  If an individual
within a group reproduces, an organism is picked at random from that
group and killed.  Organisms come in two types, which we
designate \emph{cooperators} and \emph{defectors.}  The difference
between the types manifests in a difference in reproductive
fecundities.  For simplicity, we say that cooperators reproduce at
rate 1, while defectors reproduce at an augmented rate, $1 + s$.

In order to make this an MLS-B model, we need an explicit group-level
dynamic.  So, we say that in addition to the within-group reproduction
of individuals, \emph{entire groups also reproduce.}  The offspring
group has the same proportion of cooperators and defectors as its
parent group.  The rate of group reproduction depends on that
proportion: if the size of the group is $k$ and it contains $n_c$
cooperators, then that group reproduces at a rate $1 +
u\left(\frac{n_c}{k}\right)^\alpha$.  The parameter $\alpha$ controls
the degree of nonlinearity in the dynamics.

As van Veelen~\emph{et al.}\ observe,
\begin{quotation}
\noindent Being a cooperator therefore comes at a cost---it reduces
the reproduction rate of the individual by $s$---but it has a benefit
for all group members, including itself, through an increase in the
rate at which the group as a whole reproduces.
\end{quotation}

Taking the limit of large population size and large $k$, the dynamics
of this system can be cast into a deterministic partial differential
equation.  Let $x$ be the fraction of cooperators in a group, and
denote by~$\mu(x;t)$ the relative frequency of groups having
cooperator fraction $x$ at time $t$.  Then,
\begin{equation}
\frac{d\mu(x;t)}{dt} = s \frac{d}{dx}[x(1-x)\mu(x;t)]
 + u\mu(x;t) \left[x^\alpha - \int_0^1 dy\, y^\alpha \mu(y;t)\right].
\end{equation}
The right-hand side of this PDE has two terms, reflecting two effects
at work.  The first term, with its logistic form, describes the
decrease of~$x$ within a group, due to the defectors' local
advantage.  The second term, proportional to the parameter $u$,
indicates how groups with low $x$ decrease in frequency, and those
with high $x$ increase in frequency, due to natural selection.  The
two contributions of opposite sign within the brackets ensure that the
overall population size remains constant.

Treating $\mu(x;t)$ as a time-dependent probability density function,
we can summarize the population by the moments of~$x$:
\begin{equation}
\moment{x^n(t)} = \int_0^1 dx\, x^n \mu(x;t).
\end{equation}
The overall frequency of cooperators changes at a rate that can be
computed, after some algebra:
\begin{equation}
\frac{d\moment{x}}{dt} = s(\moment{x^2} - \moment{x})
 + u(\moment{x^{\alpha+1}} - \moment{x}\moment{x^\alpha}).
\label{eq:basic-MLS-B}
\end{equation}
The simplest case is where $\alpha = 1$, and the cooperators' effect
on group reproduction rate is linear.  Then,
\begin{equation}
\frac{d\moment{x}}{dt} = s(\moment{x^2} - \moment{x})
 + u(\moment{x^2} - \moment{x}^2).
\end{equation}
The $s$ term, which reflects within-group selection, will be negative
or zero.  The $u$ term is zero or positive, and it indicates the
effect of \emph{between-group} selection.  Whether the frequency of
cooperation increases or decreases depends on which of these two
countervailing influences wins out.

Rewriting the previous equation, we find that
\begin{align}
\frac{d\moment{x}}{dt}
 &= u(\moment{x^2} - \moment{x}^2) + s(\moment{x^2} - \moment{x}) \\
 &= (\moment{x^2} - \moment{x}^2)(s+u)
    - (\moment{x} - \moment{x}^2)s \\
 &= \moment{x}(1 - \moment{x})
    \left[\left(\frac{\moment{x^2} - \moment{x}^2}
                     {\moment{x} - \moment{x}^2}
          \right)(s + u)
          - s
    \right].
\end{align}
This is reminiscent of when we turned the Price equation into a
 condition resembling Hamilton's rule, in
 Eqs.~(\ref{eq:price-hamiltonish1}) and (\ref{eq:price-hamiltonish2}).

Note that the parameter $s$ is the ``cost of cooperation,'' in that it
indicates the demotion of the reproductive rate due to being a
cooperator rather than a defector.  Likewise, the aggregate benefits
that other group members accrue thanks to the presence of a cooperator
are $b = s + u$.  The $s$ is from the reduction in the death rate, due
to the balancing of birth and death events in each group.  So, if we
write the assortment coefficient $r$ as
\begin{equation}
r = \frac{\moment{x^2} - \moment{x}^2}
         {\moment{x} - \moment{x}^2},
\end{equation}
then we can write the rate of change of $\moment{x}$ as
\begin{equation}
\frac{d\moment{x}}{dt} = \moment{x}(1 - \moment{x}) (rb - c).
\end{equation}
The sign of $d\moment{x}/dt$ depends on a factor that has the form of
Hamilton's rule.  Remember, though, that in order to achieve this
form, we had to define $r$ in terms of the current population
demographics, by way of the moments $\moment{x}$ and $\moment{x^2}$.
This means that, as the population composition changes, $r$ will vary.
We could try to apply Hamilton's rule to solve the problem, but in
order to apply Hamilton's rule, we have to solve the problem anyway!

When $\alpha = 1$, we see that we can express the change in cooperator
frequency by an equation with a natural MLS interpretation, or by a
formula with an IF flavor.  The two are freely
interconvertible.  \emph{However,} if we introduce nonlinearity by
choosing $\alpha > 1$, we still have the basic MLS expression
(\ref{eq:basic-MLS-B}), but we can no longer reshape it into an IF
condition.  Should we try to define an assortment coefficient $r$ as
we did for the $\alpha = 1$ case, we would find that it necessarily
depends on all the parameters of the dynamics.  That is, our $r$ would
include $s$, $u$ and $\alpha$, in addition to the moments of~$x$.

There's no reason why MLS-B should be the end of the story.  Earlier,
we noted that MLS-B is conceptually analogous to a pair approximation.
We have already seen multiple reasons why pair approximations, and
similar but more elaborate extensions beyond mean-field theory, fail
to capture important effects.  This was a recurring motif in
Chapter~\ref{ch:redgreen}, where competition was defined in terms of
predation on a limited resource.  We also saw it in
Chapter~\ref{ch:volunteer}, when the effects of network topology
manifested themselves.

\section{A Literature of Confusion}

In \S \ref{sec:pairapprox-altruism}, we quoted Damore and Gore's point
that errors propagate through the works in this area, thanks to
assumptions going unspoken and definitions being swept into the
Supplemental Information.  The evolutionary theory of social behaviors
is a field where there is no substitute for reading the primary
sources.

For example, Dawkins recently claimed that H\"olldobler has ``no truck
with group selection''~\cite{johnston2014}.  A 2005 piece by Wilson
and H\"olldobler proposes, in the first sentence of its abstract, that
``group selection is the strong binding force in eusocial
evolution''~\cite{wilson2005}. Later, H\"olldobler (with Reeve) voiced
support for the ``trait-group selection and individual
selection/inclusive fitness models are interconvertible''
attitude~\cite{reeve2007}. H\"olldobler's book with Wilson, \emph{The
Superorganism: The Beauty, Elegance, and Strangeness of Insect
Societies}~\cite{hoelldobler2009}, maintains this tone. Quoting from
page~35:
\begin{quotation}
It is important to keep in mind that mathematical gene-selectionist
(inclusive fitness) models can be translated into multilevel selection
models and vice versa. As Lee Dugatkin, Kern Reeve, and several others
have demonstrated, the underlying mathematics is exactly the same; it
merely takes the same cake and cuts it at different angles. Personal
and kin components are distinguished in inclusive fitness theory;
within-group and between-group components are distinguished in group
selection theory. One can travel back and forth between these theories
with the point of entry chosen according to the problem being
addressed.
\end{quotation}
This is itself a curtailed perspective, whose validity is restricted
to a narrow class of implementations of the ``multilevel selection''
idea.  Regardless, this is not at all equivalent with having ``no
truck with group selection''. The statement ``method $A$ is no better
or worse than method $B$'' is a far cry from ``method $A$ is worthless
and only method $B$ is genuinely scientific''.

We have also a 2010 solo-author piece by H\"olldobler, in a
perspective printed in \emph{Social Behaviour: Genes, Ecology and
Evolution}~\cite{hoelldobler2010}.  Quoting from page~127:
\begin{quotation}
\noindent I was, and continue to be, intrigued by the universal
observation that wherever social life in groups evolved on this
planet, we encounter (with only a few exceptions) a striking
correlation: the more tightly organized within-group cooperation and
cohesion, the stronger the between-group discrimination and
hostility. Ants, again, are excellent model systems for studying the
transition from primitive eusocial systems, characterized by
considerable within-group reproductive competition and conflict, and
poorly developed reciprocal communication and cooperation, and little
or no between-group competition, one one side, to the ultimate
superorganisms (such as the gigantic colonies of the \emph{Atta}
leafcutter ants) with little or no within-group conflict, pronounced
caste systems, elaborate division of labour, complex reciprocal
communication, and intense between-group competition, on the other
side (H\"olldobler \& Wilson 2008 [the book quoted above]).
\end{quotation}

And, a little while later, on p.\ 130:
\begin{quotation}
\noindent In such advanced eusocial organisations the colony
effectively becomes a main target of selection [\ldots\!] Selection
therefore optimises caste demography, patterns of division of labour
and communication systems at the colony level. For example, colonies
that employ the most effective recruitment system to retrieve food, or
that exhibit the most powerful colony defence against enemies and
predators, will be able to raise the largest number of reproductive
females and males each year and thus will have the greatest fitness
within the population of colonies.
\end{quotation}

H\"olldobler also says that these phenomena can be thought of as
``extended phenotypes,'' which is a Dawkinsian turn of phrase; this is
consistent with the ``MLS and IF are interchangeable'' theme.

One author who has been less often read and more often misread
third-hand is the zoologist V.\ C.\
Wynne-Edwards~\cite{wynneedwards1962, pollock1989}.  In the early
1960s, Wynne-Edwards compiled a thick volume documenting how different
species use signaling mechanisms to determine their population density
and, effectively, figure out how crowded the living situation is.
Introducing the compendium, he said that group selection would have to
be the explanation for these observations, since, in essence, that's
how cooperative behaviors are explained.  To quote an
essay~\cite{wynneedwards1963} that he wrote shortly after,
\begin{quotation}
In a recent book~\cite{wynneedwards1962} I advanced a general
proposition which may be summarized in the following way. (1)\
Animals, especially in the higher phyla, are variously adapted to
control their own population densities. (2)\ The mechanisms involved
work homeostatically, adjusting the population density in relation to
fluctuating levels of resources; where the limiting resource is food,
as it most frequently is, the homeostatic system prevents the
population from increasing to densities that would cause
over-exploitation and the depletion of future yields. (3)\ The
mechanisms depend in part on the substitution of conventional prizes,
namely, the possession of territories, homes, living space and similar
real property, or of social status as the proximate objects of
competition among the members of the group concerned, in place of the
actual food itself. (4)\ Any group of individuals engaged together in
such conventional competition automatically constitutes a society, all
social behaviour having sprung originally from this source.
\end{quotation}

Wynne-Edwards' point (2) is reminiscent of the descendant-shading
effect we saw in Chapter~\ref{ch:redgreen}.  Indeed, one can add
social signalling among consumers to the host--consumer model, and it
turns out that curtailing consumption in overcrowded situations is a
robustly evolvable trait~\cite{werfel2004}.  However, in the spatial
host--consumer model, Wynne-Edwards' point (3) is not necessary: the
consumers do not substitute any ``social status'' or ``conventional
prize'' in the place of the actual resource, but reproductive
restraint evolves anyway.

Wynne-Edwards continues,
\begin{quotation}
In developing the theme it soon became apparent that the greatest
benefits of sociality arise from its capacity to override the
advantage of the individual members of in the interests of the
survival fo the group as a whole.  The kind of adaptations which make
this possible, as explained more fully here, belong to and
characterize social groups as entities, rather than their members
individually.  This in turn seems to entail that natural selection has
occurred between social groups as evolutionary units in their own
right, favouring the more efficient variants among social systems
wherever they have appeared, and furthering their progressive
development and adaptation.

The general concept of intergroup selection is not new.  It has been
widely accepted in the field of evolutionary genetics, largely as a
result of the classical analysis of Sewall Wright~\cite{wright1929,
wright1931, wright1945}.  He has expressed the view that ``selection
between the genetic systems of local populations of a species \ldots\
has been perhaps the greatest creative factor of all in making
possible selection of genetic systems as wholes in place of mere
selection according to the net effect of alleles''~\cite{wright1945}.
Intergroup selection has been invoked also to explain the special case
of colonial evolution in the social insects~\cite{haldane1932,
sturtevant1938, williams1957}.
\end{quotation}

Others criticized Wynne-Edwards using ``group selection'' models which
we can classify as MLS-A.  These critiques, however, impose upon
Wynne-Edwards a conception of population structure that was not his
own~\cite{pollock1989}.  Indeed, the image of population structure
that Wynne-Edwards appears to have in mind is closer to that realized
in a cellular automaton lattice model, than to that seen in the
critiques by Maynard Smith~\cite{maynardsmith1964, simon2014} and
others.

We conclude this section by revisiting Fisher's fundamental theorem,
which we developed as an arithmetic identity in~\S\ref{sec:fisher}.
Fisher himself~\cite{fisher1930} stated what he called ``the
fundamental theorem of Natural Selection'' in the following way:
\begin{quotation}
The rate of increase in fitness of any organism at any time is equal
to its genetic variance in fitness at that time.
\end{quotation}
Fisher arrives at this statement after a lengthy discussion which
invokes many more particular assumptions than we did
in~\S\ref{sec:fisher}.  Moreover, he takes his fundamental theorem to
be an approximation:
\begin{quotation}
Since the theorem is exact only for idealized populations, in which
fortuitous fluctuations in genetic composition have been excluded, it
is important to obtain an estimate of the magnitude of the effect of
these fluctuations, or in other words to obtain a standard error
appropriate to the calculated, or expected, rate of increase in
fitness.
\end{quotation}
Fisher then considers a model of a panmictic population and argues
using that model that fluctuations should decrease with the square
root of the population size.

In other words, although Fisher was quite favorably impressed with the
importance of his result, he himself did not take it as universally
true.  This is in sharp contrast to more recent authors who take the
Price equation as \emph{the} way to think about evolutionary change,
thereby imparting a glow of perceived universality to Fisher's theorem.

\section{Discussion}

Consider the following statement from a recent popular-science summary
of the perennial MLS/IF dispute \cite{emj2012}:
\begin{quotation}
\noindent In the final analysis, multilevel selection is little more
than a rebranding of Hamilton's inclusive fitness (albeit the
``enhanced'' 1975 version).
\end{quotation}
That such a claim can be the punchline of a popularization is
definitely an advance from the curmudgeonly attitude that only IF is
viable and anything which sounds like MLS is old-fashioned at best or
antiscientific at worst.  However, it is only the first step in a
rhetorical journey.  We have seen that this interchangeability is
straightforwardly true in mean-field theory, the domain of MLS-A.
Nevertheless, there are more things in heaven and earth than are
dreamt of in that approximation, and it is well past time to move on
to new adventures.

If one defines MLS so narrowly that it \emph{is} indeed just IF in a
new coordinate system, than MLS will inherit all the problems which
limit IF's usefulness in modern biology.  On the other hand, if one
defines MLS broadly, then one invites misunderstanding from those who,
knowingly or not, define it narrowly.  Likewise, if we try to reserve
the term ``group selection'' for those cases which include explicit
group-level processes (MLS-B but \emph{not} MLS-A), then we engender
the same confusion.

The esteemed science communicator Larry Gonick \cite{gonick1993}
identified the key problem:
\begin{quotation}
\noindent We always overestimate the degree to which we are
understood! That is, when I talk, I assume you understand me---unless
you tell me otherwise!  And there are a lot of reasons you may not
give me this essential feedback: you're too polite; you thought you
got it, even though you didn't; you're afraid of looking stupid. The
result is a worldwide overvaluation of the level of understanding!
\end{quotation}
In this chapter, we have used the designations MLS-A and MLS-B.  A
future development might conceivably push the terminology into the
alphanumeric, perhaps introducing ``MLS-114'' and the like.  Jargon of
this kind is intimidating and appears quite nontransparent.  It is the
sort of insider-speak which fills a room with fog.  But this fog comes
with a silver lining: we may not know what ``MLS-B'' means when we
hear it, but we \emph{know that we do not know.}  This is unlike what
happens when we hear a nontechnical-sounding term like ``group
selection''---the language may seem more friendly, but that sense of
welcome is dangerous.  Its seeming hospitality leads to
miscommunication and ceaseless confusion.  Technical argot befuddles
the outsider and, all too often, the student, but thoughtfully chosen
terminology can do wonders to make one's meaning clear, at least to
fellow professionals.  (The words which doctors throw around are
intimidating, particularly when we are patients ourselves, but they do
have good reasons to use the language they learned in med school:
drawing careful distinctions which everyday speech cannot saves
lives.)  Perhaps we must accept the cost of jargon and mathematics,
until such time as we have cleared enough of that confusion that we
can make the essential points in plainer speech without error.  If our
choice is between an algebra prerequisite and another generation
wasted on Team Inclusive Fitness versus Team Multilevel Selection, we
unhesitatingly affirm our preference for the former.

\chapter{Speculations for New Mathematics}

\label{ch:speculations}

According to John Baez~\cite{baez2011b},
\begin{quotation}
\noindent The raw beginner in mathematics wants to see the solutions
of an equation. The more advanced student is content to prove that the
solution exists. But the master is content to prove that the equation
exists.
\end{quotation}
This chapter is devoted to demonstrating that equations could,
potentially, be written.

\section{More on Multiscale Information}

Our first major topic, the information-theoretic formalism for
multiscale structure, presents us with several possibilities for
future extensions.  We saw that the complexity profile $C(k)$ depends
on the amounts of information associated with all the possible subsets
of a system.  If $Q(j)$ denotes the sum of
the joint information of all collections having size $j$,
\begin{equation}
Q(j) = \sum_{i_1,\ldots,i_j} H(a_{i_1},\ldots,a_{i_j}),
\end{equation}
then the complexity at scale $k$ is
\begin{equation}
C(k) = \sum_{j = N-k}^{N-1} (-1)^{j+k-N} 
\binom{j}{j+k-N}
       Q(j+1).
\end{equation}
We saw that we can simplify this expression in the special case where
all sets of the same size carry the same amount of information.  This
is a strong symmetry requirement, and one naturally wonders whether
other, less stringent symmetries can also lead to useful and
mathematically interesting expressions for~$C(k)$.

Moreover, the Marginal Utility of Information has a conceptual
connection with a method for detecting \emph{community structure}
within networks~\cite{ziv2005}, and this relationship could
potentially be developed further.  An intuitive definition of
a \emph{module} within a network could be phrased in the following
way: a set of nodes more strongly connected among themselves than they
are to anything else~\cite{fortunato2009}.  The extent to which a
network is ``modular'' would be, then, the extent to which it consists
of multiple more-or-less independent pieces.  We could try to
formalize this intuition with some quality index for a division of a
network into putative modules.  If a partition of the vertices breaks
only a few edges, then the quality of that partition is high.  We
could become more sophisticated and compare the number of severed
connections with some null hypothesis, and then accord higher quality
rankings to those partitions which cut fewer edges than we'd expect by
random chance.  For example, if we have an undirected graph with $E$
edges and we know the degrees $k_i$ and $k_j$ of two nodes, then our
best guess for the number of edges linking the vertices $i$ and $j$ is
$\frac{k_i k_j}{2E}$.  The quality index $Q$ would then be a
comparison of this quantity to $A_{ij}$, the actual number of edges;
we could then embark on a scheme to optimize $Q$.  Such optimization
problems turn out to be {\bf NP}-hard~\cite{fortunato2009}, and while
various heuristics appear to work well enough in networks one sees in
practice, we should take care before calling their results the
``best'' partition, as they may be only first among equals in
partition space~\cite{good2009}.

An alternative, complementary perspective on modularity takes a more
dynamical view.  Suppose that each vertex in a graph comes with an
oscillator attached (perhaps a pendulum or a firefly), and that
oscillators for linked vertices are coupled.  If the dynamics tend to
synchrony, we'd expect that oscillators within a module would
synchronize with each other first; the timescale for different modules
to begin oscillating in phase would be longer.  Or, we could unleash
drunkards upon the network: a random walk starting within a module
would be more likely to stay within that module than to escape and
begin perambulating elsewhere.  Turning this intuition around, we
could \emph{define} a module as a subset of nodes which a random
walker is more apt to stay within than to leave.  We could then take
this approach and devise a procedure for partitioning a graph based on
the conditional probability distribution
\begin{equation}
G^t_{ij} \equiv p(j|i,t),
\end{equation}
denoting the probability that a random walker starting at $i$ will
be standing at~$j$ after a time period $t$.  (The time period is
typically taken to be the inverse of the smallest non-zero Laplacian
eigenvalue, but there are other ways to find a good timescale that
might be more computationally efficient.)

Because of the {\bf NP}-hardness issue mentioned earlier, we should
not expect to be able to write down a dynamical rule which, when
implemented on an arbitrary graph, will yield a steady-state
configuration revealing the optimal partition.  Instead, we ought to
expect metastable configurations, long decay timescales and fuzzy
assignments of nodes to clusters.

Ziv, Middendorf and Wiggins (ZMW) use the diffusion of random walkers
to calculate a conditional probability distribution $p(y|x)$.  Both
$Y$ and $X$ range over the vertices of the network we're studying; $y$
gives the position which a random walk started at~$x$ reaches after a
specified time.  The description variable $Z$ is an index over
modules.  The cardinality $|Z|$ indicates the number of modules we are
considering in our network partitioning, and $p(z|x)$ is the
probability that vertex $x$ is mapped into module $z$.  With these
definitions, ZMW use the information bottleneck procedure to find
optimal network partitions, \emph{i.e.,} divisions into modules which
best preserve $p(x,y)$.

One result of the ZMW procedure is a curve of information gained
versus information provided, that is, of how well a description with a
certain information content reproduces the random-walk probability
distribution of the network.  This curve is, essentially, the utility
of a description as a function of description size.  Each partition
solution $p(z|x)$ has a Shannon information content $I(Z;X)$, which
for comparison purposes we can normalize by $H(X)$.  The fidelity of
this module assignment scheme is the ratio of how well it predicts
diffusion, $I(Z;Y)$, to how the original network did so, $I(X;Y)$.
The \emph{information curve} is a plot of $I(Z;Y)/I(X;Y)$ against
$I(Z;X)/H(X)$. The ZMW paper presents such curves for some synthetic
and empirical graphs, but knowing how the fidelity curves behave for
typical classes of synthetic networks would provide much helpful
context.

The first unexpected thing happened when I looked at networks built by
the fashionable method of \emph{preferential attachment}
\cite{merton1968, price1976, barabasi1999, albert2000}.  The
parameters of this process are $N$, the total number of nodes in the
network, and $m$, the number of edges brought by each node as it is
added to the network.

\begin{figure}
\includegraphics[width=8cm]{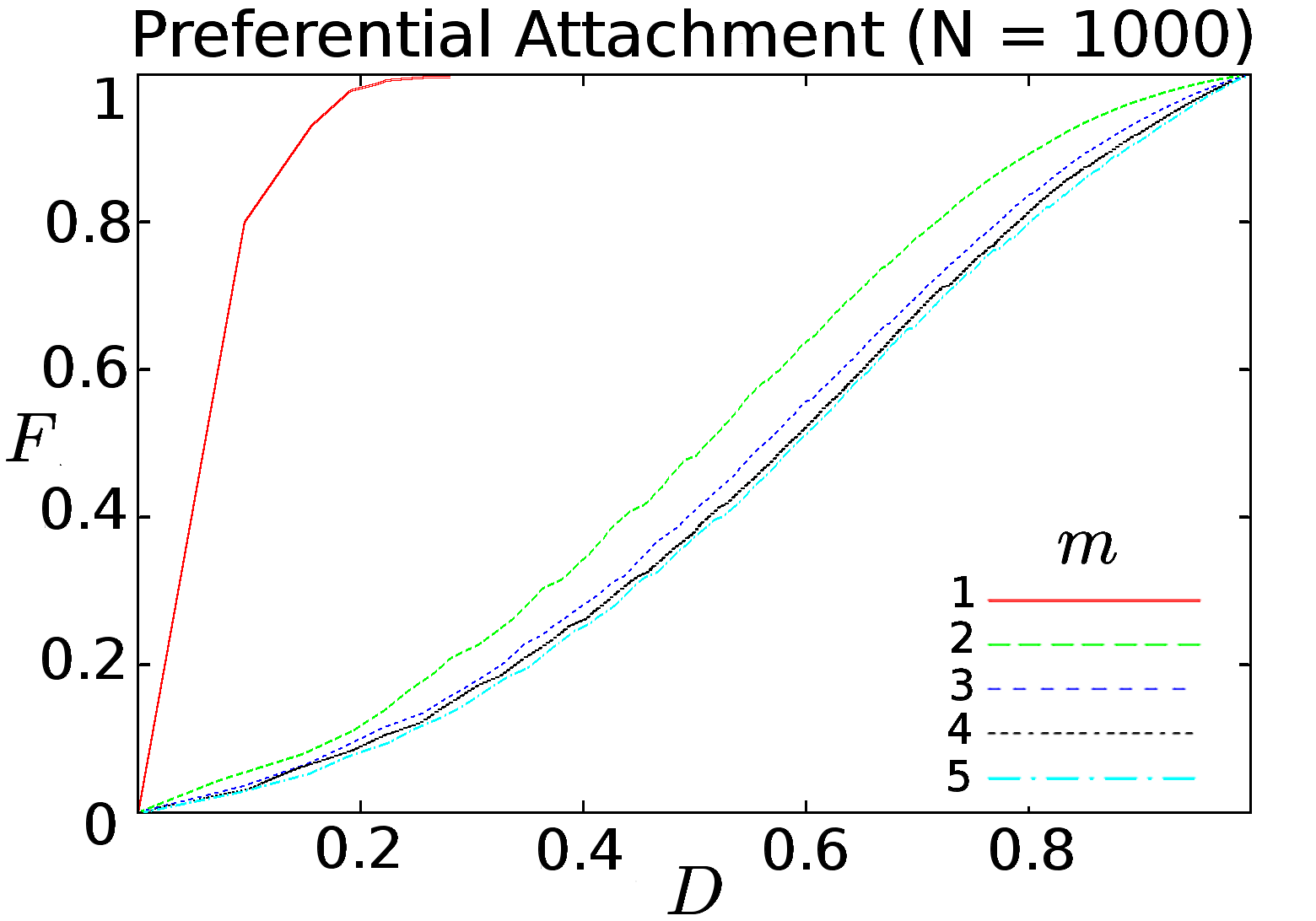}
\caption{\label{fig:ba} Fidelity curves for networks built by
  preferential attachment.  The derivatives of these curves are
  analogous to the MUI.}
\end{figure}

In ZMW, it is claimed that the fidelity curves are always concave;
however, as we can see in Fig.~\ref{fig:ba}, this is not so.  Slonim
and Tishby~\cite{slonim1999} state that for agglomerative information
bottleneck (the algorithm employed by ZMW), concavity is an empirical
result.  (In a different, though closely related implementation of the
information bottleneck concept, the fidelity curve is concave by
construction.)  From the numerical indications I've had so far, I
think this is what's happening: the marginal fidelity curve is nicely
monotonic if you plot it as a function of \emph{the number of modules
in the description.}  If you change the horizontal axis to \emph{the
Shannon entropy of the description variable,} the marginal fidelity
curve gets a bump.

All the stochastic processes we've considered in this thesis have
been \emph{classical}: nowhere have we incorporated quantum effects.
Still, it is worth asking whether our multiscale structure formalism
can be applied in quantum theory.  How should one go about
constructing a quantum analogue for our classical calculations?  The
first instinct might be to replace the Shannon index with the von
Neumann entropy, the standard information measure of quantum physics.
This choice is problematic, because while the von Neumann entropy is
strongly subadditive, it is not monotonic~\cite{mikeandike}, and so it
does not satisfy our basic axioms.  Before we consider modifying those
axioms, however, there is another possibility, one which might yield
some conceptual insight.

In quantum mechanics, we are always calculating probabilities. We get
results like, ``There is a 50\% chance this radioactive nucleus will
decay in the next hour.'' Or, ``We can be 30\% confident that the
detector at position $X$ will register a photon.'' But the nature and
origin of quantum probabilities remains obscure. Could it be that
there are some kind of ``gears in the nucleus,'' and if we knew their
alignment, we could predict what would happen with certainty? Fifty
years of theorem-proving~\cite{mermin1993} have made this a hard
position to maintain: quantum probabilities are more exotic than that.

But what we \emph{can} do is reconstruct a \emph{part} of quantum
theory in terms of ``internal gears''~\cite{spekkens2007, vanenk2007,
bartlett2012, spekkens2014}.  We start with a mundane theory of
particles in motion or switches having different positions, and we
impose a restriction on what we can know about the mundane goings-on.
The theory which results, the theory of the knowledge we can have
about the thing we're studying, exhibits many of the same phenomena as
quantum physics.  It is clearly not the whole deal: For example,
quantum physics offers the hope of making faster and more powerful
computers, and the ``toy theory'' we've cooked up does not.  But the
toy theory can include many of those features of quantum mechanics
that are commonly deemed ``mysterious.'' In this way, we can draw a
line between ``surprising'' and ``truly enigmatic,'' or to say it in a
more dignified manner, between \emph{weakly nonclassical}
and \emph{strongly nonclassical.}  Results which are weakly
nonclassical by this standard include quantum teleportation, quantum
key distribution, the no-cloning theorem, coherent superpositions
turning to incoherent mixtures by becoming entangled with the
environment, quantum discord and many more.

The ancient Greek for ``knowledge'' is \emph{epist\=em\=e}
($\epsilon \pi \iota \sigma \tau \eta \mu \eta$) and so a
restriction on our knowledge is an \emph{epistemic restriction,} or
\emph{epistriction} for short~\cite{spekkens2014}.  Finding
epistricted models for subtheories of quantum mechanics illuminates
the question of what resources are required for quantum
computation~\cite{veitch2014}.  In addition, it suggests a way to
apply our multiscale structure formalism directly, if not to the full
variety of quantum phenomena, at least to an interesting and important
subset.  We can simply treat the states of the epistricted theory as
probability distributions and use the Shannon index.

\section{Category Theory for Moment Closures}

A Petri net specifies a symmetric monoidal
category~\cite{azimuth2011}. Each truncation of the moment-dynamics
hierarchy for a system yields a Petri net, and so successive
truncations of the moment-dynamics hierarchy yield mappings between
categories. Going from a pair approximation to a mean-field
approximation, for example, transforms a Petri net whose circles are
labelled with pair states to one labelled by site states. Category
theory might be able to say something interesting here. Anything which
can tame the horrible spew of equations which arises in these problems
would be great to have. Ought we be considering, say, the strict
2-category whose objects are moment-closure approximations to an
ecosystem, and whose morphisms are symmetric monoidal functors between
them?

\section{Games with Variable Numbers of Players}

When we discussed evolutionary game theory, we treated the size of
social groups as constant.  For example, in the Volunteer's Dilemma of
\S\ref{sec:vs-panmixia}, groups of $k \geq 2$ individuals coalesced
out of the population, played the game, gathered their payoffs and
fell back into the mix.  In the network-structured version of the
Volunteer's Dilemma, the size of each group of interacting players was
fixed by the graph degree.  But what if the number of players in a
group changes from one interaction to the next?  What if some
contributions to an individual's fitness come from playing with one
partner, some from playing with two and so on?

We can begin to answer this by importing technology from statistical
physics.  In chapter~\ref{ch:prob}, we deduced the linked-cluster
theorem, which states that for a certain way of assigning numerical
values to graphs, the generating function over all graphs is the
exponential of the generating function for connected graphs.  We
developed this in the context of probability distributions for random
variables, where connected diagrams stand for cumulants, and the
generating function over all moments is the exponential of that over
cumulants:
\begin{equation}
\cQ(z) = \sum_{n=0}^\infty \frac{z^n}{n!} \moment{x^n}
       = \exp\left(\sum_{l=1}^\infty
                     \frac{\cumulant{x^l}}{l!} z^l\right).
\end{equation}

Now, it is time to deploy this machinery in evolutionary game theory.
Let the weight of a $k$-vertex connected graph be the effect upon an
individual's fitness due to interacting with $k$ other agents at once.
That is, the weight of a $k$-vertex connected graph is determined by
the payoff of a $(k+1)$-player game.  If a focal agent plays a game
with $k_1$ partners, then a game with $k_2$ partners and so on, we
represent this by a graph comprising a $k_1$-cluster, a $k_2$-cluster
and so on.  If game payoffs are mapped exponentially to fitnesses,
then the total fitness effect due to a sequence of accumulated payoffs
is the product of the fitness effects due to each payoff in the
sequence.  That is, \emph{the weight which we should assign to a graph
  comprising multiple disjoint clusters is the product of the weights
  of those clusters.}

What is the meaning of a sum over graphs in this context?  One reason
to add up a set of fitnesses is to find their average.  If each
possible sequence of cluster sizes satisfying $k_1 + k_2 + \cdots +
k_n = m$ is given equal weight, then the expected fitness effect due
to a life history involving a total of~$m$ other organisms is
proportional to the sum over all the appropriate graphs.

We have graph weights which compose in the proper manner.  Therefore,
we can use the linked-cluster theorem,
Eq.~(\ref{eq:exp-of-generating}), to deduce that the generating
function over expected fitnesses for complicated life histories is the
exponential of the generating function for the fitness effects of
individual games.

\section{Composition and a Multiscale Doi Formalism}

In Chapter~\ref{ch:doi}, we saw that classical stochastic dynamics can
be treated with notations and tools adapted from quantum theory,
specifically, raising and lowering operators (and associated entities
like coherent states).  The essential point was that these operators
satisfied the commutator relation
\begin{equation}
\comm{a}{a^\dag} = 1,
\end{equation}
and, we noted, one implementation of this abstract algebra is
differentiation and multiplication by a formal variable.  This
provides an instance of an unusual turnabout: mathematicians solve
probability problems using generating functions, the \emph{specific}
representation, while the physicist approach uses the \emph{abstract}
algebra!

One thing which the specific representation suggests that the abstract
algebra does not is the \emph{composition} of functions: with two
functions $f(z)$ and $g(z)$, it is easy to imagine evaluating $g$ at a
particular value and then plugging the result into~$f$.  We studied
this construction in \S\ref{sec:generating-functions}, where we saw
that the linked-cluster theorem is a specific example of the iterated
chain rule for differentiation.  The composition of generating
functions is used to great effect in combinatorics, as we shall now
illustrate.

Let's say we have a set with $k$ elements, and we wish to make
some arrangement of those pieces.  Perhaps we want to impose a linear
order on the elements, or fashion them into a rooted tree, or
make of them a point-set topology.  In very general terms, we'd like
to know how many ways we can do this:  given some type of mathematical
pattern $F$, how many ways can we implement $F$ on a $k$-element set?
Denote the set of all $F$-type arrangements on $k$ elements by $F_k$;
then the number of ways to implement $F$ on that many elements is the
cardinality $|F_k|$.  We can often deduce much about $F$ from the
generating function
\begin{equation}
|F|(z) = \sum_{n = 0}^\infty \frac{|F_n|}{n!} z^n.
\label{eq:species-generating}
\end{equation}
For example, the {\em vacuous structure} is just the arrangement of
``being a finite set''; its generating function is $e^z$.  Closely
related is the {\em uniform structure,} which is like the vacuous
structure except that it cannot be put on the empty set.  Its
generating function is therefore $e^z - 1$.

Consider the set $F_k$ of $F$-type structures built out of a
$k$-element set.  If each such possible structure is weighted equally,
then the ``information content'' in the statement ``there's a
structure of type $F$ on this set of $k$ elements'' is the logarithm
of $|F_k|$.  This ties the combinatorial notion of ``structure'' to
the informational one we used in Chapter~\ref{ch:multiscale}.

The bars on $|F|(z)$ in Eq.~(\ref{eq:species-generating}) suggest that
we will regard the whole generating function as the cardinality of
some object.  What kind of entity is the combinatorial species $F$?
For each $k$-element set, we have floating nearby the set $F_k$ of
$F$-type combinatorial arrangements we can perform on it; note that we
care only about the cardinality of the set we are arranging, not about
the character of its entries.  Any two sets of the same size---that
is, any two finite sets between which exists a bijection---are
equivalent for these purposes.  We thus find ourselves manipulating
${\bf FinSet}_{\bf 0}$, the category whose elements are finite sets
and whose morphisms are bijections.  The combinatorial species $F$
involves relating elements of ${\bf FinSet}_{\bf 0}$ with finite sets
of possible arrangements, {\em i.e.,} with objects in the category
${\bf FinSet}$.  Therefore, we understand $F$ as a functor between
these categories.  Specifically, $F$ send sets of structures to their
substrate sets, being a ``forgetful functor'' which forgets the
arrangement made out of the set's elements.  That is, a combinatorial
species $F$ can be represented
\begin{equation}
F: {\bf FinSet} \rightarrow {\bf FinSet}_{\bf 0}.
\end{equation}
Investing in category theory pays off when we seek relationships among
combinatorial species.  Most importantly for our current purposes, we
can compose two functors $F$ and $G$ to make a new species which
encodes the ``superstructure'' of making an $F$-type arrangement from
$G$-type structures.  For example, we could construct a linear order
out of trees.  (Or we could bake a chocolate-chip cookie in which the
chips are pieces of Oreo---or of Hydrox, which is equivalent up to
isomorphism.)  The generating function for the functorial composition
of two species is the composition of the original species'
generating functions.

The Bell numbers $\{B_n\}$ are the number of ways to partition a set
of $n$ elements into nonempty subsets~\cite{wilf1994}.  That is, $B_n$
counts the number of ways to make a finite set out of nonempty finite
sets, such that the total number of elements in the component sets
adds up to $n$.  We recognize this as making a uniform structure out
of uniform structures, so we can say immediately that the exponential
generating function for the Bell numbers is
\begin{equation}
B(z) = \sum_{n = 0}^\infty \frac{B_n}{n!} z^n = e^{e^z - 1} - 1.
\label{eq:bell-generating}
\end{equation}
Note that in Eq.~(\ref{eq:bell-generating}), we have a generating
function which is (almost) equal to the exponential of a quantity
which is itself (almost) an exponential of the formal variable $z$.
This is more than slightly reminiscent of the linked-cluster theorem,
Eq.~(\ref{eq:exp-of-generating}).  If $b_l$ were to equal 1 for all
$l$, the resemblance would be even stronger, which suggests we ought
to look into possible generalizations of the composition which gave us
the Bell numbers.  We also have those $-1$ terms in
Eq.~(\ref{eq:bell-generating}), which came from choosing the uniform
combinatorial species instead of the vacuous.  The difference between
the uniform and the vacuous species is in this case more than an
off-by-one error: if we had tried to use the vacuous species instead,
the number of possibilities would have blown up, as we could have
interleaved an arbitrary number of empty sets into our arrangement.
Composing with the vacuous species requires something more general and
robust than ordinary combinatorial species~\cite{baez2000, morton2006}
which we shall explore below.

One problem with bringing partition functions like
Eq.~(\ref{eq:exp-of-generating}) into combinatorics and category theory
is that we will sooner or later want graph weights which are not
integers.  Our graphs could be weighted by amounts of information
content, interaction energies, probabilities or even probability
amplitudes; even setting aside the quantum mechanics, we will have to
weight graphs with real or at the very least rational numbers.  We
require, then, a mathematical object whose cardinality is
nonintegral.

One candidate solution is provided by the surreal
numbers~\cite{knuth1974, conway1976, conway1996}.  Joyal observed that
the surreals, defined by Conway in terms of combinatorial games, can
be formed into a category we could call {\bf Game}, whose objects are
game positions and whose morphisms are strategies~\cite{joyal1977}.
By imposing a few extra conditions on the allowed game positions to
eliminate infinite, infinitesimal and ``pseudo-number'' quantities, we
can construct a category {\bf RealGame} whose decategorification is
precisely the real numbers $\mathbb{R}$.

Another approach, whose basic ingredients have been more fully
developed in the literature, is to replace finite sets with {\em
  groupoids,} categories in which all morphisms are invertible.  (In a
groupoid with one object, the morphisms form a group.)  Baez and
Dolan~\cite{baez2000} give the following formula for the
cardinality of a groupoid: for each isomorphism class of objects
within the groupoid, we pick a representative item and take the
reciprocal of its number of automorphisms.  The cardinality is then
the sum over isomorphism classes.
\begin{equation}
|\cG| = \sum_{[x]} \frac{1}{\hbox{aut} (x)}.
\end{equation}
When all the morphisms in the groupoid are identity morphisms,
groupoid cardinality reduces to set cardinality.

When we wrote the generating function for a combinatorial species,
Eq.~(\ref{eq:species-generating}) we considered the preimage $F_n$ of
an $n$-element object in ${\bf FinSet}_{\bf 0}$, which was a set of
$F$-type structures living in {\bf FinSet}.  If the preimage of an
object in ${\bf FinSet}_{\bf 0}$ lived instead within a groupoid
$\cG$, its cardinality would not be constrained to the natural
numbers, and could be a nonnegative rational quantity.

Take a groupoid $\cG$ and make the groupoid {\bf X} of $\cG$-colored
finite sets.  Then the functor
\begin{equation}
\Phi: {\bf X} \rightarrow {\bf FinSet}_{\bf 0}
\end{equation}
is a generalized species known as a stuff type.
Morton~\cite{morton2006} shows that the cardinality of~$|\Phi|$ is
\begin{align}
|\Phi| &= \sum_{n\in\mathbb{N}} |X_n|z^n \nonumber\\
 &=  \sum_{n\in\mathbb{N}} |\cG // S_n| z^n \nonumber\\
 &=  \sum_{n\in\mathbb{N}} \frac{|\cG|^n}{n!} z^n \nonumber\\
 &= \exp \left(|\cG|z\right).
\end{align}
Here, $S_n$ is the permutation group on $n$ elements, and $//$ denotes
the ``weak quotient'' defined as follows:

Take a category $C$ and a group $G$.  The {\em strict action} of $G$
on $C$ is a map from group elements to functors.  For every $g \in G$,
the action $A$ gives a functor $A(g): C \rightarrow C$ satisfying
$A(1) = \hbox{Id}_C$ and $A(gh) = A(g)A(h)$.  The {\em weak quotient}
of~$C$ by~$G$ is a groupoid, denoted $C//G$, whose objects are objects
of the category $C$ and whose morphisms are defined using the action
$A$.

The categorical interpretation of Eq.~(\ref{eq:exp-of-generating}) is
that we are taking the cardinality of the functorial composition of
two stuff types, one for ``being a finite set'' and the other encoding
the graph-weighting procedure.

What does all this have to do with the Doi technology for stochastic
processes?  To make the connection, we recall that the Doi formalism
is an abstract approach to probability generating functions.  So, the
question becomes, what is the meaning of the composition of two
generating functions whose coefficients are probabilities instead of
set cardinalities?

One application of generating functions in probability theory is to
problems involving \emph{randomly-sized sets of random variables.}
Let $N$ be a random variable taking nonnegative integer values, and
define
\begin{equation}
Y = \sum_{i=1}^N X_i,
\end{equation}
where each $X_i$ is a random variable defined by a probability
distribution $p_X$.  That is, we pick some number $n$ in accord with
the distribution $p_N$, and then we draw $n$ times from $p_X$ and add
the results up.  For example, we could roll a 20-sided die, obtain the
result 14, and then roll a 6-sided die 14 times.  What is the
probability distribution for the \emph{total}?

The generating function for the random variable $Y$ is the composition
of those for~$N$ and for~$X$.  Specifically,
\begin{equation}
G_Y(z) = G_N(G_X(z)).
\end{equation}

Suppose that, instead of being the outcome of a die roll, $X$ is the
number of objects present in a box.  Then, naturally, $N$ is the
number of boxes, and $Y$ is the total number of objects in all the
boxes.  Using the Doi method, we could write two vectors
\begin{align}
\ket{\phi_X} &= \sum_{i=0}^\infty p_X(i) \ket{i}, \\
\ket{\phi_N} &= \sum_{i=0}^\infty p_N(i) \ket{i}.
\end{align}
Their composition is a third vector
\begin{align}
\ket{\phi_Y} &= \sum_{i=0}^\infty p_Y(i) \ket{i} \\
 &= \sum_{i=0}^\infty \left. \frac{1}{i!}
                    \left(\frac{d}{dz}\right)^i 
                    G_N(G_X(z))\right|_{z=0} \ket{i}.
\end{align}

If either $p_N$ or $p_X$ changes over time, then $\ket{\phi_Y}$ will
change as well.  This is a way of defining a \emph{two-level}
stochastic process.  It is not, in its current form, the most useful
approach, because the way we have set things up means that the
dynamics of the two levels happen \emph{independently.}  But perhaps
there is a way to make the levels interact:  for example, for a box
containing a large number of objects to split into a pair of boxes.

Composing a probability generating function with \emph{itself} is
useful in the study of iterated processes.  If each of the objects
inside a box is itself a smaller box, then $Y$ is the total number of
small boxes.  Or, if the random variable $X$ is the number of children
produced by one individual, then $Y$ is the number of grandchildren
born to the $N$ siblings.  Setting $p_N = p_X$ amounts to declaring
that the reproductive outcomes of one generation are the same,
statistically, as those of the next.

This is a very different way of thinking about reproduction
probabilities than we used in Chapter~\ref{ch:doi}.  Perhaps the
intersection of the two can yield something interesting.

\section{Multiplayer Games and Biodiversity Indices}

In Chapter~\ref{ch:prob}, we developed the Shannon index as a measure
of information for a probability distribution.  We started with the
idea that, if $p_i$ is our probability for seeing the
$i$\textsuperscript{th} possible outcome of an experiment, then if we
repeat the experiment $N$ times, the least surprising number of times
that outcome can occur is $N_i = Np_i$.  A typical illustration of
this is the experiment of plucking a letter at random from
English-language text.  Finding the letter \emph{E} would be a less
surprising event than coming up with a \emph{Q}.

This is fine as far as it goes, but it does leave something out.  The
letters \emph{E} and \emph{Q} are, for many purposes, more dissimilar
than, for example, \emph{S} and \emph{Z}, or \emph{F} and \emph{V}.
In the first case, we have a consonant and a vowel, while the latter
two pairs are all consonants.  Furthermore, the latter two pairs are
analogous: \emph{S} and \emph{Z} both stand for sibilants, one of
which is voiced (that is, the vocal cords vibrate when the sound is
articulated).  Likewise, \emph{F} stands for a voiceless consonant,
and \emph{V} for its voiced counterpart.  From a phonetic standpoint,
the event of encountering an \emph{F} is more like the event of
encountering a \emph{V} than the event of finding an \emph{E} is like
that of meeting a \emph{Q}.  A quantitative measure of similarity
would, justifiably, assign comparable values to the \emph{F--V}
and \emph{S--Z} pairs, and a lower number to \emph{E--Q}.  Moreover,
this is a separate question from how common or how typical the
individual letters are.

For some purposes, the similarity between characters, and thus between
the events of observing those characters, is a result of their
evolutionary history.  Our letters \emph{T} and \emph{N} have been
distinct for as long as they have existed, and in a sense, longer than
that: their ancestors in the Proto-Canaanite alphabet were distinct as
well~\cite{cross1954}.  In contrast, the letter \emph{I} did not
diverge from \emph{J}, or \emph{V} from \emph{U}, until after the
Renaissance; and indeed dictionaries listed words beginning
with \emph{U} and with \emph{V} together as recently as
1837~\cite{oed-U}.

As with linguistic evolution, so with biological.\footnote{The analogy
between these two processes was already clear to Darwin, who addressed
it in Chapter~13 of the \emph{Origin.}  Later, he commented, ``The
formation of different languages and of distinct species, and the
proofs that both have been developed through a gradual process, are
curiously the same''~\cite{darwin1871}.  It is an area of ongoing
conceptual exchange~\cite{atkinson2005}.}There is a
meaningful sense in which the events of encountering two species from
the same genus are more similar events than those of finding specimens
from separate genera.

So, it is useful to consider information functions for cases in which
we have not only a probability distribution over events, but a notion
of similarity or contrast between those events as well~\cite{allen2009}.

Even if we do not have a full evolutionary history behind each type of
entity we may encounter, we can still tabulate the characteristics of
those entities.  Suppose, for concreteness, that there exists a pool
of~$n$ possible attributes, and an entity is defined by choosing
exactly $k$ of them.  The attribute sets of any two entities may be
disjoint, or they may have elements in common.  If the attribute sets
of two entities share common elements, then the events of encountering
those entities are similar.  But note that if we say ``point'' and
``line'' instead of ``attribute'' and ``entity,'' we have again the
type of finite geometries we studied in Chapter~\ref{ch:multiscale}.
This suggests a new wrinkle:  there are, very naturally, higher-order
kinds of similarity, which cannot be deduced from lower-order
relationships.  Even if any two lines meet in a common point, a set of
three lines might converge at a single intersection, or they might not.

Back in Chapter~\ref{ch:prob}, we derived the Shannon index, which, we
saw, reflects the extent to which a probability distribution is
``spread out'': the Shannon index is maximized for a uniform
distribution, and it attains its minimum value of zero when the
distribution is a delta function.  Another way to quantify the spread
of a probability distribution is an \emph{effective number.}  This is
a type of quantity, useful in mathematical ecology, which we can
motivate with the following scenario.

Imagine that we have an urn full of marbles, and we presume that when
we draw a marble from the urn, no choice is preferred over any other.
If the urn contains $N$ marbles, our probability of obtaining any
individual one of them is $1/N$.  But what if our probability
distribution is not uniform, as it would be if we thought the drawing
was rigged in some way?  In that case, we can label the marbles with
the integers from~1 to~$N$, and we say that our probability for
obtaining the one labeled $i$ is $p(i)$.

We draw one marble, replace it and repeat the drawing.  What is the
probability that we will draw the same marble both times?  Let the
result of the first drawing be $j$.  Then our probability for
obtaining that marble again is $p(j)$, and to find the overall
probability for drawing doubles, we average over all the choices
of~$j$:
\begin{equation}
p(\hbox{doubles}) = \sum_j p^2(j).
\end{equation}
For a uniform distribution, this is
\begin{equation}
\sum_j p^2(j) = \sum_j \left(\frac{1}{N}\right)^2
 = N \left(\frac{1}{N}\right)^2
 = \frac{1}{N}.
\end{equation}
That is, if all draws are equally probable, then the probability of a
coincidence is the reciprocal of the population size.  Turning this
around, we can say that whatever our probabilities for the different
draws, the effective size of the population is
\begin{equation}
N_{\rm eff} = \left[ \sum_j p^2(j) \right]^{-1}.
\end{equation}

So far, we have presumed that any pair of outcomes is as good as any
other.  For some problems, this can be good enough.  However, if we
are trying to find the effective number of organisms present in an
ecosystem, we must face the fact that some pairs of species are more
closely related than others.

Leinster and Cobbold have proposed a framework for biodiversity
indices which systematizes and extends many prior developments in that
field~\cite{leinster2011}.  The basic input data they consider is a
set of probabilities $p(i)$, which characterize the relative
preponderances of species in an ecosystem, and a \emph{similarity
  metric} which indicates how closely species $i$ resembles species
$j$.  Their diversity indices depend on a \emph{sensitivity
  parameter,} call it $q$, which indicates the relative emphasis
placed on rare species versus common ones.  The larger one makes $q$,
the less sensitive the diversity index is to improbable species.

The Leinster--Cobbold diversity index is
\begin{equation}
^q D^{\bf Z} [{\bf p}] 
 = \left[\sum_i p(i) 
         \left(\sum_j Z_{ij}\, p(j) \right)^{q-1}
   \right]^{\frac{1}{1-q}} .
\label{eq:lc}
\end{equation}
Here, we follow Leinster and Cobbold in writing ${\bf Z}$ for the
matrix of similarity values $Z_{ij}$.  If $Z_{ij} = \delta_{ij}$, we
recover the case in which distinct species are considered wholly
unrelated to one another, which is often (and
unrealistically~\cite{allen2009, chao2010}) assumed in much of the
older work on biodiversity.  We can motivate Eq.~(\ref{eq:lc}) in the
following way: if $Z_{ij}$ is the similarity between species $i$ and
species $j$, then summing $Z_{ij}\, p(j)$ over all possible values of
$j$ will give the ``ordinariness'' of species $i$.  The ``average
ordinariness'' of the whole ecosystem is then just the mean of this
taken over the probability distribution $p(i)$.  Because ``diversity''
ought to be inversely related to average ordinariness, we take the
reciprocal.  Eq.~(\ref{eq:lc}) generalizes this to \emph{power means}
of order $q - 1$.  Most of our attention will be focused on the
special case $q = 2$; that is, we will mostly consider the index
defined with the ordinary arithmetic average.

Biodiversity indices allow us to compare ecosystems.  For example,
given two test tubes full of microorganisms, one which we characterize
by a probability distribution ${\bf p}$ and the other which we
characterize by ${\bf r}$, we can compare the ``diversity profiles''
of the two microbial ecosystems by computing $^q D[{\bf p}]$ and $^q
D[{\bf r}]$ as functions of $q$~\cite{leinster2011}.  We can delve
more deeply if the two ecosystems are composed of the same species,
that is, if ${\bf p}$ and ${\bf r}$ are not just of the same length,
but defined over the same events.  Suppose that we pipette a microbe
at random from the first test tube, and that we identify this microbe
as being of type $i$ on our list of all possible microbe varieties.
What is the probability that the act of pipetting out a random microbe
from the \emph{second} test tube will yield an organism of the
\emph{same} type?  By definition, it is $r(i)$.  What, then, is the
\emph{average} probability over \emph{all possible microbe varieties}
that the results of the two experiments will ``collide'' in this way?
The answer is the expectation value of~$r(i)$ with respect to the
probability distribution ${\bf p}$, or in other words, the dot product
of the two probability vectors ${\bf p}$ and ${\bf r}$.  As before, we
would like a measure of diversity to be inversely related to a
tendency toward coincidence.  We can therefore define a \emph{cross
  diversity} as
\begin{equation}
^2 D[{\bf p}; {\bf r}] = \frac{1}{\sum_i p(i) r(i)}.
\label{eq:cross-diversity}
\end{equation}
We can also include a distance metric ${\bf Z}$ in the cross
diversity, as we had with the Leinster--Cobbold index:
\begin{equation}
^2 D^{\bf Z} [{\bf p}; {\bf r}] = \frac{1}{\sum_{ij} Z_{ij} p(i) r(j)}.
\end{equation}
As before, this reduces to the ordinary cross diversity of
Eq.~(\ref{eq:cross-diversity}) in the limiting case that $Z_{ij}$ is a
Kronecker delta, that is, when a species resembles only itself and is
distinct from all others.

For reasons stemming from biology, Leinster and Cobbold prefer to set
up their indices as ``effective numbers'' of species present, rather
than as entropies.  This has become a standard practice in
mathematical ecology.  In just about any circumstance, it's reasonable
to say that an island with four equally abundant, unrelated species
is only half as biodiverse as an island with eight equally abundant,
unrelated species.  Effective numbers preserve this desirable scaling
property, while entropies of the forms familiar from information
theory do not~\cite{jost2006}.  Relating these effective numbers to
various entropies people have defined is not conceptually difficult.
Generally, we expect an entropy to be logarithmically related to an
effective-number diversity measure, because an entropy should count
the number of questions needed to specify an item in a set.  (For
example, in the study of complex networks or graphs, the ``effective
degree'' of a vertex is the exponential of the entropy of its
edge-weight distribution~\cite{sayama2012}.)

We can think of these diversity indices in another way, which suggests
a natural generalization.  The key comes from \emph{game theory.}
Imagine a game in which each player's goal is to match the move made
by the other.  The score earned by a player who makes move $i$ is 1 if
the other player also makes the move $i$, and 0 otherwise.  If the
players make their moves \emph{randomly,} in a way characterized by
the probabilities $p(i)$, then the \emph{expected payoff} obtained by
either player is just ${\bf p} \cdot {\bf p}$.  Now, suppose the
matching is not such an all-or-nothing affair:  perhaps there are wild
cards in the deck, so that the ace ($i = 1$) can match any other.  Or,
perhaps matching one card with another of the same suit is almost as
good as matching it with a copy of the same card.  Then the expected
payoff will include cross terms, since the score of an $i$ matched
against a $j$ is no longer just $\delta_{ij}$.  Diversity indices,
then, are \emph{measures of expected welfare in games whose goal is
  agreement.}

This new perspective hints at a generalization:  what about games
played by more than two players?

Leinster and Cobbold think in terms of \emph{distance} between
species, and distance is naturally a pairwise thing.  However, as they
point out, some of the earlier work in the area considered
\emph{inter-species conflict} instead~\cite{ricotta2006}.  Conflicts,
or interactions more generally, do not have to break down into
pairwise relationships.  In human affairs, what Alice does in the
company of Bob and Carol does not have to be a linear combination of
what she does when alone with Bob and when alone with Carol.  In game
theory, the payoff in a multiplayer game is not restricted to being a
linear sum of pairwise games (recall Chapter~\ref{ch:volunteer}).  Or,
consider a parasite species with multiple hosts in its life cycle: the
total effect on humanity due to \emph{Anopheles} mosquitos
and \emph{Plasmodium} microbes depends on both species taken together.

We imagine, therefore, an ``interaction tensor'' of the form
$Z_{ijk}$, which tells us how the presence of species $i$ and $j$
taken in combination affects a focal species, $k$.  The natural
modification of Eq.~(\ref{eq:lc}) is
\begin{equation}
^q D^{\bf Z} [{\bf p}] 
 = \left[\sum_k p(k) 
         \left(\sum_{ij} Z_{ijk}\, p(i) p(j) \right)^{q-1}
   \right]^{\frac{1}{1-q}} .
\label{eq:lc2}
\end{equation}

Biodiversity measures, generalized to multiplayer games as we have
done here, have an application in quantum information theory, of all
places.  This is one of the oddest bits of interdisciplinary
boundary-jumping which I have seen, so I think it's worth talking
about for a while.

In quantum physics, we take what we think we know about a system, roll
it into a density operator $\rho$, and use that density operator to
make statistical predictions about what the system might do in
particular experiments.  But presenting that information as a matrix
operator is not always the most illuminating choice.  We can actually
rewrite any finite-dimensional density matrix as a probability
distribution, using the idea of \emph{informationally complete
measurements} \cite{prugovecki1977, dariano2004}.  These are
generalized measurement procedures (positive operator valued measures,
or POVMs) which have an appealing ability: given a probability
distribution over the possible outcomes of an informationally complete
POVM, we can compute all the statistics which we could have gotten
using the density matrix.  Such POVMs can be constructed in any
finite-dimensional Hilbert space~\cite{caves2002, fuchs2002}.  The
nicest variety are the \emph{symmetric} informationally complete
POVMs, known familiarly as SICs~\cite{zauner1999, renes2004}.  A SIC
for a $d$-dimensional Hilbert space is a set of $d^2$ operators $\{E_i
= \frac{1}{d} \Pi_i\}$ where the rank-one projection operators
$\{\Pi_i\}$ satisfy
\begin{equation}
\tr (\Pi_k \Pi_l) = \frac{d\delta_{kl} + 1}{d + 1}.
\label{eq:sic-trace}
\end{equation}
An arbitrary density matrix $\rho$ can be decomposed in terms of SICs.
If $p(i)$ is the probability that performing a SIC measurement on the
system yields the outcome labeled by $i$, then
\begin{equation}
\rho
 = \sum_{i=1}^{d^2} \left((d+1) p(i) - \frac{1}{d}\right) \Pi_i
 = (d + 1) \sum_{i=1}^{d^2} p(i) \Pi_i - \mathbb{I}.
\label{eq:rho-to-sic}
\end{equation}
Exact expressions for SICs have been found for dimensions 2--16, 19,
24, 28, 31, 35, 37, 43 and 48~\cite{fuchs2013}.  High-precision
numerical approximations have been discovered for dimensions
2--67~\cite{scott2010}, and more recently, E.\ Schnetter has claimed
numerical solutions for dimensions 68--76, 78--81, 83--85, 87, 89, 93
and 100 \cite{fuchs2013}.  It is not known whether SICs exist for all
values of~$d$, but it has become commonplace to assume that they
do.\footnote{Following a conjecture about a way to reduce the number
of equations which must be solved to obtain a SIC \cite{appleby2007,
fuchs2014b}, I've found low-precision evidence that such a structure
exists in $d = 77$ as well.}

The extremal states in the space of density matrices are the ``pure''
states, which satisfy the condition $\rho^2 = \rho$.  Thanks to a
theorem of Flammia, Jones and Linden \cite{flammia2004, jones2005,
  appleby2011b}, we can also characterize pure states as those
Hermitian matrices satisfying
\begin{equation}
\tr\rho^2 = \tr\rho^3 = 1.
\end{equation}
This result is well worth calling a remarkable theorem: it is simple,
powerful and easy to prove once asserted, but it was apparently missed
completely until 2004~\cite{fuchs2014b}.  In turn, this definition of
purity yields the following two conditions on the probability
distribution $p(i)$ \cite{fuchs2009, appleby2011, fuchs2013}.  First,
\begin{equation}
\sum_{i=1}^{d^2} p(i)^2 = \frac{2}{d(d+1)},
\label{eq:purity1}
\end{equation}
and second,
\begin{equation}
\sum_{ijk} c_{ijk}\, p(i) p(j) p(k) = \frac{d+7}{(d+1)^3},
\label{eq:purity2}
\end{equation}
where we have defined the real-valued, symmetric three-index tensor
\begin{equation}
c_{ijk} = \hbox{Re}\, \tr (\Pi_i \Pi_j \Pi_k).
\label{eq:c-tensor}
\end{equation}
The second condition, Eq.~(\ref{eq:purity2}), has been nicknamed the
\emph{QBic equation} \cite{fuchs2014b}.  The full state space is the
convex hull of probability distributions which meet
Eqs.~(\ref{eq:purity1}) and (\ref{eq:purity2}).  It would be
interesting to be able to motivate these equations from something
other than the pre-existing quantum formalism: is there a reason,
independent of quantum physics, to care about functionals of
probability distributions like we see on the left-hand sides of
Eqs.~(\ref{eq:purity1}) and (\ref{eq:purity2})?

Formally, the conditions defining quantum-state purity,
Eqs.~(\ref{eq:purity1}) and (\ref{eq:purity2}), become
\begin{equation}
^2 D^\delta [{\bf p}] = \frac{d(d+1)}{2}
\label{eq:purity1lc}
\end{equation}
and
\begin{equation}
^2 D^{\bf c} [{\bf p}] = \frac{(d+1)^3}{d + 7}.
\label{eq:purity2lc}
\end{equation}
This form of the QBic equation is, of course, still a cubic
constraint, in that it employs three instances of the probability
vector ${\bf p}$.  However, as far as the sensitivity parameter $q$ is
concerned, it is a ``second order'' equation, as we have kept to the
special case $q = 2$.

What can we learn from these restatements of the pure-state
conditions?  Amusingly, assigning a pure state to a quantum system
means that the \emph{effective number} of possible outcomes for a SIC
experiment which one is willing to contemplate is just
$\binom{d+1}{2}$.  The fact that an effective number works out to be a
combinatorial quantity hints strongly, at least to me, that this is a
promising avenue to explore: combinatorics is, after all, the art of
counting cleverly.  (Thinking in terms of effective numbers has at
least a few pence of ``cash value'' \cite{fuchs2014b}, in that it means
I can remember what goes on the right-hand side of the quadratic
purity condition.)  Another way to think of this is that when all SIC
outcomes are judged as equiprobable, that is to say $p(i) =
\frac{1}{d^2}$, the effective number of experimental outcomes is the
total number which comprise the SIC: $^2 D^\delta = d^2$.  So, if we
focus on the quadratic constraint, ascribing a pure state means
neglecting $\binom{d}{2}$ possible outcomes of a SIC experiment.
Entertainingly, this is also the best known upper bound on the number
of entries which can be zero in a quantum-state assignment ${\bf p}$
\cite{appleby2011}.  This is not a coincidence: we can deduce that
bound by starting with the normalization of ${\bf p}$ and squaring to
find
\begin{equation}
\left( \sum_i p(i) \right)^2 = 1.
\end{equation}
We then apply the Cauchy--Schwartz inequality to find, writing $n_0$
for the number of zero-valued entries in ${\bf p}$,
\begin{equation}
(d^2 - n_0) \sum_{{\rm nonzero}} p(i)^2 \geq 
 \left( \sum_{{\rm nonzero}} p(i) \right)^2 = 1.
\end{equation}
We see the inverse of the quadratic diversity appearing on the
left-hand side.  Consequently,
\begin{equation}
n_0 \leq d^2 - {}^2 D[{\bf p}],
\label{eq:weak-zeros-bound}
\end{equation}
and from Eq.~(\ref{eq:purity1}) we know the right-hand side equals
$d(d-1)/2$, as advertised.

This bound is perhaps not the tightest possible.  In fact, it is
suspected \cite{fuchs2013} that the actual upper bound on the number
of zeros permitted in a quantum state is just the dimension, $d$.

In addition, quantum mechanics implies a constraint on \emph{pairs} of
probability vectors \cite{appleby2011, fuchs2013, fuchs2014}.  If we
begin with state assignments written as density matrices $\rho$ and
$\sigma$, then using Eq.~(\ref{eq:rho-to-sic}) we can deduce that the
Hilbert--Schmidt inner product of those state assignments is
\begin{equation}
\tr\, \rho\sigma = d(d+1) \sum_i r(i) s(i) - 1.
\end{equation}
Because the Hilbert--Schmidt inner product is always nonnegative, two
quantum state assignments defined on Hilbert spaces of the same
dimension $d$ ``can never be too nonoverlapping''~\cite{fuchs2013}:
\begin{equation}
\sum_i r(i) s(i) \geq \frac{1}{d(d+1)}.
\label{eq:never-too-nonoverlapping}
\end{equation}
Comparing this to Eq.~(\ref{eq:cross-diversity}) above, we have a
bound on the cross diversity of ${\bf r}$ and ${\bf s}$:
\begin{equation}
^2 D[{\bf r}; {\bf s}] \leq 2 \binom{d+1}{2}.
\end{equation}
From the first purity condition, Eq.~(\ref{eq:purity1}), we know that
all valid probability vectors lie within a ball.  The dot product of
any two state assignments will therefore obey the bound
\begin{equation}
\sum_i r(i) s(i) \leq \frac{2}{d(d+1)},
\end{equation}
meaning that their cross diversity is also bounded from below:
\begin{equation}
\binom{d+1}{2} \leq {^2 D[{\bf r}; {\bf s}]} \leq 2 \binom{d+1}{2}.
\end{equation}
The lower bound is saturated if ${\bf r} = {\bf s}$ and ${\bf r}$ is a
pure state.

Tabia \cite{tabia2012, tabia2013} has discovered a fascinating
simplification of the QBic equation, Eq.~(\ref{eq:purity2}), in the
case of a qutrit, a system whose Hilbert space has dimension $d = 3$.
In this special case, Eq.~(\ref{eq:purity2}) can be reduced through a
clever choice of SIC to
\begin{equation}
\sum_i p(i)^3 - 3 \sum_{(ijk) \in S(9)} p(i) p(j) p(k) = 0.
\label{eq:qbic-qutrit}
\end{equation}
Here, $S(9)$ denotes the \emph{Steiner triple system} of order 9, a
set of 12 elements which can be found by cyclically tracing all the
horizontal, vertical and diagonal lines in the array
\begin{equation}
\begin{array}{ccc}
1 & 2 & 3\\
4 & 5 & 6\\
7 & 8 & 9
\end{array}
;\qquad\hbox{that is to say,}\qquad
S(9) =
\begin{array}{ccc}
(123) & (456) & (789) \\
(147) & (258) & (369) \\
(159) & (267) & (348) \\
(168) & (249) & (357)
\end{array}.
\label{eq:steiner}
\end{equation}
The order-9 Steiner triple system also rejoices in the name of the
\emph{Hesse configuration} \cite{bengtsson2010, tabia2012, dang2012,
  tabia2013}.  This construction is also an example of a dual affine
plane, a type of incidence geometry which is known to be relevant to
SICs more generally \cite{wootters2006, appleby2008, baron2009}, and
which we saw all the way back in Chapter~\ref{ch:multiscale}.

We consider ``striations'' of the $3 \times 3$ array: we can carve it
up into horizontals, verticals, left-leaning diagonals or
right-leaning diagonals, and each of these four striations divides the
array into three parallel sets of numbers.  That is, each striation
produces one row of the table in Eq.~(\ref{eq:steiner}).

We can easily check the simplified QBic equation,
Eq.~(\ref{eq:qbic-qutrit}), in one case where we know it should hold
true: for the SIC states themselves.  In a SIC representation, the
$k$\textsuperscript{th} vector making up the SIC has a $1/d$ in the
$k$\textsuperscript{th} slot and $\frac{1}{d(d+1)}$ everywhere else.
Specializing to $d = 3$, any of the nine SIC states appears in exactly
four of the index triples listed in Eq.~(\ref{eq:steiner}).  This
makes checking that all nine SIC states satisfy
Eq.~(\ref{eq:qbic-qutrit}) a straightforward arithmetic problem.

Note that we can recast the $d = 3$ QBic equation,
Eq.~(\ref{eq:qbic-qutrit}), as
\begin{equation}
\sum_i p(i)^3 + 3 \sum_{(ijk) \in S(9)} p(i) p(j) p(k)
 = 2 \left(3 \sum_{(ijk) \in S(9)} p(i) p(j) p(k)\right)
 = 2 \left(\sum_i p(i)^3\right).
\label{eq:qbic-qutrit-recast}
\end{equation}
On the right-hand side of Eq.~(\ref{eq:qbic-qutrit-recast}), we have
the inverse of the ternary diversity we defined back in
Eq.~(\ref{eq:lc2}), with experimental outcomes treated as wholly
dissimilar.  On the left-hand side, we have something which is
starting to look like a ternary diversity with some sets of outcomes
distinguished as more similar than others.

Let $Y_{ijk}$ be the completely symmetric tensor which is 1 if $(ijk)
\in S(9)$ and 0 if no permutation of $(ijk)$ is in $S(9)$.  Then
\begin{equation}
\sum_{ijk} Y_{ijk}\, p(i) p(j) p(k)
 = 6 \sum_{(ijk) \in S(9)} p(i) p(j) p(k).
\end{equation}
Now, let $Z_{ijk}$ be 1 if all subscripts are equal and $\frac{1}{2}
Y_{ijk}$ otherwise.  Then the $d = 3$ QBic equation reads
\begin{equation}
^2 D^{\bf Z} [{\bf p}]
 = \left(\sum_i p(i)^3 
         + 3 \sum_{(ijk) \in S(9)} p(i) p(j) p(k)
   \right)^{-1}
 = \frac{1}{2}\, {^2 D^\delta [{\bf p}]}.
\label{eq:qbic-qutrit-diversity}
\end{equation}
For qutrits, pure states are those for which treating as similar the
proper sets of SIC outcomes reduces the diversity of possible outcomes
by 2.

The Hesse configuration has other interesting properties in relation
to qutrit SICs, including a result I found about ``mutually unbiased
bases'' and compatibility criteria for quantum-state
ascriptions~\cite{stacey2014b}.  This is a result of quantum
information theory---actually, a correction to others' earlier work in
that field---which I proved thanks to an interest in mathematical
biology and complex systems.  These calculations are, if taken just as
mathematics, neutral on philosophical questions about quantum
physics.  From that perspective, they are matters of complex
projective geometry.  However, the research tradition they derive
from, and which they might feed back into, is a philosophy which
treats quantum physics itself in evolutionary terms, as a tool which
agents immersed in the creative profusion of the world can use to make
the best of life's Darwinian contest~\cite{fuchs2013, fuchs2014b}.

\section{Gauge Theory and Evolution}

The study of evolution has a certain conceptual intersection with the
subject of economics.  Both are concerned with the effects of limited
resources.  And, both make use of game theory.  On the evolutionary
side, however, we are not concerned with whether agents are
``rational''; nor do we start with an assumption that a system is
``efficient'' or that the participants in it are by any standard
well-informed.  This is a significant difference in outlook between
evolutionary theory and economics, and arguably, economics needs to
catch up~\cite{fox2009, harmon2011, krugman2011, lagi2011, lagi2012,
lagi2015}.

I can think of one idea, though, which is rather in the borderlands of
economics, which might fruitfully be transposed over into mathematical
biology.  That is the application of \emph{gauge theory}
and \emph{differential geometry} to economic indices.  This subject
was inaugurated by Malaney~\cite{malaney1996}, in collaboration with
Weinstein.  More recently, it has been promoted by
Smolin~\cite{smolin2009} and discussed by Baez~\cite{baez2011c}.

A big concern in the Malaney thesis is how to define a cost-of-living
index when the goods which are relevant to daily life change over
time.  If Richard III regarded a horse as a fair trade for his
kingdom, how much of England should we be able to swap for a Prius?
We can answer this kind of question quantitatively, as long as old
goods remain in circulation at least temporarily as new ones are
introduced.  During the period of overlap, we can evaluate the price
at which agents participating in the economy will trade an old good
for a new one and vice versa.  By chaining overlap intervals, we can
gradually eliminate all the goods in the initial ``basket,'' while
maintaining an unbroken sense of what counts as a decent standard of
living.

An evolutionary analogue would be the origin of new traits.  A common
mechanism for this is gene duplication: a whole stretch of DNA
accidentally gets copied twice, so the offspring carries two copies of
the same gene.  This doesn't make much immediate difference, but it
does provide redundancy: over the generations, mutations accumulate,
and because mutations which knock out one gene leave the other intact,
the species is more resilient.  Over time, one of the gene copies can
gain a new function, while the other keeps doing the old.  This is
studied in the context of gene interaction networks, where it's known
as the ``duplication-divergence model''~\cite{ispolatov2005}.

Adapting the Malaney--Weinstein economics stuff might provide a way to
talk about what ``fitness remaining constant'' means in this context,
and to write equations for population dynamics.

To lay the groundwork for this project, we can establish a dictionary
from economics to evolution.  Instead of \emph{goods and services,} we
can speak of \emph{traits.}  In economics, one considers
the \emph{amount of a good} held by an agent; the counterpart in
evolutionary dynamics is the \emph{expression level}
or \emph{quantitative value} of a trait.  Rather than a \emph{pricing
system} that assigns prices to baskets, we have \emph{fitness
functions,} given in the simplest case as a linear combination of
trait values with coefficients (\emph{e.g.,} the $\beta$ values of
Chapter~\ref{ch:price-mls}).

The Malaney thesis~\cite{malaney1996} defines a ``barter'' as a set of
debts and possessions whose net monetary value is zero.  For example,
if six cupcakes are worth ten cookies, then a debt of ten cookies can
also be paid by six cupcakes.  The evolutionary analogue is a mutation
which does not affect the reproductive fitness of the individual.  On
the molecular level, this happens all the time: replication errors
swap out nucleotides so that the offspring end up carrying different
DNA sequences, but those sequences still code for the same protein.
In the mathematical setup we discussed in the previous chapter, a
neutral mutation could be one trait diminishing in value while another
one increases, so the $w_i$ we get by taking the dot product doesn't
change.

On the economic side, we have phenomena like inflation, in which
the function from baskets to prices changes over time.  This has a
natural counterpart in evolution; for example, shifts in external
environmental conditions can be represented as variations in the
parameters of the fitness function.

By carrying over the basic notions of Malaney~\cite{malaney1996} to
the new terminology, we have our first result: \emph{If the covariant
derivative of an evolutionary history vanishes, then that history
consists of neutral mutations.}

So far, in fashioning this map from one jargon to another, we have
neglected population structure.  It would be interesting to push
further, into models where the fitness function is not constant across
the population.  Moreover, the work on the economics side presumes
that price scales linearly with quantity, and we know that doing
mathematical biology means facing up to nonlinearity sooner rather
than later.

Gauge theory provides a platform for understanding the phenomenon of
moving through a circuit and not ending up exactly how you began.  In
Chapter~\ref{ch:redgreen}, we saw an evolutionary dynamic which is
reminiscent of this: a consumer strain can go from rare and successful
to rare and dwindling, while local information indicates no difference
between the environmental conditions.  Can we express this phenomenon
in terms of fiber bundles?

\chapter{Conclusions}

\label{ch:conclusions}

\section{Review}

\begin{quotation}
\noindent Whatever universe a professor believes in must at any rate 
be a universe that lends itself to lengthy discourse.

---William James~\cite{james1907}
\end{quotation}

We began, many chapters ago, with the idea of a \emph{complex system}
being one which exhibits organization at multiple scales.  Formalizing
this concept mathematically, we saw that indices of multiscale
structure can describe the patterns which result from evolutionary
processes.  Then, we saw how the emergence of multiscale structure can
control the evolution of trait values in an ecosystem, and we explored
how we can expand game theory to multiplayer scenarios, providing
another angle on the question of scale.  With tools from probability
theory in hand, we brought adaptive dynamics into the stochastic
regime, and we found a way to quantify the interplay between mutation
and selection, revealing how the outcome depends upon population
structure.

Academic writing traditionally patterns itself rather like the Twelve
Labors of Hera\-kles~\cite{avalos2007}.  First, there is the statement
of the problem, and then, a litany of failed attempts to solve it,
which we call a ``literature review.''  Next, a new and better
solution is proposed, and its triumph is proclaimed.  I elected to
deviate from this template, as far as the overall layout of this
thesis is concerned.  This choice hinged upon a tradeoff that I think
is important enough to warrant discussing explicitly.  A literature
review is a specialized history in miniature, and so it is at this
juncture worth thinking about historical expositions of science more
generally.

This thesis has addressed several variations on the theme of
multiscale structure in evolutionary dynamics.  I have tried to make
the material flow as smoothly as possible, and to build up the ideas
in a sequence which helps them come across transparently.  It may be,
however, that in doing so, I have clouded the distinction between new
research and old.  The first glimpse of a discovery is not always the
most clear, after all, and later treatments of a topic can have the
advantage of experience over the early ones.  Initial reports are
commonly soaked in the confusions of the time, which require hard work
(and, sometimes, as the proverb says, funerals) to overcome.
Consequently, an exposition which aims to express the current
synthesis is apt to be ahistorical.

Any field of science which has reached a sophisticated level of
development is susceptible to this problem.  Suppose I want to teach a
classful of college sophomores the fundamentals of quantum
mechanics. There is a standard ``physicist's history of
physics''~\cite{feynman1985} which goes along with this, one that
progresses through a familiar litany of famous names: Planck,
Einstein, Bohr, de Broglie, Heisenberg, Schr\"odinger, Born. We like
to go back to the early days and follow the development forward,
because the science was simpler in its initial stage---or so we tend
to believe.

The problem is that all of these men were highly trained, professional
physicists who were thoroughly conversant with the knowledge of their
time.  But this means that any one of them knew more classical physics
than a modern college sophomore. They would have known Hamiltonian and
Lagrangian mechanics, for example, in addition to techniques of
statistical physics. Unless you know what they knew, you can't really
follow their thought processes, and we don't teach big chunks of what
they knew until after we've tried to teach what they figured out! For
example, if you aren't fairly conversant with thermodynamics and
statistical mechanics, you won't be able to follow why Planck proposed
the blackbody radiation law he did~\cite{pais1979}, and a crucial step
of the development will be lost, without your even knowing it.

Consequently, any ``historical'' treatment at the introductory level
will probably end up conventionalized. One has to step extremely
carefully! Strip the history down to the point that students just
starting to learn the science can follow it, and you might not be
portraying the way the people actually did their work. That's not so
bad, as far as learning the facts and formul\ae{} is concerned, but
you open yourself up to all sorts of troubles when you get to talking
about the process of science. Are we doing physics differently than
folks did $N$ or $2N$ years ago? If we aren't, is that a problem?
Well, we sure aren't doing it like they did in the textbooks we
learned out of.

Carelessly repeating a ``scientist's history'' instead of teaching a
history of science leads to a kind of inadvertent myth-making that
Stephen Jay Gould designated ``textbook cardboard''~\cite{switek2008}.
An example from biology would be the assertion that no one put
genetics and natural selection together until the
1930s~\cite{stoltzfus2014}.  In physics, to name one of many
possibilities, we have the canard that in the last years of the
19\textsuperscript{th} century, physicists thought that the only
remaining task was to calculate answers to more decimal
places.\footnote{This sentiment is often attributed to Kelvin, but
never with an actual pointer to a primary source.  In a 1900 lecture,
expanded the next year to an essay, Kelvin described two ``clouds over
the dynamical theory of heat and light''~\cite{thomson1901}.
Dispelling the first cloud turned out later to require special
relativity, and removing the second was a task for quantum mechanics.
Kelvin concludes his discussion of the first problem by the remark,
``I am afraid we must still regard Cloud No.\ I.\ as very dense.''
Figures as prominent as Tait~\cite{tait1886}, Gibbs~\cite{gibbs1901}
and Maxwell himself~\cite{maxwell1860} all pointed out that classical
physics fails to grasp the specific heats of gases.  A text as widely
admired and merchandised as the \emph{Feynman Lectures} laid out the
historical situation~\cite{feynman1964, stacey2011}.  Even so, the
assertion of \emph{fin-de-si\`ecle} physicists' na\"\i{}ve folly lives
on, helping thinkpieces to be glib and cocktail talk to be smug.}  The
``physicist's history'' of twentieth-century physics is replete with
textbook cardboard~\cite{howard2004, camilleri2009, stacey2014c,
camilleri2015}, and no doubt we'll keep this tradition going in the
twenty-first.

Having argued that there exists a cause for concern, I am now in the
unheroic position of admitting I have no way to solve it.  The only
way out seems to be flexibility, sacrificing one objective for another
as the circumstances permit.

The field of complex-systems research has an additional challenge.
Many of the models it employs are defined computationally: we begin
with a specification of the model, and then we implement it as a
computer program.  Not infrequently, that implementation is within the
range of a fairly novice programmer.  The primary challenge lies not
in the coding, but in deciding which model would be interesting to
explore, and in knowing how to investigate it systematically.  The
\emph{mathematical} prerequisites for working with adaptive-network
models, say, are less demanding than they are for many topics at the
modern physics frontier.  Furthermore, when analytical treatments are
possible for complex-systems models, they often only apply to special
cases.  Nonequilibrium phase transitions provide a good example: a
basic implementation of directed percolation is easy to
code~\cite{henkel2008}, but the statistical field theory which
describes it is difficult to obtain (Chapter~\ref{ch:doi}), and the
elaborate mathematics is most useful near the critical point.

It may be that the special case of a problem, the case amenable to
analytical work, was historically discovered first, and computational
methods which made the general case accessible only came later.
However, as we have discussed, computational methods can admit easier
expositions than densely mathematical ones.  Thus, an explanation
which moves from the familiar to the esoteric would invert the
historical order.

As a result, presenting all of the literature review
first~\cite{cham2006} would have been a dissatisfying scheme for this
thesis.  Moreover, such a sequence would have required muddling
through the remarkably confused MLS/IF literature we surveyed in
Chapter~\ref{ch:price-mls}, before seeing the simulation results in
Chapters~\ref{ch:redgreen} and \ref{ch:volunteer}, which are
significantly easier to appreciate.  Likewise, for a reader trained in
physics, the Price equation we met in Chapter~\ref{ch:price-mls} is
likely to be unfamiliar and, indeed, to appear arcane and even
overwrought.  In contrast, the dynamical stability analysis presented
in Chapter~\ref{ch:volunteer} is a technique more routine to a physics
student, albeit applied to a less common-or-garden problem.  Again,
choosing to begin with the approachable material means that we turn
history on its head.

That said, optimizing towards one goal can impede the achievement of
another.  With this in mind, I will use this section to revisit the
previous chapters and more cleanly separate research from review.

Chapter~\ref{ch:multiscale} discussed a general formalism for
multiscale structure, based on information theory.  We saw two indices
of structure, the complexity profile and the Marginal Utility of
Information (MUI).  The complexity profile was introduced some years
ago by Bar-\!Yam~\cite{baryam2004a}.  More recently, Allen developed
an axiomatic foundation upon which the complexity profile and related
quantities could be constructed.  The three of us coauthored a paper
on the subject~\cite{allen2014}.  During the course of that project, I
derived the binomial transform equation for computing the complexity
profile in the special case of exchange-symmetric systems,
Eq.~(\ref{eq:D-of-k-binomial}).  I also worked out the illustrative
examples using combinatorial geometry, which our three-author paper
was already long enough without.  The complexity profiles for the
imitation dynamics (noisy voter model) and the frequency-dependent
Moran process are new calculations for this thesis.

Chapter~\ref{ch:redgreen} is based on a paper by Gros, Bar-\!Yam and
myself, as I indicated in its concluding note.  The organism-swapping
test and the use of survival-probability curve intersections were new
contributions of that paper.  In addition, the use of percolation
arguments to find the critical thresholds in certain limits had not
been done for that model before, and the discrepancy between
predator-prey and epidemic dynamics (\S\ref{sec:percolation}) had not
been commented upon.  Relating the correlation lengths to the evolved
transmissibilities is new, as are the crossover in the perimeter-area
relationship and the 99\textsuperscript{th}-percentile curves.  The
scaling argument which shows the inability of pair approximation to
handle the correlation length is a new application of earlier
semi-numerical work.  So is the combination of the pair approximation
with the coagulation and fragmentation model to predict how $\tau$
will evolve.

The next chapter, on multiplayer games, contains several novelties.  I
encountered the Volunteer's Dilemma in papers which argued that it was
an understudied type of scenario~\cite{nowak2010, archetti2011}.
While writing Chapter~\ref{ch:volunteer}, however, I found that I had
to treat the game in ways which I hadn't read about.  The
continuous-time, well-mixed dynamical systems, with their baseline
growth and death rates, logistic forms, pessimistic view of the cost
of Volunteerism and unfixed total population size, are unusual.  The
lattice models are, likewise, atypical with respect to the literature.
The analytical computations which conclude the chapter are an
extension to multiplayer games of ideas which had only been applied to
dyadic interactions.

Chapter~\ref{ch:prob} is largely a review of material which would
have otherwise required references to selected chapters in a
scattering of textbooks.  The part about an evolutionary analogue of
the Jeffrey rule is, to my knowledge, not covered elsewhere.  (I've
mentioned the possibility over the past few years to people
better-read than I, and prior art never came up.)  Anomalous
cross-diffusion terms, as seen in \S\ref{sec:variations-on-diffusion},
are known in the literature~\cite{lugo2008, biancalani2010}, but that
literature arrives at them in a much more complicated way.

I learned at school the physicist's justification of the
Fokker--Planck equation~\cite{kardar2005} which now appears in
Chapter~\ref{ch:adaptive}, and I read about the deterministic limit of
adaptive dynamics~\cite{allen2013d}.  The applications to the
continuous Prisoner's Dilemma and Snowdrift games are new developments
for this thesis, as is the connection to random matrix theory.

Both Chapters~\ref{ch:invasion-fitness} and \ref{ch:price-mls} are
pedagogical reviews, for the most part.  The argument that the
``modified mean-field model'' is insufficient for eco-evolutionary
considerations (\S\ref{sec:mmf}) is my contribution.  My approach to
the interconversion between MLS-A and neighbor-modulated fitness
calculations (\S\ref{sec:mls-and-inclusive}) is more abstracted and
generalized than Bijma and Wade's~\cite{bijma2008}, which is where I
began.

The work on which this thesis builds is, in many places, fairly new
itself.  This adds an extra layer to the challenge of exposition: the
ideas are sometimes recent enough that their significance has yet to
be fully hashed out.  I hope this thesis can be a part of that process.

In 1996, John Horgan published a book titled \emph{The End of
Science.}  This attracted some serious criticism at the
time~\cite{hoffman1996}, but the passing of years has brought an even
more strongly negative review.  To wit: in 1997, the first \emph{E.\
coli} genome was published~\cite{blattner1997}, and in 1998, the
expansion of the Universe was found to be
accelerating~\cite{riess1998, perlmutter1999}.  Accomplishments like
these are not the final fits of a dying enterprise; nor are they the
concluding, semi-senescent huzzahs before a quiet retirement.  They
are new beginnings, which raise up new opportunities and which all the
work that comes after must acknowledge.  These are the kind of
discovery which go beyond providing answers: they change the questions
which we are able to ask.  And to these examples we could add many
more.

I mention this because the bulk of the references on which I leaned
most directly date to the post-\emph{End-of-Science} period.  The
complexity profile, for example, emerged in 2004~\cite{baryam2004a},
and the axiomatic formalism of multiscale structure only last
year~\cite{allen2014}.  Chapter~\ref{ch:volunteer}'s analytical
calculations depend on a scheme which was likewise published in
2014~\cite{allen2014b}, and the Fokker--Planck treatment of stochastic
adaptive dynamics is an extension of research reported the year before
that~\cite{allen2013d}.  The theory of the structure coefficient
$\sigma$, which provides a convenient way to discuss many mathematical
models of evolution in a unified fashion, was put forth in
2009~\cite{tarnita2009}, and it is perhaps still not as well known as
it should be.

\section{Just One More Thing}

Another way in which my organizational scheme is a little heterodox is
that I decided to move the Acknowledgments to the end.  Partly, this
is to soothe my conscience:  having all those pages come before now
makes me a little more inclined to feel that this work is substantial,
and thanking those who helped it along is not entirely damnation by
faint praise.  And, partly, this arrangement is to benefit those who,
like me, skip to the end to see what the thing is about.

I am indebted to my collaborators, first of all to those with whom the
relation is made official by coauthorship: Marcus de Aguiar, Benjamin
Allen, Yaneer Bar-\!Yam and Andreas Gros.  Many portions of this
thesis are my attempts to build on something Ben has done.  It would
be a much slimmer document without his achievements as wellsprings.
Without our conversations, it might not exist at all.

Special mention must also be given to my thesis committee: Yaneer
Bar-\!Yam, Aparna Baskaran and Albion Lawrence.  Their input
throughout the stages of this project has been consistently helpful.

John Baez had very kind things to say about the work which became
Chapter~\ref{ch:redgreen}, and also stimulated parts of
Chapters~\ref{ch:prob} and \ref{ch:doi}.  Karla Z.\ Bertrand read many
iterations of what is now Chapter~\ref{ch:redgreen}, and in addition
helped test my presentation in Chapter~\ref{ch:multiscale}.  Chris
Fuchs got me to thinking about some of the matters which inform
Chapters~\ref{ch:prob} and \ref{ch:speculations}.

Credit must go to my far-flung correspondents, who kept my curiosity
active and my sense of honesty engaged by reminding me of questions in
varied fields of science.  A large fraction of our interactions these
past few years have taken place through a strange one-to-many
telegraph.
\begin{quotation}
\noindent @archymck \\
@arikia \\
@artologica \\
@arutherfordium \\
@AstroKatie  \\
@bengoldacre \\
@biochembelle \\
@bug\_gwen \\
@carlzimmer \\
@cgranade \\
@cryptogoth \\
@csferrie \\
@cuttlefishpoet \\
@dabacon  \\
@Dhunterauthor \\
@DrEugeniaCheng \\
@DrMathochist \\
@DrMRFrancis \\
@drskyskull \\
@elakdawalla \\
@ErinPodolak \\
@evelynjlamb \\
@FlyingTrilobite \\
@HirokiSayama \\
@HowardBarnum \\
@Laelaps \\
@mandaYoho  \\
@MarkCC \\
@michael\_nielsen \\
@mikethemadbiol \\
@Myrmecos  \\
@nervous\_jessica \\
@phylogenomics \\
@planet4589 \\
@pleunipennings \\
@PsiWavefunction \\
@pzmyers \\
@RaquelHRibeiro \\
@rebeccawatson \\
@rf \\ 
@sc\_k \\
@scicurious \\
@seelix \\
@shanley \\
@TomLevenson \\
@TRyanGregory \\
@vihartvihart
\end{quotation}

Eric Downes was my gateway to the people and the puzzles with whom and
which this thesis began.

My housemates put up with me.  I'm not entirely sure why I deserved
it.

For my family, all my love.

Lieutenant Columbo, Avatar Korra and Daria Morgendorffer were always
there for me in the small hours of the morning.

{\small \bibliography{thesis.bib}}

\end{document}